\documentclass[a4paper,10pt,twoside]{report}
\usepackage[left=1.5cm,right=1.5cm,top=2cm,bottom=2cm]{geometry}
\usepackage[final]{pdfpages}
\usepackage{setspace}
\usepackage{ulem}
\usepackage{soul}
\usepackage{color}

\usepackage{blindtext,graphicx}
\usepackage[absolute]{textpos}
\usepackage{array}
\usepackage{algorithm}
\usepackage{algorithmic}
\usepackage{attrib}
\usepackage{mathtools}

\usepackage{pifont}% http://ctan.org/pkg/pifont
\usepackage[n,
advantage,
operators,
sets,
adversary,
landau,
probability,
notions,
logic, ff, mm, primitives, events, complexity, asymptotics, keys,
]{cryptocode}

\usepackage{multirow}
\usepackage{colortbl}
\definecolor{lightgray}{gray}{0.9}
\usepackage{enumitem}
\setlistdepth{20}
\renewlist{itemize}{itemize}{20}
\setlist[itemize]{label=$\bullet$}

\usepackage{etoolbox}
\makeatletter
\preto{\@verbatim}{\topsep=2pt \partopsep=2pt }
\makeatother

\usepackage{cite}
\usepackage{amsmath,amssymb,amsfonts}
\usepackage{textcomp}
\usepackage{bmpsize}
\usepackage{xcolor}
\usepackage{lipsum}
\usepackage{sidecap}
\sidecaptionvpos{figure}{t}
    
\newcommand{\TX}[1]{$TX^{#1}$}
\usepackage{graphicx}
\usepackage{verbatim}
\usepackage{latexsym}
\usepackage{mathchars}
\usepackage{setspace}

\input{blocked.sty}
\input{uhead.sty}
\input{boxit.sty}
\input{icthesis.sty}

\newcommand{\QQQ}{{\sf C\kern-0.48emQ}}   % Rational numbers
\newcommand{\normallinespacing}{\renewcommand{\baselinestretch}{1.5} \normalsize}

\newcommand{\syncc}{~\stackrel{\textstyle \rhd\kern-0.57em\lhd}{\scriptstyle L}~}

\usepackage{hyperref}
\begin{document}

\title{}

\author{\LARGE Jacob Tyge Göker Swambo}
\submitdate{December 2022}

\normallinespacing
\maketitle

\preface
\addcontentsline{toc}{chapter}{Abstract}

\begin{abstract}

The broad topic of this thesis is the design and analysis of \textit{Bitcoin custody systems} -- systems to safeguard, manage, and control Bitcoin assets. Both the technology and threat landscape are evolving constantly. Therefore, custody systems, defence strategies, and risk models should be adaptive too. We consider designing for resilience against accidents and attacks, and using extensible and modular risk models. This way, we support a cyclic process of design and analysis, where feedback from live deployments can be integrated, and custody systems can \textit{evolve}. We begin with a motivating discussion about the relevance of Bitcoin as a new monetary system with unique features; censorship-resistance and pseudonymous participation. We explain that custody systems are the foundation of the Bitcoin economy, upon which all economic activity depends. 

We introduce Bitcoin custody by describing the different types, design principles, phases and functions of custody systems. We review the technology stack of these systems and focus on the fundamentals; key-management and privacy. We present a perspective we call the \textit{systems view}. It is an attempt to capture the full complexity of a custody system, including technology, people, and processes. We review existing custody systems and standards. We discuss how various risk modelling methods have been applied previously with custody systems and their components. 

We explore \textit{Bitcoin covenants}, a promising innovation. This is a mechanism to enforce constraints on transaction sequences. Although previous work has proposed how to construct and apply Bitcoin covenants, these require modifying the consensus rules of Bitcoin, a notoriously difficult task. We introduce the first detailed exposition and security analysis of a deleted-key covenant protocol, which is compatible with current consensus rules. We demonstrate a range of security models for deleted-key covenants which seem practical, in particular, when applied in autonomous (user-controlled) custody systems. We conclude with a comparative analysis with previous proposals.

Covenants are often proclaimed to be an important primitive for custody systems, but no complete design has been proposed to validate that claim. To address this, we propose an autonomous custody system called Ajolote which
uses deleted-key covenants to enforce a \textit{vault} sequence. Withdrawal from a vault is subject to a delay period, during which it can be rejected by the user or automated servers (watchtowers). We argue that this affords a novel expression of an inherent security-accessibility trade-off that standard multi-signature custody systems do not support. We evaluate Ajolote with; a model of its state dynamics, a privacy analysis, and a risk model covering each phase (setup, operation, and recovery). We propose a threat model for custody systems which (we argue) captures a realistic attacker for a system with offline devices and user-verification. We perform \textit{ceremony analysis} to construct the risk model. 

The thesis offers a synthesis of concepts and methods from sources across academic literature, industry articles, risk-management methodologies, and technological practice, and offers several original contributions as mentioned above. Through integrating whole-system designs with adaptive risk modelling methods, this thesis commences down a research path towards \textit{evolving Bitcoin custody}. 
\end{abstract}
\cleardoublepage

\addcontentsline{toc}{chapter}{Acknowledgements}

\begin{acknowledgements}

I would like to express my gratitude to my supervisor, Prof. Peter McBurney, who made my journey into computer science not only possible, but a pleasure. To my early collaborators, Spencer Hommel, Bob McElrath and Bryan Bishop; our regular discussions propelled my understanding and kept me on the path of pragmatism, thank you. To my recent collaborators, Antoine Poinsot, Kevin Loaec, and Edouard Paris; I could not have produced this thesis without the practical experienced gained from working together with you, merci. I would also like to thank Dr. Steve Phelps and Patrick McCorry for insightful feedback in the early stages of my studies.  

To my wife, Yaprak, whose boundless support and love kept me inspired and grounded. Her 
 faith in me made this journey all the more enriching and fulfilling. To my family, whose encouragement never wavered, I hope to make you proud. I thank my friend, Arnau, for providing the initial spark that lead me down this path of study and for being an outstanding example to follow. I want to thank Wil for encouraging me towards cultivating a practice of wisdom and ethics, which spurred my well-being alongside this intellectual pursuit. To Aggelos, whose genius musicianship was an inspiration for many evenings of respite from our academic endeavours, efcharistó. To the whole community of friends, colleagues and staff at King's College London, thank you for helping me to feel at home.   
\end{acknowledgements}
\clearpage

\begin{center}

\vspace*{\fill}
\textit{`To be free of all authority, your own and that of another, is to die to everything of \\ yesterday, so that your mind is fresh, young, innocent, full of vigour and passion.'}\\
Jiddu Krishnamurti

\vspace{1cm}

\textit{`If you know the enemy and know yourself you need not fear the results of a hundred battles.'}\\
Sun Tzu

\vspace{1cm}

\textit{`Amateurs hack systems, professionals hack people'}\\
Bruce Schneier

\vspace*{\fill}
\end{center}

\body
\chapter{Introduction}
\label{ch:introduction}

This opening chapter seeks to contextualize and outline the trajectory of our research into Bitcoin custody systems. These are systems which ensure the safekeeping and operation of Bitcoin assets, combining cybersecurity measures with procedural standards. Starting with an examination of Bitcoin's disruptive potential and an overview of monetary theory and Bitcoin, we highlight the compelling economic forces driving the need for development in Bitcoin custody. This is followed by an explanation of the thesis title. The key objectives of the thesis are then articulated, accompanied by a statement of originality and a breakdown of the thesis structure.

\section{Motivation}

\subsection{Bitcoin, the Disruption}

The control structure of money may be characterized by three dimensions \cite{ControlStructure}; decentralization of money issuance, decentralization of transaction handling, and virtualization of representation. The technological innovation that was introduced to the world with Bitcoin \cite{Nakamoto2008} and cryptocurrency technology was the creation of a form of money with a previously unseen control structure. Comparisons are often made between Bitcoin and commodity money (e.g. gold), cash, and commercial bank deposits, however it is fundamentally different to each of these. The decentralization of money issuance corresponds to a competitive market between issuers; cash in the fiat system is created in a monopolistic environment, whereas gold is mined in a competitive industry. If the transaction handling is decentralized then the asset being transacted is a bearer instrument; cash is a bearer instrument, while commercial bank deposits are not. Cash transactions are peer-to-peer, while commercial bank transfers occur in hierarchical systems. Finally, the representation of a money ranges from virtual to physical; commodity money and cash are physical, while commercial bank deposits are virtual. What was previously unseen, that is now seen in the world of cryptocurrencies, is a form of money that is at the same time vitrually represented, competitively issued, and whose transactions are handled by a decentralized network. 

Until recently, governments and central banks have been able to maintain a monopoly over the issuance of currency within their jurisdiction. Their ability to define what is and what isn't legal tender has made it difficult for new currencies to be issued by any entity other than the central banks or their delegates in lieu of the state. A private company issuing a new currency, despite for example having a more favourable monetary policy, is susceptible to coercive pressures which render the currency unattractive for economic use \cite{Hayek.1978}. With the advent of Bitcoin, the technology to privately create new forms of money has become available to everyone, as is demonstrated by the plethora of alternate cryptocurrencies available now. What allows these systems to persist despite efforts to shut them down or otherwise manipulate them is the decentralization of the security of the system. A lack of centralized authority in these systems makes it costly and difficult both technically and legally for a coordinated attack to cease the operation of a cryptocurrency.

As society continues on its transition from the industrial age into the information age, it is plausible that the cyber-economy will continue to grow, and escape beyond the jurisdiction of nation-states \cite{Davidson:1996:SIS:548392}. Privacy technology will make it ever more difficult to hold entities accountable for avoiding tax, sanctions, and other forms of capital controls, further incentivising participation in the cyber-economy. Whether for good or for ill, entities (individuals, corporations, states) may claim monetary sovereignty by participating in the cyber-economy, which is available to them regardless of where they are geographically based so long as they have an internet connection. While such entities will acquire great freedom in their cyber-economy participation, they will also be responsible for maintaining their own security. So understanding the nature of cryptocurrencies; to what extent they are secure relative to each other and to traditional financial assets, and the technicalities of securing cryptocurrencies against theft, loss, and death, will become increasingly important. The fact that so many universities and financial institutions now have dedicated research and development teams working with cryptocurrency technology is testament to these arguments.

This technology is inherently a tool for anarchists \cite{bitcoinAnarchist}. What is meant here by anarchy is not chaos, but rather anti-government. Anarchy in the form of a network protocol that is used to enable coordination between free and willing participants, despite perhaps facing coercive pressures against participation. This technology is radically permission-less; hence it may be used by revolutionaries, dictators, entrepreneurs, institutions, and democrats alike. The existence of this technology limits the ability of national gatekeepers to suppress financial commerce among entities within and across jurisdictions; it brings optionality to those entities. For example, if a business is operating within the bounds of a jurisdiction with unfavorable tax laws or capital controls, they may temporarily store some wealth in a form of money which transcends national jurisdictions, until they can claim the `jurisdictional arbitrage' that exists when the business is re-located to a more appealing jurisdiction. The allure of such a powerful tool will inevitably find its way into the hands of powerful players.   

\subsection{Money, Currency and Bitcoin}

The following is a brief analysis of the functions of money, and describes to what extent these are expressed by Bitcoin in its current state of development. The purpose is to underline the economic motivations behind the need for further research and development in Bitcoin custody. \textit{Custody}, within the realm of Bitcoin, describes the practices and mechanisms that ensure the safe holding and secure management of the digital currency, be it for personal ownership or for an organization or even an application. For non-technical individuals and organizations to leverage Bitcoin's benefits, we must significantly lower the barriers to setting up and using secure custody systems.

To understand the utility of Bitcoin, one may look at a general framework on the utility of money, or the `functions' of money, as is commonly described in economic theory \cite{Hayek.1978}. In 2014 work from the Bank of England \cite{BoEcrypto} an analysis of the extent to which Bitcoin expresses each of the following functions was given; a store of value, a medium of exchange, and a unit of account. Since then, there has been significant progress in the development of Bitcoin, in associated network applications \cite{Sidechains101,Poon2016, StrongFederations, decker2018}, in the regulatory landscape which concerns the bridges between the Bitcoin economy and traditional asset classes, and the institutionalisation of Bitcoin. In order to illustrate some open areas of research which require more dedicated work, we present a brief analysis of how Bitcoin expresses each of the functions of money. The functions of money are commonly thought to operate in a hierarchy \cite{BoEcrypto}. It should be noted that the expression of these functions is not binary (on or off), rather, they are expressed along a continuum, which is determined by the size of a collective of people utilising the asset for these functions. First, an asset which is to be money must function as a store of value. Subject to many people using this asset as a store of value, it may then begin to function as a medium of exchange. Finally, predicated on its use as a medium of exchange, an asset may function as a unit of account, that is, an asset whose price is used as a reference price for other goods, commodities and services. 

A \textbf{unit of account}, with which to measure the value of commodities and services. It is important to realize that valuations are subjective, and so an asset may function as a unit of account for some people, and not for others. For those who are accustomed to using fiat currencies as a measure of their wealth, bitcoin seems like a very volatile asset and thus a poor unit of account. For those who are disillusioned with the fiat money system, with beliefs that, in future, scarce digital assets will appreciate in value relative to other commodities and services, Bitcoin functions as a reliable measure of their wealth and a unit of account. The question of the extent to which various people hold such beliefs is out of the scope of this thesis.  To increase the extent to which Bitcoin is used as a unit of account requires increasing the extent to which it behaves as a medium of exchange and as a store of value. 

\sethlcolor{green}

A \textbf{medium of exchange}, with which to make payments and contracts for deferred payments. Bitcoin is a system which comprises both a distributed ledger of the history of all transactions, and the payment processing network itself \cite{Nakamoto2008}. There are some features of this payment network which differentiate it from traditional payment networks.

Bitcoin is a permissionless payment network, that is, transactions may not be censored provided appropriate measures are taken regarding privacy and anonymity. National borders are irrelevant to the processing of Bitcoin transactions. The availability of the payment network is dependent on a market mechanism - when there is significant demand for usage of the payment system the price of transacting (fees) increases. Transactions can be of arbitrary values, the network is suitable for small (of the same order as fees) and very large payments. Given the virtual nature of Bitcoin, it is understandable that the transactions may be augmented with code such that they are subject to conditions based on authorisation, time, secret knowledge, and combinations thereof.

The throughput of Bitcoin transactions is limited by design. Transactions are processed in batches, where one block of transactions is produced, on average, every 10 minutes. Further, merchants accepting bitcoin payments are encouraged to wait for multiple blocks (typically six) before considering the transaction to be immutable. Thus, the settlement properties of Bitcoin are presumably insufficient for large numbers of very low value transactions which require immediate finality. There has been significant research into so-called layer 2 applications which leverage a slightly weaker security model for high frequency, low fee transactions processed externally to the Bitcoin system. These applications can use the settlement properties of Bitcoin blocks when disputes arise between users. Examples of such protocols are; the Lightning Network \cite{Decker2015,Poon2016}, eltoo \cite{decker2018}, and Liquid \cite{Sidechains101, StrongFederations}. Another interesting area of research is in the potential for the \textit{atomic} exchange of bitcoin for other digital resources \cite{DELGADOSEGURA2020832}. Here, an atomic exchange means that either the payment is made and the digital resource is sent, or no payment is made and no digital resource is sent. There is no way for either party to cheat the other.

The innovation in payments and contracts is clear. Predicated on beliefs of Bitcoin functioning as a secure store of value, one may argue that the technological capability for it to function as a medium of exchange is unlike any form of money seen before. While there has been significant research and development in enabling better functionality as a medium of exchange, there is an argument to be made that the technical research in the design and anlysis of Bitcoin custody systems is lacking, especially when considering; the usability requirements of different user groups, the lack of mature open-source software for resilient cryptographic key management strategies, and the combination of these strategies with a privacy framework. Presumably, the asymmetry of knowledge between the public and the technically-literate third-party Bitcoin custodians is an opportunity for profit, and, so far, these for-profit companies haven't made a significant offering of free tools for self-managed custody systems, which would undermine their business model. However, Bitcoin behaves as censorship resistant money only if third-party custodians are not necessary. Self-managed custody enables free participation in the growing cyber-economy. Individuals, businesses and other organizations should be capable of self-managed custody to retain Bitcoin's property of censorship-resistance as it continues to be adopted globally.

A \textbf{store of value}, with which to transfer ‘purchasing power’ (the ability to buy goods and services) from today to some future date. The value of bitcoin at any point in time is based on the aggregate actions of the market participants in all bitcoin markets. Valuations are subjective, and are often based on projections into the future about the level of demand that is anticipated for an asset's various utilities relative to its supply. Bitcoin has a finite supply which aids in forecasting the projected value. A store of value may be expected to retain its purchasing power better under conditions of high liquidity. 

Each holder is in fact utilizing Bitcoin for the function of a store of value. The bearer nature of bitcoin means that holders are responsible for their own security. They may  outsource this security provision to a third party, such as a digital asset exchange or web wallet, at the risk of being censored or defrauded by the operators of those systems.
The design space of potential custody systems is large, given the abundance of operational security controls, key-management techniques and privacy services available. Defining specific solutions and enhancing their usability would better enable self-managed custody and drive adoption of Bitcoin. This thesis aims to examine how to design and analyse Bitcoin custody systems, to reduce custodial risk and better enable it to function as a store of value for individuals, businesses, and institutions. There is a wide array of possibilities to intelligently use Bitcoin's programmable nature which (to the best of the authors' knowledge) has yet to be reviewed in detail. 

\subsection{Evolving Bitcoin Custody}

 In this section I will explain the title of this thesis, `Evolving Bitcoin Custody'. To begin, custody of Bitcoin is a foundational practice for any application of the technology. A suite of financial services built on Bitcoin would not be possible without considering the operational processes for how the underlying is protected in custody. Applications (to name a few) in payment networks, international trade, open-timestamps, and as a reserve asset in the cyber-economy all critically depend on how custody is enacted. A weak custody system can greatly facilitate compromised services and fraudulent activity. 

This begs the question; what is a strong custody system? A strong custody system prevents and resists compromise and enables Bitcoin applications to operate as intended. For the purpose of this thesis, my conjecture is that a static solution is inappropriate; there is no solution with fixed protections and mitigating risk controls that can indefinitely remain uncompromised. This is because of the simple fact that adversaries can and will adapt their methods. Security is a game of cat and mouse. It is a dynamic arms-race between the defence and the offence. The defence must out-manouver the offence, must adapt to new information and learn from mistakes and incidents, or risk being compromised by the offence. Moreover, as the environment in which this security game takes place changes (e.g. as Bitcoin's rules and capabilities are developed), affording new defence strategies and tactics, the defence must adapt and utilise new techniques. All this points to the conclusion that a strong custody system must \textit{evolve}. 

My methods and objectives are thus aimed at contributing to the evolution of Bitcoin custody systems, and enhancing the capacity for Bitcoin custody systems to evolve. How? One priority is to design and analyse in the open, and to elicit feedback from as many interested parties as possible. If a diverse collection of commercial enterprises use common open tools and technologies, there will be an incentive for them to collaborate on upkeeping and developing that foundation. An open process is amenable to a cyclic process of design and analysis. Open-source technology and open-access research empower and educate others towards adopting self-managed custody systems. A second priority is to select methods which support differential analysis for adapting custody systems and operating environments. For example, selecting a formalism for an attack model that is modular and can be extended or reduced as needed. In essence, this thesis encourages open collaboration and promotes dynamic, adaptable strategies.

\section{Objectives}
\label{sec:thesis-objectives}
The objectives of this thesis are as follows:

\begin{itemize}
    \item[O1] Establish categorical definitions within the domain of Bitcoin custody. Evaluate existing technology, documentation, standards and analyses. Write an informed overview to bring coherence to this emerging field of study.
\end{itemize}

This objective is crucial as it sets the foundation for the rest of the research. The rapidly evolving nature of the Bitcoin ecosystem necessitates a comprehensive review and categorization of Bitcoin custody methods and technologies to ensure subsequent research is built upon an accurate and up-to-date understanding of the field. This knowledge foundation will be built and laid out in Chapter 3.

\sethlcolor{green}

The next objective is central to the research as it seeks to develop and evaluate novel solutions for managing and mitigating the inherent risks in Bitcoin custody. One such promising innovation is the use of Bitcoin \textit{covenants}, which are specific conditions that constrain bitcoins transactions. As will be shown, they can be used to both limit fraudulent transactions and delegate limited operational control to automated systems, and thus enhance custody systems.

\begin{itemize}
    \item[O2] Design and analyze a novel covenant protocol that works with state of the art Bitcoin features. Provide a comparative analysis of alternative covenant constructions. 
\end{itemize}

As we'll see, existing covenant constructions rely on features that Bitcoin does not have, at least not yet. Constructing a covenant with existing Bitcoin features can open new avenues to improve the security and efficiency of Bitcoin custody systems. It can also serve as a benchmark and testing ground when evaluating alternative covenant constructions which may one day be enabled. This work will be presented and elaborated in Chapter 4.

\begin{itemize}
    \item [O3] Design and analyse a novel, self-managed custody system for an individual.
\end{itemize}

Building upon the insights and innovations of the previous chapters, this objective focuses on the practical implementation of these concepts into a novel, self-managed Bitcoin custody system called \textit{Ajolote}. This system, designed for individual users, will be a demonstration of the practical application of the theoretical constructs discussed earlier. This objective is aimed at contributing directly to the field of Bitcoin custody with a realistic system design, highlighting this research's practical relevance. The design and analysis of this system will be presented in Chapter 5.

In summary, the objectives are closely interlinked, each building upon the previous one, taking us from a comprehensive understanding of the existing landscape (O1), through innovation and design of new solutions (O2), to the practical application of these ideas in a novel system (O3). With these, we contribute to the evolution of Bitcoin custody.

\section{Statement of Originality}
\label{intro:statement-originality}

The initial research for chapters \ref{ch:bitcoin-covenants} and \ref{ch:vault-custody} were undertaken as a collaboration between myself, Bob McElrath, Bryan Bishop and Spencer Hommel. The collaboration began in June 2019 and ended in June 2020 with the release of our papers titled `Bitcoin Covenants: Three ways to Control the Future' and `Custody Protocols using Bitcoin Vaults'. Upon later reflection, I found that these papers had become out-dated and the presentation and argumentation could be improved significantly. With the permission from my collaborators I have re-used many ideas and descriptions from the original papers. However, I have updated previous work to reflect the current state-of-the-art, added new insights throughout, improved the methodology, conducted new analyses and evaluation, and moved from the passive voice to the active voice. The original papers will remain available online \cite{Swambo2020cov,Swambo2020vault}.

The work found in appendix \ref{ch:risk-framework-revault} was undertaken as a collaboration between myself and Antoine Poinsot. This project was part of a broader collaborative effort to design and build the Revault custody system. Our work together on this project began in August 2020 and ended in December 2021. This work is tangential to the primary content of the thesis, offering another relevant case study of a Bitcoin custody system. The results may be of independent interest.

\section{Publications}

Appendix \ref{ch:risk-framework-revault} was published as presented herein.

Swambo, J., Poinsot, A. (2021). \textbf{Risk Framework for Bitcoin Custody Operation with the Revault Protocol.} In: Bernhard \textit{et al.}, Financial Cryptography and Data Security. FC 2021 International Workshops. FC 2021. Lecture Notes in Computer Science, vol 12676. Springer, Berlin, Heidelberg. https://doi.org/10.1007/978-3-662-63958-0\_1

\section{Thesis Structure}

The remainder of this thesis is structured as follows. In chapter \ref{ch:background} we introduce the background theory in sufficient detail to understand the thesis. In chapter \ref{ch:custody} we offer an introduction to Bitcoin custody with reference to existing and related work. This includes a technology primer, a systems view perspective (people, process and technology), and modelling risk. Chapter \ref{ch:bitcoin-covenants} is our detailed presentation of a mechanism to construct Bitcoin covenants with key-deletion. We also describe therein alternative proposals and share a comparative analysis. With chapter \ref{ch:vault-custody} we put forward a custody system design specification called Ajolote, which uses the deleted-key covenant mechanism. We evaluate Ajolote, primarily by constructing a detailed risk model and demonstrating the attack and failure modes. In chapter \ref{ch:conclusions} we offer concluding remarks which summarise the thesis contributions and highlight directions for future work. In Appendix \ref{ch:risk-framework-revault} there is associated work on another custody system called \textit{Revault}. We provide a risk framework, as a formal attack model, for that system's operation. The methodology and findings therein are complementary with the approach taken in constructing the risk model for Ajolote.
\chapter{Background Theory}
\label{ch:background}

This chapter describes the foundational theory underpinning the entirety of this thesis. It commences with an exploration of cryptographic functions utilized in Bitcoin, alongside an examination of their intrinsic security properties. Subsequently, Bitcoin itself is analyzed, with a broad discussion of the blockchain, peer-to-peer protocol, mining, and governance. Transactions, addresses and outputs are then scrutinized in detail due to their pivotal role in the technical contributions of this thesis. Moreover, abstract security strategies and concepts are presented, including the secure data deletion concept. This is relevant to the covenant protocol detailed in chapter \ref{ch:bitcoin-covenants} and later applied in the Ajolote custody system design in chapter \ref{ch:vault-custody}.

\section{Cryptography}
\label{sec:cryptography}

\textbf{Hash functions} are one-way functions that take an input message $m$ and produce a fixed-length output as a bit-string called the \textit{hash value}, $h(m)$ \cite{Understandingcryptography}. The hash value is typically interpreted as a unique fingerprint for the message. Hash functions are considered \textit{secure} if they possess three properties:
\begin{quote}
    \textit{Pre-image resistance.} This refers to the function being one-way. Given $h(m)$, it is infeasible to compute $m$. \\
    \textit{Second pre-image resistance.} This refers to the the function having weak collision resistance. For a \textit{given} message $m_1$ with hash value $h(m_1)$, it is infeasible to compute another message $m_2$ such that $h(m_1)=h(m_2)$. \\
    \textit{Collision resistance.} This refers to the function having strong collision resistance. For \textit{any} message $m_1$ with hash value $h(m_1)$, it is infeasible to compute another message $m_2$ such that $h(m_1)=h(m_2)$. 
\end{quote}

We will see hash functions used frequently as we explore different pieces of Bitcoin. The hash function used most commonly in Bitcoin is `secure hash algorithm 256', or simply SHA256 \cite{SHA256}. It outputs a 256-bit string. Another hash function used in Bitcoin is RIPEMD-160 \cite{RIPEMD-160} which outputs a 160-bit string. 

A \textbf{commitment scheme} is another cryptographic primitive that enables us to commit to a message, $m$, but keep the message itself hidden until sometime later when we want to prove what our message was \cite{CommitmentSchemes}. A commitment scheme is considered \textit{secure} if the commitments it produces, $C(m)$, are;
\begin{quote}
    \textit{Binding. } The creator of the commitment $C(m)$ cannot compute a different message $m'$ such that they have the same commitment value, $C(m) = C(m')$. \\
    \textit{Hiding. } The commitment value $C(m)$ reveals no information about the message $m$.
\end{quote}

A \textbf{Merkle tree} (also called a hash tree) is an efficient commitment scheme that repeatedly uses a hash function. It is a data structure that compactly commits to large data sets. The root of the tree is a commitment (a single hash value) to each leaf node of the tree. Leaf nodes are hash values from data that will be committed to. A binary Merkle tree structure is generated by computing the hash of pairs of leaf and branch nodes until there is a single root hash value. Any leaf nodes can be revealed and it can be demonstrated that they were used to produce the tree. 

\textbf{Proof-of-work} is a way for a user (the prover) to prove that they have expended a certain amount of computational effort to others (the verifiers) \cite{Back2002}. In Bitcoin, proof-of-work is achieved by generating a hash value with a specific number, $n$, of zeros at the beginning of the bit-string. The higher $n$ is, the more difficult it is to generate a valid hash value because (on average) it takes more guesses at messages to compute a bit-string with more leading zeros. The security of proof-of-work depends on the pre-image resistance of the hash function. The verification of a proof-of-work only requires computing the hash value given the message, which is trivial.

\textbf{Public-key cryptography} (or asymmetric cryptography) is the basis on which bitcoin is securely held and transacted. Contrary to \textit{symmetric} cryptography schemes, where entities share a secret key to perform tasks such as encryption and message authentication, in an asymmetric cryptography scheme each entity has both a \textit{public key} (which may be shared) and a \textit{private key} (which must be kept hidden). Public-key cryptography is based on the intractability of certain mathematical problems. Key-pairs are generated using a one-way function. It is easy to compute the public key given the private key, but infeasible to compute the private key given the public key. 

In Bitcoin, public-key cryptography is used for \textbf{digital signatures} based on the mathematics of elliptic curves over finite fields. The approach leverages a hard mathematical problem that arises from the algebraic structure of elliptic curves over finite fields. Specifically, the security of the digital signature schemes used in Bitcoin depend on the hardness of the Elliptic Curve Discrete Lograithm Problem (ECDLP). Let us briefly consider how to derive a Bitcoin key-pair.

Bitcoin uses a specific standardized set of constants which define the elliptic curve, known as `secp256k1' \cite{secp256k1}. A private key, $k$, corresponds to a random integer. The public key, $K$, corresponds to a point on the elliptic curve. It is computed by elliptic curve point multiplication of the private key with the generator point of the curve, $G$,
\begin{equation}
    K = kG
\end{equation}

Multiplication on an elliptic curve over a finite field is easy to calculate one-way, but difficult to invert. Thus, deriving the public key from the private key is simple, but deriving the private key from the public key is assumed to be an intractable problem (given sufficiently large numbers), called the Elliptic Curve Discrete Logarithm Problem.

Bitcoin supports two digital signature schemes that rely on the same elliptic curve. Key-pairs are compatible with both schemes. These are the Elliptic Curve Digital Signature Algorithm (ECDSA) \cite{ECDSA} and the Schnorr Signature algorithm \cite{Schnorr1991}. These signature schemes provide powerful \textit{security services}. Specifically, they provide:

\begin{quote}
    \textit{Message integrity. } A signed message cannot be changed without invalidating the signature. \\ 
    \textit{Message authentication. } A signature generated with a private key owned by a user shows that the message was sent by that user.\\ 
    \textit{Nonrepudiation. } The signer of a message cannot deny that they have signed the message.
\end{quote}
Schnorr signatures, in particular, offer several advantages over ECDSA due to their mathematical linearity. This property of Schnorr signatures simplifies and enhances several cryptographic operations, which we will now discuss:

\begin{itemize}
    \item[] \textit{Public Key Aggregation:} Schnorr signatures allow for the aggregation of multiple public keys into a single public key that can be used to verify a signature that represents a group of signers.

    \item[] \textit{Signature Aggregation:} Similarly, Schnorr signatures allow for the aggregation of multiple signatures into a single signature. This is beneficial in scenarios when multiple entities need to sign a single transaction.
\end{itemize}
A concept that is particularly relevant for Bitcoin custody is \textbf{key derivation}. A key derivation function $KDF$ is simply a method to take an established key, $K_1$, and compute from it a new key $KDF(K_1,r) = K_2$ where $r$ is some random value. Key derivation can be used for private and public keys. By using a one-way function in the implementation of $KDF$, the key derivation cannot be reversed. This is useful when re-using public keys could cause a problem, as is the case in some cryptography protocols which were designed for single-use session keys. As a general principle in cryptography, it is always better to use fresh keys where possible. Re-used keys can create statistical correlations that enable cryptanalysis techniques to weaken or break, for example, digital signatures.  

For a detailed exposition of cryptography engineering, readers are referred to the classic textbook `Understanding Cryptography' by C. Par and J. Pelzl \cite{Understandingcryptography}. For an overview of cryptography in Bitcoin readers are referred to `Mastering Bitcoin' by A. Antonopoulos \cite{Antonopoulos:2014:MBU:2695500}.

\section{Bitcoin}

Bitcoin is a system built on a combination of numerous information technology mechanisms. It was introduced by S. Nakamoto \cite{Nakamoto2008} and has since been described in systematic detail by A. Antonopoulos \cite{Antonopoulos:2014:MBU:2695500} and others. In the following, let us prepare for the main content of this thesis by reviewing how it works.

\subsection{Blockchain Data Structures}

\begin{figure}
    \centering
    \includegraphics[width=\textwidth]{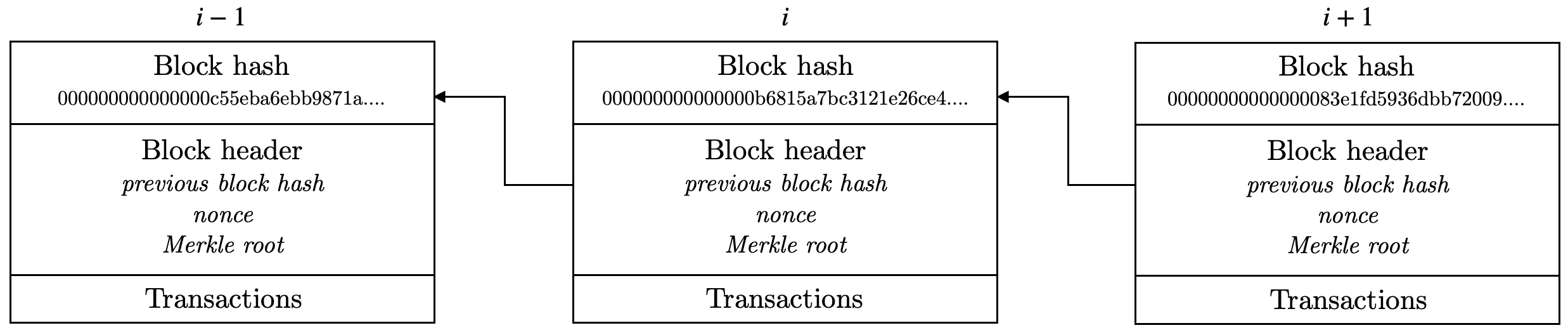}
    \caption{Blockchain data structure and the block data type.}
    \label{fig:blockchain}
\end{figure}

A \textbf{blockchain} is a data structure that holds a historical record of transactions. It is used as a public ledger from which the balances of users can be derived and new transactions can be verified. As the name indicates, the data structure is a chain of blocks and is shown in figure \ref{fig:blockchain}. Blocks are ordered and timestamped. Each block contains a block header and batch of transactions. A block is uniquely identified by a block hash, which is the SHA256 hash of its block header. Blocks are given a deterministic ordering by including the previous block hash in the block header. The transactions contained in a block are set as leaves in a Merkle tree. The root of this tree, called the Merkle root, is included in the block header. This is a binding commitment between the complete batch of transactions and this particular block in a manner that allows users to verify which transactions are included. Transactions can't be modified, added or removed without altering the Merkle root, and consequently the block hash. Finally, a nonce is included in the block header as a variable field to allow block creators to generate a proof-of-work using the block hash. Bare this in mind when we consider mining and the consensus mechanism.   

\begin{figure}
    \centering
    \includegraphics[width=0.6\textwidth]{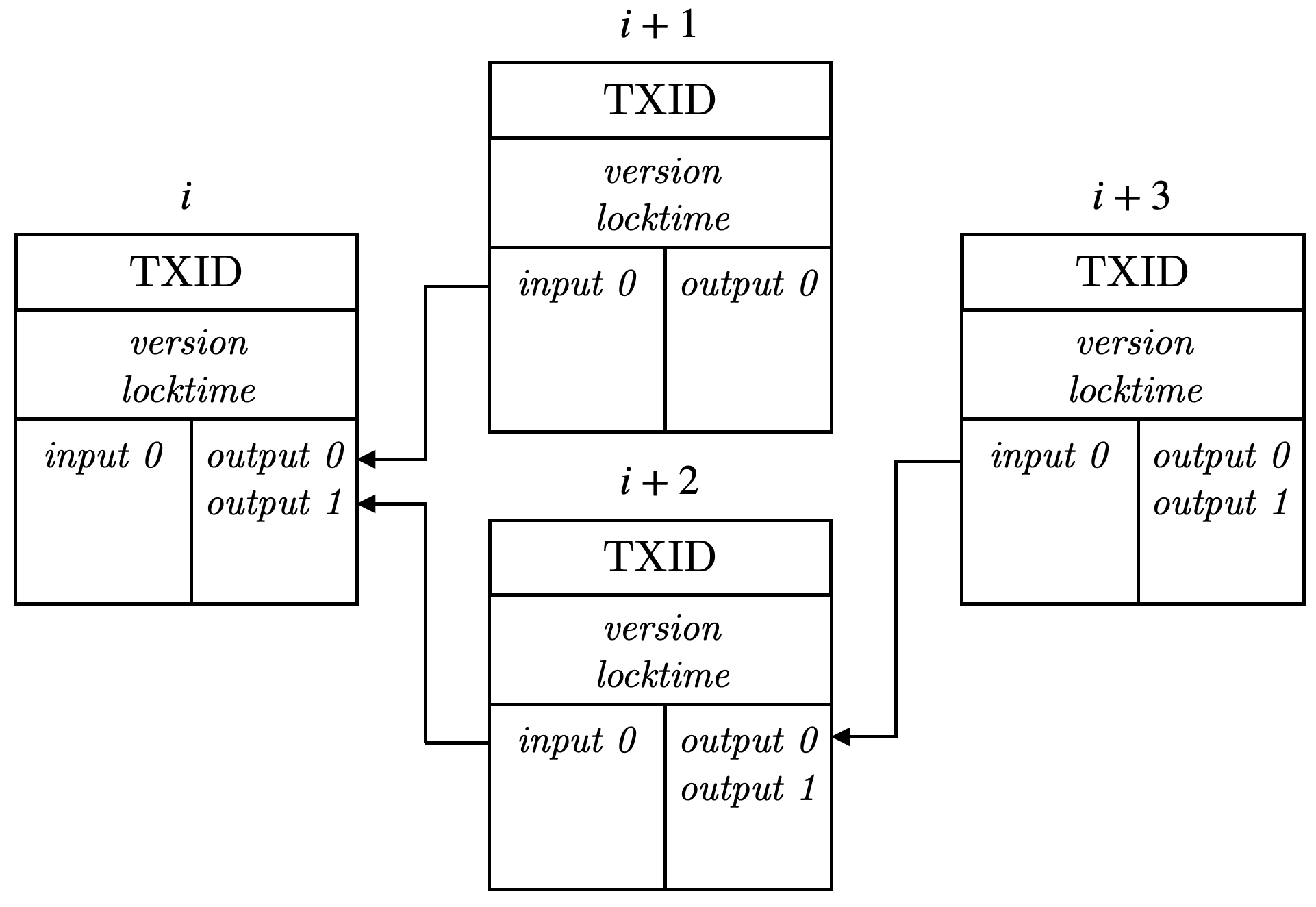}
    \caption{An example transaction graph. Each input must reference an about from a previous transaction. Some outputs remain unspent.}
    \label{fig:transaction-graph}
\end{figure}

Users create, digitally sign, and broadcast \textbf{transactions} to send bitcoin among each other. Transactions are objects with a specific data structure, described in table \ref{tab:transaction-type}. Note that the intuitive concept of accounts and balances is not directly encoded into the blockchain nor into transactions. Instead, unspent transaction outputs (\textbf{UTXOs}) are tracked and designated to a specific address controlled by users. Each transaction has a set of inputs and a set of outputs. The inputs reference existing UTXOs created in previous transactions, to be consumed in the current transaction, as shown in figure \ref{fig:transaction-graph}. The outputs set the target address and amount for each new UTXO being created. The total amount of the inputs must be greater or equal to the total amount of the outputs. The difference is offered as a transaction fee. Transactions will be further detailed in \ref{subsec:Txs-Script}.

\begin{table}
    \centering
    \begin{tabular}{|l|p{9cm}|}
    \hline
    \rowcolor{lightgray}  \textbf{Transaction Field} & \textbf{Meaning} \\
    \hline
         Transaction ID & Unique identifier \\
    \hline
         Version & Transaction data format version \\
    \hline     
         Inputs & A list of 1 or more transaction inputs  \\
    \hline     
         Outputs & A list of 1 or more transaction outputs \\
    \hline     
         Witnesses & A list of witnesses (unlocking scripts), up to one for each input (may be empty) \\
    \hline 
         Locktime & The block number or timestamp at which this transaction becomes valid \\
    \hline     
    \end{tabular}
    \caption{Fields of a transaction and their meanings.}
    \label{tab:transaction-type}
\end{table}

At any given block, the complete set of UTXOs can be derived by stepping through the historical record up to and including that block while tracking which UTXOs have been consumed and which haven't. The UTXO set is required to validate new transactions; to avoid double-spending attempts and to prevent creating bitcoin out of nothing. Users need only track the UTXO set for the current block to determine the validity of new transactions. A user need only download and verify the historical record once (this is called \textit{synchronising}), and the resulting UTXO set can be relied on from then on. 

\subsection{Peer-to-Peer Protocol}

The Bitcoin network consists of a distributed set of peers who establish connections among each other to send and receive block data. Moreover, each peer validates and verifies block and transaction data according to \textit{consensus rules}, rejecting data when the rules are violated. What emerges from the actions of individual peers is a distributed consensus protocol to manage the blockchain state. Any internet connected device can join the peer-to-peer network. The network is not hierarchical, each peer acts as both client and server to others. 

Network discovery messages, transaction and block announcements, requests and delivery of transactions and blocks are all part of the protocol. Peers will track the transactions they receive if they are valid, and forward them to other peers. The pool of transactions tracked by a peer is called the \textit{memory pool}. The protocol includes a multitude of methods to mitigate denial-of-service attacks on peers, such as limiting the size of their memory pool and temporarily banning connections with peers that violate consensus rules. 

\subsection{Mining}

A certain type of node in the network will participate as a \textit{miner}. A miner's role is to collect transactions from the peer-to-peer network (and elsewhere in some cases), build a valid block containing them, and propose the block to the network. Miners must produce a proof-of-work with the block hash for the block to be accepted. At any given time there is a \textit{difficulty parameter} in the global state which determines the difficulty of producing the next valid proof-of-work.  

Since producing proof-of-work is costly for the miner (computational effort requires energy expenditure), they require an incentive to perform this service for the network. The incentive is payment in bitcoin. If a block constructed by a miner is accepted by the network, all transaction fees and a \textit{block subsidy} will be paid to the miner. The block subsidy is newly issued bitcoin units (on a schedule that will eventually decay to zero). 

Anyone can participate as a miner. If the payment is greater than the cost, a rational player would start mining. The difficulty parameter is increased if throughput over a span of 2016 blocks is faster than the target; \textit{one block per ten minutes}. Conversely, it is decreased if the throughput is less than the target. The difficulty adjustment makes the system resilient. Why?

It is necessary to impose a throughput limit that gives sufficient time for the peer-to-peer network to distribute and verify block data. The limit generally prevents distributed denial-of-service attacks wherein the network peers are overwhelmed by new blocks. Moreover, if throughput was too high, it would result in consensus instability with many \textit{forks} \cite{InfoProp}. A fork is when more than one  valid and competing blocks are observed. A fork can persist, with several blocks competing against several others. The consensus rules determine how to select from multiple versions of the blockchain if a peer is presented with conflicting but valid states. The \textit{fork-selection} consensus rule is to consider the blockchain with the most cumulative proof-of-work as the \textit{canonical chain}. Nonetheless, blockchain forks can be frustrating to users because new transactions become less reliable. Thus, mechanisms to reduce the frequency of forks are important. 

On the other hand, if the difficulty makes the cost of producing a proof-of-work higher than the reward, rational players will stop mining. This will reduce throughput below the target. This is also frustrating for users whose transactions will take more time. By systematically reducing the difficulty once the throughput is low, the system incentives become compatible once more for rational miners. 

\subsection{Transaction Process}

In this example exchange, we detail the process by which Alice sends Bob 1 BTC. This example is designed to elucidate the interactions between users, their bitcoin applications, and the underlying peer-to-peer network, building on the concepts discussed so far.

\begin{enumerate}
    \item \textbf{Setup:}
    \begin{enumerate}
        \item Bob opens his bitcoin application and ensures it is synchronized with the blockchain.
        \item He observes a balance of 0 BTC.
        \item Bob generates a new receiving address and securely shares it with Alice.
    \end{enumerate}
    
    \item \textbf{Initiation of Transfer:}
    \begin{enumerate}
        \item Alice opens her bitcoin application and verifies synchronization with the blockchain.
        \item Her balance reads 1.5 BTC.
        \item Using the `Send' feature, she initiates a transfer of 1 BTC to Bob's provided address.
        \item Her application constructs and signs the transaction, then broadcasts it to her 8 connected peers.
    \end{enumerate}

    \item \textbf{Peer Validation:}
    \begin{enumerate}
        \item Each of Alice's peers checks the transaction against consensus rules.
        \item Upon validation, they verify Alice's signature.
        \item Valid transactions are retained in memory and relayed to each peer's respective connections, allowing the transaction to propagate through the network.
    \end{enumerate}

    \item \textbf{Mining and Block Inclusion:}
    \begin{enumerate}
        \item Miners, also peers, receive the transaction and perform the same validation process.
        \item Considering transaction fees, miners decide whether to include Alice's transaction in the next block they attempt to mine.
        \item A successful miner, one who generates a valid proof-of-work, includes Alice's transaction in a valid block. The miner broadcasts this block.
    \end{enumerate}

    \item \textbf{Block Propagation:}
    \begin{enumerate}
        \item Peer nodes, upon receiving the new block, validate its contents and the embedded transactions.
        \item Nodes append validated blocks to their blockchain copies and relay the block.
        \item The block diffuses throughout the network, reaching all nodes, including those of Alice and Bob.
    \end{enumerate}

    \item \textbf{Balance Update:}
    \begin{enumerate}
        \item Alice's application reflects her new balance, reduced by 1 BTC and the associated transaction fee.
        \item Bob's application updates his balance, showing 1 BTC.
    \end{enumerate}
\end{enumerate}

This example captures the intricacies of peer-to-peer exchanges within the Bitcoin network, shedding light on the multiple processes that ensure secure and accurate transactions.

\subsection{Governance}

The consensus rules of Bitcoin can only be changed given widespread support among participants in the system. There needs to be sufficient consensus about changing the rules of consensus. This is a difficult process that is not strictly formalised. As Bitcoin has become more widely adopted and the economy building on it has grown, it seems to have become increasingly difficult to change the consensus rules. There are two types of consensus rule changes. A \textit{hard-fork} is a change that makes previously invalid blocks valid. Without unanimous agreement to abide by the new rules the network would separate into two factions because any peers abiding by the old rules will reject a certain type of block as invalid, while peers following the new rules will accept them. A \textit{soft-fork} is a change that makes previously valid blocks invalid. A soft-fork is considered to be backwards-compatible because peers following the old rules will accept blocks made according to the new rules. A soft-fork still puts peers following old rules at risk because they may accept a block that is no longer accepted by those following the new rules. However, due to the fork-selection rule, if a majority of miners are producing blocks according to the new rules, this non-consensus block would be discarded once the chain of blocks that are valid for all peers becomes longer. Soft-forks are the preferred method for updating the consensus rules but they do not come without risk. Proposals for soft-forks should be well specified (for example, as a Bitcoin Improvement Proposal (BIP) \cite{BIP1}) and their implementation should be widely and thoroughly reviewed. 

\subsection{Inputs, Outputs and Script}
\label{subsec:Txs-Script}

In preparation for later chapters    we will need to understand how transactions are verified. Note that \textit{coinbase} transactions used by miners to issue new bitcoin will be ignored for this discussion. We saw the transaction data structure in \ref{tab:transaction-type}. Now, tables \ref{tab:output} and \ref{tab:input} describe an input and output record respectively. Transactions are entries to the public ledger that reassign ownership of Bitcoin. The strict controls over reassigning ownership of Bitcoin are based on digital signatures and additional programmable constraints. Inputs and outputs contain code written in a language specifically designed for Bitcoin called Script. 

\begin{table}
    \centering
    \begin{tabular}{|l|p{9cm}|}
    \hline
        \rowcolor{lightgray} \textbf{Output Fields} & \textbf{Meaning} \\
        \hline
        Amount  & Number of Satoshis being assigned to the public key script\\
        \hline
        Script length & Length of the public key Script \\
        \hline
        Public key Script & Expression of conditions (or a commitment to them) that need to be satisfied to consume this output \\
        \hline
    \end{tabular}
    \caption{Fields of an output record and their meanings. A \textit{Satoshi} denotes $10^{-8}$ bitcoin units.}
    \label{tab:output}
\end{table}

\begin{table}
    \centering
    \begin{tabular}{|l|p{9cm}|}
    \hline
        \rowcolor{lightgray} \textbf{Input Fields} & \textbf{Meaning} \\
        \hline
        Output hash & The hash of the referenced transaction containing the output \\
        \hline
        Output index & The index of the specific output in the referenced transaction \\
        \hline
        Script length & The length of the signature Script \\
        \hline
        Signature Script & Script for confirming transaction authorization  \\
        \hline
        Sequence & Optionally signals `replace-by-fee' \cite{BIP125},  optionally encodes a relative time-lock \cite{BIP112}\\
        \hline
    \end{tabular}
    \caption{Fields of an input record and their meanings. The  output hash and output index are often simply referred to as an `outpoint'.}
    \label{tab:input}
\end{table}

Script is a stack-based language \cite{ScriptWiki}. It was designed to support stateless verification of Bitcoin transactions. Here, `stateless verification' refers to the fact that no state information, or contextual data about the system, is required prior to the execution of Script and no such information is retained post-execution. Importantly, Script is not Turing complete, as it lacks the ability to perform loops or complex conditional logic (e.g. go-to).  This inherent limitation allows for easier reasoning about the potential termination of a Script program, be it successful or otherwise \cite{btc-spendable}. To clarify, Script is constructed such that it cannot diverge or run indefinitely, further reducing the risk of computationally intensive executions that could potentially lead to denial-of-service attacks on network peers and miners as they verify transactions. A subset of relevant Script instructions is provided for reference in table \ref{tab:opcodes}.

\renewcommand{\arraystretch}{1.25}
\begin{table}
\centering
\begin{tabular}{|l|l|p{8cm}|}
\hline
\rowcolor{lightgray} \textbf{Script Instruction} & \textbf{Operation Type} & \textbf{Description} \\
\hline
OP\_CHECKSIG & Cryptography & Verifies a signature against a public key and pushes the result (1 if the check succeeded, 0 if it did not) onto the stack\\
\hline
OP\_CHECKSIGVERIFY & Cryptography & Verifies a signature against a public key and if the result is not true, the script fails\\
\hline
OP\_CHECKSIGADD & Cryptography & Verifies a signature against a public key and adds a numerical result to the stack \\
\hline
OP\_CHECKMULTISIG & Cryptography & Verifies multiple signatures against multiple public keys \\
\hline
OP\_HASH160 & Cryptography & Hashes the top stack item with SHA-256, then with RIPEMD-160 \\
\hline
OP\_CHECKLOCKTIMEVERIFY & Time Lock & Verifies if the transaction's lock-time is greater than or equal to the top stack item \\
\hline
OP\_CHECKSEQUENCEVERIFY & Time Lock & Verifies the relative lock-time (set using the sequence number) of the transaction based on the top stack item \\
\hline
OP\_NUMEQUAL & Numeric & Checks if the top two stack items are numerically equal, pushes the result to the stack \\
\hline
OP\_DUP & Stack Operations & Duplicates the top stack item \\
\hline
OP\_EQUALVERIFY & Stack Operations & Checks if the top two stack items are equal, fails if not \\
\hline
OP\_EQUAL & Stack Operations & Checks if the top two stack items are equal, pushes the result to the stack \\
\hline
OP\_0 & Push Value & An empty array of bytes is pushed onto the stack \\
\hline
OP\_1 & Push Value & The number 1 is pushed onto the stack \\
\hline
OP\_IF & Control & If the top stack value is not False, the statements are executed \\
\hline
OP\_VERIFY & Control & Marks transaction as invalid if the top stack value is not true \\
\hline
OP\_RETURN & Control & Marks transaction as invalid \\
\hline
\end{tabular}
\caption{Table listing relevant Bitcoin Script instructions along with their operation type and a simple description of what it does.}
\label{tab:opcodes}
\end{table}
\renewcommand{\arraystretch}{1}

The Sequence field in a Bitcoin transaction input has a dual role. It can be used to signal `replace-by-fee' (RBF) or to encode a relative time-lock. Introduced in BIP 125 \cite{BIP125}, RBF is a method that allows an unconfirmed transaction in a node's memory pool to be replaced with a different transaction that spends at least one of the same inputs and pays a higher transaction fee\footnote{If the Sequence number of an input is less than {\tt 0xffffffff - 1}, then it signals that the transaction can be replaced.}. Introduced in BIP 112 \cite{BIP112}, \textit{relative time-locks} allow an input to specify a minimum time or block height difference from when the output being spent was included in a block. The transaction output can't be spent until a certain time period has passed since the output was included in a block\footnote{If the Sequence number of an input is less than {\tt 0xfffffffe}, and the version of the transaction is 2 or higher, then the Sequence number is interpreted as a relative lock-time.}. To create a relative lock-time, the sequence field is used in conjunction with the {\tt OP\_CHECKSEQUENCEVERIFY} instruction. 

Although Script's syntax -- the set of rules that parse byte strings into instructions -- is consistently applied, its semantics, or the consensus rules that govern instruction interpretation and execution procedure, can vary based on the context. `Context' here is referring to the specific output type, as will be described in the following subsection. One method to make these contextual differences more manageable is through the concept of `lock' and `unlock' Scripts. `Lock' Scripts can be understood as conditions set in a Bitcoin transaction output, which must be fulfilled for the transaction output to be spent. Conversely, `unlock' Scripts, or scripts provided in the transaction redeeming the locked output, serve to fulfill these conditions by providing the required data or signatures. During the process of verifying a transaction, the unlock script is executed first, followed by the lock script, with the remaining stack elements from the execution of the unlock script. The transaction is considered valid if the lock script execution is completed successfully, without any errors. The `context' determines how the lock script is committed within an output and outlines where both the lock and unlock scripts are provided within the transactions.

Each output specifies an amount and a \textit{public key script}\footnote{In the reference implementation this is called the {\tt scriptPubKey}.}. The public key script will either be a lock script, or a commitment to a lock script. In the latter case, the lock script has to be provided in a subsequent transaction that consumes this output. The locking conditions are represented as a Script program with data and/ or instructions to manipulate stack elements during execution. The owner of an unspent transaction output (UTXO) is anyone or any group who can satisfy its locking conditions with a valid unlock script which typically depends on secret information. The unlock script is provided in the subsequent transaction redeeming the output balance. Transactions have two containers for the unlock script. The first is the \textit{signature script}\footnote{In the reference implementation this is called the {\tt scriptSig}} in the input, and the second is the witnesses field in the transaction. Next, we will describe when and why these different containers are used for Script verification.

\subsection{Output Types and Addresses}

There are 8 output types that are considered \textit{standard} (as of Bitcoin version 23.0). The default policy for nodes in the network is to relay transactions with standard outputs only. Note that this is not a consensus rule, but it is broadly enforced. These output types each have a different purpose and are specified as a public key script template, with fixed instructions but variable data elements. These are shown in table \ref{tab:scriptpubkeys}. The corresponding script template in a transaction spending from the given output types is shown in table \ref{tab:scriptSigs} with the location where the script is contained. These templates may encompass both unlock and lock scripts.

\begin{table}
    \centering
    \resizebox{\textwidth}{!}{
    \begin{tabular}{|l|l|}
    \hline
     \rowcolor{lightgray} \textbf{Output Type} & \textbf{Public Key Script Template} \\
    \hline 
    pay-to-public-key (P2PK) & {\tt <public key> CHECKSIG} \\
    \hline
    pay-to-public-key-hash (P2PKH) & {\tt DUP HASH160 <public key hash> EQUALVERIFY CHECKSIG} \\
    \hline
    pay-to-script-hash (P2SH) & {\tt HASH160 <lock script hash> EQUAL} \\
    \hline
    pay-to-multi-signature (P2MS) & {\tt 2 <public key A> <public key B> <public key C> 3 CHECKMULTISIG}\\
    \hline
    null-data & {\tt RETURN <data>} \\
    \hline
    pay-to-witness-public-key-hash (P2WPKH) & {\tt 0 <20-byte public key hash>} \\
    \hline
    pay-to-witness-script-hash (P2WSH) & {\tt 0 <32-byte script hash>}  \\
    \hline
    pay-to-taproot (P2TR) & \textit{key path}: {\tt 1 <32-byte public key>} \\
    
    & \textit{script path}: {\tt 1 <32-byte witness program>} \\
    \hline
    \end{tabular}}
    \caption{Public Key Script templates for standard output types. Instructions are denoted without the {OP\_} prefix for brevity. Variable data elements are denoted between angled brackets such as {\tt <data>}.}
    \label{tab:scriptpubkeys}
\end{table}

The output types P2PK, P2MS, and null-data are incorporated in the tables for the sake of comprehensiveness. However, it's important to note that P2PK and P2MS have been largely replaced by more recent standard output types. Additionally, the null-data outputs, while part of the Bitcoin protocol, offer limited practical utility for our purposes and hence, will not be elaborated upon in subsequent discussions.

An \textit{address} is derived from the public key script in a particular way depending on the output type. It is a string of characters that users must share in order to receive a payment and it includes version information, a commitment to the lock script and a checksum for error detection.  There are several types of address corresponding to different output types, each with a different purpose. Addresses are created by users and can be controlled independently or collaboratively. Now, let's consider common address/output types and elaborate on how `ownership' of bitcoin is defined. 

\begin{table}
    \centering
    \begin{tabular}{|p{3cm}|p{9cm}|l|}
    \hline
     \rowcolor{lightgray} \textbf{Output Type}  & \textbf{Template} & \textbf{Location} \\
    \hline 
    pay-to-public-key (P2PK)  & {\tt <signature>} & Signature Script \\
    \hline
    pay-to-public-key-hash (P2PKH) & {\tt <signature> <public key>} &Signature Script \\
    \hline
    pay-to-script-hash (P2SH)  & {\tt <arbitrary unlock script> <serialised lock script>} &Signature Script \\
    \hline
    pay-to-multi-signature (P2MS) & {\tt 0 <signature A> <signature B>} &Signature Script\\
    \hline
    null-data  & & \\
    \hline
    pay-to-witness-public-key-hash (P2WPKH) & {\tt <signature> <public key>} &Witnesses\\
    \hline
    pay-to-witness-script-hash (P2WSH)  & {\tt <unlock script> <serialized lock script>}  &Witnesses\\
    \hline
    pay-to-taproot & \textit{key path}: {\tt <signature>}  &Witnesses\\
    (P2TR) & \textit{script path}: {\tt <tapleaf unlock script> <serialised tapleaf lock script> <control block>}  &Witnesses \\
    \hline
    \end{tabular}
    \caption{Signature Script and Witness templates for standard output types. }
    \label{tab:scriptSigs}
\end{table}

A P2PKH output type commits to a public key by including the public key hash in the public key script. A user who has received payment to their P2PKH address can spend these coins by providing both their public key and a digital signature in the signature script of their transaction. To create a valid digital signature, the user must have access to the private key associated with the public key. Users can openly share their P2PKH address. Their public key will remain hidden until they spend from the address, at which point they must reveal their public key.  

A P2SH output type commits to a lock script by including the hash of the serialised lock script in the public key script \cite{BIP13}. This output type was made standard in 2012 and it allows the receiver to define and share the locking script without revealing what that script is (at least until spending from it) \cite{BIP16}. Spending a P2SH output requires providing the exact script that produced the hash as well as a valid unlock script (typically with signatures) in the signature script of a new input.  The execution logic of Script when spending from a P2SH address is modified. First, the provided lock script is checked against the script-hash commitment in the address. Subsequently, standard execution proceeds wherein the unlock script is executed first, followed by the execution of the lock script, utilising any remaining stack elements left by the unlock script.

With the soft-fork called the `Segregated Witness soft-fork', several consensus rules were added to Bitcoin \cite{BIP141}. These rules enabled the separation of a transaction's data representing its \textit{effects} (on blockchain state) from its `witness' data (the unlock script) required for \textit{validation}. Consequently, a new class of `Segregated Witness' (SegWit) transactions were enabled. These transactions offer enhanced storage and bandwidth efficiency and make use of an updated semantics of Script. This is the reason for having two containers for the unlock script. Standard SegWit outputs can be redeemed by setting an empty signature script and instead placing signatures and the unlock script in the witness field. There are two version 0 SegWit output types that are  equivalent to P2PKH and P2SH. These are called pay-to-witness-public-key-hash (P2WPKH) and pay-to-witness-script-hash (P2WSH). 

The implementation of the `Taproot' soft-fork introduced several additional consensus rules to Bitcoin. For all output types, with the exception of P2TR, the Elliptic Curve Digital Signature Algorithm (ECDSA) is employed. However, in the case of P2TR, the Schnorr digital signature algorithm is utilized \cite{BIP-Schnorr}\footnote{In both instances, the elliptic curve {\tt secp256k1} is used, and private keys are compatible.}. The linearity of Schnorr signatures simplifies both public key aggregation and signature aggregation. This means that collaborative control of outputs, which previously required multiple {\tt OP\_CHECKSIG} instructions in Scripts, can now be achieved more privately, reducing storage and bandwidth costs for the network. 

Furthermore, the linearity property enables the commitment of lock scripts by adding them to a public key, a process known as a `public key tweak'. This results in the output type being represented as a public key, effectively concealing the fact that a script was committed to, at least until the output is spent. In essence, P2TR output types amalgamate behaviors akin to P2PK, P2MS, and P2SH into a single, more efficient output type. The technical constructions in chapters \ref{ch:bitcoin-covenants} and \ref{ch:vault-custody} both utilise P2TR outputs and so let us explore these in more detail.  

\subsection{Taproot and TapScript}

This section provides an overview of the key concepts from the Taproot \cite{BIP-Taproot} and TapScript \cite{BIP-Tapscript} specifications that are pertinent to this thesis. While many low-level details are omitted for brevity, interested readers can find comprehensive information in the cited specifications. 

Recall from Table \ref{tab:output} that an output record contains a `public key script' field, which holds a commitment to the lock script. More specifically, the structure of the values in the public key script determines the output type, such as P2PKH, P2WSH, and others. In this section, we focus on P2TR output types. A Taproot output type is structured as follows:

\begin{equation*}
    public\text{ }key\text{ }script:\text{ } <witness\text{ }version>\text{ } <taproot\text{ }output\text{ }key>
\end{equation*}

Here, the witness version is 1, and the Taproot output key is a 32-byte public key as defined in the Schnorr specification for Bitcoin \cite{BIP-Schnorr}. The Taproot output key, denoted as $q$, encodes a commitment to both a key-path spend option and script-path spend options. It is the $x$ component of $Q$, where

\begin{equation}
    Q = P + int(t)G
\end{equation}

In this equation, $P$ is an internal public key, $G$ is the elliptic curve base point, and $t$ is the \textit{TapTweak} hash cast to an integer. The internal public key is used for key-path spending. To ensure domain separation, so-called `tagged hashes' are used. The notation is 

\begin{equation}
    hash_{tag}(x) = SHA256(SHA256(tag)||SHA256(tag)||x)
\end{equation}

for arbitrary data $x$ and a string $tag$. Here, $||$ denotes concatenation. This approach theoretically reduces the likelihood of hash collisions within the system \cite{BIP-Taproot}. The TapTweak hash contains a commitment to a tree of TapScripts that can be used in script-path spending. It is a hash of the root $k_m$ of a Merkle tree \cite{Merkle1987ADS}, known as the TapTree, and $P$,

\begin{equation}
    t = hash_{TapTweak}(P||k_m)
\end{equation}

\begin{figure}
    \centering
    \includegraphics[width=0.7\textwidth]{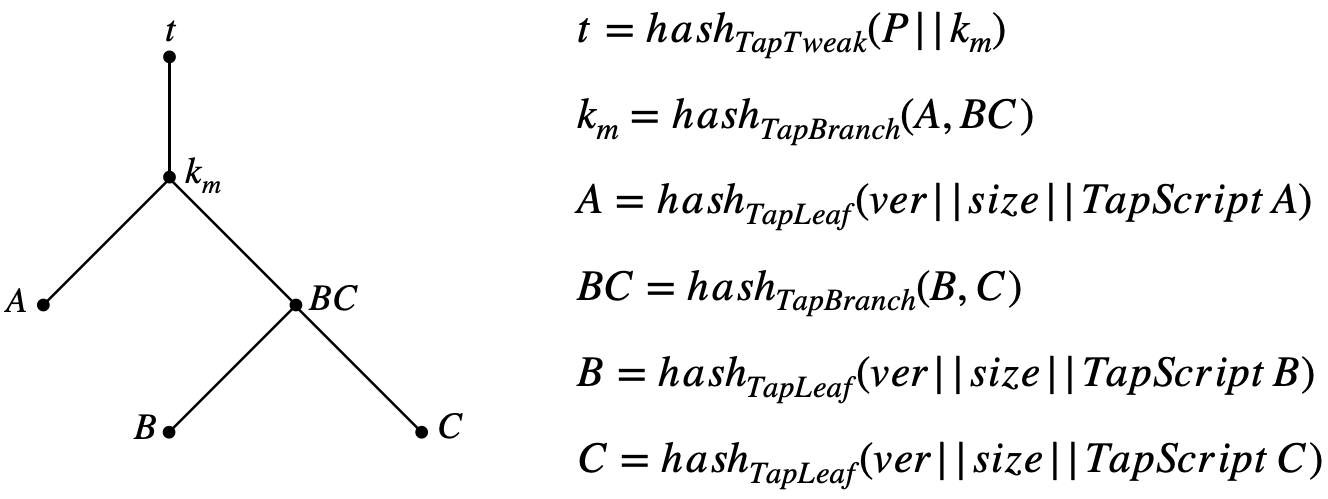}
    \caption{An abstract TapTree with three leaf nodes $A$, $B$, and $C$ which are hashes of three TapScripts. Leaf nodes, branch nodes and the TapTweak are each tagged hashes of data as shown on the right. The TapScript version and size are represented by $ver$ and $size$, respectively. The TapTweak $t$ is the hash of the internal public key $P$ concatenated with the root hash of the Merkle tree $k_m$.}
    \label{fig:abs-taptree}
\end{figure}

To comprehend how script-path spending operates, we need to understand the TapTree data structure. An example is depicted in Figure \ref{fig:abs-taptree}. The structure's purpose is to compactly commit to distinct data sets and enable existence proofs for a data set without revealing the others. It is a Merkle tree, also known as a Hash tree. A TapTree is a binary tree that may be unbalanced. Each leaf node is the hash of a ($\tt leaf$ $\tt version$, $\tt TapScript$) tuple. Each TapScript is a distinct lock script that can be satisfied to spend the output. Each branch node is the hash of its two child nodes. The root node is the Merkle-root hash, $k_m$, which is a deterministic derivation from the given leaf nodes. 

{}

In this thesis, we are primarily interested in script-path spending. Key-path spending does not enable the complex access policies or privacy guarantees we seek. Therefore, we must use an internal public key with an unknown private key to prevent key-path spending. A method for this was described in the Taproot specification \cite{BIP-Taproot}, stating that one should use `the hash of the standard uncompressed encoding of the secp256k1 base point $G$ as $X$ coordinate'. The intuition here is that the hash function produces a random value, based on a `nothing-up-my-sleave' number, and so the resulting public key was not generated from a known private key in the standard way, by computing $P = pG$. Given the internal public key, one cannot find the private key without solving the elliptic curve discrete logarithm problem which is assumed to be hard. 

A transaction that consumes a Taproot output as its input must provide a witness with a specific structure that satisfies the chosen TapScript. The witness structure has the form

\begin{equation*}
    script-path\text{ }witness:\text{ } <initial\text{ } stack\text{ } elements>\text{ } <TapScript> \text{ }<control \text{ }block>
\end{equation*}

The initial stack elements are the unlock script (typically containing all signatures). The TapScript is the chosen lock script that must be present as a leaf node in the TapTree. The control block contains the internal public key $p$ and a proof of the existence of the TapScript in the TapTree. The Merkle proof contains all necessary hashes to derive $k_m$, but doesn't reveal unused TapScripts. For example, in Figure \ref{fig:abs-taptree}, the inclusion proof to spend with TapScript C includes the hashes B and A.

When the script-path witness is broadcast as part of the transaction, it becomes public knowledge. The control block reveals a minimum depth of the complete TapTree. The TapScript reveals public keys and system parameters such as the length of time-locks.

\subsection{Miniscript and Policy}

Miniscript is a language designed to be decoded into Bitcoin Script \cite{miniscript}. Its primary aim is to simplify the process of writing Bitcoin Script as generic access policies, which are composed of common subpolicies such as signature verification, time-locks, and hash-locks. Miniscript enhances the analysis of spending cost, script size, and correctness. 

In practical terms, an engineer-readable form is used to write a policy, which is then compiled to generate the corresponding Miniscript. This Miniscript directly maps to Script, which is subsequently used in Bitcoin transactions. A reference of the Miniscript policies utilized with the Ajolote custody system is provided in Table \ref{tab:miniscript}.

\renewcommand{\arraystretch}{1.25}
\begin{table}
    \centering
     \begin{tabular}{|l|p{8cm}|}
        \hline
        \rowcolor{lightgray} \textbf{Policy} & \textbf{Description} \\
        \hline
        pk(NAME) & Require public key named NAME to sign. NAME can be any string up to 16 characters. \\
        \hline
        older(NUM) & Require that the nSequence value is at least NUM. NUM cannot be 0. \\
        \hline
        and(POL,POL) & Require that both subpolicies are satisfied.\\
        \hline
        or([N@]POL,[N@]POL) & Require that one of the subpolicies is satisfied. The numbers N indicate the relative probability of each of the subexpressions (so 9@ is 9 times more likely than the default).\\
        \hline
        thresh(NUM,POL,POL,...) & Require that NUM out of the following subpolicies are met (all combinations are assumed to be equally likely).\\
        \hline
    \end{tabular}
    \caption{Description of Miniscript policies extracted directly from the Miniscript website \cite{miniscript}.}
    \label{tab:miniscript}
\end{table}
\renewcommand{\arraystretch}{1}

\section{Foundations of Security}

Throughout this thesis we will rely on strategies and concepts derived from the history of warfare and conflict. As many others have done before, these concepts will be applied to the modern domain of information security. Let us describe each of these concepts because they appear regularly throughout. 

\textbf{Defence-in-depth} is an approach to system security wherein multiple layers of controls are used to guard an asset. An attacker must exploit a vulnerability in each control layer to reach their target. The redundancy of controls increases the energy and time required from an attacker to be successful. Often, this can give the defender an opportunity to react with reinforcements. Defence-in-depth should be deployed across each system layer: people, process, and technology. Ideally, the defender would set-up diverse controls that require different kinds of expertise and or specialist equipment for an attack sequence to be successful. Ideally, the defender would weaken the attacker as the sequence progresses, by gathering information and where possible counter-attacking. For example, by determining the attacker's location and calling the police. 

\textbf{Compartmentalisation} is an approach to segregate critical information or assets from one another by limiting access to them with, for example, role-based policies, physical environments, or cryptographic keys. This strategy helps a defender to limit the scope and damage of successful attacks. Interactions between compartments should be limited for maximum effectiveness. While defence-in-depth forces an attacker to complete a sequence of exploits, compartmentalisation forces an attacker to complete multiple separate attack sequences to compromise the system. 

\textbf{Choke points} are used to force the attacker through a specific channel controlled and monitored by the defender. Traditionally, a castle would be constructed with high walls and a moat, with only a bridge and a gate readily accessible. Together these act as a choke point, where intruders must pass through the gate to enter. Defenders can inflict devastating damage while the attacker passes through the choke point, or deter attacks from happening at all due to their asymmetric advantage here. A firewall is an example of a choke point in network security. 

\textbf{Least privilege} is a design constraint that states that any agent (human, machine or process) should only have access to the information or assets they require in order to carry out their legitimate responsibilities, and only for the minimum amount of time necessary. It is unnecessary to give more privileges than required, and doing so increases the fragility of a system. For example, an administrative intern does not need root access to the company database to update employee attendance sheets, they merely need read and write access for that sheet. 

\textbf{Weakest links} are components within systems that have the most exploitable vulnerabilities. By definition, a system is only as secure as its weakest link. When reinforcing a system, add countermeasures to mitigate attacks on the weakest components first. 

\textbf{Fail-secure} components are crucial, particularly with complicated systems with many components. If a component fails, the system should not permit insecure behaviour. This is generally achieved by applying a deny-by-default policy and not permitting any behaviour while the component is inactive. Using fail-secure components may not always be possible if continuity and availability are strict requirements. 

\section{Key-deletion}
\label{subsec:key-deletion}

The successful enforcement of deleted-key covenants--the technical construct to be introduced in chapter \ref{ch:bitcoin-covenants}--is contingent upon the effectiveness of private key deletion. Here we review this topic and explore the practical implications of different approaches. 

The secure erasure of data from a physical medium is a critical security process in many contexts where the disclosure of privileged or confidential information must be avoided. As defined in a survey of the state of knowledge of secure data deletion \cite{SOKSecureDeletion}; `data is securely deleted from a system if an adversary that is given some manner of access to the system is not able to recover the deleted data from the system'. The techniques for securely deleting data differ when using a magnetic hard drive or flash memory. Moreover, the techniques differ when attempting to delete through interfaces at different levels of the system; through a controller at the hardware layer, through the file system at the operating system layer, and through software at the user-interface layer. If the private key's entire life-cycle can be contained in volatile memory (RAM) alone, then its deletion will be dependent on the physical properties of SRAM or DRAM cards. However, if the private key is ever written to disk, its erasure will be dependent on the physical storage medium being used too, typically it will be EEPROM or Flash storage. 

Kissel \textit{et al.} discuss guidlines for media sanitization in \cite{NIST}. They define four types of media sanitization which, in the following order, become more thorough and resistant to information recovery; disposal, clearing, purging, and destroying. Disposal here means simply discarding the storage media and is very susceptible to information recovery. Clearing is a sanitization process for which information cannot be re-gained using data, disk, or file recovery utilities. Overwriting information with random data is an example of clearing. Purging is a sanitization process that resists specialized laboratory attacks. An example of purging from magnetic drives is to use a degausser, which generates a strong magnetic field to remove the information contained in the drive's magnetic domains. Finally, destroying media is the most secure form of media sanitization. This could be done by incinerating, shredding, melting or pulverising the storage media. Factors such as cost must be considered along with effectiveness when deciding which media sanitization processes are suitable. In the context of an organization, the sanitization process should be conducted on a significantly representative sample of storage media to verify the process, ideally employing a non-biased expert entity to conduct the verification.

Valamehr \textit{et al.} \cite{InspectionResistantMemory} discuss physical inspection attacks for determining information contained on storage devices, stating that there are two classes of attacks; \textit{passive} and \textit{intrusive}. Passive attacks will probe the interface of the device to discern timing or electrical differences. Active attacks will breach the boundaries of the device to probe, scan and alter the device internally.  According to Valamehr \textit{et al.} both EEPROM and Flash memory store charge on a floating gate. When their memory cells are overwritten, some residual information is retained as a bias on the substrate. This is the basis for physical inspection attacks with an electron microscope. Similarly for volatile memory, SRAM memory cells can retain some level of information and are subject to similar analyses. Their work explores how to construct device architectures that resist information leakage even when subjected to physical inspection attacks. 

Other work, which also relies on the imperfect nature of recovering data, attempt to incorporate an additional layer of security by using key hiding techniques based on leakage-resistant cryptography. Examples include  exposure-resilient functions and all-or-nothing transforms \cite{ERFAON} and distributed public key schemes secure against continual leakage \cite{DPKSSecureAgainstLeakage}. This exemplifies only a small portion of the work being done to improve physical inspection security of devices, but also demonstrates that the current state-of-the-art fails to handle the totality of the complexity of secure key deletion. These and similar methods may prove useful in future if the theory and implementations advance to support better performing systems and hardware device manufacturers can integrate them into their products.

Other work focuses on \textit{provable} deletion for embedded devices \cite{ProofsOfSecureErasure, EPoSE, DeletingSecretData}. The `black box' nature of secure deletion functions, which are often carried out in Trusted Platform Modules (the standard, for example, of key-management in the financial industry \cite{Anderson:2008:SEG:1373319}) is unsatisfactory as it requires complete trust in the correct implementation of software inside the module. Provable deletion on the other hand is a scheme to enable users to verify the deletion operations. The provable deletion schemes of \cite{ProofsOfSecureErasure, EPoSE} work by requiring the embedded device to compute a function on a data-set that fills its storage capacity, and return the result as a proof that it stored that data-set which necessitated overwriting all other data previously contained on the device. These techniques may prove very useful for a hardware device manufacturer who wishes to enable deleted-key covenant functionality. 

Trusted erasable memory (using secure deletion tools and physical device security measures) and disposable erasable memory (renders physical inspection useless through destruction) differ in their security guarantees and cost effectiveness. Even while using state-of-the-art in trusted erasable memory, one can never be totally certain that no important confidential information has been retained. However, the fact that these devices can be re-used indefinitely for as many secure deletions as is desired makes them more conducive to use in covenant-based applications. Disposable memory offers maximally secure key deletion, but requiring a signing device to be destroyed for each activation of a covenant is impractical. An example of how to minimize wasted resources here would be to create batches of covenants on a simple, cheap and `bare bones' signing device such as a smart card, and to destroy it, activating multiple covenants at a smaller cost than one per card.  

No single method will ever enable perfect key deletion. However, the key deletion process can be distributed across numerous heterogeneous devices which rely on different methods. Each new device and deletion method acts as an additional countermeasure that increases the cost for an attacker. Thus, the security of key deletion can be enhanced by the set of enforcers and the effort they put in to mitigate information recovery attacks. For an enforcement policy of $m$-of-$n$, the key deletion process can resist up to $n-m+1$ private keys being recovered. The trade-off is that more security costs more money and time.

Thus it has been shown that there is a range of possible designs for secure key deletion, each with a different security model and set of assumptions. If security is paramount, then disposable (destructible) memory should be used. However, if probabilistic security is sufficient, a distributed set-up of trusted erasable memory devices may be used. 

\section{Protocols}
\label{sec:protocols}

\subsection{Network Protocols}
\label{subsec:network-protocols}

A network protocol is a formal set of rules, conventions and data structures that governs how computers and other network devices exchange information over a network. Kurose and Ross give the following definition \cite{kurose2013computer}:
\begin{quote}
    ``A protocol defines the format and the order of messages exchanged between two or more communicating entities, as well as the actions taken on the transmission and/or receipt of a message or other event.''
\end{quote} 

Some well-known network protocols include the Transmission Control Protocol (TCP) \cite{RFC793}, Internet Protocol (IP) \cite{RFC791}, Hypertext Transfer Protocol (HTTP) \cite{RFC2616}, and File Transfer Protocol (FTP) \cite{RFC959}. Network protocols are often architected and implemented in a hierarchical manner, forming what is known as the protocol stack. In this paradigm, each layer provides services to the layer above it and relies on the services of the layer beneath it. This layered approach allows for modularity, making design, implementation, and testing more manageable. Higher-level protocols deal with abstract data and system-level functions, whereas lower-level protocols handle more concrete tasks such as data transmission and reception. Examples include the TCP/IP model, which encapsulates protocols like IP, TCP, and HTTP in different layers.

In this thesis, we focus on designing and analyzing new protocols that leverage existing common internet protocols such as TCP/IP, as well as Bitcoin network protocols. These new protocols aim to provide advanced transaction functionalities in the Bitcoin ecosystem while maintaining compatibility with the existing network infrastructure.

\subsection{Ceremonies}
\label{subsec:ceremonies}

A ceremony, as introduced by Ellison \cite{Ellison2007CeremonyDA}, is an extension of a network protocol which broadens a system specification to include humans and their interactions alongside networked devices. Humans and their actions are considered `out-of-band' to a network protocol, but not to a ceremony. Often, failing to consider the human(s) operating a system can create security vulnerabilities. Moreover, humans have capabilities that network devices do not (e.g. strong authentication for someone they know) and can thus enable otherwise ignored countermeasures that mitigate vulnerabilities \cite{TransformingWeakestLink}. However, humans also introduce complications. Humans have numerous weaknesses including non-deterministic behaviour, memory limitations, bounded attention, lack of knowledge of security, and so on. A taxonomy of human weaknesses was presented by Carlos and Price \cite{HumanWeakness}. They propose design recommendations to minimise human-based vulnerabilities: 

\begin{quote}
    1. Do not give users an unfeasible task.\\
    2. Prevent inappropriate interactions from being performed.\\
    3. Consider that the expected behaviour might change.\\
    4. Do not rely on user's authentication capabilities (of digital objects or strangers).\\
    5. Integrate security into the main workflow (not as a secondary task).\\
\end{quote}
For Bitcoin custody systems, humans are critical components. There is no escaping human-mediated inputs when making payments. A custody system is owned and managed by one or more humans and their decisions are essential to ongoing operations. 

\subsection{Threat Model}

A threat model serves as a structured and systematic blueprint that represents an array of potential threats capable of compromising the security of a system. This blueprint encapsulates diverse elements such as potential attack vectors, profiles and capabilities of prospective attackers, as well as the specific vulnerabilities these attackers could exploit.

When it comes to conducting a risk analysis for a ceremony, traditional threat models designed for network protocols fall short. The evolution of threat models has seen a progression from passive attacks, which merely eavesdrop on communication channels, to active attacks, where messages are intercepted, modified, replayed or otherwise manipulated. This notion of active attackers was first considered by Needham and Schroeder \cite{NeedhamSchroeder} and later formalized by Dolev and Yao \cite{DolevYao}. The Dolev-Yao threat model has become the \textit{de-facto} standard for network protocol analysis \cite{Bella2007FormalCO}. To encapsulate the Dolev-Yao model succinctly, it posits an attacker who controls all network traffic and can perform any message operation, barring cryptanalysis. The underlying assumption is that if a network protocol can withstand such a potent attacker, it will remain secure against any realistically feasible attacker.

However, this model neither embodies realism nor sufficiency for a ceremony. Certain communication channels are not strictly confined to a network. For instance, a {\tt human} $\leftrightarrow$ {\tt offline device} channel might comprise of a display and input buttons. A remote attacker, absent from the room, would not have the ability to intercept this channel, rendering it unrealistic to assume that an attacker can control that traffic and manipulate those messages. Designing a ceremony to secure against an omnipotent, yet unrealistic, attacker would necessitate additional countermeasures and complexities. This could impose an undue burden on the user, potentially leading to erroneous enactment of the ceremony. Consequently, an adaptive threat model was proposed by Martina \textit{et al.} for ceremony analysis \cite{AdaptiveThreatCeremony}. They introduced a concise notation to define an attacker's capabilities for any given channel. We will utilize the same notation as depicted in Table \ref{tab:attacker-capabilities}. A Dolev-Yao threat model characterizes an attacker possessing all capabilities across all channels, whereas a passive attacker threat model describes an attacker who can only eavesdrop on channels.

\begin{table}
    \centering
    \begin{tabular}{|l|l|l|}
    \hline
         \rowcolor{lightgray} & \textbf{Capability} & \textbf{Meaning} \\
        \hline
         E & Eavesdrop & Attacker can learn message contents \\
         I & Initiate & Attacker can initiate messages with channel entities \\
         A & Atomic Breakdown & Attacker can decompose messages into distinct data elements \\
         C & Crypto & Attacker can perform cryptographic operations \\
         B & Block & Attacker can block messages from being received\\
         F & Fabricate & Attacker can fabricate messages \\
         S & Spoof & Attacker can spoof the identity of one entity to the other \\
         O & re-Order & Attacker can delay messages to a receiving entity and thus re-order them \\
        \hline
    \end{tabular}
    \caption{Summary of attacker capabilities for a ceremony threat model.}
    \label{tab:attacker-capabilities}
\end{table} 

\chapter{Understanding Bitcoin Custody: Technologies, Systems, and Risk Models}
\label{ch:custody}

In this chapter we review the field of Bitcoin custody and introduce necessary concepts and perspective in preparation for later chapters. We begin by introducing terminology, motivating principles, and functional elements in section \ref{sec:custody-intro}. In section \ref{sec:custody-tech-primer} we investigate the broad space of technology used in custody systems, covering the theory and practice of key-management and privacy in Bitcoin. Technological components form only a part of a custody system. In addition, there are people and processes. We consider the comprehensive view which integrates technology, people and processes in section \ref{sec:custody-systems-view}. Finally, in section \ref{sec:custody-modelling-risk}, we consider how to model risk, and review examples in the field. This chapter addresses thesis objective \textbf{O1} (see \ref{sec:thesis-objectives}).

\section{Introduction to Bitcoin Custody}
\label{sec:custody-intro}

Bitcoin custody, put simply, is the act of holding and controlling bitcoin. A custody system is comprised of technologies, processes, and people. The ideal system would be safe (preventing failure) and secure (preventing attack). Of course, an ideal system is not realistic. In designing Bitcoin custody systems we can strive for safety and security by building tolerance to failure and attack and by designing methods to recover from partial failure and compromise. 

Custody systems are the foundation of all applications and uses of Bitcoin. Custody systems can be very simple or complex. Let us elaborate on types, design principles, phases, and functions to build an intuition about custody systems. 

\subsection{Types}

A custody system can be controlled by one or more people. The custody system may also be operated by some people on behalf of others who are the legal owners of the bitcoin contained in the system. We say that a custody system is \textit{autonomous} if it is controlled and legally owned by the same entities. An autonomous custody system is self-governing. Autonomous is often referred to as \textit{self-custody} or \textit{sovereign} in the literature and industry. The ability for anyone to use autonomous custody systems in Bitcoin is a critical difference from state-backed fiat monetary systems, which are typically controlled by privileged entities (banks and payment network operators). Autonomous custody systems cannot be censored from receiving and sending bitcoin. The catch is that autonomous custody systems come with responsibility to actively maintain safety and security. We will call a custody system that is not autonomous a \textit{heteronomous} custody system. An example is a trust or an exchange. The same processes and technologies can be used for both types of system. The existence of autonomous custody systems for bitcoin are crucial to maintain the property of censorship-resistance. Heteronomous custody systems inherently give another entity power to prevent the user from making transactions, i.e. to censor the user. 

Another distinction to make is about \textit{single-party} and \textit{multi-party} custody systems. A single-party custody system is completely controlled by one person. That one person has full access to the system except in situations of failure or attack. In multi-party custody systems, multiple people participate in controlling how bitcoin is spent. Access policies can enforce dual control (e.g. person A \textit{and} person B), joint control (e.g. person A \textit{or} person B), or can enforce arbitrary control hierarchies with separation of duties. An example is a large company where the finance team and compliance team both need to authorise expenses requested from different product teams. 

Custody systems can have components or functions which are outsourced to a \textit{third-party}. Examples include a fiduciary co-signer for payments or a fiduciary recovery assistant in situations of failure or compromise. A third-party could assist with data-availability for a custody system, or with reliable access to the Bitcoin network, or could provide price information about bitcoin. 

Finally, let us describe the often-used jargon \textit{cold} versus \textit{hot} storage. Cold storage refers to keeping private keys on offline devices or media. Cold storage can include a purpose-built hardware device (e.g. a hardware security module), paper, engraved metal, or even human memory. This minimises network-based attacks at the expense of making the keys less available, requiring manual action from the user to access the bitcoin. Hot storage refers to keeping private keys on networked devices. While being online significantly amplifies the potential attack surface, it simultaneously enhances the availability of the bitcoin, thus facilitating easier access. It is particularly useful for automated processes.  

\subsection{Design Principles}

Using design principles is important to guide the development of custody systems. It helps in setting objectives and defining system constraints. Moreover, it helps by communicating to users the intent of system architects, so that users can evaluate if the system is suitable for them. 

An example set of design principles are given by a custody services company called Casa in their `wealth security protocol' \cite{Casa-WSP}. Let us paraphrase them here. 
\begin{quote}
    \textit{Minimal knowledge.} Users should disclose minimal information to others, including trusted counter-parties and other third-parties. \\
    \textit{High security.} There should be no single point of compromise. Use defense-in-depth to increase the difficulty for attackers to compromise the custody system. \\
    \textit{Usability is security.} Poor usability can create vulnerabilities. Poor usability will result in nobody using the custody system. \\
    \textit{Expert support.} Users are generally not security experts. A custody system can make use of expert support to serve a wider range of users, to integrate feedback and develop better systems. \\
    \textit{Redundancy.} There should be no single point of failure. A custody system can be resilient despite failure of some of its components. \\
    \textit{Sovereignty.} Bitcoin was designed to give users autonomy. Heteronomous custody systems require trusting others, which creates vulnerabilities. \\
    \textit{Incentive alignment.} Creating an alignment of commercial incentives with the success of the custody system will result in better development of the custody system. \\
    \textit{Bitcoin First.} Bitcoin is the most secure and valuable digital currency. The same custody systems that are used for Bitcoin could eventually be used for other digital assets.
\end{quote}

Blockchain Commons, a not-for-profit benefit corporation working with developer communities in the area of digital assets and digital identity \cite{BlockchainCommons} proposed similar design principles. Their stated goal is to ``put the user first and enable responsible key-management''. They have designed and built a custody system called Gordian. Below, the Gordian principles are quoted;
\begin{quote}
``\textit{Independence.} Gordian improves user freedom from involuntary oversight or external control. \\
\textit{Privacy.} Gordian protects against coercion with non-correlation, privacy, and pseudonymity. \\
\textit{Resilience.} Gordian protects users to decrease the likelihood of them losing their funds via any means. \\
\textit{Openness.} Gordian supports open infrastructure that allows developers to create their own applications."
\end{quote}

Keep these principles in mind to understand why, sometimes, Bitcoin custody systems are designed with an abundance of cryptographic, hardware, and physical controls. They help to explain why, often, Bitcoin custody systems are inconvenient for users in comparison to the simple mobile applications used in retail banking. 

\subsection{Phases}

Let us define three phases for a Bitcoin custody system. In the \textit{setup} phase the components of the custody system are initialised and the physical environments that store components are prepared. If there are third-party service providers, some data exchange may be required to initialise their system too. The setup phase will consist of generating keys that will control bitcoin, preparing key-storage devices, preparing device storage locations, preparing other computation devices, and testing the entire system. Once the setup phase is complete, the \textit{operation} phase can begin. This includes receiving and sending bitcoin payments, and checking the balance and health of the system. If the system experiences partial failure or suffers an attack, the \textit{recovery} phase is designed to replace components and consolidate the system. Recovery will often involve key backups and a transition of control to newly generated keys controlled by remaining and new components. 

\subsection{Functions}

Custody systems are comprised of components that perform one or more of several common functions. Let us summarise these now. 

Bitcoin private keys are 256-bit random numbers and are a fundamental data type for custody systems. True randomness is required to generate private keys securely. Many key-storage devices come with a built-in `true random number generator'. Many devices also allow the user to enter random data, or `entropy', manually. The user can roll dice to generate true random numbers. The common function is \textbf{private key generation}. 

Private keys are often stored on digital hardware; either general purpose (e.g. a mobile or laptop), or purpose-built (e.g. a hardware security module). To improve the resilience of a custody system against loss of private keys, they are often backed up on analogue media. Examples include engraved metal and archival paper. Lopp has a detailed study of different types of analogue backup media \cite{lopp-seed,lopp-metal}. When using analogue media, the private key will often be encoded as a human-friendly mnemonic seed, usually 24 words, as specified in BIP-39 \cite{BIP39}. The common function is \textbf{private key storage}.

To operate the custody system, private keys must be accessed and used to sign transactions. This function requires computation, and many purpose-built signing devices exist \cite{commercial-HWs}. General purpose devices can function as a signing device too. The common function is \textbf{signing}.

Bitcoin payments are encoded as transactions. Transactions define how control of bitcoin transfers from one entity to another. Transaction messages are authorised with signatures generated using private keys. A \textbf{transaction coordinator} is a device (often general purpose) that can create transactions and share them with signing devices. 

The counterpart to signature generation is \textbf{signature verification}. Typically, signing devices and transaction coordinators will be equipped with this functionality.

Another common function is a \textbf{Bitcoin network server}. This is a server running Bitcoin software to be a peer in the peer-to-peer network. This function consists of downloading new blocks of transactions, verifying blocks and transactions, and coordinating the broadcast of transactions. 

As previously noted, the current state of bitcoin is encapsulated by the set of unspent transaction outputs, also referred to as \textit{coins}. Similarly, the state of a custody system is the set of coins which it controls. The total balance is the sum of available coins. Knowing which coins exist requires a Bitcoin network server. Knowing which coins are in control of the system requires knowledge of private key storage (though not necessarily direct knowledge of the private key). The common function is \textbf{state tracking} (the custody system state).

The term \textbf{wallet} is often used to describe software that can perform one or several of these common functions. It is an ambiguous term. Sometimes it is used to refer to a complete custody system. Sometimes it is used to refer to a set of keys controlled by the user. Sometimes it is used to describe a key-storage and signing device. We will avoid any confusion by not using this term except with an explicit description of the functions supported or in reference to other work that uses it. The term is particularly unhelpful for modular custody systems which partition functionality across separate components.

\section{Technology Primer}
\label{sec:custody-tech-primer}

There is a broad array of technologies that can be used in custody systems to achieve security, safety, usability, and privacy. Herein we will present the concepts, discuss their utility, and refer to noteworthy research. 

Technology can be used to create custody systems based on a user-friendly yet resilient \textit{key-management strategy} coupled with a \textit{privacy framework}. The former relates to control of bitcoin, while the latter relates to what information is revealed as assets and technologies are used. Moreover, the two are inter-related; maintaining privacy helps to reduce the risk associated with compromised keys, and conversely, key-management techniques may reveal operational information to external observers. Financial privacy may be paramount for success in business, and is critical for the safety of users with significant holdings.

\subsection{Key-management in Theory}

\subsubsection{Security Objectives}

Each key has a life-cycle. First, a key is generated. Then, it has a period wherein it is used to perform cryptography tasks. Then it is decommissioned (no longer used). Finally, it may be deleted. The user must control the key for its entire life-cycle. 

Key-management is a long-standing issue for the usability of cryptography schemes. For an autonomous custody system, users are in control of their own bitcoin and have no possibility for recourse in the event of loss or theft, therefore, methods for dealing with various failure modes in key-management are necessary. Au \textit{et al.} used actuarial techniques to estimate that, by June 2020, \$7.3 billion worth of bitcoin had been lost since the system launched in 2009 \cite{CCC}. 

In Bitcoin, coins are specified as unspent transaction outputs (UTXOs) and are publicly linked to addresses. In the simplest case, an address is generated by taking the hash of the relevant public key\footnote{Addresses may also represent more complex associations between UTXOs and, for example, multiple public keys.}. In order to move coins, a transaction message (which meets the validation criteria of consensus rules) must be signed with the private key that corresponds to the address (public key) linked to those coins and be broadcast to the network. 

Thus, \textit{confidentiality} of the private key is a critical security objective. If a person or machine successfully learns the private key, they will be able to use it to sign transactions to move any bitcoin from the address associated with that private key to any other address.

\textit{Confidentiality} is also important for public keys, in order to maintain privacy. Achieving this while interacting with a public ledger of records can be challenging.  Re-using public keys can result in several transactions being identified as from the same user. Observers can simply match addresses across different transactions. An address (derived from one or more public keys) is a pseudonym for a user. The use of pseudonyms in Bitcoin enables optional audit-ability. In some cases, it may be desirable to hold and transact bitcoin in a publicly verifiable way. In other cases, it may be desirable to remain anonymous. 

A custody system ought to be designed to support the user's expected usage behaviour. The security objective here is to enable sufficient \textit{availability} for private and public keys to enable their use. Some users may need low frequency and irregular access to their private keys, while others may require continual access. With the public key, the user can check their balance and create an address to receive payments. We may expect higher availability requirements for public keys. 

\textit{Integrity} of one's private and public keys is an important security objective. A user may demonstrate the integrity of their private key by using it to sign a message or a test transaction. The user must take care to ensure the integrity of their public key when sharing an address (derived from the public key) with their counter-party in order to receive a payment. Together, a valid signature on a message and a public key can be used to demonstrate to a third-party that one controls bitcoin; this is referred to as proof-of-reserves \cite{BIP127}. 

\subsubsection{Key-trees}

Bitcoin keys are typically generated as a \textit{deterministic wallet} (in this case, `wallet' is referring to a collection of keys). A deterministic wallet starts with a mnemonic seed phrase, as standardised with BIP-39 \cite{BIP39}. From this, a key derivation function is used to produce a master key-pair and subsequently to produce a chain of descendant key-pairs. Most wallets adhere to the standard introduced in 2012, BIP-32 \cite{BIP32}, referred to as \textit{hierarchical deterministic} (HD) wallets in which an initial mnemonic seed phrase is used to derive a \textit{tree} of (public key, private key) pairs. 

We will refer to HD wallets as \textit{key-trees}. A key-tree simplifies the management of using many keys. If designed well, a key-tree can enhance privacy, improve coordination among devices and regular counter-parties, optimise storage of large key-sets, and simplify the backup of keys. Now we consider how a key-tree is constructed. 

Recall that one of the inputs to a key derivation function is a random number. In BIP-32 both public and private keys, $(k,K)$, are appended with a random number $c$. This is an extra 256-bit string referred to as the \textit{chaincode} which is an argument in the key derivation function. Together, the key and chaincode are called an \textit{extended key}. A private and public key-pair share the same chaincode, such that the extended private key is $(k, c)$ and the extended public key is $(K,c)$. 

There is a key derivation function to go from;
\begin{quote}
    parent private key $\rightarrow$ child private key, \\
    parent private key $\rightarrow$ child public key, and\\
    parent public key $\rightarrow$ child public key.
\end{quote}
Deriving a private key from a public key is not possible because of the ECDLP. Deriving a parent key from a child key is not possible because the key derivation function is one-way. 

A \textit{derivation index} is included as an argument in the key derivation function and is used to separate key-chains as branches in the key-tree. Each extended key-pair can be the root of a sub-tree with its own derived branches. 

There are two types of derivation; \textit{hardened} and \textit{non-hardened}. Hardened derivation includes the private key as an argument in the key derivation function (whether the parent key is private or public). One can't compute a hardened child key without knowledge of the parent private key. Conversely, non-hardened derivation includes the public key as an argument in the key derivation function. Non-hardened derivation doesn't require knowledge of the private key.

As a consequence, knowing an extended private key is sufficient to compute a complete chain of descendant private and public keys. Whereas knowing an extended public key is sufficient to compute all descendant non-hardened public keys. 

It is useful to share an extended public key with a regular counter-party. This way, they may derive a fresh public key each time they make a payment to you without any interaction. Note, however, that given a hardened extended public key, $(K_1,c_1)$, and \textit{any} extended private key in the descendent chain, $(k_i,c_i)$, it's possible to compute the hardened extended private key $(k_1,c_1)$ \cite{Buterin-Deterministic-wallets}. Generally, one must assume that a compromised hardened extended private key results in a compromised sub-tree. 

Observe that the user can backup their entire key-tree with a single mnemonic phrase. Similarly, a signing device may optimise storage by only storing the master extended key-pair, the derivation index, and the current chain depth instead of the entire key-tree. 

Das \textit{et al.} presented a formal model for BIP-32 HD wallet systems and demonstrated precisely the security properties it achieves \cite{Exact-bip32}.  Several other works have proposed variations of the standard with slight improvements or additional functionality \cite{gutoski2015hierarchical,arcula,hd-privilege,Groth2022}. 

\subsubsection{Pseudonymity}

The motivation for using fresh public keys with each payment is to maintain privacy. Recall that a public key is a pseudonym for the user. Once a pseudonym is linked to a real identity, we consider the privacy to be broken. 

\begin{quote}
    \textbf{Privacy objective 1.} Avoid creating links between real identities and pseudonyms. 
\end{quote}

This could mean avoiding `Know Your Customer' regulated transactions, protecting the IP when broadcasting transactions to the network, or not giving away geo-location data when interacting with Bitcoin applications and services. It should be obvious too that sharing information about public keys and addresses with people (family, friends or colleagues) or online sites (forums or exchanges) compromises privacy.

Unfortunately, transacting in bitcoin for goods or services creates a scenario where the seller and buyer must reveal a pseudonym to each other. Revelation of pseudonyms will always be a risk. 

\begin{quote}
    \textbf{Privacy objective 2.} Manage a diverse set of un-linked pseudonyms. 
\end{quote}

Compartmentalising bitcoin across multiple pseudonyms (as UTXOs with different addresses) is a risk-management strategy that mitigates the impact of accidental or forced identity leaks through de-anonymization methods.

Key-trees enable users to avoid re-using public keys and enhance privacy. They help users maintain maximal un-linkability of their pseudonyms. If privacy is paramount, then hardened key derivation can be used. A hardened descendant key-pair can only be computed with access to the parent private key. Provided that the private key is not compromised, it is not feasible for an attacker to derive the chain of public keys even if any individual public keys are revealed.

\subsubsection{Transaction graph analysis}

Methods to cluster ownership of UTXOs are not only based on the public keys used to derive addresses, but also on the structure of connected transactions. 

An UTXO created with a prior transaction becomes a spent input for a new transaction. A single transaction may contain several inputs and outputs. It is even possible to aggregate inputs from various signatories in the same transaction and to distribute new outputs among addresses controlled by different recipients. 

This way, transactions are linked. We can represent this as a \textit{transaction graph}. This graph can be derived from the public blockchain data structure to analyse which addresses are likely associated with the same custody system, service, or person. When a UTXO is spent it is totally consumed by the transaction; any excess value that is not intended for the recipient must be used to create a \textit{change output} with an address controlled by the sender. It is advisable to generate fresh change addresses with each transaction, to adhere to the objective of maintaining a diverse set of un-linked pseudonyms. 

Following this method eventually leads to a situation where the total value of funds is thinly distributed across many addresses in small amounts. If there is not enough value in any individual unspent output for some intended transaction, then these change outputs must be aggregated to form a new UTXO with sufficient value. Doing this creates a public correlation between the change outputs and the transactions that created them. 
\begin{quote}
\textbf{Privacy objective 3. } Minimise cross-transaction correlation.     
\end{quote}
We will consider privacy services that can obfuscate transaction graph analyses in section \ref{subsec:privacy-tools}.

\subsubsection{Access control}
\label{subsec:custody-access-control}

A seed (or the master private key) is a single point of compromise for a custody system using a single key-tree. For single-party and multi-party custody systems, researchers and practitioners have instead turned to using systems with keys (or key-shares) distributed across hardware devices, locations, software and people. This is how to build defence-in-depth and how to avoid weak links at the key-management layer. 

With multiple keys and key usage controls (cryptography and physical), arbitrary access control structures can be defined which determine who or what devices can authorise sending and receiving payments. 

When keys are contained to distinct devices or media, they can be stored in distinct physical locations. Each location will have its own physical controls that define the \textit{physical security environment}; safe, combination lock, alarm, cameras, security guards, \textit{etcetera}. The design of physical security environments is beyond the scope of this thesis. It is common for home owners and organisations to implement physical controls to safeguard their assets. Each control forms part of the access control structure.

The access control structure is also determined through cryptography and Bitcoin engineering. The general approach is to use a \textit{multi-signature scheme}. An example access control structure is quorum authorisation; a threshold $m$ of signatures are required from a set of $n$ private keys where $1 \leq m \leq n$. Another example is a hierarchy; a signature is required from private key $k_A$, or two signatures are required from private keys $k_B$ and $k_C$. 

Access structures, both hierarchical and quorum-based, can be implemented via \textit{on-chain} or \textit{off-chain} mechanisms. On-chain controls, which embed data directly into transactions, inherently leave a trace on the public ledger. While this provides an inherent audit-ability, it comes at the cost of privacy. Conversely, off-chain controls do not necessarily leave a trace on the public ledger, thereby offering a higher degree of privacy. However, it's worth noting that audit-ability mechanisms can still be incorporated into off-chain controls if necessary.

Bitcoin has a built-in programming language called Script that can be used to construct on-chain controls and enforce arbitrary multi-signature access structures. We will simply refer to these as \textit{multi-signature} access structures. This will be explored in more detail in the following chapter, section \ref{sec:Preliminaries}. With this approach, the minimal information that will be revealed on-chain is a sufficient set of signatures and their associated public keys. For example, with a $3-$of$-5$ quorum structure, $3$ signatures and their public keys must be revealed to access the bitcoin\footnote{We are assuming the use of a privacy-optimised Taproot Tree, a concept we shall expand upon further in section \ref{sec:ax-taptrees}.}. 

\textit{Threshold-signature schemes} have been studied and implemented for both ECDSA \cite{GGN16,DS16,GG18,LN18,DKLS18,DKLS19,BGG19,CCLST19,DOKSS20,GJDM20,BLMS20,CGGMP21,YCX21,P21,WMDZW21,DJNPO20,CCLST21,ANOSS21,GS22,PCCY22,L21} and Schnorr signatures \cite{SS01,GJKR05,KG21,L22,GKMN21,MuSig2,JPS22,CKM21,RRJSS22}. In this case, cryptography protocols among multiple participants are required to generate a distributed set of keys, to generate signatures, and to verify signatures. The common pattern is that several public keys can be aggregated into a single public key, and signatures for each public key can be aggregated into a single signature. The on-chain component is the aggregate public key and aggregate signature. In principle, arbitrary access control structures could be enforced off-chain. In practice, these schemes are still being heavily researched and quorum access structures have primarily been focused on in this literature ($m-$of$-n$).

Multi-signature structures and threshold-signature schemes have different security assumptions. The latter are nascent and although significant progress has been made recently, using a nascent cryptography protocol in a custody system adds risk. Both theoretical and implementation vulnerabilities can emerge. Adding dependence on new and complex code repositories (in addition to Bitcoin core itself) increases the attack surface for the foundational key-management layer \cite{attacking-thresh-wal}. Several formal security models have been proposed for threshold signature schemes and in time their guarantees will become more reliable, with more realistic assumptions (such as the trend towards modeling communication over an asynchronous network \cite{RRJSS22,JPS22}). Multi-signature structures are a composition of; the security of the signature scheme (ECDSA or Schnorr), the enforcement of consensus rules for Script validation, and the properties of the Script itself (such as, will it terminate, are all the expected execution paths reachable, \textit{etc.}).

ECDSA threshold-signature schemes are generally considered to be less efficient, and harder to scale than Schnorr threshold-signature schemes \cite{RRJSS22}. This is because of non-linearity present in the algebraic structure of an ECDSA signature, where a term requires a field inversion.  In contrast, Schnorr signatures have a linear algebraic structure and will likely be amenable to more efficient and practical threshold-signature schemes. 

Rather than dive into the details of each proposal, since the designs are developing quickly, let us consider the features that are most relevant to the design of custody systems. 

Ideally, the key-generation procedure, signing procedure, and verification procedure will each be \textit{non-interactive}; each participant can enact the procedure independently with asynchronous message passing, requiring receiving only one message and sending only one message. After key generation, it will be the case that each participant must receive public keys and share their public key with others. For signing, the transaction message must be communicated to a participant first, and subsequently the signatures must be sent to be combined to form a valid witness, but additional communication rounds should not be necessary. For verification, the verifier needs the combined signature(s) and the transaction message, but no additional communication. This non-interactivity is especially important to simplify usage with offline signers. Script-based multi-signature is non-interactive in this way, while threshold signature schemes are progressing towards being non-interactive. 

The signing session should be \textit{robust}, as defined by Ruffing \textit{et al.} \cite{RRJSS22}. This means, for an $m-$of$-n$ scheme, the signing session succeeds (outputs a valid signature) if $m$ honest signers participate. Other signers may be malicious and attempt to disrupt the session, but will fail to deny this service to the honest signers. 

The public nature of the ledger presents an opportunity for fiduciary transparency. The custody system would benefit from  \textit{audit-ability}. Whether the custodian is demonstrating solvency with proof-of-reserves, or an individual is seeking insurance for their holdings, the ability support auditing the access control structure is beneficial.

In a similar vein, integrating accountability into the custody system can provide an additional layer of security. Accountability, in this context, pertains to controls ensuring that activities on the network and within the system can be attributed to a distinct individual \cite{rhodes2013information}. The distinguishability of actions of key-holders can deter malicious behaviour, thereby fortifying the reliability and integrity of the custody system.

Bitcoin transaction size grows linearly with the number of required signatures when using multi-signatures. This is a problem for the user because larger transactions cost more in fees. However, threshold-signature schemes always result in transactions of the same size because of public key aggregation and signature aggregation. They will be cheaper for the user. For the same reasons, multi-signature structures suffer worse privacy than threshold-signature structures, with (at least a part of) the access structure being revealed on-chain. As we shall see in section \ref{subsec:privacy-tools}, multi-signature structures  also hinder the use of privacy tools by making them impractical.
 
Finally, a feature that is only possible with threshold-signatures is to change or update the access structure without a transaction. New keys (or key-shares) can be authorised by the signers with an updated access control structure. This is achieved with a \textit{key refreshment protocol}. Achieving this with multi-signature requires a transaction, and for the same reasons as above, this leaks information and costs the user in transaction fees. 

It is also worth noting that a combination of on-chain and off-chain controls can be used. Together, new access control structures with different privacy and efficiency properties can be achieved. 

\subsection{Transaction Coordination}

A transaction coordinator will handle transaction construction and coordination of communication with signers and the Bitcoin network. Transaction construction requires; knowledge of the UTXO set, an ability to determine which UTXOs are owned by the user, transaction template information, selection of inputs, details of outputs and a fee. Once constructed, the coordinator will send the transaction to a sufficient number of signers, and receive signatures. Typically, the coordinator will verify these signatures. The coordinator must support the communication protocols of each of the signing device types.

The Partially Signed Bitcoin Transaction (or PSBT) format \cite{BIP174} was created to simplify transaction coordination. It is a binary transaction format that contains all the necessary information for the signer to produce a signature. Typically, the coordinator will send the PSBT to each signer who will respond with signatures. The coordinator combines all the signatures and finalises the signed transaction before broadcasting it to the Bitcoin network. A second version, PSBT 2, has also been proposed to simplify the process of adding new inputs and outputs to the transaction \cite{BIP370}.

The transaction coordinator may also communicate with signers to make other requests, such as for public keys, output information, or otherwise. Blockchain Commons have created a specification for \textit{Uniform Resources} \cite{UR-crypto-commons} in an attempt to standardise a self-identifying encoding of cryptographic data to improve interoperability between custody system components being developed across the industry.  

Constructing transactions requires user input. User-friendly abstractions and interfaces are essential for the transaction coordinator to mitigate the risk of attacks based on deceiving users. Parts of transaction construction which can be automated should be automated. The user, however, must authorise critical fields directly on the signing device. For example, by manually checking the transaction fee and the destination addresses and amounts of all outputs, the user restricts a compromised transaction coordinator from creating arbitrary theft transactions. A labelled address book can be very useful for the user to avoid handling long and seemingly random alphanumeric strings. Output addresses do not have meaningful representations for users. One way to improve read-ability is to construct higher-level representations of the access control structure. Miniscript policies are the leading example \cite{miniscript}. However, while Miniscript is legible to an engineer, it may not be for the general user. 

Constructing transactions requires selecting UTXOs to fulfil the payment amount that has been input by the user. Typically, this is achieved automatically. The algorithm for selecting coins has implications for the cost to the user, and the user's privacy (recall the discussion of transaction graph analysis). Some studies have analysed how to optimise coin-selection and note the trade-off between minimising transaction fees and maximising privacy by avoiding correlating transactions \cite{FormalWallet,CoinSelection,LoppCoinSelection,bip126,stonewall}.

\subsection{Bitcoin Network}

Bitcoin nodes participate in a peer-to-peer network, maintaining a set of connections to other nodes in order to push and retrieve information to construct and validate their local view of the blockchain and to broadcast transactions. Unfortunately the peer-to-peer network in bitcoin is not designed for anonymity. It is a diffusion broadcast network, designed for low latency and for consistency (every node should receive each broadcast transaction). As a communication protocol, the network stack does not encode consensus rules and is amenable to new developments. Prior to 2015, the peer-to-peer network relied on a trickle mechanism for disseminating transactions to peers. Some privacy breaking methods were presented \cite{DeanonymisationP2P, AnonymitybitcoinP2P} and subsequently the networking stack upgraded to the current diffusion broadcast system.

Following the taxonomy of methods to break anonymity and privacy through the network stack of bitcoin, as presented by \cite{PrivacySurvey2018}, there are four methods known in the literature; utilizing anomalously relayed transactions, utilizing first relayer information, utilizing an underlying network graph, and setting an IP address cookie. The outcome of these methods is to link bitcoin addresses with IP addresses. Setting an address cookie (by providing a fake `fingerprint' IP address to a peer) can also result in linking bitcoin addresses with each other. In practice these methods aim for large-scale deanonymization by, for example, creating a super-node which makes in-coming connections with all peers in the network, or by creating a botnet of many nodes who in aggregate connect to all peers in the network, and performing traffic analyses to classify peers through different heuristics. More targeted deanonymization approaches are also possible. 

An interesting development in this domain is Dandelion \cite{Dandelion} and its newer iteration Dandelion++ \cite{Dandelion++} which are attempts to redesign the network stack to improve privacy properties for users. The central idea here is to pre-fix broadcasted messages with an anonymity phase based on randomized routing. In addition to low latency and consistency, it is claimed that this would offer better anonymity by default, mitigating some of the above methods for breaking anonymity, without significantly increasing the latency in network communications. 

Until a carefully audited implementation of a network stack like Dandelion++ is merged into bitcoin-core, users must use methods that aren't specific to Bitcoin to maintain network privacy. The most popular anonymization network is The Onion Router (TOR) \cite{TOR}. Bitcoin can be configured to communicate through TOR by default, but it has been demonstrated that other vulnerabilities emerge when users rely on TOR \cite{BitcoinTor}. Other anonymization networks such as the Invisible Internet Project (I2P) \cite{I2P} may be useful too, but given that the design goals of these anonymization networks don't align completely with those of Bitcoin's peer-to-peer network (they are not generally meant for broadcasting messages to all peers), they will likely not be the most scalable nor secure solutions for bitcoin nodes. Users are advised to consider employing a Virtual Private Network (VPN), which can provide an additional layer of protection. Furthermore, it is prudent to restrict the exposure of Bitcoin-related information only to sites that establish a secure HTTPS connection, as a measure to counter potential network adversaries.

The complexity of maintaining privacy increases when bitcoin users interact with other networks and systems, a fact that was demonstrated by Goldfeder \textit{et al.} in \cite{CookieBlockchain} where it was shown how third-party web trackers enable privacy breaking threat vectors for cryptocurrency users. Goldfeder \textit{et al.} recommend common web-browser extensions such as uBlock  Origin,  Adblock  Plus,  or  Ghostery  to  block  trackers but warn that these are not perfect solutions \cite{TrackerBlockerSurvey}. 

One final point to bear in mind while participating in the peer-to-peer network is that bitcoin nodes which verify the entire blockchain state locally, rather than outsourcing this verification, are able to participate without leaking information on the UTXOs managed in their local custody system. Custody system software which doesn't maintain its own view of the blockchain state must query other nodes or services to retrieve the information relevant to their managed UTXOs. In querying another machine for a subset of transactions from the blockchain, information is revealed which can be used to correlate users with UTXOs. Different proposals suggest the use of privacy-preserving filters to obfuscate the information being revealed when sending queries. However, the Bloom filter technique \cite{BIP37} introduced in 2012 was soon shown to offer little privacy benefits \cite{Gervais:2014:PPB:2664243.2664267}. More recent proposals for privacy-preserving filters include a draft BIP \cite{CompactFilter} which relies on a different filter construction called a compact filter, and a construction which relies on trusted execution environments \cite{Matetic2018BITEBL}. These may be relevant in the context of some custody protocols which rely on resource constrained devices which cannot independently verify the blockchain state but should be avoided where possible if privacy is a major concern. 

\subsection{Key-management in Practice}

\subsubsection{Key generation}

Key generation requires randomness. The best sources of randomness involve high entropy physical processes such as the roll of a die or the flip of a coin. A 256-bit private key can be generated randomly by flipping a balanced coin 256 times and noting a $0$ for heads or a $1$ for tails. Completely analogue methods are possible, but usually a small program will be used for computation in key derivation. Typically, to improve usability, the random bits are encoded as mnemonic seed words such as those standardised in BIP-39 \cite{BIP39}. There are several examples of guides to generate bitcoin seed phrases with dice documented online \cite{BIP39-diceware,glacier-dice,swanson-dice,offline-bip39-dice,vault12-dice-seed,coinmonks-dice-seed}.

Key-management hardware devices include a `true random number generator' (TRNG). This is built with an integrated circuit that samples data from a physical process (usually an electric sine wave signal generated with a free oscillator) \cite{Ledger-threat-model}. However, it is difficult to distinguish the output of a pseudo-random number generator (PRNG) and a TRNG, especially from a single data point. A PRNG uses a deterministic function to generate numbers that appear random, but the same input will produce the same output. By forcing the use of a PRNG, an attacker could compromise key generation and deceive a user into using a key which was chosen by the attacker. There will always be a risk that a device manufacturer is corrupt or compromised and subverts the key-generation process by using a PRNG. Laboratory tests can reveal weaknesses in a device's random number generator through statistical analysis. Device manufacturers will seek certification from audit laboriatories to enhance their claim that their devices have TRNGs.

In a multi-key setting, devices or users must generate their key-pair and share public keys among each other. This process happens during the setup phase. This critical phase of a custody system has been under-studied to-date. 

Bitcoin Secure Multisig Setup (BSMS) has been proposed as a standard and adopted by some of the industry's software and hardware providers \cite{BIP129}. It is an important first step towards specifying and analysing the setup process for a multi-signature custody system. It attempts to enable participating signers to verify the configuration (the set of shared extended public keys and the access structure), and prevent leakage of public keys during the setup process. It uses a combination of message authentication and encryption, based on symmetric key cryptography. The process depends on a trusted coordinator, and mitigates passive attackers (who may eaves drop on communication). However, it is vulnerable to active attackers who may engage in man-in-the-middle attacks, or compromise the coordinator itself. 

\subsubsection{Key storage and signing}

Let us classify the types of key-storage: keys in local storage of a networked device, encrypted keys in local storage of a networked device, offline storage of keys (both digital and analogue), and third-party key-storage (such as professional custodians or digital asset exchanges).  

Commercial products for hardware devices (supporting; key-generation, key-storage, signing, and verification) have emerged as the preferred key-management option for the average user with a non-trivial amount of bitcoin who desire autonomy \cite{commercial-HWs}. Devices support one of a variety of communication channels; wired, usb, blue-tooth, camera and QR codes, near-field communication (NFC), and SD cards. Devices will typically have a screen to communicate with the user, and buttons to allow the user to input data. Devices are designed with several countermeasures to physical and remote attacks. Arapinis \textit{et al.} presented a formal specification of a hardware wallet  and demonstrated how a theoretical and idealised protocol can be secure \cite{FormalHardware}. They go on to state realistic attacks that compromise the user or the client device interacting with the hardware. Attacks are often discovered for hardware devices, including fault injection and side-channel attacks \cite{Volokitin-HW-attacks,MOBT,BeatCoin}. In general, it is safer to assume that if a motivated attacker gets physical access to a hardware device, they can extract all sensitive material contained on it, or inject malware \cite{Donjon-trezor-attack,Donjon-ellipal-attack,Donjon-extracting-seeds,Donjon-trezor-sc-attack,Donjon-laser,Trezor-hack,Trezor-glitch-hack,Volokitin}. Remote attacks on hardware devices are more difficult, but not unheard of \cite{Donjon-ellipal-attack,low-level-hw-attacks}. It would also require compromising the client device that communicates with the hardware device. Finally, supply-chain security is a growing concern, with counterfeit hardware devices being inserted into the supply-chain running malware \cite{HW-sc-security}. 

Software applications with encrypted keys in local storage tend to be used only for small amounts. User passwords and passphrases are relatively weak countermeasures to prevent unauthorised access. They are susceptible to human error (choosing weak passwords) and to remote attacks by malware injection and key-loggers \cite{SecurityRefArchitecture,Android-wallet-attacks,SmartphoneWalletThreats}. Physical attacks on software applications that store keys locally are also practical \cite{Forensic-Electrum-Core,Donjon-software}. More often, software applications are used as a transaction coordinator and network server and are used in combination with a hardware device. 

Third-party providers for key-management will often use more sophisticated custody systems than users. The problem is that these providers are large and public targets for attackers. The potential payoff results in highly sophisticated attack sequences involving social engineering, bribery, coercion, weapons and more. Malicious insiders are also a critical threat for these providers to contend with. There is a long history of exchanges and custodians being hacked or defrauding customers \cite{Exchange-risk-empirical,ExchangeClosure,PreventingExchangeHeist,mtGox}.

A key-storage device or medium can be used actively or as a backup. A backup is a way to build resilience for the custody system. If the active key-management device is faulty or lost, the user can recover their bitcoin from the backup.  Often, hardware devices will come with a paper document which can be filled in with the user's seed phrase and used as a backup of keys. Software `wallets' encourage the user to write down their seed phrase during the on-boarding process. As mentioned previously, engraved metal and archival paper can offer better resilience. Offline backups have the benefit of not being susceptible to remote attacks. However, they are subject to trivial physical inspection. The risk of an `evil maid' attack is ever-present. A second hardware device can be used as a backup by importing the seed phrase into it. That way, the backup is no longer trivially read by an attacker because the device requires a PIN to unlock. Also, the backup is more accessible for the user in order sign transactions if their active device was compromised. 

In other work, types of `wallet' systems are broadly classified \cite{eskandari2015usability,Crypto-vs-Fiat-Wallet,ResilientCustody,gazibtd513088}; mobile wallet, desktop wallet, third-party web wallet, paper wallet, and hardware wallet. These solutions are a useful classification of single, whole-key set-ups. However, each of them in isolation must deal with an apparent security-usability trade-off. There is a tension between maintaining \textit{confidentiality} of the key while also ensuring it is sufficiently \textit{available} for use in signing transactions. For example, a dedicated hardware device may be very effective at protecting keys from extraction since it significantly reduces network-based attacks, but it requires the user to manually connect the device and authenticate before being able to access the keys. As another example, a mobile-based software wallet would be highly accessible to the user, but would be susceptible to a vast attack surface given all of the interfaces through which it can be accessed, resulting in stolen keys.  

\subsubsection{Usability}

Usability is a critical property for key-management. If key-management is too difficult and error-prone, and loss of funds is too common, users will tend to either rely on heteronomous custody systems or will abandon the use of Bitcoin altogether. The current state of user experience in Bitcoin is characterised by immature `wallet' applications with a tedious on-boarding process (generating and backing up the seed), un-intuitive mental models (e.g. addresses and UTXOs) and exposed details of how the blockchain functions (waiting for transactions to be included in the blockchain before payments are processed).   

Fröhlich \textit{et al.} performed a systematic literature review of human-computer interaction in blockchain and cryptocurrency \cite{SoK-HCI-crypto}. Since the inception of Bitcoin, researchers have pursued various methods; quantitative data analysis, interviews, questionnaires, lab studies, field studies, and workshops. One of the major themes they identified was the usability of cryptocurrency wallets, wherein 16 papers were identified which highlight challenges and issues in that domain, and recommend potential solutions. Little work, they note, actually implements and evaluates proposals to improve the usability of wallets. They show that we can consider usability from the perspective of the tasks that the user has to perform; creating a new account (e.g. setting up user authentication with the application), creating a new a collection of keys, buying cryptocurrency, receiving and sending cryptocurrency, purchasing goods, reviewing the value of cryptocurrency, and backing up or restoring the keys. These demonstrate a more broad context than what we have considered as part of a custody system. For example, how will the user acquire bitcoin to begin with? How will the user purchase goods or services with bitcoin? Fröhlich \textit{et al.} describe how types of custody system are suitable for different user groups. It's also likely that one user will have several types; a hardware device for secure long-term storage, exchanges for trading, a mobile key-management application for payments, and a browser based key-manager for interacting with decentralised applications. Another important insight is that users may begin with simpler custody systems, and as they gain experience and education through interacting with Bitcoin, they can graduate towards more advanced systems that are more resilient to attack and failure.  

Perhaps most relevant for the direction of this thesis was the work of Mangipudi \textit{et al.} \cite{usability-multi-device}. They investigate the perceptions of users regarding multi-device custody systems, and demonstrate how a little amount of education can alter users' perceptions. They surveyed 255 users, and classified these into `newbies' and `non-newbies' based on initial questions. Most users weren't aware of multi-device custody systems to begin with. The second phase of questions was conducted after showing the users two videos describing the security benefits of multi-device custody systems, and performing a brief knowledge-check to ascertain how well they absorbed this new information. Users were asked if they would be willing to transition to a multi-device custody system, and 71.9\% said yes. 

\subsection{Mixers}
\label{subsec:privacy-tools}

Analysing the transaction graph that is derivable from the public blockchain data structure enables de-anonymisation attacks \cite{AnonymityTxGraph, Reid2013, QuantitativeAnalysisTxGraph,HF16,JBWD18,MPJLMVS16}. Obfuscating the transaction graph is the basis of a popular privacy service methodology called `mixing'. While the general objective of maintaining a diverse set of un-linkable pseudonyms does obfuscate the transaction graph somewhat, successfully using a mixing service will reduce correlations among users' mixed UTXOs and other UTXOs they control. Ghesmati \textit{et al.} present a systematic literature review of privacy tools in Bitcoin, as well as a comparative evaluation \cite{SoK-privacy-21}. 

The most popular mixing technique, implemented in several software `wallets' is CoinJoin \cite{CoinJoin-Maxwell}. The basic idea is that a single transaction can have many inputs from different users, and many outputs to different users, and if the amounts are equal then an observer won't know which output corresponds to which input. The number of users (or input-output pairs) represents an \textit{anonymity set}. If the anonymity set is small, the technique is less effective. Moreover, even with a large anonymity set, mixing is not a perfect solution because other methods exist that can correlate UTXOs and addresses and link them to user identities. CoinJoin-based methods do not require trust in a centralised party.

To protect the mixed UTXOs from becoming correlated again, one must employ a privacy preserving coin-selection algorithm. This may limit the speed at which one may transact, since the total value of a privacy preserving transaction will be limited by the size of the available (previously mixed) UTXOs. If the desired amount of coins to be moved is larger than any single mixed UTXO, then some form of mixing technique may be used again. 

Naive centralised mixers require complete trust that the central coordinator will not steal all the bitcoin. However, several centralised mixers added mechanisms to mitigate this risk. \textit{MixCoin} \cite{BNMCKF14} and \textit{BlindCoin} \cite{VR15} required the mixer to commit to details of a mixing transaction, before users send their bitcoin, so that they can be held accountable. If the mixer doesn't fulfil their obligation, users will have proof that the mixer is corrupted. \textit{LockMix} \cite{BSKKC20} uses a $2-$of$-2$ multi-signature Script between a user and the mixer as an escrow. A game is created between user and mixer such that if they cooperate and complete the mix, the user gets their bitcoin back and the mixer takes a fee. Conversely, if they can't cooperate, they both lose. \textit{Obscuro} \cite{TLKBS18} uses a trusted execution environment to perform the role of the mixer. This merely shifts trust from a server-side website to the implementation on a security-focused hardware device, and ultimately still requires the user to trust that the service will not cheat them.

To-date, the usability of privacy tools is cumbersome and un-intuitive, and users don't understand the risks they may be taking in using them \cite{coinjoin-usability}. Moreover, these tools typically rely on sufficient collaboration among single-key custody systems and struggle to bring together enough users for an efficient service. These tools are not yet practical for multi-device custody systems unless threshold signature schemes are used at the key-management layer. In the future, as this technology progresses and matures, easy-to-use mixers and multi-device custody based on threshold signature schemes may offer the optimum privacy and security for users. 

 \section{Systems View}
\label{sec:custody-systems-view}
Let us refer to the most comprehensive level of abstraction of a custody system as the \textit{systems view}. The systems view covers the totality of the custody system, including; all components, all functionality, all processes, the users, and each phase. Little academic work exists on the systems view of custody.  Practice is leading theory. However, as shown, there is a growing literature of the underlying cryptography, forensic analyses of software and hardware in custody systems, Bitcoin's Script language and transaction protocols, of usability, and algorithm design. Moreover, there is a broad literature studying Bitcoin; the consensus protocol, the peer-to-peer network, the economics and incentive structure.

\subsection{Existing Systems}

A custody system enables the protection of assets and their functionality and is comprised of software and documentation that supports enacting a set of procedures for each phase; setup, operation and recovery. Herein we give an overview of existing Bitcoin custody systems. We review those which are sufficiently specified. These vary in form; some are system specification documents, some are software repositories, others are user-manuals, and some are a combination. Ideally, a custody system would have all three; from theory to engineering to practice.  

While the design space of possible custody systems is large, in practice only a few are used and most rely on similar techniques such as the distribution of private keys and offline devices. The design space is continually evolving as the Bitcoin protocol advances and as software and hardware mature. In recent years, new Script capabilities, new signature schemes, and new transaction validation semantics have become available in Bitcoin. Meanwhile, there is an array of software developers and hardware manufacturers striving to build competitive products to better serve users for their custody systems. 

Each custody system we will review was designed for particular users in a context. Each system must approach trade-offs such as security versus convenience and security versus system cost. Herein we review the approach of the custody systems, critique aspects that are weak (from system design to the methodology for describing and analysing `security'), and highlight strengths. The custodial architecture outlined in Chapter \ref{ch:vault-custody}, called `Ajolote', seeks to amalgamate the strengths and mitigate the shortcomings of each system discussed herein.

A simple custody system that is commonplace now is a hardware device (for key-storage and signing) with a companion application on mobile or desktop (for transaction coordination and Bitcoin network services). This system is designed for an individual with a non-trivial amount of bitcoin. It is intended to be an autonomous system, though the user trusts the hardware device manufacturer to have not, for example, embedded a backdoor access mechanism. The user requires little experience in dealing with computer systems and can be guided by simple instructions to use the system securely. A user guide is typically documented on the commercial website of the hardware device producer. The master seed phrase is backed up on paper or metal to mitigate the loss or failure of the hardware device. Hardware device manufactures often document known attacks and vulnerabilities for their products, as we will see in section \ref{sec:custody-modelling-risk}.

A prominent, open-source example of an autonomous custody specification is the Glacier Protocol \cite{Glacier}. It is suitable for an individual with a significant portion of their wealth in Bitcoin and some experience with the Linux operating system. It involves a slow and tedious procedure to withdraw funds. It does not support multi-party custody nor does it handle inheritance planning. Glacier relies primarily on Script-based multi-signature to enforce an $m-$of$-n$ access policy where each of $n$ private keys are written on paper and stored in separate secure locations. Users enact the setup and operation procedures with two separate laptops. With Glacier, users make minimal trust assumptions, often assuming that software, hardware, or networked devices can be exploited. To overcome this, the specification makes use of parallel hardware stacks and eternally quarantined hardware (which has never been connected to a network). This specification is somewhat out-dated in its choice of key-management techniques, not making use of purpose-built signing devices, modern Bitcoin Script nor metal-inscribed seed storage. However, it will still work given the backwards-compatibility of all Bitcoin upgrades since it was written. A minimal software implementation called GlacierScript is used to automate deposit and withdrawal procedures and computations. Bitcoin Core software is used for all critical cryptography routines. An array of attack and failure vectors are defined, though without a clear elaboration of a methodology for enumerating them. The specification also documents the basic privacy properties for its users, who will reveal their addresses when paying or receiving from counter-parties, as is usual in Bitcoin. Network privacy through the use of Tor is suggested but not included in procedures. Since Glacier is a `do it yourself' specification, users can remain private about using it. 

Another approach is given by Casa who have publicised their `wealth security protocol' \cite{Casa-WSP}. Here, the company offers custody services to clients including; professional assistance, custom (not open-source) key-management software, an emergency key recovery process, and strong data protection guarantees for privacy. This system was designed for users with significant wealth in Bitcoin, with a focus on usability. The withdrawal process is simple thanks to a transaction coordination app that guides users, yet may take hours of travelling between locations to enact. Users don't need experience in key-management (and in fact can be assisted where necessary). The system uses multi-signature Script to enforce a $3-$of$-5$ access policy, where each key is stored on different hardware running different software and stored in a different secured location. One of those keys is controlled by Casa, who can then assist with `emergency key recovery' in the event that the user has lost access to $2$ of their keys. In normal circumstances, Casa can not block a user accessing their funds, nor can Casa steal user funds. A fundamental difference in approach to back-ups is recommended here, which is to use `seedless' hardware signing devices. They suggest the process of writing a seed, and storing it separate from the hardware signing device creates a larger attack surface through adding complexity to the overall system. Instead, this system has simplified key-rotation significantly through automation. Rather than recreating a hardware signing device from a back-up, the user would simply need to authorise the invalidation of the lost/compromised device and initialise a new one. This procedure requires one or more transactions to sweep all funds, signed with 3 of the 4 remaining active keys. Several other features are used: PIN requirements for all devices, `sovereign recovery instructions' so that users can access funds in the event that Casa vanishes, an emergency application lockdown button in case of a physically coercive attack, and a device health-check procedure with historical logs. The `wealth security protocol' documents several attack vectors and failure modes for components of the system, though no methodology for how these were enumerated was given. Mitigating techniques were discussed for many, while some attack vectors remain unmitigated. The latter include address spoofing, malicious insider key theft (if retrieving all software and hardware from Casa), extreme disaster scenarios, and extortion. Ajolote draws inspiration from this custody system in that it also uses a 3-of-5 access policy with no additional backups, instead relying on a recovery process that can replace key-management devices as needed.

With their `Smart Custody Book', Allen and Appelcline have laid critical groundwork for making autonomous custody systems accessible with a specification for a simple system titled `Cold Storage Self-Custody Scenario' \cite{SmartCustody-cold}. The book contains more; we will review the contents on helping users understand the risks involved with their custody system in the section \ref{sec:custody-modelling-risk}. The specification is for an individual with a sizeable amount of their wealth in bitcoin ($>5\%$), who is not an experienced security practitioner and who doesn't intend to make transactions often. The access policy is a single signature, where the private key is redundantly stored in two locations; a home safe, and a bank safe deposit box. The system includes a plan for inheritance; written letter(s) for their heir(s) stored in both locations. The user also stores a metal seed backup in the safe deposit box which prevents certain types of extreme disaster events from resulting in lost bitcoin. They describe generic `adversaries' that include safety issues (like disaster and incapacitation) as well as threats (like casual physical theft and personal network attacks). They include an analysis of risks with a clear methodology. The procedures are extensible, with options that can enhance a user's custody system if they are particularly concerned with a certain type of risk. The presentation of the specification is user-friendly, with lists and check-boxes to guide each step. It is a complete example of how to approach designing for autonomous custody, with the user in-mind.  

Allen and Appelcline have recently provided another example of an open-source custody specification titled `Multisig Self-Custody Scenario' \cite{SmartCustody-multisig}. This is a step-by-step guide for an individual with significant wealth in Bitcoin. It expects the user to be planning for inheritance. It is reasonably convenient, enabling active use of funds. The specification relies on Bitcoin Script to enforce a $2-$of$-3$ access policy. Moreover, one of those keys is designated as the recovery key and is split into three shares via Shamir Secret Sharing. Keys and key-shares should be distributed across three physical locations and one cloud storage. An open-source application called Gordian Seed Tool is used primarily for generating cryptographic seeds, deriving bitcoin public and private keys, communicating with a transaction coordinator through QR codes (communication via an air-gap), and signing PSBTs. The specification is presented with the following default choices; use Sparrow software \cite{sparrow-wallet} as the transaction coordinator, use Foundation's Passport \cite{foundation-passport} as an air-gapped signing device, and use two distinct iOS or MacOS devices with Gordian Seed Tool \cite{gordian-seed-tool} (one as an air-gapped signing device and the other as a recovery key manager). A template inheritance letter is provided to share with a close family member or friend in the event of the user becoming disabled or deceased. The specification provides a list of adversaries for both attacks and failures. It provides a description of failure modes and attack vectors for single and multiple components. Strong privacy protections are not considered within. The approach taken by Allen and Appelcline to simplify the user experience, by creating an accessible user-guide and elucidating risks and mitigations, is vital for any autonomous custody system. We adopt this methodology with Ajolote as well, seeking to empower users and draw on their unique strengths.

The company Square, Inc. published open-source software and documentation for their custody system called `Subzero' \cite{Subzero}. It is a system they use in their payments business. They describe it as  `HSM-backed Bitcoin Cold Storage', where HSM stands for hardware security module. In their system, they use `FIPS 140-2 certified' HSMs. Federal Information Processing Standards (FIPS) is a standard developed by the National Institute for Standards (NIST) that must be complied with when dealing with sensitive data in, for example, banking and healthcare. The documentation does not describe objectives nor design principles. The document covers the setup and operational phase, but has no information about recovery. They don't divulge precisely their access control policy, except that it requires a multi-signature. They don't present any risk model or threat analysis. They provide their code for review and offer a bug bounty. For the procedures `wallet initialisation' and `signing ceremony' they use message sequence diagrams which explicitly show human-device interaction. These are candidates for a more rigorous analysis. However, overall the details are too sparse to perform a comprehensive analysis of the system. 

Revault is an open-source multi-party custody system with a detailed specification and software implementation \cite{practical-revault,revault-repos}. It is intended to be an autonomous custody system for institutions and organisations with a large amount of bitcoin. The architecture distinguishes two types of role for users in the system; stakeholders and managers. Stakeholders are the primary controllers, and they delegate limited access rights to managers. The access control policy is implemented with a combination of Script-based multi-signatures and pre-signed transaction sequences. The system is unique in that it uses servers to monitor the Bitcoin network and the custody system, to detect misbehaviour and to respond to events. The system also has an emergency fall-back feature as a countermeasure for physical attacks on the stakeholders. We present a risk framework for the operation of this system in Appendix \ref{ch:risk-framework-revault}. The usage of pre-signed transaction sequences in Revault serves as a mechanism to predetermine specific actions. In this instance, deviation from this predetermined path necessitates endorsements from all primary stakeholders. This bears resemblance to the method adopted by Ajolote, although in that case there is no provision for an alternate route. Both systems utilise servers for vigilance over the Bitcoin network and for initiating countermeasures against perceived threats.

Other custody systems exist but lack the type of specification we are interested in for the purposes of performing system view analyses. Examples include; Electrum \cite{electrum-multisig}, Nunchuck \cite{nunchuck-multisig}, Blockstream Green \cite{blockstream-multisig}, and Wasabi \cite{wasabi}.  Similarly, other user-guides exist but lack the depth required for rigorous analysis \cite{tordl,yeticold,rusty-guide}.

\subsection{Custody System Standards}
\label{subsec:custody-system-standards}

Standards are being drafted for the custodianship of digital assets which provide models and terminology for custody and set minimum requirements for technical operations, secret generation, secret recovery, development and maintenance. Let us consider the current set of available standards.

The CryptoCurrency Security Standard (CCSS) is an attempt to create a uniform set of requirements for digital asset storage systems, primarily custody systems for organisations \cite{ccss}. These requirements cover; key generation, key storage, key usage, key compromise policy, key-holder grant/revoke policies and procedures, data sanitisation policy, proof-of-reserves, and audit logs. That is to say, they primarily cover key-management. These requirements are consistent with our overview of key-management herein. These requirements can be used as a baseline benchmark for analysing custody systems. 

The Digital Asset Custody Standard (DACS) is an alternative to CCSS. It also aims to be a baseline upon which users and auditors can analyse custody systems. They propose requirements and recommendations that are split into two streams: operations and infrastructure. The operations stream is beyond the key-management layer. It includes the choice of `custody model', which refers to the accounting method used by custody service providers to track customer accounts. It requires the specification of a threat model, risks, and countermeasures from physical security environment controls, to software and hardware controls, and to personnel controls. However, specifics are omitted. Rather, they merely require these controls are documented. One of their requirements is that all key-storage devices are offline. Another is that a proof-of-reserves procedure should be documented. It appears to lack requiring procedures to handle situations of compromise. The infrastructure stream concerns the key-management layer. In particular, recommendations for how secrets are generated.  

Another document titled `General Security Considerations for Cryptoassets Custodians' (GSCCC) reviews technical and operational risks for custody service providers \cite{CryptoassetCustody}. It presents a model of a custody system architecture including a data flow diagram that depicts; customer inputs from a web interface, communication with the blockchain network, a transaction signing module, and various data stores for customer credentials and asset-related data. It describes the functional components, the process to send a transaction, key types, key-generation process, and key-usage process. Of note, there is no process for recovery in the event of compromise. Risks to the custodian are broadly considered; from the custody system to external factors such as the internet infrastructure, user environment, blockchain, and reputation. Finally, the document describes security controls for the custodian; to use a combination of hot and cold key-storage, to require two administrators for key-usage, to backup keys, to use hardware signing devices and apply strict physical controls on their access. 

Altogether, the requirements and recommendations proposed with CCSS, DACS and GSCCC are consistent with the overview of custody within this chapter. 

Taking a broader look to the practice of information security, there are several cybersecurity frameworks including `The US National Institute of Standards and Technology (NIST) Framework for Improving Critical Infrastructure Cybersecurity (NIST CSF)' \cite{NIST-CSF}, `The Center for Internet Security Critical Security Controls (CIS)' \cite{CIS-CSC}, and `The International Standards Organization (ISO) frameworks ISO/IEC 27001 and 27002.' \cite{ISO-IEC} which specify some high-level characteristics and guidance that would apply to custodianship of digital assets. This perspective can help situate custody protocols in the wider context of an organization's objectives and capabilities, and ensure that integrating a custodial operation doesn't create unseen risk factors that aren't present in a technical analysis. 

\section{Modelling Risk}
\label{sec:custody-modelling-risk}

\subsection{Overview}

Throughout this thesis the terms `threat model', `attack model', and `risk model' are used precisely and it is important that we distinguish them here. A \textit{model} is an abstraction to aid in thinking about a system. Models are often incomplete or inaccurate but they can be useful. We are attempting to design, analyse, and improve Bitcoin custody systems in order to avoid theft and loss. We can achieve this by constructing models of what things can go wrong and how they can go wrong, and then find ways to mitigate those events.

We will use the term \textit{risk} as defined in `How to measure anything in Cybersecurity Risk' by Hubbard and Seiersen \cite{Hubbard}.
\begin{quote}
    ``\textbf{Risk:} A state of uncertainty where some possibilities involve a loss, catastrophe, or other undesirable outcome." \cite{Hubbard}
\end{quote}
Uncertainty is quantifiable even if subjective belief is involved. Using standard techniques from mathematics such as Bayesian inference we can measure changes in risk and state whether or not risk has been reduced.
\begin{quote}
    ``\textbf{Measurement of risk:} A set of possibilities, each with quantified probabilities and quantified losses." \cite{Hubbard}
\end{quote}
In this way, the reduction of risk requires a measure-able reduction of \textit{impact} (losses) and \textit{likelihood} (probability of the event occurring). 

\textit{Threats} are a person or thing likely to cause undesirable outcomes. Notice that this term is vague. This is in part why `threat models' can take many forms. Shostack states that there are several approaches to constructing a threat model \cite{Shostack}; asset-centric, attack-centric, and software-centric. The act of constructing a threat model is a process to enumerate threats. To develop a risk model from a threat model, one can quantify the impact and likelihood of each threat. 

Focusing on assets, that is, things of value, is a way to structure an exercise in enumerating threats. What is valuable to the user? Of course, their bitcoin holdings, but also less tangible things such as their privacy or their communication credentials. With a list of assets in hand, one can systematically consider what types of threat can compromise those assets. Considering assets is compatible with quantifying the \textit{impact} of lost assets. 

Focusing on attacks is another way to structure a threat enumeration. Common threats exist across systems; denial-of-service attacks, theft, information disclosure, \textit{etcetera}. An attack model can help in understanding what actions an attacker must take in order to compromise some asset, or otherwise cause some chaos. By understanding the complexity of an attack, or the expertise required, or the cost, we can make an assessment of its \textit{likelihood}. Attacks are often modeled with attack trees \cite{AttackTrees} or attack graphs \cite{SurveyAttackModeling}. Attacks can be better understood with an experimental approach too. For example, attempting to physically extract private keys from a mobile device in a laboratory setting. 

Focusing on software is yet another way to structure a threat enumeration. By formally specifying how a software system works (for example, with data-flow diagrams for each process and component), we are forced to consider every entry-point into that system. By considering each computation and each action in turn, we may systematically determine threats. 

Often, a combination of approaches will yield a more complete model. A \textit{vulnerability} is a weak point in a component or process that will grant an attacker leverage towards compromising an asset. Once vulnerabilities are discovered in a system, then \textit{countermeasures} can be developed and applied to mitigate the risk associated with that vulnerability. Countermeasures are typically a new component or process in the system. As such, the risk model must be updated to account for the countermeasure, including the benefits and possible unwanted side-effects it yields. 

A custody system exists in a complex environment of nested layers consisting of people, physical spaces, physical devices, internet networks, mathematical spaces, and so on. Exhaustively enumerating threats is not possible. However, the objective in constructing a threat model is to enumerate threats sufficiently to defend against the attacks and failures which pose the most risk. 

Risk analysis is intended to support decision making, and in this case improve the design of custody systems. In general, less risk is equivalent to more `security'. We can understand if a risk analysis is valuable by asking the following questions. Does the risk analysis improve the identification and management of risks? Does it help us to measurably reduce risk? Constructing and using such a model is a strategic process that considers attack and failure scenarios, vulnerabilities within the system, the likelihood and impact of attack and failure scenarios, and thus the risk of using the system. 

\subsection{Examples in Custody}

In their `Smart Custody Book', Allen and Appelcline present a guided approach to help users construct a risk model which is user-friendly and compatible with a modular custody system (with assets distributed across devices and sub-systems) \cite{SmartCustodyBook}. The process starts by listing assets, quantifying their value, and drawing a diagram representing each asset's environment (e.g. hardware device at home, bitcoin on an exchange) and the transfer paths between them. This is an asset-centric approach. The second phase of the process attempts to enumerate threats to the assets. It begins with considering the transfer paths between the user's asset environments, then considering the environments themselves. The reader is prompted with lists of possible vulnerabilities but is also instructed to brainstorm alternatives. Finally, non-physical assets (such as privacy and availability) are considered. With each vulnerability, the user must quantify the impact and likelihood. It is suggested to compute risk as $risk = impact$ x $likelihood$ and to visualise it by making a chart of \textit{impact} versus \textit{likelihood}. This helps the reader to determine where the highest risks are, and those components should be addressed first. The third and final phase of the process is to resolve risks. In this, they provide a long (but not exhaustive) list of `adversaries' which include accidental (e.g. flood) and active (e.g. violent thief) threats. Readers are instructed to map adversaries to the ranked list of risks. This is an attack-centric approach. Then, the reader can consider how to reduce that risk. Any countermeasure that is used will change the system and require an updated risk model. The process should be repeated with each change until the assessment of risks is acceptable. This is an exemplary process for constructing a risk model for a custody system. It is targeted more towards users than system designers. Allen and Appelcline demonstrate that it is possible to educate users sufficiently to take responsibility for their risk management.  

Homoliak \textit{et al.} propose a `security reference architecture' for blockchains \cite{SecurityRefArchitecture}. They present a stacked model of the architecture of a blockchain consisting of a network layer, consensus layer, replicated state-machine layer, and an application layer. They describe known vulnerabilities, threats and defenses at each layer (obtained from broadly reviewing the literature). They suggest that application designers should consult the security reference architecture layer by layer when assessing risks. Indeed, if we are to take the systems view of Bitcoin custody, we would want to understand risks that emanate from each layer of the technical environment. This work would benefit from being written as an open-source repository which can be updated by the broader community as new research is generated.  This collection of threats vulnerabilities were reviewed and considered when conducting the risk analyses for the covenant protocol introduced in chapter \ref{ch:bitcoin-covenants} and for Ajolote in chapter \ref{ch:vault-custody}.

Jain \textit{et al.} consider risk management with a custody system for a bitcoin exchange \cite{ExchangeReservesRisk}. They analyse a system design that combines two distinct access control structures; a hot wallet and a cold wallet. They work at a layer of abstraction above cryptographic primitives and key-management technologies. They assume that lower layers are imperfect, that system components can fail and that theft will occur. They inquire how to optimise risk management by carefully managing flows in their two tier system and setting limits to the amount in the hot wallet. They model deposits, withdrawals, and thefts as Poisson processes and derive an equation to compute the expected balance of the system over a given period. 

Eyal proposed a formal model of multi-signature access policy designs \cite{Eyal2021OnCW}. For each key in the policy, probabilities are assigned to the event of loss, leakage, or theft. They evaluate optimal designs for small state-spaces, with 5 or less keys as part of the policy. Assuming the probabilities are uncorrelated for each key, then failure drops exponentially with the number of keys. This points to the significant benefit that derives from multi-location and heterogeneous device storage. The model illuminates the importance of understanding whether a particular key is more likely to be lost, leaked, or stolen, and helps to understand how that may impact the inherent trade-off between security and safety. A key held on a server, for example, is more likely to be leaked or stolen than one held on an air-gapped signing device. However, the server key may be much less susceptible to loss. So a safety-focused user may prefer holding a key on a server, while a security-focused user may prefer holding that key on an air-gapped signing device. While specifying concrete probabilities is not realistic, the model still provides useful insight into methods for evaluating a custody system's resilience to attacks and to failures. 

Developers of the Enno mobile wallet published their threat model \cite{enno-TM}. They take the software-centric approach to enumerating threats and applying controls to mitigate threats. They follow the STRIDE methodology publicised by Microsoft \cite{STRIDE}, which guides the analyst to consider common threat categories for sensitive data. They also present an attack-tree towards compromising the seed contained on the mobile. While the custody system itself is not resilient to threat nor failure, the analysis demonstrates how a systematic threat enumeration process can produce valuable insights about the custody system. 

Bulut and Sertkaya use a structured approach to modeling the security of several types of wallet (online, mobile, desktop, hardware, paper) \cite{wallets-TM}. They follow the `common criteria' framework \cite{cc-framework}. They formally define the `security problem' as a combination of; threats to assets (e.g. eves dropping), assumptions about environment (e.g. secure platform), and organizational security policies (e.g. authentication). They elaborate on each to obtain a well-defined security problem and match threats, assumptions and policies to each wallet type. They go on to define security objectives and controls that mitigate threats. Putting it all together, they evaluate each wallet type by matching threats and policies with security objectives and summarising how each wallet type addresses its security problem. While these custody systems are simplistic (single-key systems), Bulut and Sertkaya have demonstrated how to apply the common criteria framework in a logical way to evaluate and compare those custody systems.   

Kozak presents a thesis on the security of hardware wallets (that is, key-management hardware devices) \cite{thesis-HW-security}. They assess the threat model commonly used for hardware wallets \cite{FormalHardware} and present a list of common attack vectors for that model. They go into more detail on the theory of physical attacks on hardware devices such as power consumption analysis (a side-channel attack vector). They describe in detail the software and hardware of the Trezor One hardware device and experiment with physical attacks. This is an attack-centric study towards evaluating the risks present with hardware devices. The practical, experimental work is particularly valuable because it demonstrates the likelihood of attack, and enables an analyst to consider risk (as the combination of impact and likelihood). They conclude that if a hardware device is stolen, it is only a matter of time before a physical attack can completely compromise the device. 

In other work, Dabrowski \textit{et al.} consider hardware device security in a context of an attack model in which supply-chain attacks are possible \cite{HW-SPOF}. This has typically been ignored throughout the industry on the assumption that vendors' device authenticity checks are sufficient. In reality, Dabrowski \textit{et el.} argue, many supply-chain attacks have been observed. What this results in is hardware devices which are susceptible to active attacks (man-in-the-middle attacks), through malware injection or hardware implant. They summarise features and security engineering practices of hardware wallets. The critical point is that hardware devices are not a panacea for security, and that treating them as if they are creates a single point of failure/compromise. They advocate removing the single point of failure/compromise by combining hardware authenticity checks with mutual verification between the hardware device and the accompanying software client, making use of collaborative signatures and key generation.  

Aumasson and Shlomovits construct an attack model for the custody system of a digital asset exchange based on a threshold signature scheme \cite{attacking-thresh-wal}. The system has two tiers; a hot tier with a $2-$of$-2$ access policy between the customer mobile and the exchange's server, and a cold tier with a $2-$of$-3$ access policy across offline devices in distinct physical security environments. They consider active attacks that can intercept and modify data in transmission. In sum they describe three separate attacks, the exploitation, and a method to mitigate the attack. They show how failure of the implementation to comply precisely with the specification from the literature creates vulnerabilities. They note also that complex multi-round protocols involving advanced cryptography primitives (verifiable secret sharing, homomorphic encryption, commitments and zero-knowledge proofs) are prone to vulnerabilities. Detailed descriptions of attack can be very insightful for an analyst attempting to model risk in various custody system designs. 

Fröhlich \textit{et al.} take a different approach to constructing a threat model \cite{HCI-threat-model}. They elicit information from a group of expert security and cryptocurrency researchers, and through several rounds of iteration they broaden and validate the model (following the Delphi method \cite{Delphi}).  The scope of their model is limited to threats relevant for users of general cryptocurrencies, and does not consider specific technical threats associated with any given implementation, nor legal and regulatory threats. The result is a threat model with six categories; accidental, privacy, phyiscal, financial fraud, social, and technical threats. `Threat agents' are proposed (inclusive of attackers and accidents); non-target specific (e.g. computer viruses), employees, organised crime, corporations, human (unintentional), human (intentional), and natural (e.g. flood, fire, \textit{etcetera}). `Potential consequences' are listed as a summary of impact to the user; disclosure of personal data, complete loss of cryptocurrency, partial loss of cryptocurrency, temporary loss of access, endangered personal health, loss of reputation, and reduction of value (price of the cryptocurrency). With this, threats are detailed and mapped to threat agents and impact, along with countermeasures for each threat. This systematic, user-centric approach is very instructive, and shows the difficulty of constructing a comprehensive threat model. With a panel of experts and several iterations, the model is not exhaustive and the countermeasures presented must be analysed further. To improve the user-experience (to achieve better privacy and security), they recommend three directions for future work; educating users, building assistive systems, and user-interface guidelines. 

A threat model has been presented for Ledger hardware devices \cite{Ledger-threat-model}. Therein, they decompose the model into layers; device, OS, application. They propose security objectives (such as \textit{confidentiality} of user seeds), and security mechanisms to meet those objectives (such as random number generation). They briefly describe their security mechanisms and note, in some cases, that vulnerabilities are still present. From their model, it is clear that physical attacks, either while the device is in the supply chain or after it is delivered to the user, are highly effective against the hardware device. Sensitive data can be extracted through many attacks; side-channels (e.g. power consumption analysis), laser and electromagnetic interference, glitch fault injection, and SRAM extraction. Moreover, the user can be deceived or coerced to authorise theft transactions. 

While an admission of imperfect security is difficult for the marketing department to spin, any hardware manufacturer that is claiming perfect security is being dishonest. Other hardware device manufactures also present documentation describing threats and countermeasures. Trezor documentation has a list of threats and countermeasures \cite{Trezor-threat-model}. An attack-centric threat model (device attacks, computer attacks, service attacks, and user attacks) is provided for Bitbox, and again describes how they mitigate the risk of each attack with various countermeasures \cite{Bitbox-threat-model}. Cold card discusses threats and countermeasures across several user-guides (each designed for a different type of user, from `quick-start' to `paranoid') \cite{Coldcard-hsm-threat-model}. A detailed security model has been documented for Foundation's `passport' device \cite{foundation-TM}. Combined, this collection of models presents a coherent picture of the strengths and weaknesses of hardware devices. In each case, the user must trust the vendor. While a single-key custody system would have to trust one vendor, that trust requirement can be distributed across multiple vendors with a custody system that implements a resilient access control structure. 

Knox Custody, a Bitcoin custody service provider, has published a set of risk management controls. This is similar in form to the custody standards seen in section \ref{subsec:custody-system-standards}. They structure their controls in accordance with the key life-cycle; general considerations, generation and de-archiving, storage and transport, and signing. Each section has a set of controls. Each control is described and the rationale for applying this control is given; the impact and what is protected. They have designed a custody system with an appreciation for the systems view. Their controls include people, process and technology, and each phase of custody. The list of controls indicate this is a custody system with multi-signature access control policies based on keys that are stored in geographically distributed facilities on hardware security modules. The article explicitly attempts to achieve clarity in communication, and provide transparency about the risk management practices. However, without an open specification document for the precise technology components used, it is still difficult to make a judgement about this system. At least, the risk controls can be considered as design constraints when designing custody systems. 

\section{Conclusion}

In this chapter we introduced Bitcoin custody. We defined types of custody system; autonomous and heteronomous, single-party and multi-party, cold key-storage and hot key-storage. We considered design principles from other work which may act as guidelines. We described three phases of custody; setup, operation, and recovery. We decomposed the functions of custody systems. 

We provided an overview of the suite of technologies that can be applied in Bitcoin custody. The foundations, we learned, are a resilient key-management strategy coupled with a privacy framework. Resilience to failure and attack can be obtained by using a multi-key access structure and distributing keys across different physical security environments, hardware and software stacks. We considered objectives for maintaining privacy, and learned of the numerous ways in which a user's privacy can be compromised. Then we reviewed mixers, which can help to obfuscate de-anonymising transaction graph analysis.

With a clear grasp of the fundamentals underpinning Bitcoin custody, we turned towards a  comprehensive perspective we referred to as the `systems view'. We reviewed the majority of existing, well-documented and well-specified Bitcoin custody systems. Each one with a different approach, targeting a particular user group and using a different composition of technologies. We noted the state of user-guides and supporting documents which (in the best cases) help to understand the risk model. Then we took account of emerging standards for Bitcoin (and, more generally, digital asset) custody. Taken altogether, the standards are a reasonable benchmark for custody system designers. 

Finally, we defined the nuances of risk-related terms and presented how they will be used for the remainder of this thesis. In brief, a risk model is constructed by mapping a set of threats with some measure of their likelihood and impact. We can understand likelihood by examining the attack sequences. Impact depends on the assets in the system and how the user values them. Risk is the uncertainty of something `going wrong' and is the composition of impact and likelihood. This was necessary context for our review of examples of risk modelling from the literature and industry. These take a range of approaches and provide valuable input about the risks associated with people, processes and technology.

We are now prepared to study in detail a new technology mechanism which has yet to be applied to Bitcoin custody; covenants. This is the topic of chapter \ref{ch:bitcoin-covenants}. Following that, chapter \ref{ch:vault-custody} introduces a new, single-party, autonomous custody system that utilises covenants. We specify, evaluate, and construct a risk model for the system. In Appendix \ref{ch:risk-framework-revault} we consider Revault, a similar custody system in a multi-party context and construct a risk framework; a formalised attack model with a methodology to perform context-specific risk analyses. 
\chapter{Bitcoin Covenants}
\label{ch:bitcoin-covenants}

In the ensuing chapter, we investigate the intricacies of Bitcoin Covenants, relying on the necessary background of Bitcoin transactions, inputs, outputs and Script from chapter \ref{ch:background} to comprehend various covenant proposals. We present a deleted-key covenant protocol, constructed using advanced Bitcoin features like Schnorr signatures and TapScript, aimed at enabling a novel technology-based risk control within the Ajolote custody system to be introduced in the following chapter. Furthermore, we scrutinise the practical aspects of applying deleted-key covenants and explore potential caveats and their solutions. Crucially, our approach is the only known way to construct Bitcoin covenants without altering the consensus rules. This chapter addresses thesis objective \textbf{O2} (see \ref{sec:thesis-objectives}).

\section{Introduction}

\subsection{Motivation}

In traditional finance parlance, a financial covenant is a commitment in a debt agreement (or other indenture) that certain activities will or will not be carried out \cite{CovInvestopedia}. What we describe herein only applies to the enforcement of the \textit{commitments} in the covenant, and not to dispute resolution in the case of a default on the \textit{debt agreement}. When there is a default on the debt agreement, the resolution could be dependent on a legal framework or on some collateral liquidation process in a peer-to-peer loan system. Traditional mechanisms for dispute resolution of broken commitments may also apply to debt agreements denominated in bitcoin. However, with bitcoin it's possible to enforce a class of covenants automatically with a mechanism that is secure under a well-defined set of assumptions. 

Traditionally, there are two classes of covenants; affirmative (where the conditions of the covenant require some specific action to be performed) and negative (where conditions of the covenant require refraining from specific actions). Typically, these covenants will be related to operational capacity of a company - for example to ensure adequate levels of insurance, or to maintain an interest coverage ratio. A simpler example of a covenant is a restriction of the set of possible recipients for future payments. An ability to enforce arbitrary covenants would increase the viability of Bitcoin as a financial system capable of supporting functions required in modern economies. Even a restricted class of bitcoin covenants could enable new opportunities for novel internet-based  business models. Applications of bitcoin covenants are being explored in the design of custody (see chapter \ref{ch:vault-custody}), payment processing \cite{BIP119}, decentralised exchange protocols \cite{Bitmatrix} and more. 

In general, applications that use covenants can benefit from enforcing a transaction sequence and mitigating risks associated with the compromise of private keys. An ability to prove the enforcement of a covenant may increase the incentive for the counter-parties to continue with a transaction protocol since the covenant restricts them from altering the transaction sequence and they may otherwise lose access to the funds in the protocol. Moreover, the transaction sequence can encode arbitrary state machines, albeit with memory limited by amount of state that can be embedded in transactions. 

\subsection{Contributions}

This chapter introduces a novel protocol for a covenant mechanism termed \textit{deleted-key} covenants. While other covenant mechanisms have been proposed, deleted-key covenants stand out as the only mechanism currently operable in Bitcoin, as alternatives necessitate a soft-fork upgrade. This chapter contributes to the ongoing debate on the value of such soft-forks by providing a direct comparison with what is currently feasible. It should be emphasized, however, that the contributions of this chapter are primarily theoretical in nature, given that no live or test deployment of the deleted-key covenant protocol has been undertaken as of the time of writing.

We introduce necessary terminology and concepts related to covenants in section \ref{sec:Preliminaries}. Therein, we also offer a summary of the internal details of cryptographic signature operations in Script in section \ref{sec:Preliminaries}. Although this information is publicly accessible, it is dispersed across multiple informational websites, Bitcoin Improvement Proposals (BIPs), and code implementations. Our consolidated overview serves as a valuable introduction for Bitcoin contract designers.

We review existing proposals that enable covenant mechanisms in section \ref{sec:related_work}, explaining the changes to the consensus rules they require. Moreover, we present an overview of the related literature including analyses of Bitcoin Script and how to appropriately model it with covenants, and covenant applications. 

In section \ref{sec:Deleted-Key-Covenants}, we present a multi-party protocol for enforcing a deleted-key covenant and provide an analysis of its relevant security properties in section \ref{sec:DKC-security-analysis}. Subsequently, in section \ref{sec:designing-for-applications}, we discuss several crucial aspects for designing Bitcoin contracts that utilize deleted-key covenants, including composability, the class of possible covenants, fee allocation strategies, safety concerns, and requirements for a proof-of-reserves protocol.

Finally, in section \ref{sec:covenant-comparison}, we compare deleted-key covenants with three covenant mechanisms based on separate soft-fork proposals. We elucidate the benefits and drawbacks of each, demonstrating that these proposals could significantly enhance the fundamental security and practicality of covenant-based applications in Bitcoin. We also elaborate on the subtle differences in their practical applications. 

\section{Preliminaries}
\label{sec:Preliminaries}

In this chapter we will assume the reader is familiar with Bitcoin, with a high-level understanding of blockchain data structures, the peer-to-peer network, the consensus protocol and the economic incentives that underpin the system's security. Furthermore, we will assume the reader is familiar with concepts from cryptography including hash functions, Merkle trees, public key cryptography and digital signatures. We will  consider formal definitions relevant to covenants, see how covenants are constructed, and explore internal details of Signature operations in Script, which are used to commit to covenant constraints. 

\subsection{Definitions for Covenants}

This section introduces a series of definitions pertinent to Bitcoin covenants. These definitions provide a crucial framework for the discussion and analysis of covenants in subsequent sections. Bitcoin covenants refer to a commitment to a certain set of constraints that govern the transfer of control of the coins. Such transfers occur through transactions, and thus mechanisms for Bitcoin covenants revolve around the creation of a commitment to a proper subset of transaction data to be broadcasted in the future. These mechanisms impede transactions that do not contain that subset of data. Cryptography and the Script language underpin the functioning of covenant mechanisms.

\vspace{0.1cm}
\noindent \textbf{Definition 1.} A \textit{secure bitcoin covenant} is an unbreakable commitment to a specific set of constraints that apply to one or more transactions. 

Examples of specific constraints might include limiting the addresses to which coins can be sent, or the number of coins that can be transferred to certain addresses.

\vspace{0.1cm}
\noindent \textbf{Definition 2.} A \textit{covenant-bound output} is one that can only be consumed in a transaction that adheres to the constraints specified in the covenant.

\vspace{0.1cm}
\noindent \textbf{Definition 3.} A \textit{covenant transaction} is one that consumes at least one covenant-bound output.

\vspace{0.1cm}
\noindent \textbf{Definition 4.} A \textit{deposit transaction} is one that produces a covenant-bound output.

\vspace{0.1cm}
\noindent \textbf{Definition 5.} A covenant mechanism's \textit{enforcement conditions} are the set of conditions required to create a covenant-bound output.

\vspace{0.1cm}
\noindent \textbf{Definition 6.} An \textit{active} covenant is one for which all of its enforcement conditions have been satisfied.

\vspace{0.1cm}
\noindent \textbf{Definition 7.} \textit{Custodial power} is the capacity to spend coins. Custodial power can be \textit{distributed} according to arbitrary access control structures between several parties.
\vspace{0.1cm}

For each covenant mechanism discussed herein, custodial power can be delegated to an arbitrary access control structure. The covenant-bound output should commit to a lock script that encodes the covenant \textit{and} a set of public keys that determine who is granted custodial power. To distribute custodial power among $k$ participants one can use a $j$-of-$k$ multi-signature lock script where each of $k$ participants provide a public key $Q_{i}$ for $i \in \{1, ..., k\}$. A threshold of $j$ participants must collaborate by generating signatures with their associated private keys $q_{i}$ in order to authorise the covenant transaction.

\noindent \textbf{Definition 8.} A \textit{recursive covenant} begins with the creation of a covenant-bound output that commits to a subset of transaction data. Each covenant transaction that follows creates a covenant-bound output that commits to the same subset of transaction data.  The constraints of the covenant that are being enforced apply to all subsequent covenant transactions. 

A recursive covenant may be conceptualised as a state machine where each transaction in the sequence is the state transition, and the state is stored in some part of the transaction data. 

\noindent \textbf{Definition 9.} A Bitcoin \textit{smart contract} is a computer protocol that allows users to exchange bitcoins according to (simple or complicated) pre-agreed rules. No trusted third-party is required since the rules are enforced by the consensus protocol for maintaining and extending the blockchain. 

\subsection{Covenant Creation and Verification Process}

Informally, the creation and verification of a Bitcoin covenant transaction, hypothetically enforced with a Script instruction, involves three main steps:

\begin{enumerate}
    \item \textbf{Commitment:} As the covenant creator, you formulate the constraints of the covenant and commit these to the public key script of an output. This commitment sets the rules that any transaction attempting to spend this output--the \textit{redeeming transaction}--must follow.
    \item \textbf{Push Transaction Data:} You use a Script operation to push the redeeming transaction data onto the execution stack. This can be done either directly or through a cryptographic commitment, which securely binds to the transaction data.
    \item \textbf{Verification:} This way, when the lock and unlock Scripts are executed, one can verify whether or not the redeeming transaction complies with the constraints committed to in the output's public key script. This step confirms or refutes the validity of the covenant transaction and its adherence to the defined rules. If the transaction does not comply, it is considered invalid and will not be included in the blockchain.
\end{enumerate}

A cryptographic commitment, by definition, is uniquely bound to a message and keeps the message hidden until it is revealed \cite{Menezes:1996:HAC:548089}. In the context of Bitcoin covenants, we use such commitments to bind an output to one or more covenant transaction types, while keeping the covenant transaction hidden until it is revealed. This is achieved by embedding either a hash or a signature of the covenant transaction data into an output's public key script, creating what we refer to as a covenant-bound output.

The lock script of the covenant-bound output needs to access the relevant transaction fields to verify that a transaction spending the output adheres to the committed constraints. Currently, the only instructions capable of reading transaction data are the signature operations. Without additional script operations, it appears impossible to transfer verified transaction data onto the execution stack in any other way. This limitation has led to numerous proposals to update the Script language to enable consensus-enforced covenants.

\subsection{Signature Message Types}

\label{subsec:sig_msg_types}

Now we will inspect \textit{signature message types}. By understanding what subset of transaction data are committed to with digital signatures, i.e. what is used as the \textit{message} for a signature, we will come to understand the details and differences of covenant proposals.

The Script instruction {\tt OP\_CHECKSIG} is used to verify a digital signature. It removes from the Script execution stack two elements, expecting a {\tt public key} and {\tt signature}. The signature message is the serialization of a subset of current transaction data and previous output data. Valid signatures can be produced to endorse different subsets of current and previous transaction data. The type of signature is controlled by setting a {\tt SIGHASH} flag to one of six types. The {\tt ALL}, {\tt SINGLE} and {\tt NONE} types refer to including all, one, or none of the outputs in the signature message, respectively. Each of these types may be concatenated with {\tt ANYONECANPAY} (or {\tt A1CP}), giving a total of six types. {\tt A1CP} is used to indicate that only the input which the script is associated with should be included in the signature message. A single transaction may have many signatures of different types, for example if there are several inputs, or if one of the input's unlock script has multiple signatures. 

Knowing the signature message types, we can appreciate better the risks involved with \textit{transaction malleability}. Transaction malleability refers to the modification of transactions without invalidating them and their accompanying signatures. Designers of transaction protocols must be weary of malleability as a potential attack vector, but may also use malleability as a feature in certain contexts. This topic will be discussed in detail in section \ref{sec:safety-concerns}.

The signature message types for the semantics of signature operations (e.g. {\tt OP\_CHECKSIG}) in `legacy' output types (P2PK, P2PKH, P2SH, P2MS) are distinguished in table \ref{tab:legacy_sighash}. We show which data types are committed to and the source of those data.

\begin{table}
    \centering
\resizebox{\linewidth}{!}{
    \begin{tabular}{|l|l|l|l|l|l|l|l|}
    \hline
    \rowcolor{lightgray} \textbf{Data Source} & \textbf{Data Type} & {\tt ALL} & {\tt SINGLE} & {\tt NONE} & {\tt A1CP||ALL} & {\tt A1CP||SINGLE} & {\tt A1CP||NONE} \\
    \hline 
    Current Tx & Version &Y&Y&Y&Y&Y&Y \\
    \hline 
    Current Tx & Number of Tx inputs &Y&Y&Y&N&N&N \\
    \hline 
    Current input & Index & Y&Y&Y&N&N&N \\
    \hline    
    Current input & Previous output outpoint & Y&Y&Y&Y&Y&Y \\
    \hline
    Current input & Subscript length & Y&Y&Y&Y&Y&Y \\
    \hline
    Current input & Subscript &Y&Y&Y&Y&Y&Y \\
    \hline
    Current input & Sequence &Y&Y&Y&Y&Y&Y \\
    \hline
    Other inputs & Index & Y&Y&Y&N&N&N \\
    \hline  
    Other inputs & Previous output outpoint &Y&Y&Y&N&N&N \\
    \hline
    Other inputs & Subscript length &Y&Y&Y&N&N&N \\
    \hline
    Other inputs & Subscript &Y&Y&Y&N&N&N \\
    \hline
    Other inputs & Sequence &N &N&Y&N&N&N \\
    \hline
    Other inputs & Number of Tx outputs &Y&Y&Y&Y&N&N \\
    \hline
    Output at current input index & Value &Y&Y&N&Y&Y&N \\
    \hline
    Output at current input index  & Public key script length &Y&Y&N&Y&Y&N \\
    \hline
    Output at current input index  & Public key script  &Y&Y&N&Y&Y&N \\
    \hline
    Other outputs & Value &Y&N&N&Y&N&N \\
    \hline
    Other outputs & Public key script length &Y&N&N&Y&N&N \\
    \hline
    Other outputs & Public key script  &Y&N&N&Y&N&N \\
    \hline
    Current Tx & Locktime  &Y&Y&Y&Y&Y&Y \\
    \hline
    \end{tabular}}
    \caption{Data committed to by signatures for each signature hash type when legacy Script semantics for {\tt OP\_CHECKSIG} are used. This includes spending P2PK, P2PKH, P2MS and P2SH output types.}
    \label{tab:legacy_sighash}
\end{table}

The semantics for signature operations is different when spending version 0 SegWit outputs (P2WPKH and P2WSH). In particular, the algorithm to derive the signature message has changed some of the data types and their ordering. The precise subset of data committed to in this case is presented for each signature hash type in table \ref{tab:segwit_sighash}. 

\begin{table}
    \centering
    \resizebox{\textwidth}{!}{
    \begin{tabular}{|l|l|l|l|l|l|l|l|}
    \hline
       \rowcolor{lightgray}  \textbf{Data Source} & \textbf{Data Type} & {\tt ALL} & {\tt SINGLE} & {\tt NONE} & {\tt A1CP||ALL} & {\tt A1CP||SINGLE} & {\tt A1CP||NONE} \\
    \hline
        Current Tx & Version & Y&Y &Y &Y &Y &Y \\
    \hline
        Current Tx & Hash previous outpoints (of all inputs) &Y &Y &Y &N &N &N \\
    \hline
        Current Tx & Hash sequence (of all inputs) & Y& N &N &N &N &N \\
    \hline 
        Current input & Index & Y&Y(I) &Y(I) &N &N &N \\
    \hline
        Current input & Previous output outpoint &Y &Y &Y &Y &Y &Y \\
    \hline
        Current input & ScriptCode &Y &Y &Y &Y &Y &Y \\
    \hline
        Current input & Previous output value & Y & Y & Y& Y& Y&Y \\
    \hline
        Current input & Sequence & Y &Y &Y & Y& Y&Y \\
    \hline
        Output at current input index & Output hash & Y & Y & N & Y&Y&N \\
    \hline
        Current Tx & Hash outputs &Y &N &N &Y &N &N \\
    \hline
        Current Tx & Locktime  & Y & Y&Y &Y &Y &Y \\
    \hline
        Current input & SIGHASH type of signature & Y&Y &Y &Y &Y & Y\\
    \hline
    \end{tabular}}
    \caption{Data committed to by signatures for each signature hash type when Segregated Witness version 0 semantics for {\tt OP\_CHECKSIG} are used. This includes spending P2WPKH and P2WSH output types. Note that the ScriptCode is the public key hash for P2WPKH outputs or script hash for P2WSH outputs. (I) means that the commitment is implicit by reference to other explicit commitments.}
    \label{tab:segwit_sighash}
\end{table}

The semantics for Script changed significantly with the Taproot soft-fork, and is referred to as TapScript. Here we are particularly interested in the semantics for signature operations and the signature message. The precise subset of data committed to in this case is presented for each signature hash type in table \ref{tab:taproot_sighash}.

\begin{table}
    \centering
    \resizebox{\textwidth}{!}{
    \begin{tabular}{|l|l|l|l|l|l|l|l|}
    \hline
        \rowcolor{lightgray} \textbf{Data Source} & \textbf{Data Type} & {\tt ALL} & {\tt SINGLE} & {\tt NONE} & {\tt A1CP||ALL} & {\tt A1CP||SINGLE} & {\tt A1CP||NONE} \\
    \hline
        Current Tx & Version & Y&Y &Y &Y &Y &Y \\
    \hline
        Current Tx & Hash previous outpoints &Y &Y & Y&N &N &N \\
    \hline
        Current Tx & Hash previous output amounts &Y &Y &Y &N &N &N \\
    \hline
        Current Tx & Hash previous output scriptPubKeys &Y &Y &Y&N &N &N \\
    \hline
        Current Tx inputs & Hash sequence (of all inputs) &Y &Y &Y &N &N &N \\
    \hline 
        Current Tx & Hash all outputs &Y &N &N &Y &N &N \\
    \hline
        Current input & Spend type & Y & Y&Y &Y &Y &Y  \\
    \hline
        Current input & Previous output outpoint & Y(I) & Y(I)&Y(I) &Y &Y &Y \\
    \hline
        Current input & Previous output amount & Y(I) & Y(I)&Y(I) &Y &Y &Y \\
    \hline
        Current input & ScriptPubKey & Y(I) & Y(I)&Y(I) &Y &Y &Y \\
    \hline
        Current input & Sequence & Y & Y&Y &Y &Y &Y \\
    \hline
        Current input & Index & Y &Y &Y &N &N &N \\
    \hline
        Current input & Annex (if present) & Y & Y & Y & Y & Y & Y \\
    \hline
        Output at current input index & Output hash & Y(I) & Y & N& Y(I) & Y& N \\
    \hline
    \end{tabular}}
    \caption{Data committed to by signatures for each signature hash type when TapScript semantics (Segregated Witness version 1) for {\tt OP\_CHECKSIG} are used. This includes spending from P2TR addresses. Note that `spend type' and `annex' are specific to Taproot spending rules and can be used in future upgrades to extend the data committed in signature messages. (I) means that the commitment is implicit by reference to other explicit commitments.}
    \label{tab:taproot_sighash}
\end{table}

These tables are relevant to covenant mechanisms because, as will be shown in \ref{sec:Deleted-Key-Covenants}, signatures can be used to commit to transaction data and enforce the constraints of a covenant. Understanding precisely what the signature message includes shows what can and can not be altered without invalidating the signature, and this helps us to appropriately encode restrictions in transactions. This is also relevant because other proposals (described in \ref{subsec:related_covenants}) aim to modify signature message types to enable consensus-enforced covenants.

\section{Related Work}
\label{sec:related_work}

\subsection{Bitcoin Covenant Mechanisms} 
\label{subsec:related_covenants}

An early proposal from Möser, Eyal, and Sirer \cite{moeser2016bitcoin} described {\tt OP\_CHECKOUTPUTVERIFY}, an operation which checks if the script of a specific output (in the covenant transaction) matches a template provided in the lock script committed to in the covenant-bound output. The template is specified for an output at a given index and ensures that it has a certain \textit{value} (in satoshis) and a \textit{pattern}. The pattern consists of operations and placeholders, where placeholders can represent either variable data (public keys, public key hashes) or a script pattern. The use of placeholder keys gives application designers the ability to dynamically insert new keys. The pattern placeholder enables recursive covenants. Note that this is a covenant mechanism that restricts only the output but not the rest of the transaction. This allows the covenant transaction to be modified in many ways without invalidating it. This can be both useful and dangerous, as will be discussed in section \ref{sec:safety-concerns}. What is not clear is whether or not {\tt OP\_CHECKOUTPUTVERIFY} introduces new denial-of-service attack vectors on Bitcoin nodes. If a script can be constructed that is very costly to execute, then attackers can use it to drain resources from nodes who are validating transactions. A systemic denial-of-service attack like this is critical to consider for new Script operations. 

More recent work proposed a covenant mechanism that requires introducing an operation {\tt OP\_CHECKSIGFROMSTACK} \cite{Covenants2} which checks a signature for a message found on the stack, rather than the usual signature message of a subset of transaction data. The trick here is to input the transaction data to the stack with the signature by including it in the unlock script and if the same signature can be verified using both {\tt OP\_CHECKSIG} and {\tt OP\_CHECKSIGFROMSTACK}, then the transaction data message must be the same. The covenant restrictions are committed to in the public key script. Given verified transaction data on the stack, we can ensure it has certain properties to enforce a covenant. This operation also enables recursive covenants by requiring the covenant-bound output's public key script to match the public key script of the covenant transaction's output. This work was implemented on a Bitcoin side-chain platform called Elements. The original proposal requires the concatenation operation, {\tt OP\_CAT}, which has long been disabled in Bitcoin due to concerns about it enabling unbounded resource consumption \cite{btc-spendable}.

Another Script extension to enable covenants is called {\tt OP\_CHECKTEMPLATEVERIFY} \cite{BIP119}. This operation will compute the hash of a subset of transaction data and check if it matches a 32-byte hash on the execution stack. The 32-byte hash is the commitment contained in the covenant-bound output. The subset of transaction data, the \textit{template}, is similar to the message types used for signatures but with slight differences. More templates may be added in future proposals. The proposal makes a conjecture that the default template is safe. This operation does not enable recursive covenants. In the default template, the commitment hash covers the entire transaction, except for the transaction ID. That means the covenant transaction can not be modified without invalidating it. The public key scripts in the covenant transaction's outputs must be completely specified up front. Compared with {\tt OP\_CHECKOUTPUTVERIFY} this mechanism has much less functionality, while being safer. There is a possibility, in future, to add new standard templates which commit to different subsets of transaction data. The current default template is designed for low validation costs, to avoid denial of service attacks and centralisation pressures. 

A signature can only be used as a commitment and contained in the output if the signature itself does not depend on that output. For all signature message types described in section \ref{subsec:sig_msg_types}, there is a circular dependency between the output and the signature that makes constructing a covenant impossible. However, one proposal is to optionally remove the references to the previous output from the signature message algorithm \cite{BIP-anyprevout} and this can be used to construct covenants. This alternative proposal for a covenant mechanism is based on the introduction of an {\tt ANYPREVOUTANYSCRIPT} signature message type \cite{BIP-anyprevout} as a TapScript upgrade. This means signature operations (e.g. {\tt OP\_CHECKSIG}) can verify a signature over a transaction message where the reference to the specific previous output is omitted. This means the output referenced by the transaction input can be changed without invalidating the signature, hence `any previous output'. Critically for covenants, that means a covenant-bound output can commit to a specific signature message and avoid creating a circular dependence between the commitment and the message being signed. 

Another research direction is unfolding wherein new operations are proposed that can push specific transaction data (hashed, in some cases) to the stack. This approach enables \textit{granular} covenants - committing only to single field values rather than to transaction signature message subsets or templates. Noteworthy proposals include {\tt OP\_PUSHTXDATA} \cite{BIP-PUSHTXDATA}, {\tt OP\_TXHASH} \cite{TXHASH}, and {\tt OP\_TX} \cite{OPTX}. However, most recently, the Bitcoin side-chain platform called Elements has implemented and activated a set of transaction introspection operations \cite{tapscript_elements} in its modified version of Bitcoin Script. There is an operation for each field of the transaction and each field of its inputs and outputs. We will refer to these operations as {\tt INSPECT\_X} where {\tt X} is a placeholder for the field whose value is pushed to the stack by the operation. A covenant mechanism using {\tt INSPECT\_X} works as follows. A covenant-bound output must commit to inspecting transaction data, and verifying only if the field value satisfies a constraint (e.g. it matches a target value). An example lock Script is {\tt INSPECT\_X <target\_value> EQUALVERIFY}. The `granular' property gives flexibility to application designers and enables recursive covenants. These operations were designed to not introduce unbound resource use, new disk access requirements, or complex control flow. However, the operations add complexity to the analysis of modified Bitcoin Script. It is an open question how to formally model and verify properties of this version of Script where a dependence on global transaction state has been introduced \cite{tapscript_elements_blog}. 

More generally, the design of Bitcoin's programming language is an active area of research and there is not broad consensus on the best approach to improving it. Some argue that the language should consist only of primitive building blocks, while others would prefer more complicated procedural operations. Numerous proposals for incremental improvements (operation by operation) have been discussed above, while other proposals aim for a total overhaul of the Script language. Simplicity is the primary example, purpose-built to be a constrained turing-incomplete language that supports static analysis and formal verification \cite{Simplicity}. 

Each of the previously proposed methods for implementing a covenant require some non-trivial modification to Bitcoin-core. Gathering community approval for changes to the consensus rules (even backwards compatible changes) has proven to be a slow process. These proposals demand an analysis of how the required modifications broaden the attack surface for Bitcoin users and to determine whether this is an acceptable cost for the additional functionality of a new financial tool. These proposals have not concretely presented the benefits they offer since no comparison has been made with what is already possible currently with bitcoin. Deleted-key covenants, the primary topic of this chapter, are currently possible and are appropriate for a baseline comparison.

The concept of a deleted-key covenant, the central theme of this chapter, was first introduced by McElrath \cite{P2TST}, although without formalization or a comprehensive security analysis. Recognizing its potential, I joined McElrath to further explore and develop the idea. Bishop, working on a similar prototype for signed transaction coordination \cite{BishopVaults}, and Hommel, with an investigation into how to integrate hardware signers with covenant functionality \cite{HommelVaultmbed}, also contributed to this collaborative endeavor. Our joint work resulted in two papers on Bitcoin covenants \cite{Swambo2020cov} and vault-based custody \cite{Swambo2020vault}. However, the content of this chapter goes beyond our initial collaboration. As detailed in section \ref{intro:statement-originality}, I have revisited and extensively revised our original findings with the agreement of my collaborators. This included updating the research to mirror current advances, refining the methodology, incorporating new analyses and evaluations, and adjusting the writing style \cite{Swambo2020cov,Swambo2020vault}. The enhancements in this chapter provide a richer and more nuanced insight into deleted-key covenants.

\subsection{Analysis and Applications of Covenants}
\label{subsec:anal-app-covenants}

Other work focuses on modeling, formal specification and verification for smart contracts implemented with Bitcoin Script and transactions. Tolmach \textit{et al.} provide a thorough and broad survey on the topic for several smart contract platforms including Bitcoin \cite{SurveySC}. Hu \textit{et al.} cover similar ground with their comprehensive survey on smart contract construction and execution \cite{SurveySC-21}. Bartoletti and Zunino provide a survey of formal models of Bitcoin contracts \cite{SurveyFormalBitcoinContracts}. 

A relevant direction for research is to extend these formal models to handle various covenant mechanisms, and verify properties in the design of Scripts and contracts that use those mechanisms. To this end, Bartoletti \textit{et al.} extend their high-level contract language, BitML \cite{BitML}, to express {\tt OP\_CHECKOUTPUTVERIFY}-like covenants \cite{CovenantsUnchained}. They claim that model-checking techniques based on exploring the complete state-space are not suitable for analysing relevant security properties of contracts with \textit{recursive} covenants. In general, formal verification tools will be more complicated for recursive covenants. Given a formal model of bitcoin Script with covenants enabled, it would be more feasible to answer critical questions, such as; `can we verify that this unspent transaction output is spendable?' \cite{btc-spendable}, or, `can we ensure the script execution always terminates so that denial-of-service attacks are not unwittingly enabled?'.

Depending on the breadth of functionality of any covenant mechanism that is added to Bitcoin, numerous new applications will become possible or practical. This includes advanced custody functionality (for individuals and groups), new payment protocols to improve network throughput and privacy, fungible and non-fungible token and exchange protocols, and advanced oracle protocols.  

An interesting covenant for custody (introduced in \cite{moeser2016bitcoin}) is called a \textit{vault}. The basic idea is to create a covenant-bound output that can be spent in one of two ways; a time-locked spend to release funds from the wallet, or an immediate spend to a highly secured destination address. A thief trying to release funds from the wallet would be forced to wait out the time-lock, giving the wallet owner an opportunity to thwart the attack by pushing their bitcoin immediately to a secured address. This particular use-case will be explored in detail in Chapter \ref{ch:vault-custody}.

Another example use case within custody is `secure dynamic access structures', which were shown to be possible with certain types of covenant mechanisms \cite{ParalysisProofs}. This is a way to change, in a secure manner, the access control structure of some bitcoins when a participant is provably absent. For example, if there is a 4 out of 4 multi-signature company custody system but one of the participants resigns, then the policy can be converted to be 3 out of 3 by providing a proof that the fourth participant has been absent for sufficient time. 

An infinitely recursive covenant would diminish the fungibility of the bitcoin it controls. The bitcoin it controls could be permanently subjected to restrictions in how they are transferred. In the worst case, this would make the coins unacceptable for payments. Since users must share a receiving address for payments, it is not feasible to trick a user to accept coins bound by a recursive covenant. On the other hand, this could be viewed as a feature as it would enable distinct tokens to be represented on the blockchain as discussed by Bartoletti \textit{et al.} \cite{SoundBitcoinTokens}. Moreover, it would be possible to enable an exit-path in a recursive covenant. This would mean the fungibility could be regained.

A notable design for a covenant-based application could bring \textit{decentralised exchange} to the Bitcoin network. `Bitmatrix' is an automated market maker based on recursive covenants \cite{Bitmatrix}. In addition to recursive covenants, the design requires support for asset issuance provided by the Liquid Network, a Bitcoin sidechain, which Bitcoin itself does not have \cite{Liquid}. The application is a smart contract which controls assets in liquidity pools, sets the prices of those assets, and is prepared to trade with users. For users, they need not relinquish custody of their assets in order to trade; the application is \textit{non-custodial}. 

Another popular layer 2 protocol within Bitcoin is Discreet Log Contracts (DLCs) \cite{DLC}, which could see 30 to 300 times efficiency improvements if non-recursive covenants were enabled \cite{DLC-covs}. Put simply, DLCs are a way to facilitate conditional payments where the conditions are based on data external to the blockchain, provided by data oracles. This could become a critical infrastructure piece for executing financial smart contracts underwritten with bitcoin.

Within the domain of payment protocols, which generally aim to improve throughput and privacy of payments on the bitcoin network, speculative proposals where covenants could be useful are in CoinPool \cite{CoinPool} and Eltoo payement channels \cite{decker2018}.

\section{Deleted-key Covenants}
\label{sec:Deleted-Key-Covenants}

\subsection{Overview}
The \textit{Deleted-key Covenant} protocol, introduced here, offers a unique approach which differs from consensus-enforced covenants. Distinctly, the covenant constraints are enforced by procedures external to consensus, rather than with Script. The protocol involves the creation of a covenant transaction with stipulated constraints, the generation of a signature for this transaction using an ephemeral private key, and subsequent deletion of this key. The covenant transaction is only verifiable with this signature, which we call the \textit{enforcement signature}. Once the covenant-bound output has been funded by the deposit, control is dictated solely by the signed covenant transaction since no further signatures can be produced for transactions involving the same output. This transaction may be stored until circumstances permit its broadcast. Although the protocol specification in Section \ref{subsec:protocol-spec} is applicable to multiple participants, the underlying concept remains consistent. 
 
\subsection{Roles and Notation}
\label{subsec:roles-notation}

The protocol comprises three participant roles:

\begin{quote}
     \textbf{Depositor:} Responsible for the construction of the covenant-bound output and its funding.\\
     \textbf{Enforcer:} Duties involve the creation of the off-chain commitment to the covenant transaction through its signature and the deletion of the associated private key.\\
     \textbf{Custodian:} Holds the covenant transaction and, when it is time to broadcast it, participates in finalising its authorisation, thereby rendering it valid for inclusion in the blockchain.
\end{quote}

There may be one or more participants with the role of \textit{enforcer} and \textit{custodian}. For simplicity, we will consider only a single participant with the role of \textit{depositor}\footnote{An extension to the protocol with multiple depositors is possible but the added complexity brings new attack vectors which require mitigating. This obfuscates what is \textit{necessary} in the single depositor case. Moreover, multiple instances of the protocol (as presented) can be enacted in order to fund covenants from multiple depositors.}. A single participant may take multiple roles.

Let there be $k$ custodians, labelled by $i \in \{1,2,...,k\}$, each with a public key $Q_i$ in the set of custodial public keys $\mathcal{Q}$. Let there be $n$ enforcers, labelled by $l \in \{1,2,...,n\}$, each with an ephemeral private key $p_l$ and corresponding public key $P_l$ in the set of enforcement public keys $\mathcal{P}$.

\begin{figure}
    \centering
    \includegraphics[width=0.35\linewidth]{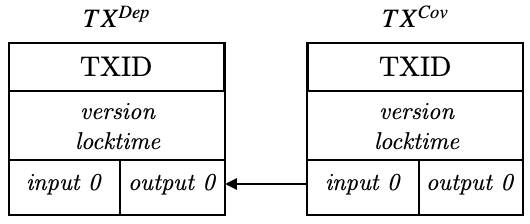}
    \caption{Diagram depicting a simple covenant. The deposit transaction, $TX^{Dep}$, creates an output with a lock script, $l_{cov} = l_{enf}||l_{cust}$, which contains the set of enforcement public keys $\mathcal{P}$ and custodial public keys $\mathcal{Q}$. The input of the covenant transaction, $TX^{Cov}$, references the output from the deposit. }    
    \label{fig:simple-covenant}
\end{figure}

Let the covenant transaction and deposit transaction be called $TX^{Cov}$ and $TX^{Dep}$, respectively. These are depicted in figure \ref{fig:simple-covenant}. The covenant-bound output in $TX^{Dep}$ commits to a lock script $l_{cov}$ which is the concatenation (denoted by $||$) of two subscripts; the enforcement script $l_{enf}$ and the custodial script $l_{cust}$ such that $l_{cov} = l_{enf}||l_{cust}$. Generally, $l_{enf}$ represents a multi-signature policy over $\mathcal{P}$ while $l_{cust}$ represents multi-signature access policy over $\mathcal{Q}$. 

Recall that there are two approaches to implementing a multi-signature lock with Taproot outputs; one using an aggregate public key for key-path spending, and another using a multi-signature TapScript for script-path spending. To avoid confusion with the security model of threshold multi-signature schemes let us use script-path spending. Thus, $l_{enf}$ should take the form

\begin{equation}
\label{lenf}
    l_{enf} = P_1 \textit{ } {\tt CHECKSIG } \textit{ } P_2 \textit{ } {\tt CHECKSIGADD ...}\textit{ } P_n\textit{ } {\tt CHECKSIGADD}\textit{ } m \textit{ }{\tt NUMEQUAL}
\end{equation}

This is an $m$-of-$n$ multi-signature lock script with the enforcement public keys. Similarly, $l_{cust}$ should take the form

\begin{equation}
\label{lcust}
    l_{cust} = Q_1 \textit{ } {\tt CHECKSIG } \textit{ } Q_2 \textit{ } {\tt CHECKSIGADD ...}\textit{ } Q_k\textit{ } {\tt CHECKSIGADD}\textit{ } j \textit{ }{\tt NUMEQUAL}
\end{equation}

which represents a $j$-of-$k$ multi-signature lock script with the custodial public keys. 

\subsection{Enforcement Conditions}
\label{subsec:enf-cond}

The objective of the protocol is to yield an \textit{active} covenant transaction. A deleted-key covenant is deemed to be active if the following enforcement conditions hold:

\begin{enumerate}
    \item There exists a covenant transaction, denoted $TX^{Cov}$, which encodes the covenant constraints. Furthermore, the witnesses field of $TX^{Cov}$ is populated with enforcement signatures that satisfy the enforcement subscript $l_{enf}$.
    \item Each enforcement private key, denoted $p_l$, is deleted. As such, no new signatures can be generated to satisfy the enforcement subscript.
    \item The deposit transaction, denoted $TX^{Dep}$, is included in a block that is sufficiently deep in the blockchain, such that it is highly improbable for the transaction to be removed or modified\footnote{The generally accepted depth in the Bitcoin network is six blocks. This convention is predicated on the principle that it would be highly improbable for a malicious entity to amass enough computational power to overwrite six blocks simultaneously. This depth thus provides sufficient assurance of the transaction's permanence.}.
\end{enumerate}

\subsection{Protocol Specification}
\label{subsec:protocol-spec}

\begin{figure}
    \centering
    \includegraphics[width=\textwidth]{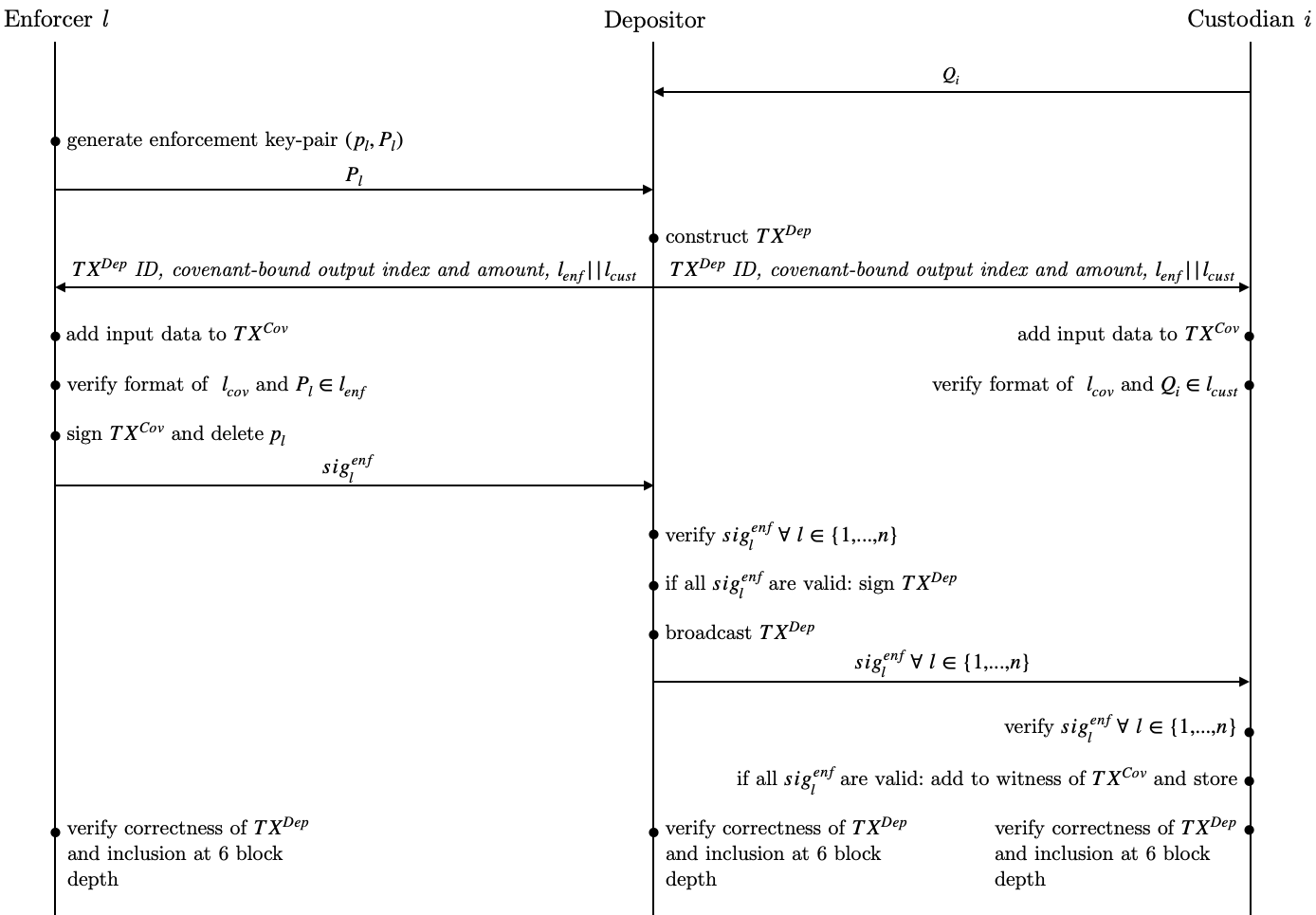}
    \caption{Deleted-key covenants protocol. The enforcer and custodian roles here represent multiple participants while there is only a single depositor. Each enforcer and custodian act independently, though performing the same actions as others with the same role. For example, each custodian will send their public key $Q_i$ to the depositor in the first step, and each enforcer will generate their ephemeral key pair $(p_l,P_l)$ in the second step. At the termination of this protocol, the custodians have control over a covenant-bound output.}
    \label{fig:deleted-key-protocol}
\end{figure}

Figure \ref{fig:deleted-key-protocol} depicts a Message Sequence Chart (MSC)\footnote{A Message Sequence Chart is a form of representation for communicative interactions between autonomous entities formally defined and provided with a semantics under International Telecommunications Union (ITU) Standard Recommendation ITU-T Z.120 \cite{MSC}.} representing the protocol for constructing a deleted-key covenant among multiple participants. In this figure, the enforcer and custodian roles represent multiple participants, with a single depositor for simplicity. Each enforcer and custodian act independently, performing the same actions as their counterparts. 

The protocol involves a series of clearly defined steps:

\begin{enumerate}
    \item Each custodian communicates their custodial public key $Q_i$ to the depositor.
    \item Each enforcer generates an enforcement key pair $(P_l, p_l)$ and dispatches $P_l$ to the depositor.
    \item The depositor assembles $TX^{Dep}$, ensuring that it creates and spends only Taproot output types. The covenant-bound output pays to an address derived from a lock script $l_{cov}=l_{cust}||l_{enf}$.
    \item The depositor communicates the deposit transaction details (transaction ID, covenant-bound output index and amount, and $l_{cov}$) to each enforcer and custodian.
    \item Each enforcer and custodian uses this `previous output' data to fill the input field of $TX^{Cov}$.
    \item Each enforcer and custodian verify the Script format of $l_{cov}$ and the inclusion of their public key in the correct location of $l_{cov}$. 
    \item Each enforcer produces an enforcement signature $sig_l^{enf}$ for $TX^{Cov}$, deletes $p_l$, and sends the signature to the depositor.
    \item The depositor verifies each enforcement signature. If all are legitimate, the depositor signs $TX^{Dep}$ and broadcasts it.
    \item The depositor transmits the set of enforcement signatures to the custodians.
    \item Each custodian verifies the set of enforcement signatures. Once validated, the custodian adds them to the witnesses field of $TX^{Cov}$ and stores the transaction.
    \item All participants verify the correctness of $TX^{Dep}$ and await its inclusion in a block with a depth of 6 in the blockchain. Only then they will recognise $TX^{Cov}$ as an active covenant transaction. 
\end{enumerate}

\subsection{Potential Applications}

This section presents some potential real-world applications of the deleted-key covenant protocol. In each use case, the covenant constraints and the roles of participants are discussed. Note that these applications have not been deployed. We sketch them here to demonstrate the potential utility of the protocol. 

\subsubsection{White-list}

In this application, we utilise a `pay-to' covenant which is characterized by a constraint that dictates a specific amount of bitcoin to be strictly directed to a predetermined address. This can be achieved by assigning the public key script field in the output of $TX^{Cov}$ to the desired address, and likewise setting the amount field in the output. 

An instance where this covenant can prove advantageous is within a firm operating an autonomous custody system, which seeks to impose stringent rules on its outflows. Broadening the application, a collection of `pay-to' covenant transactions could be constructed that spend the same covenant-bound output, thereby enabling the firm to establish a white-list of acceptable recipient addresses. In this scenario, the firm would assign the role of enforcers and custodians to multiple internal entities, which could be either employees or servers. The depositor would be one or more internal entities. This application considerably mitigates the risks associated with the custodian role, as compared to a situation where the white-list is not enforced. 

The principal advantage of this application is the heightened control over fund allocation, which significantly reduces the risk of malicious or erroneous transactions by imposing a rigid white-list of recipient addresses. However, this stringent control could potentially be seen as a limitation, as it restricts the ability to transact with entities outside of the white-list. Despite this, in scenarios where security is of paramount importance—such as in large corporations or institutions managing substantial assets—the trade-off could be deemed acceptable.

\subsubsection{Insured Custody}

In a slightly distinct variation of the previous application, a firm desiring access to more cost-effective insurance solutions can setup a white-list and invite external entities such as an insurance company and an independent auditor to assume the role of enforcer. The firm, meanwhile, fulfills the roles of the depositor and custodian. With the enforcers ensuring adherence to the white-list covenant constraints, an additional layer of security is added to the firm's custody operation. This additional security provides the insurance company with sufficient assurance to underwrite an insurance policy for the firm at a competitive rate. The incorporation of an independent auditor further augments the checks and balances, offering a crucial resource in the event of a dispute between the firm and the insurance company. 

\subsubsection{Time-Locked Vault}

The `vault covenant' concept, discussed in section \ref{subsec:anal-app-covenants} during the review of related work, has been promoted as an appealing application for covenants with the potential to significantly enhance custody systems. 

For implementation using a deleted-key covenant, the output for $TX^{Cov}$ must commit to a specific type of lock script. This lock script is designed to allow spending of the output via one of two distinct methods. The first method enables immediate spending of the output by utilizing a specifically designated offline key. Conversely, the second method grants spending access to a designated online key, subject to a time-lock constraint, meaning the funds can only be moved after a specified period.

One of the pivotal characteristics of the time-locked vault covenant is its capacity to effectively deter theft attempts. Given a sufficiently robust network monitoring system and a swift response procedure, attempted thefts—which are typically aimed at the online key—can be detected promptly. The detected theft can then be prevented by utilizing the priority granted to the offline key. The offline key is designed to have precedence over the online key and can be used to spend the funds before the expiration of the time-lock, effectively circumventing the theft attempt.

This promising application serves as the focus for chapter \ref{ch:vault-custody}. In that chapter, we will conduct a comprehensive investigation into the advantages and disadvantages of integrating a vault covenant into a custody system. 

\section{Security Analysis}
\label{sec:DKC-security-analysis}
\subsection{Security Analysis Methodology}

We adopt a rigorous, structured approach to define the concept of `security' for the deleted-key covenant protocol. We delineate four specific security objectives, providing the foundations for a comprehensive analysis of conditions under which they can be satisfactorily achieved. The analysis is premised on explicit assumptions (refer to \ref{subsec:covenants-assumptions}) and a comprehensive threat model (refer to \ref{subsec:covenants-threat-model}). Our objectives and corresponding analyses span the entire life-cycle of a deleted-key covenant, including its establishment, active period, and its execution.

\begin{quote}
\textbf{Objective 1:} Ensure the integrity of protocol enactment.
\end{quote}
Under this objective, we seek to ascertain the risk, characterized by both impact and likelihood, associated with incorrect enactment of the protocol. The focus is directed toward the procedures undertaken up to the potential activation of the covenant. In particular, we reveal the level of assurance participants have in knowing whether or not the protocol was enacted with integrity. Our analysis is systematic and considers potential malicious activities of participants and external entities that might endanger the protocol enactment.

\begin{quote}
\textbf{Objective 2:} Protect honest participants against losses.
\end{quote}
With this objective we are still focused on risks that materialise during the protocol enactment. We aim to ensure that honest participants are shielded from any potential harm. More specifically, we seek to preclude scenarios where either external attackers or malicious participants can steal from or prevent access to honest participants' Bitcoin or sensitive data. Even in the event where an attack perpetrated by a malicious entity halts the protocol, our aim is to prevent any losses from befalling honest participants.

\begin{quote}
\textbf{Objective 3:} Guarantee the irrevocability of covenant enforcement.
\end{quote}
The focus of this objective lies in the active period of the covenant. We will argue for the security of the covenant enforcement predicated on the successful enactment of the protocol. The analysis evaluates the complexity and feasibility of breaching the enforcement conditions once they are fully established.

\begin{quote}
\textbf{Objective 4:} Maintain honest custodial control for covenant execution.
\end{quote}
The ultimate purpose of the deleted-key covenant is to ensure that the custodians retain the ability to execute $TX^{Cov}$ when a sufficient threshold of them choose to do so. This objective maintains that the practical utility of the covenant is preserved and provided to the custodians, according to their decision-making process.

\subsection{Assumptions}
\label{subsec:covenants-assumptions}

We categorize our assumptions into five sections: Communication, Cryptography, consensus protocol, key-management, and transaction data.

\textbf{Communication:}
We assume participants have authenticated and private communication channels with each other where necessary, for every message. While secure communication is a well-understood problem with numerous solutions, note that this assumption hides some of the burden of preparing to enact the protocol. With many participants that burden can be cumbersome, but the same set of secure channels can be re-used for multiple instances of deleted-key covenants.

\textbf{Cryptography:} 
Our cryptographic assumptions rest on the perceived security of the cryptographic primitives employed in the Bitcoin network, which were described in Section \ref{sec:cryptography}. Key among these assumptions is the intractability of the Elliptic Curve Discrete Logarithm Problem (ECDLP) in the context of the secp256k1 curve used by Bitcoin. This hardness underpins our confidence that digital signatures cannot be forged and that private keys cannot be inferred even when the associated public key is known. We further assume that the hash functions employed within the Bitcoin protocol, particularly those used in deriving public key scripts from Bitcoin Script, exhibit strong collision resistance, precluding the viable generation of distinct inputs that hash to the same output.

\textbf{Consensus protocol:}
We assume that each participant runs their own Bitcoin server which can verify blocks and transactions according to the consensus rules. In particular, they can verify proof-of-work and the inclusion of transactions in the blockchain.

\textbf{Key-management:}
We assume that all custodial keys are `fresh', i.e., that they have not been used before in other transactions or any other protocol. This ensures that if these keys leak, or when they are revealed on-chain once custodians broadcast $TX^{Cov}$, no additional sensitive information will be disclosed such as other public keys they control.

We assume that the depositor's bitcoin private keys have not been compromised before the covenant is active. This is a reasonable assumption because an attacker that captures the depositor's signing device can simply steal all of the depositor's funds, rendering any analysis of the covenant protocol useless.

\textbf{Transaction data:}
We assume that the templates of $TX^{Cov}$ and $l_{cov}$ are agreed upon by all participants prior to enacting this protocol. The procedures for this will depend on the application in which the protocol is used. The specific input data which references the covenant-bound output and the keys used in the locking script $l_{enf}||l_{cust}$ are shared during the protocol. All protocol participants will know the public key scripts for each output of $TX^{Cov}$. 

We assume that the lock scripts for each output of $TX^{Cov}$ are known in advance by the entities that will eventually unlock those outputs.

We further assume that $TX^{Dep}$ spends outputs that are not correlated with others in the depositor's custody system.

\subsection{Threat Model}
\label{subsec:covenants-threat-model}

We scrutinize the Deleted Key Covenant protocol within the context of both a network protocol and a security ceremony, applying the robust Dolev-Yao threat model to the network protocol component, while considering endpoint compromise of user devices (e.g., through physical attacks) in the security ceremony aspect.

This adversarial model endows the attacker with all the capabilities delineated in Table \ref{tab:attacker-capabilities}. Essentially, the attacker can eavesdrop, initiate, atomically break down messages, conduct cryptographic operations, block, spoof, and reorder messages. This comprehensive set of capabilities allows for a systematic exploration of possible attack vectors and helps to outline distinct risks to which the protocol may be susceptible.

Despite the broad set of capabilities we assign to the attacker, there is a pivotal assumption: communication channels among participants are authenticated and encrypted. This assumption implies that while an attacker can intercept messages, they cannot manipulate or tamper with the content of these messages in transit without detection. Traditional man-in-the-middle attacks are thus possible only if the attacker compromises a participant or their device, thereby gaining access to their secure communication credentials. In such circumstances, the attacker could spoof the identity of the compromised participant. This is why we also consider the security ceremony aspect, extending the capabilities of the adversary beyond the traditional Dolev-Yao model. Consequently, our analysis considers the potential risks and vulnerabilities should one or more participants act with malevolent intent.

\subsection{Ensure the Integrity of Protocol Enactment}

In our approach to analyzing protocol enactment integrity, we systematically examine the potential for corruption or compromise of the participant types: Depositor, Enforcers, and Custodians. Additionally, we explore the implications of collusion between compromised participants. 

To structure this analysis, we put forth several claims associated with aspects of protocol enactment integrity. Each claim is followed by an evaluation of corresponding vulnerabilities and potential attacks that might undermine the verification process, culminating in the minimum viable attack that invalidates the claim. This methodology aligns with the `proof by contradiction' approach: we assume each claim to be valid and then demonstrate its invalidity under specific adversarial conditions. By structuring our analysis in this manner, we ensure a comprehensive and systematic examination of the protocol's robustness against a defined set of adversarial scenarios, thereby underlining the protocol's enactment integrity.

A fundamental aspect of our analysis pertains to the verification of the permanence of $TX^{Dep}$, which contains the covenant-bound output. The means through which each participant ensures the permanence of $TX^{Dep}$ is a crucial piece for the analysis. Specifically, a participant's Bitcoin server maintains synchronization with the Bitcoin network. Proof-of-work incurs a substantial cost, thus staying synchronised mitigates the adversary's ability to deceive about the blockchain state, and in particular of $TX^{Dep}$. The server actively searches for $TX^{Dep}$ within each block. Upon detection, it verifies the proof-of-work for the subsequent 5 blocks.

\textbf{Claim:} Each participant's Bitcoin server can verify that $TX^{Dep}$ is included sufficiently deep in the blockchain (6 blocks), or detect that they are suffering an attack.

The subsequent two tables enumerate possible vulnerabilities and attacks related to this claim.

\begin{table}[H]
\centering
\begin{tabular}{|>{\columncolor{lightgray}\raggedleft\arraybackslash}p{3.5cm}|p{11.5cm}|}
\hline
    \textbf{Vulnerability A(i)}& A Denial-of-Service (DoS) attack could disrupt a participant's connection to the Bitcoin network, thereby inhibiting their server's ability to maintain synchronization. This disruption would prevent the participant from verifying the inclusion of $TX^{Dep}$ in the blockchain. \\
    \textbf{Attack A(i)}& In this attack, the adversary obstructs all incoming messages to a participant, consequently causing a state of network isolation.\\
\hline
\end{tabular}
\end{table}

\begin{table}[H]
\centering
\begin{tabular}{|>{\columncolor{lightgray}\raggedleft\arraybackslash}p{3.5cm}|p{11.5cm}|}
\hline
    \textbf{Vulnerability A(ii)}& A Sybil attack could lead the participant to receive counterfeit updates from the Bitcoin network, thereby disrupting their server's ability to maintain synchronization. This disruption would prevent the participant from verifying the inclusion of $TX^{Dep}$ in the blockchain. \\
    \textbf{Attack A(ii)}& In this attack scenario, an adversary spawns multiple fake nodes on the Bitcoin network, thereby inundating the participant with falsified network information. \\
\hline
\end{tabular}
\end{table}

If we evaluate the aforementioned vulnerabilities and corresponding attacks, it's apparent that even these potent threats do not invalidate our claim. In the case of a Denial-of-Service attack, the participant's connection to the Bitcoin network might be interrupted, but they would likely recognize the disruption, thus preventing them from mistakenly attributing `integrity' to the protocol enactment. Similarly, under a Sybil attack, while the participant may be inundated with falsified network information, a participant could recognize insufficient proof-of-work associated with falsified blocks and detect the attack.

Therefore, considering the substantial hurdles for an adversary to succeed in these powerful attacks, coupled with the likely detection of such attacks by vigilant participants, we uphold our claim.

\subsubsection{Part 1: The custodians}

\textbf{Procedure summary:} Upon receipt of $l_{cov}=l_{enf}||l_{cust}$, each custodian verifies the inclusion of their public key $Q_i$ within $l_{cust}$ and checks if $l_{cov}$ matches with the correct subscripts \ref{lenf} and \ref{lcust}. They subsequently receive and validate the enforcement signatures $sig^{enf}_l$ for $TX^{Cov}$. Next, the custodian validates $TX^{Cov}$, which spends the covenant-bound output generated by $TX^{Dep}$, ensuring compliance with consensus rules. Finally, they verify the permanence of $TX^{Dep}$.  

\textbf{Claim:} Each custodian is capable of verifying the inclusion of their custodial public key in the covenant-bound output's script, and the correctness of the format of $l_{cov}$. 

\begin{table}[H]
\centering
\begin{tabular}{|>{\columncolor{lightgray}\raggedleft\arraybackslash}p{3.5cm}|p{11.5cm}|}
\hline
    \textbf{Vulnerability B}& A custodian might be deceived regarding the inclusion of their public key $Q_i$ within $l_{cust}$ or the format of $l_{cov}$. \\
    \textbf{Attack B(i)}& An attacker impersonates the depositor, altering the contents of $l_{cov}$ in $TX^{Dep}$, but then forwards the custodian a correct version of $l_{cov}$ that includes $Q_i$. This assault necessitates compromising the depositor's authentication and encryption credentials. \\
    \textbf{Attack B(ii)}& Alternatively, the depositor themselves could alter the contents of $l_{cov}$ in $TX^{Dep}$ but provide the custodian with a correct version of $l_{cov}$ containing $Q_i$. \\
\hline
\end{tabular}
\end{table}

Based on our evaluation, even in the face of the potential attacks B(i) and B(ii), our claim remains valid. This is because, even though a custodian might be deceived temporarily, the permanent record of $TX^{Dep}$ in the blockchain ensures that any deceit is eventually uncovered. Thus, each custodian is indeed capable of confirming the inclusion of their custodial public key in the covenant-bound output's script, supporting the integrity of the protocol enactment.

\subsubsection{Part 2: The enforcers}

\textbf{Procedure summary:} An enforcer generates a key-pair, $(P_l, p_l)$, and transmits $P_l$ to the depositor. Upon receiving the transaction ID, covenant-bound output index and amount, and $l_{cov}$ from the depositor, the enforcer verifies the inclusion of $P_l$ in $l_{enf}$ and checks if $l_{cov}$ matches with the correct subscripts \ref{lenf} and \ref{lcust}. The enforcer then generates $sig_l^{enf}$ and deletes $p_l$. Finally, they verify the permanence of $TX^{Dep}$.  

\textbf{Claim:} Each enforcer is capable of validating the inclusion of their own enforcement public key, $P_l$, in the covenant-bound output's script, and the correctness of the format of $l_{cov}$.

\begin{table}[H]
\centering
\begin{tabular}{|>{\columncolor{lightgray}\raggedleft\arraybackslash}p{3.5cm}|p{11.5cm}|}
\hline
    \textbf{Vulnerability C}& An enforcer might be deceived regarding the inclusion of their public key $P_l$ within $l_{enf}$ or the format of $l_{cov}$. \\
    \textbf{Attack C(i)}& An attacker may impersonate the depositor and modify the contents of $l_{cov}$ in $TX^{Dep}$. However, they send the enforcer a correct version of $l_{cov}$ that includes $P_l$. This attack requires the compromise of the depositor's authentication and encryption credentials. \\
    \textbf{Attack C(ii)}& Alternatively, the depositor could alter the contents of $l_{enf}$ in $TX^{Dep}$ and provide the enforcer with a correct version of $l_{cov}$ that includes $P_l$. \\
\hline
\end{tabular}
\end{table}

In creating an enforcement signature, $sig_l^{enf}$ for $TX^{Cov}$, the enforcer relies on a version of $TX^{Dep}$ which has a covenant-bound output that commits to $l_{enf}$ inclusive of $P_l$, with the correct format for $l_{cov}$. This signature is only valid for the said transaction. Therefore, the impact of such attacks is that a compromised or corrupt depositor could exclude the enforcer from the protocol. However, when comparing the version of $TX^{Dep}$ in the blockchain to the expected version, the enforcer would become aware of this discrepancy.

This analysis indicates that, despite potential attacks, the enforcer is capable of validating the inclusion of their enforcement public key, $P_l$, in the covenant-bound output's script. Therefore, the claim holds true, reinforcing the integrity of the protocol.

\subsubsection{Part 3: The depositor}

\textbf{Claim}: the depositor can exclude participants, but the misbehaviour will be discovered, thus the integrity of protocol enactment can be ensured. 

\textbf{Scenario:} The depositor acts as a man-in-the-middle, specifically targeting enforcers within the protocol.

\begin{table}[H]
    \centering
\begin{tabular}{|>{\columncolor{lightgray}\raggedleft\arraybackslash}p{3.5cm}|p{11.5cm}|}
    \hline
        \textbf{Vulnerability G}& Depositor can manipulate inclusion and exclusion of enforcers within the protocol. \\
        \textbf{Attack G}& The depositor may deceive enforcers by impersonating others, replacing $m$ genuine enforcement public keys $P_l$ with malicious counterparts $P'_l$. \\
        \textbf{Impact G}& This allows the depositor to forge enforcement signatures, effectively nullifying the covenant activation. Honest, but excluded enforcers are aware of exclusion, but included ones may misattribute integrity to the protocol. \\
        \textbf{Countermeasure G}& Enforcers could communicate out-of-band to detect the subversion, ensuring the integrity of protocol enactment. \\
    \hline
    \end{tabular}
\end{table}

Countermeasure G is pivotal in safeguarding the integrity of protocol enactment against deceptive practices by the depositor. Taking the example of insured custody, with a 2-of-2 enforcement subscript involving an auditor and an insurance firm, we see that this countermeasure is plausible and straight-forward to implement. If any discrepancies are detected, the parties need only communicate with each other, negating the impact of the attack. Therefore, despite being a straightforward attack, it is mitigated by collaboration and vigilance among participating enforcers.

\textbf{Scenario:} The depositor acts as a man-in-the-middle, specifically targeting custodians within the protocol.

\begin{table}[H]
    \centering
    \begin{tabular}{|>{\columncolor{lightgray}\raggedleft\arraybackslash}p{3.5cm}|p{11.5cm}|}
    \hline
        \textbf{Vulnerability H}& Depositor can manipulate inclusion and exclusion of custodians within the protocol. \\
        \textbf{Attack H(i)}& The depositor may deceive custodians by impersonating others, replacing $j$ genuine custodial public keys $Q_i$ with malicious counterparts $Q'_i$. \\
        \textbf{Impact H(i)}& The depositor controls the final authorization of the covenant transaction. Honest but excluded custodians are aware, but included ones may misattribute integrity to the protocol. \\
        \textbf{Attack H(ii)}& The depositor may deceive custodians by impersonating others, replacing $k-j+1$ genuine custodial public keys $Q_i$ with malicious counterparts $Q'_i$. \\
        \textbf{Impact H(ii)}& The depositor can block the finalisation of the covenant transaction. Honest but excluded custodians are aware, but included ones may misattribute integrity to the protocol. \\
        \textbf{Countermeasure H}& Custodians can communicate out-of-band, allowing any excluded party to complain and ensure that all are included. \\
    \hline
    \end{tabular}
\end{table}

Consider a collaborative custody setting where custodians want to set-up a multi-signature access policy with a covenant that stipulates that funds must be sent to a specific address. Communication is expected among custodians, and countermeasure H proves to be a simple and effective way to ensure the integrity of the protocol enactment. Any excluded custodian would recognize the exclusion and initiate communication with others, thereby flagging the anomaly. This reinforces the idea that the outlined attack, although plausible, is not necessarily detrimental. The success of countermeasure H relies on the cooperative nature of the participating custodians and the facility for out-of-band communication.

The presented analysis demonstrates that while the depositor has the ability to manipulate inclusion and exclusion of participants, the defined countermeasures can effectively neutralize these attacks. By enabling out-of-band communication, collaboration among enforcers and custodians can ensure the claim is upheld. This is critical to consider in the design of applications based on deleted-key covenants. 

\subsubsection{Part 4: Collusion}
\label{subsubsec:collusion}

Now, let us highlight a limitation inherent to the system: despite the ability of each custodian and each enforcer to independently verify the integrity of their inclusion in $l_{cust}$ and $l_{enf}$ respectively, they lack the means to substantiate the integrity of other participants. Similarly, while enforcers can be confident in the deletion of their private enforcement key, $p_l$, they possess no mechanism to provide irrefutable evidence of such deletion to other participants. Consequently, this necessitates certain trust assumptions regarding custodial control and covenant enforcement, which we shall further elucidate.

\textbf{Claim:} The integrity of the protocol enactment relies on a sufficient number of custodians and enforcers remaining uncompromised. Specifically, their authentication and encryption keys should stay secure. The protocol can withstand the compromise of up to (but not including) $min(j, k-j+1)$ custodians, and $m$ enforcers, without the integrity of the protocol enactment being undermined.

\begin{table}[H]
    \centering
    \begin{tabular}{|>{\columncolor{lightgray}\raggedleft\arraybackslash}p{3.5cm}|p{11.5cm}|}
    \hline
        \textbf{Vulnerability D}& Custodians deceiving the depositor and enforcers. \\
        \textbf{Attack D(i)}& Compromise authentication and encryption keys of $j$ custodians. Replace $j$ custodial public keys $Q_i$ with attacker-owned public keys $Q'_i$. \\
        \textbf{Impact D(i)}& The attacker gains custodial control. However, $TX^{Cov}$ remains bound by the covenant enforcement. The attacker can prematurely broadcast $TX^{Cov}$, disrupting a time-sensitive protocol.\\
        \textbf{Attack D(ii)}& Compromise authentication and encryption keys of $k-j+1$ custodians. Replace $k-j+1$ custodial public keys $Q_i$ with attacker-owned public keys $Q'_i$. \\
        \textbf{Impact D(ii)}& The attacker can refuse to authorise $TX^{Cov}$, resulting in the deposited funds being effectively burned.\\ 
    \hline
    \end{tabular}
\end{table}

\begin{table}[H]
    \centering
    \begin{tabular}{|>{\columncolor{lightgray}\raggedleft\arraybackslash}p{3.5cm}|p{11.5cm}|}
    \hline
        \textbf{Vulnerability E}& Enforcers deceiving the depositor and custodians \\
        \textbf{Attack E}& Compromise authentication and encryption keys of $m$ enforcers. Replace $m$ enforcement public keys $P_l$ with attacker-owned public keys $P'_l$ and covertly refuse to delete the associated enforcement private keys.\\
        \textbf{Impact E}& The attacker obstructs the covenant enforcement. The attacker can generate new signatures to satisfy $l_{enf}$ for transactions that are not constrained by the covenant. However, the attacker would not be able to satisfy $l_{cust}$, and would thus not be able to spend from $TX^{Dep}$. \\
    \hline
    \end{tabular}
\end{table}

Upon examining the attacks D(i), D(ii) and E and their potential impacts, we find that the integrity of the protocol's enactment can be preserved provided a sufficient number of custodians and enforcers are not compromised. While the attacks undermine the integrity of the protocol enactment, they are the minimum viable attacks. Any less than that would not undermine the integrity of the protocol. Hence, the claim holds.

As the enforcers lack a mechanism to demonstrate to other participants whether they have truly deleted their keys, the enforcement of the covenant requires a level of trust among participants. To avoid this reliance on trust, each participant should act as an enforcer and configure $l_{enf}$ as an $n-$of$-n$ multi-signature script. This would grant each participant the unilateral authority to prevent the fabrication of malicious enforcement signatures. Otherwise, participants need to trust that a substantial number of enforcers remain honest. Particularly, if $l_{enf}$ is set as an $m-$of$-n$ multi-signature script, they must trust that at least $m$ enforcers are honest. 

\subsubsection{Summary}

In conclusion, our analysis of the integrity of protocol enactment identifies potential vulnerabilities and the corresponding attacks that may be employed by adversaries. Through the methodology of `proof by contradiction', we have analyzed the minimum viable attacks that might invalidate the claims associated with various aspects of the protocol enactment integrity. We will discuss how the identified limitations, particularly related to trust assumptions regarding custodial control and covenant enforcement, and how out-of-band communication among participants should be considered in practical applications in section \ref{sec:designing-for-applications}. Here, we summarise the salient points:

\begin{itemize}
    \item Each participant can assess the integrity of the protocol enactment with respect to their own inclusion, but cannot overcome the trust threshold requirements. 
    \item Applications which involve no out-of-band communication are likely to suffer attacks from the depositor. 
    \item Applications wherein participants communicate with each other can effectively dissuade a depositor from malicious actions, or else the depositor will be discovered. In these cases, the depositor cannot undermine the integrity of protocol enactment without colluding with a sufficient threshold of the enforcers or custodians.
    \item The integrity of the protocol enactment can only be ensured if up to (but not including) $m$ enforces and $min(j, k-j+1)$ custodians are corrupt or compromised.
\end{itemize}

\subsection{Protect Honest Participants Against Losses}
\label{subsec:safeguarding-honest-participants}

In this section of the analysis we assess how protocol participants are protected from potential harm. We systematise the analysis with an asset-centric approach, considering in turn bitcoin assets, identity information, and custody system information. 

\subsubsection{Protection of Bitcoin}

Custodians and Enforcers, who do not input Bitcoin are trivially protected against Bitcoin loss resulting from their participation. The Depositor, however, could potentially face theft or loss of Bitcoin. The critical issue here is whether participating in the protocol increases the risk of such theft or loss for the depositor. Examination of the collusion attacks presented in section \ref{subsubsec:collusion} reveals that the depositor could be tricked into transferring Bitcoin to an attacker. This attacker might, at the very least, aim to render the depositor's Bitcoin inaccessible or, in the worst-case scenario, plan to execute a theft transaction, thereby gaining control of the Bitcoin from the covenant-bound output. 

\begin{table}[H]
    \centering
    \begin{tabular}{|>{\columncolor{lightgray}\raggedleft\arraybackslash}p{3.5cm}|p{11.5cm}|}
    \hline
        \textbf{Vulnerability F}& The Depositor can be misled into creating a malicious custodial subscript $l_{cust}$ without including a sufficient threshold of honest custodian's public keys. \\
        \textbf{Attack F(i)}& An attacker compromises the authentication and encryption credentials of $k-j+1$ custodians, replacing $k-j+1$ custodial public keys $Q_i$ with malicious ones $Q'_i$. The attacker then refuses to provide signatures to satisfy $l_{cust}$. \\
        \textbf{Attack F(ii)}& $k-j+1$ malicious custodians collude and refuse to provide signatures to satisfy $l_{cust}$. \\
        \textbf{Impact F}& Honest custodians are unable to finalise authorisation of $TX^{Cov}$, making funds inaccessible. The Bitcoin deposited by the Depositor is lost. \\
    \hline
    \end{tabular}
\end{table}

\begin{table}[H]
    \centering
    \begin{tabular}{|>{\columncolor{lightgray}\raggedleft\arraybackslash}p{3.5cm}|p{11.5cm}|}
    \hline
        \textbf{Vulnerability G}& The Depositor can be misled into creating a malicious custodial subscript $l_{cust}$, in which a sufficient threshold of malicious custodial public keys are included. The Depositor can also be deceived into creating a malicious enforcement subscript $l_{enf}$, in which a sufficient threshold of malicious enforcement private keys are included. \\
        \textbf{Attack G(i)}& An attacker compromises the authentication and encryption keys of $m$ enforcers and replaces $m$ enforcement public keys $P_l$ with attacker-owned public keys $P'_l$. The attacker secretly refuses to delete the associated enforcement private keys. The same attacker also compromises the authentication and encryption keys of $j$ custodians, replacing $j$ custodial public keys $Q_i$ with attacker-owned public keys $Q'_i$. \\
        \textbf{Attack G(ii)}& There is collusion among $m$ malicious enforcers, who collectively refuse to delete their enforcement private keys, and $j$ malicious custodians.  \\
        \textbf{Impact G}& The attacker circumvents covenant enforcement by generating new signatures to satisfy $l_{enf}$ for transactions that are not constrained by the covenant. The attacker can also satisfy $l_{cust}$. Consequently, the attacker gains control of the output created by $TX^{Dep}$. The bitcoin deposited by the Depositor is stolen. \\
    \hline
    \end{tabular}
\end{table}

Consequently, the depositor's security relies on an underlying trust assumption that there exists an honest threshold of custodians and enforcers, which serves as a safeguard against potential loss. Note however that this is no different than typical payments made with bitcoin transactions. The fact that the payment was part of the deleted-key covenant protocol doesn't increase the risk associated with the depositor making the payment. Rather, the presence of the enforcement subscript is an additional mitigation to theft.

\subsubsection{Protection of identity and custody system information}

In our protocol, all public keys transmitted are fresh, meaning they are not correlated with other public keys within the custody systems of the participants. Since all communication occurs on pairwise channels with the depositor, the depositor's system will be able to associate public keys with authentication credentials of the enforcers and custodians. The corruption or compromise of the depositor can result in disclosed public key information being linked to the authentication credentials of participants. However, if the authentication details are anonymous and do not contain information about the real identity of the participants, no identity nor custody system information would be revealed.  

An attacker that has spoofed (accessed and used the authentication credentials of) a custodian or enforcer will receive all other participant's public key data in a single message from the depositor, encoded in the covenant-bound output script $l_{enf}||l_{cust}$. This data will thus not be linked to other participants' authentication credentials.

An adversary gaining control over a custodian or enforcer can associate the depositor's authentication credentials with the data in $TX^{Dep}$. However, since the depositor uses fresh public keys and only spends outputs that are uncorrelated with others in their custody system, no further information about the depositor's custody system is revealed.

\subsubsection{Summary}

To conclude, in our analysis for the objective `protect honest participants against losses' we find that most participants are protected from any losses, including theft of bitcoin or other sensitive data. The only exception is the potential for the depositor to be deceived by collusion between custodians and enforcers. However, this vulnerability is not a flaw within the protocol design but rather an inherent aspect of making any Bitcoin payment. Therefore, the objective of protecting honest participants is achieved with the assumption that less than $m$ enforcers and $min(j, k-j+1)$ of custodians are malicious. 

\subsection{Guarantee the Irrevocability of Covenant Enforcement}

To guarantee the irrevocability of the covenant enforcement, it is pertinent to dissect each enforcement condition separately and thoroughly assess the scenarios in which they could potentially be compromised.

\begin{quote}
    a) There exists a covenant transaction, denoted $TX^{Cov}$, which encodes the covenant constraints. Furthermore, the witnesses field of $TX^{Cov}$ is populated with enforcement signatures that satisfy the enforcement subscript $l_{enf}$.
\end{quote} 

This enforcement condition underscores the critical necessity for the custodian(s) to persistently store $TX^{Cov}$ and the accompanying enforcement signatures. A loss of either entity would render the covenant-bound output irredeemable. This predicament necessitates robust redundancy measures in storage; this can either be achieved by entrusting multiple custodians with the storage of $TX^{Cov}$ or, in a single custodian scenario, by preserving $TX^{Cov}$ across various backup storage media.

\begin{quote}
    b) Each enforcement private key, denoted $p_l$, is deleted. As such, no new signatures can be generated to satisfy the enforcement subscript.
\end{quote}

To break this enforcement condition, an attacker would need to either recover deleted keys, or forge signatures associated with the enforcement public keys in some other way. The difficulty of recovering deleted keys depends on the method used by the enforcer to delete them. As discussed in section \ref{subsec:key-deletion}, many deletion methods exist and a wide range of security models can be selected from. These have been summarised in table \ref{tab:key-deletion}. Applications which use covenants often will benefit from re-using storage media and making an efficient process for purging it (e.g. through de-gaussing) on-demand. One-time applications with a lot of bitcoin at risk might favour destroying the storage media.

\begin{table}
    \centering
    \begin{tabular}{|l|p{9cm}|l|}
    \hline
        \rowcolor{lightgray} \textbf{Method} & \textbf{Attack Vector} & \textbf{Relative Difficulty} \\
        \hline
        Disposing & Find the storage media & Low \\
        \hline
        Clearing & Analyse residual charge on storage media & Medium \\
        \hline
        Purging & Exploit failures to completely remove residual charge on storage media with advanced laboratory techniques & High \\
        \hline
        Destroying & None & Impossible \\
        \hline
    \end{tabular}
    \caption{Description of attack vectors and their relative difficulty with four generic methods to deleting sensitive data.}
    \label{tab:key-deletion}
\end{table}

Note, this qualitative description of the relative difficulty of an attack is provided to give an intuition to application designers. We did not find it worthwhile to model the human-computer interaction for enacting the deletion process as it would not result in a precise quantification of risk. We have provided sufficient information for users of covenant-applications to select a suitable method and understand its limitations. 

Fortunately, Schnorr signatures have received significant attention from security researchers and its security properties have been formalised rigorously. Schnorr signatures are ``\textit{strongly unforgable under chosen message attacks} in the random oracle model assuming the hardness of the elliptic curve discrete logarithm problem (ECDLP)'' \cite{BIP-Schnorr, SchnorrSecurity}. Informally, this means that attackers can not forge new Schnorr signatures without having access to the relevant private key, even if they are given (signature, message) pairs for arbitrary messages. If an efficient method for solving the ECDLP emerges, then not only would this enforcement condition be broken, but so too would the basis of security for bitcoin transactions and more broadly much of the security infrastructure for modern internet technologies. 
 
\begin{quote}
    c) The deposit transaction, denoted $TX^{Dep}$, is included in a block that is sufficiently deep in the blockchain, such that it is highly improbable for the transaction to be removed or modified.
\end{quote}

Once it is included in the chain, $TX^{Dep}$ cannot be modified or removed without a chain reorganisation. A chain reorganisation attack is very expensive, requiring the input of a significant amount of energy. Even in the case where a highly motivated attacker out-competes the rest of the mining network and removes $TX^{Dep}$ from the chain, there is nothing to stop it from being re-broadcast and included in a future block. Only a sustained 51\% attack where $TX^{Dep}$ is continually censored would revert this enforcement condition indefinitely.

The analysis of the irrevocability of covenant enforcement, broken down into separate enforcement conditions, suggests that the proposed covenant scheme maintains strong resistance against potential breaches. Redundant storage of the covenant transaction ($TX^{Cov}$) can mitigate loss of funds risks. Further, the deletion of private enforcement keys, paired with the robustness of Schnorr signatures, reduces the feasibility of key recovery or signature forgery. Lastly, the inclusion of the deposit transaction ($TX^{Dep}$) deep within the blockchain, along with the prohibitive costs of a chain reorganization attack, ensures the near immutability of $TX^{Dep}$. Altogether, these measures significantly enhance the irrevocability of covenant enforcement under the proposed scheme, given adherence to the described protocol.

\subsection{Maintain Honest Custodial Control for Covenant Execution}

The final security objective is to ensure that honest custodians maintain control for covenant execution. Since custodial power is distributed using a $j$-of-$k$ multi-signature Script, the scenarios under which this objective is achieved are readily identifiable.

\textbf{Claim:} The multi-signature custodial subscript ensures the authorization of the covenant transaction $TX^{Cov}$ by a minimum of $j$-of-$k$ custodians. If at least $j$ custodians are honest, the final authorization will be possible. Conversely, if fewer than $j$ custodians are malicious, premature authorization will not occur.

\begin{table}[H]
    \centering
    \begin{tabular}{|>{\columncolor{lightgray}\raggedleft\arraybackslash}p{3.5cm}|p{11.5cm}|}
    \hline
        \textbf{Vulnerability I}& Custodians' key-management systems are subject to compromise. \\
        \textbf{Attack I}& An attacker may compromise $j$ custodians' systems to extract or destroy their private keys. \\
        \textbf{Impact I}& The compromised private keys can authorize $TX^{Cov}$ prematurely or withhold authorization in a denial-of-service attack, potentially disrupting time-sensitive applications or locking the deposited funds. \\
    \hline
    \end{tabular}
\end{table}

However, it is crucial to note that this type of attack is not exclusive to the deleted-key covenant protocol but is inherent to multi-signature custody systems. The critical difference in this context is that the impact is mitigated since the attacker cannot modify $TX^{Cov}$. This distinctive feature, and its potential to enable new risk controls for Bitcoin custody systems, will be explored in detail in chapter \ref{ch:vault-custody}.

The enforcement of the access policy is ingrained within the Bitcoin consensus rules. All nodes within the network participate in the verification process to ensure that the signatures satisfy the custodial lock script. This mechanism of strongly enforced access control facilitates an unambiguous understanding of the security objective. The integrity of the process is maintained through decentralized validation, leaving little room for interpretation or deviation.

It is pertinent to note that the description provided herein does not delve into the intricacies of custodians' key-management systems, nor does it address the broad spectrum of potential attacks that custody systems might be susceptible to. These attacks can generally be classified into two categories: remote network attacks and physical device or person attacks. While a comprehensive exploration of these vulnerabilities is beyond the scope of this section, it is essential to emphasize that custodians must adhere to best-practice key-management. This includes, for example, the utilization of hardware device signers and the adoption of covert operations, measures that contribute substantially to the robustness and resilience of the system.

\subsection{Analysis Conclusion}

In summarizing the extensive analysis presented herein, it should be emphasized that a methodical and comprehensive approach was taken to scrutinize the security aspects of the deleted-key covenant protocol. Four specific security objectives were defined, each corresponding to stages of the covenant's life cycle, encompassing its establishment, the active period, and the final execution. Additionally, we laid down explicit assumptions to ground our analysis, while characterizing the potential adversary's capabilities, which encompass all the traits of a Dolev-Yao attacker, enhanced with the capability to compromise the end-point devices of participants. Through this examination, the conditions under which each objective could be satisfactorily met were revealed. Key findings of the analysis include the following:

\begin{itemize}
    \item An honesty assumption regarding enforcers is requisite for the protocol; specifically, for an $m-$of$-n$ enforcement Script, at least $n - m + 1$ enforcers must be honest.
    \item Participants have the option to remove the aforementioned honesty assumption by personally enacting the role of enforcer and setting the multi-signature threshold as $m=n$.
    \item Though custodians can be compromised, the inherent constraints of the covenant significantly mitigate the potential rewards for the attacker when compared to a regular payment lacking the covenant's restrictions.
    \item The disclosure of sensitive information does not pose a risk to honest participants, particularly concerning authentication credentials and public keys.
    \item It is recognized that the depositor, similar to any entity handling and transacting in Bitcoin, remains vulnerable to theft due to potential compromises in key management systems, as well as payment attacks where an attacker fraudulently replaces recipient's identity and public keys.
\end{itemize}

This conclusion affirms the robustness of the deleted-key covenant protocol, acknowledging its strengths and identifying potential weaknesses, all grounded in the methodological framework established at the outset of this inquiry.

\section{Designing for Applications}
\label{sec:designing-for-applications}

In this section we will introduce several more details that are relevant for practical applications using deleted-key-covenants.

\subsection{Composability of Covenant Deposits and Commitments}
\label{subsec:Composability}

The design space for covenant-based protocols can be extended from the simple concept depicted in figure \ref{fig:simple-covenant} by increasing the number of inputs and outputs in a covenant transaction and by creating multiple dependent covenant transactions. Additional covenant commitments can be \textit{joint} in the sense that each commitment must be satisfied, or can be \textit{disjoint} such that only a subset of commitments must be satisfied.

\begin{figure}
    \centering
    \includegraphics[width=0.4\linewidth]{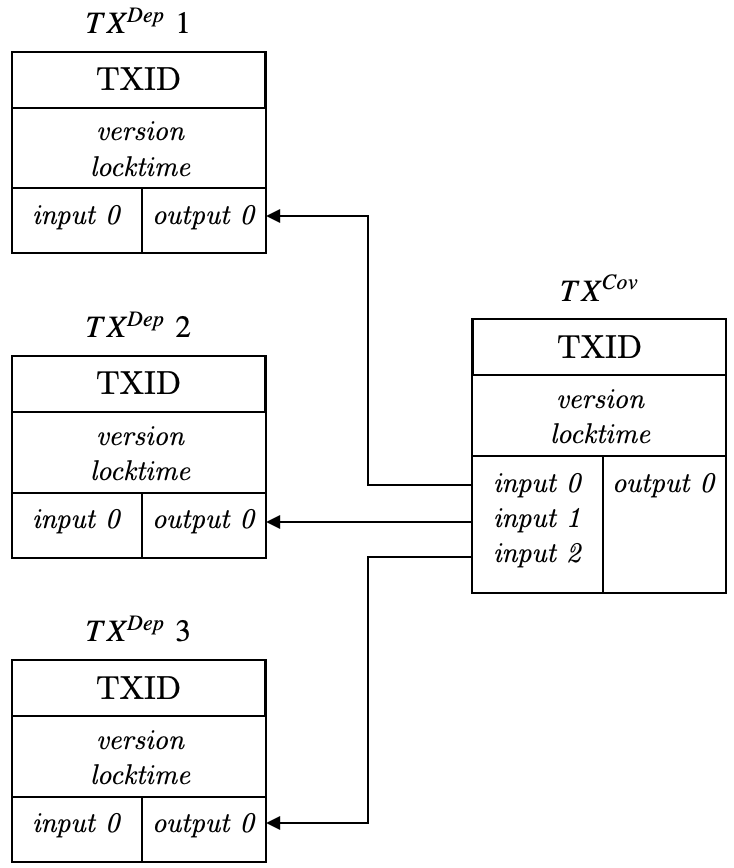}
    \caption{A covenant transaction with three inputs that spend outputs in different deposit transactions. Each output commits to an enforcement script $l_{enf}$ and requires sufficient enforcement signatures to be spent. \textit{All} deposits must be confirmed for the signed covenant transaction to become active.}      
    \label{fig:multi-deposit-covenant}
\end{figure}

Figure \ref{fig:multi-deposit-covenant} depicts a covenant transaction with multiple deposit transactions, each of which participate in enforcing the covenant by committing to an enforcement script $l_{enf}$ within an output. While this is possible, there is a risk that if \textit{any} of the deposits fail to be confirmed on the blockchain, the funds could be lost. With this design, the enforcement condition should instead read

\begin{quote}
    c) \textit{All} deposit transactions have been included in a block that is sufficiently deep in the blockchain to not be removed or modified.
\end{quote}

\begin{figure}
    \centering
    \includegraphics[width=0.6\linewidth]{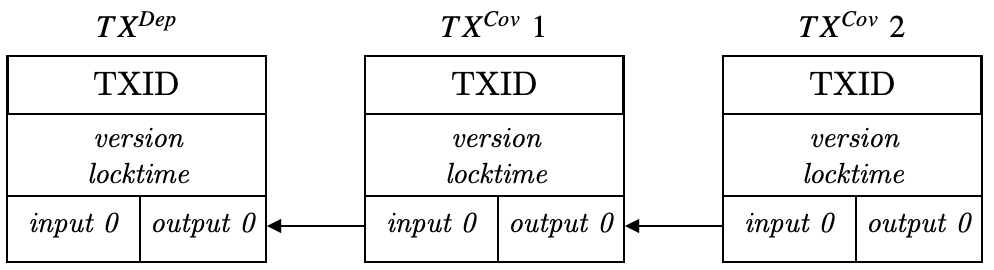}
    \caption{A chain of covenant transactions whose constraints are \textit{joint} in that each must be satisfied to spend the funds which they control. All covenant transactions with dependent transactions must be non-malleable to protect their dependent transactions from being invalidated.}      
    \label{fig:chain-joint-covenant}
\end{figure}

Figure \ref{fig:chain-joint-covenant} depicts joint covenant commitments by constructing a chain of dependent covenant transactions. Recall that transactions that spend only P2TR output types ensure that transactions are not malleable, such that their transaction ID can not be changed. This is critical because references to previous outputs in the chain of dependent transactions use the transaction ID, and if this could be modified then the subsequent transactions would be invalidated. Since the enforcement private keys are deleted, no alternative transactions can be finalised to spend those outputs, and they would be indefinitely locked. Here, \TX{Cov} $1$ must be broadcast before \TX{Cov} $2$ would be considered valid. Therefore, the constraints encoded by both transactions must be met in order to satisfy the full covenant and regain arbitrary spending capability with the funds. Another way to achieve joint commitments would be to add each constraint to the lock Script of an output in a single covenant transaction. 

\begin{figure}
    \centering
    \includegraphics[width=0.4\linewidth]{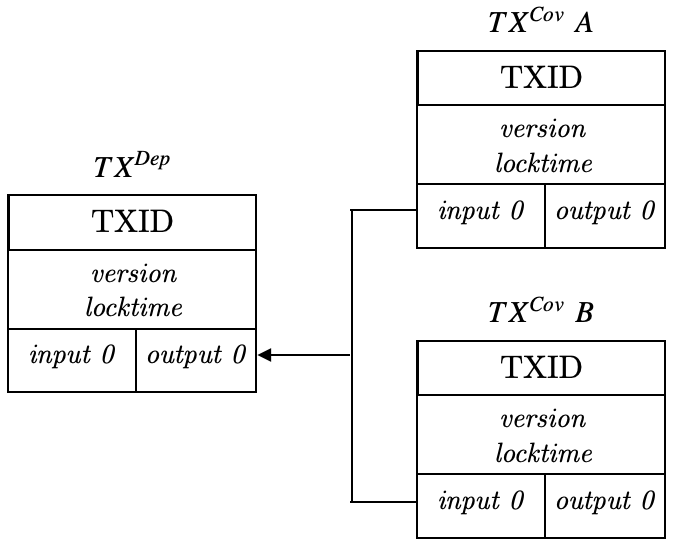}
    \caption{Two covenant transactions which spend from the same deposit output are considered to have \textit{disjoint} commitments in that either \TX{Cov} $A$ or \TX{Cov} $B$ must be satisfied while the other is invalidated.}      
    \label{fig:disjoint-covenant}
\end{figure}

Figure \ref{fig:disjoint-covenant} depicts disjoint covenant commitments. There is a choice with which set of constraints to adhere to, either those encoded by \TX{Cov} $A$ or by \TX{Cov} $B$. The confirmation of either transaction invalidates the other. Another way to achieve a disjoint commitment would be to use conditional logic in the lock Script of an output in a single covenant transaction. 

Each of these concepts (multiple deposits, joint and disjoint covenant commitments) can be combined arbitrarily to produce multiple dependent chains of covenant transactions in a tree-like structure. Moreover, covenant transactions can be constructed with arbitrary numbers of inputs and outputs provided that transaction size limits are adhered to. In this way, application designers can construct state-machines from transaction sets. Custodians control the transition choices and timing by authorising the broadcast of the next transaction in the tree. State can be encoded in a transaction with various methods, described and empirically analysed by Bartoletti \textit{et al.} \cite{BitcoinMetadata}.

While infinitely recursive covenant transactions are not possible (with deleted-key covenants), a chain of covenant transactions of arbitrary length is possible. The chain must be pre-defined since each covenant transaction depends on a previous transaction. Creating the unsigned transactions requires generating the enforcement public keys and knowing the custodial public keys. Each enforcer must be involved from the beginning and the amounts for each transaction and the specific constraints of each covenant commitment must be known in advance.

\subsection{Class of Possible Covenants}
\label{subsec:ClassOfCovenants}

It should be clear that the class of possible covenants is predominantly restricted by the parameters of a transaction template. This means that traditional covenants such as those based on the operational capacity of a business could only be implemented by use of an \textit{oracle} who would validate data \textit{external} to the blockchain and then either authorise (or refuse) the covenant transaction. Without oracles, deleted-key covenants can have commitments that only require validating \textit{internal} blockchain and transaction data, since each Bitcoin node has the capacity to independently validate transactions (and transaction chains). Constructing a condition that relies on external validation requires participation from an oracle but this introduces some trust assumptions that weaken the security model of the covenant. Script does not enable explicit requests to external servers, nor does it enable importing mutable state variables. A basic mechanism for enabling an oracle is to give them a necessary signing key in the enforcement Script, and design them to only provide the signature to authorise the transaction once some condition in the external world has been met. Some issues that arise with the use of oracles include; they can be hacked, bribed, or be poorly designed and return incorrect outputs, or they can go offline and a required private key can vanish leading to a loss of funds. Well designed oracles may rely on a consortium of independent signatories to introduce Byzantine fault-tolerance. Moreover, signatories may be instantiated on secure hardware for trusted computing. A more advanced oracle mechanism is \textit{Discreet Log Contracts} \cite{DLC} and it is being actively researched. Integrating external validation methods is a topic for future research.  

Covenants that require only internal validation are our focus here. Bitcoin transaction signatures can be set to commit to different signature message types, allowing for some selection for which subset of transaction data are included. The subsets were shown explicitly in section \ref{subsec:sig_msg_types}. Recall that chains of dependent transactions must avoid malleability to prevent the risk of indefinitely locking bitcoin. However, the final transaction in a chain can safely be modified because no subsequent transactions depend on it having a static transaction ID. Covenant transactions may be defined long before they are broadcast, and enabling some modification of the transaction without invalidating the signatures could be useful in some contexts. For example, to help with dynamically allocating fees, or to adjust how much change is returned to the one paying. 

A minimally constrained deleted-key covenant transaction would use the {\tt A1CP||NONE} signature type and, referring to table \ref{tab:taproot_sighash}, the enforcement signature(s) would commit to the transaction version and the data for a single input (except its index). The enforcement signature(s) would be valid for any transaction that shares that subset of data; the outputs and other inputs are free to change. This particular usage is unlikely to be broadly useful since the only restriction the covenant could have is on the transaction version and the input's sequence and annex fields.

A maximally constrained deleted-key covenant transaction would use the {\tt ALL} signature type and the enforcement signature(s) would commit to every field of the transaction, except the witnesses field. Most of the power of this type of covenant comes from the locking Script associated with each output. One can specify that a given set of inputs can only be spent to a set of output addresses whose amounts are well defined and whose unlocking scripts are any set of constraints specifiable in Bitcoin Script.

A partially constrained deleted-key covenant transaction could use the {\tt A1CP||ALL} signature type and the enforcement signature(s) would commit to every field of the transaction except for other inputs and the witnesses field. This would allow additional inputs to be added by custodians without invalidating the enforcement signature(s). Since the outputs are fixed, adding new inputs increases the fee paid to miners. Since the fee-market is driven by current demand and is dynamic, this method can be useful to application designers. 

For enforcement signatures with the type {\tt SINGLE}, the enforcers commit to all inputs and a single output. New outputs can be added to the transaction. This means that custodians can update the output set to handle any excess bitcoin that is not directed to the fixed output. Again, an example use-case for this is to adjust the fee by adding a change output immediately prior to broadcasting the transaction. 

Thus we have seen why signature message types play an important role in defining what class of covenant commitments are possible. Note that all signature message types are possible: {\tt ALL}, {\tt A1CP||ALL}, {\tt SINGLE}, {\tt A1CP||SINGLE}, {\tt NONE}, and {\tt A1CP||NONE}. This approach is a hack because the signatures message types were not designed for this purpose. With a more efficient and flexible way to get verified transaction data onto the execution stack, the class of possible covenants would be more broad. Other covenant mechanism proposals differ from what is presented here. 

\subsection{Dynamic Fee Allocation}
\label{subsec:cov-dyn-fee-alloc}

There are two mechanisms in Bitcoin for increasing the fee paid in a transaction if the transaction becomes stuck in the memory pool due to having a low fee-rate (in \textit{satoshis/kb}) relative to others in the pool; `replace-by-fee' (RBF) \cite{BIP125} and `child-pays-for-parent' (CPFP). Application designers should consider both CPFP and RBF, and the best approach might involve a combination of the two.

RBF is a signalling mechanism that alerts miners that a transaction may be replaced by an equivalent one with a higher fee. Miners and nodes opt-in to a policy of detecting RBF-enabled transactions and updating their memory pool if certain conditions are adhered to. For example, the replacement transaction must have a higher absolute fee, and the fee must be sufficiently high to cover the additional bandwidth cost imposed on the network by the replacement. See the BIP for a precise specification of conditions \cite{BIP125}. Miners have an incentive to claim the higher fee replacement if they are aware of it. Note that the replacement conditions are only policies set by each node but they are not consensus rules. A replacement transaction may be consensus-valid but nodes may still reject to propagate it. The replacement conditions are designed to protect the peer-to-peer network from denial-of-service attacks. However, if a replacement transaction is sent directly to a miner, that miner may include it in their block despite it not adhering to the replacement conditions.

CPFP became the default transaction selection method for block building in Bitcoin Core $v0.13$. Bitcoin's consensus rules enforce a strict ordering of transactions in the blockchain. However, transactions which depend on an unconfirmed transaction can be included in the same block as its ancestor (parent) as long as it appears later in the block. Thus, if an unconfirmed transaction has a small fee, but its child transaction has a fee large enough to cover itself and the unconfirmed parent transaction, miners have an incentive to include both in a block. If there is a chain of unconfirmed covenant transactions in the memory pool, the final transaction can pay for the whole chain, but the whole chain will appear in the same block. If a transaction chain is to be broadcast across multiple blocks, then so-called `CPFP outputs' should be added to each transaction. CPFP outputs are additional outputs that are controlled by a key the user has for fee management. The CPFP output itself need not have any value (in \textit{satoshis}), but it can be consumed in a `fee-bump transaction' that pays for itself and the unconfirmed parent transaction. This allows the transaction chain to progress even if the market fee-rate hikes. 

\subsection{Safety Concerns}
\label{sec:safety-concerns}

There are three classes of attack that application designers must be aware of when using different signature message types; transaction ID malleability attacks, pinning attacks, and theft. If a transaction is consensus-valid but none of the \textit{required} signatures used {\tt ALL}, then some part of the transaction can be modified without invalidating it, in some cases enabling these attack vectors. 

Generally, `pinning' is a class of attacks in which the attacker takes advantage of an aspect of network topology, transaction relay and/ or memory pool policies to prevent timely inclusion of a transaction in a block. An example is when a user broadcasts a transaction with an {\tt A1CP||ALL}-type signature and with RBF enabled, with the intention of adding an additional input to increase the fee if the transaction is low-priority in the transaction queue. However, an attacker can add low-value (in \textit{satoshis}) inputs with large size (in \textit{kb}) to the transaction and re-broadcast it. If the absolute fee is sufficient, nodes in the network (by default in bitcoin-core \textit{v23.0}) will accept the replacement and propagate it. However, the fee-rate (fee per \textit{kb}) will be very low, making it low priority for miners to include in the blocks they produce. Pinning attacks create risks in many off-chain transaction protocols \cite{Pinning,Pinning-Riard,Pinning-Bastien}.

Transaction ID malleability attacks are disruptive for chains of dependent covenant transactions. An attacker modifies any part of an intermediate transaction in the chain, resulting in a different Transaction ID. In doing so, all subsequent covenant transactions become invalid, and the bitcoin they control is effectively burned. 

Theft is a possible attack vector if a transaction is signed with {\tt SINGLE} and broadcast. An attacker could add an output to the transaction to redirect to them-self any residual value that is implicitly allocated as a fee. Miners that detect such a transaction will not gain from this type of attack since the fee is already directed to them-self. However, with RBF enabled, any listener on the network can broadcast a replacement transaction and the theft is likely to succeed. 

Without RBF enabled, attackers would not be able to broadcast a replacement transaction. However, modified transactions would still be consensus-valid. The thief would have to collude with a miner to get it included. An attacker would have to offer something sufficient to the miner in return. However, the miner that creates the next block is probabilistic, not deterministic. An attacker would have to collude with several powerful miners to have a reasonable likelihood of success. Thus we see how, in general, not enabling RBF is a strong mitigation for these classes of attack. 

\subsection{Fee Strategy}
\label{subsec:cov-fee-strat}
Given the general approaches and safety concerns discussed in previous sections for increasing transaction fees, we now move on to discuss those methods in the context of deleted-key covenant transaction applications.  

To be safe, partially constrained covenant transactions should be protected by a non-empty custodial script $l_{cust}$ for which custodians provide {\tt ALL}-type signatures. Application designers should determine their tolerance for malleability or pinning attacks and act accordingly. A more secure design, where {\tt ALL}-type signatures are used would be less flexible. A more flexible design in which, for example, new inputs or outputs can be added, would be less secure to rely on. Thus there is a security-flexibility trade-off for \textit{satisfied} (fully-signed, consensus-valid) covenant transactions. 

On the other hand, for covenant transactions that are not completely satisfied (and not yet consensus-valid) that trade-off is not present. For example, enforcers can safely use any signature types, enforcing only the constraints which they require, relying on custodians to satisfy covenant transactions with {\tt ALL}-type signatures. Enforcers can grant flexibility to custodians, but custodians must not grant flexibility to potential attackers. Moreover, some custodians can grant flexibility to others, but a sufficient threshold of custodians must use {\tt ALL}-type signatures to maintain safety when broadcasting. 

In table \ref{tab:FeeMethods}, we summarise several methods for increasing the fee with deleted-key covenants. The signatures $sig_l^{enf}$ and $sig_i^{cust}$ are created by enforcers and custodians respectively. The pre-broadcast method highlights how to increase the fee before the transaction is transmitted to the peer-to-peer network. The post-broadcast method highlights how to increase the fee for transactions that have been broadcast but are low-priority in the memory pool. If the method is safe to use given that the covenant transaction has dependent covenant transactions, then `works with dependent $TX$' is marked `Yes', otherwise `No'. Finally, we show whether or not pinning and malleability attacks are possible.

The Fee input(s) and Change output(s) strategies are safe to use for single covenant transactions, but not for chains of covenant transactions (except for the final one). 

The CPFP output(s) strategy is probably the most adaptive, since it works for chains of covenant transactions across multiple blocks, and can in principle handle any market fee-rate scenario. It requires broadcasting a second `fee transaction', relying on the CPFP mechanism to pay for itself and the parent covenant transaction. The fee transaction can also enable RBF in case it needs to be replaced with a higher fee. If the CPFP output is not required and not consumed, this could lead to an increase in the set of unspent transaction outputs, which has to be maintained in memory by all nodes in the network. 

The `prepare fee range' strategy requires the user to sign several variants of a covenant transaction with a range of fees, anticipating broad changes in the market fee-rate. The more variants that are signed within a range, the more accurate the fee-rate will be to current market conditions. However, each signature is a burden for the user to generate (especially if they are using an offline signing device). For a chain of covenant transactions of length $t$, if each one is prepared with $p$ variants, then $p^t$ covenant transactions must be prepared. This is because each covenant transaction is bound to a specific variant that it is spending from and so $p$ variants must be created for each new dependent covenant in the chain \textit{for all} $p$ variants of the previous covenant. This scales exponentially poorly in both storage and computation required for signature generation.   

If none of the required signatures from the custodians use the {\tt ALL} type, as with the `unsecured fee input(s)' method, then the covenant transactions will be at risk of both pinning and malleability attacks. It is not suitable for a chain of dependent transactions. It is reasonable to enable RBF in a case where the timeliness of including a transaction in a block is not critical. While the transaction would be at risk of being pinned, it would likely be included in a block eventually.

\renewcommand{\arraystretch}{1.25}
\begin{table}
    \centering
    \resizebox{\textwidth}{!}{
    \begin{tabular}{|l|l|l|l|l|l|l|l|}
    \hline
         \rowcolor{lightgray} \textbf{Strategy} & \textbf{$sig_l^{enf}$ type} & \textbf{$sig_i^{cust}$ type} & \textbf{Pre-broadcast} & \textbf{Post-broadcast} & \textbf{Works with} & \textbf{Pinning} & \textbf{Malleability} \\
         \rowcolor{lightgray}\textbf{Name}& & & \textbf{method} & \textbf{method} & \textbf{Dependent}& \textbf{attacks} & \textbf{attacks} \\
         \rowcolor{lightgray} && &&& \textbf{Txs} &&\\
    \hline
         Fee & {\tt A1CP||ALL} & {\tt ALL} & Add inputs, & Use RBF, add inputs, & No & No & No \\
         input(s) & &  & new $sig_i^{cust}$ & new $sig_i^{cust}$ & & & \\
    \hline 
        Change & {\tt SINGLE} & {\tt ALL} & Add/ modify & Use RBF, add/ modify & No & No & No \\
        output(s) & & & unenforced outputs,& unenforced outputs, & & & \\
        && & new $sig_i^{cust}$ & new $sig_i^{cust}$  & & & \\
    \hline 
        CPFP & {\tt ALL} & {\tt ALL} & Fee TX & Fee TX with RBF & Yes & No & No \\
        output(s) && & & & & & \\
    \hline 
        Prepare & {\tt ALL} & {\tt ALL} & Prepare TXs with & Use RBF, prepare TXs & Yes & No & No \\
        fee range & &  & range of fees & with range of fees & & & \\
    \hline 
        Unsecured & {\tt A1CP||ALL} & {\tt A1CP||ALL}  & Add input(s) & Use RBF, add input(s) & No & Yes & Yes \\
        fee & &  & and witness(es) & and witness(es) & & & \\
        input(s) && & & & & & \\
    \hline 
    \end{tabular}}
    \caption{Comparative summary of strategies for allocating fees to deleted-key covenant transactions.}
    \label{tab:FeeMethods}
\end{table}
\renewcommand{\arraystretch}{1}

\subsection{Proof-of-Reserves}
\label{sec:ProofOfReserves}

Financial auditing often requires verifying claims that an entity has access to funds. With Bitcoin, so-called `proof-of-reserve' protocols exist for this purpose. In the simple case, the claimant aims to demonstrate their ability to spend an unspent transaction output, $u$. As per the proof-of-reserves protocol described in \cite{BIP127}, the claimant can construct an invalid transaction (e.g. one which creates more bitcoin than it consumes) that consumes $u$ and generate a valid signature to satisfy the associated lock script. The invalid transaction and signature together can be safely shared with the auditor for verification, demonstrating that the claimant can generate similar but valid transactions that consume $u$. 

For funds that are bound by deleted-key covenants, the proof-of-reserves protocol must be adapted slightly. The custodians have control of $TX^{Cov}$ with witnesses that satisfy $l_{enf}$, but they cannot generate signatures to satisfy $l_{enf}$ with a different transaction. However, as long as $l_{cust}$ is not empty, $sig_l^{enf}$ can be shared with an auditor without granting them the ability to finalise $TX^{Cov}$ for broadcast. Given this, the custodian must:

\begin{enumerate}
    \item Enact the simple proof-of-reserves protocol for a fake output with $l_{cust}$ as its lock script,
    \item Share the enforcement signature(s), the lock script $l_{cov}=l_{enf}||l_{cust}$, and $TX^{Cov}$ with the auditor.
\end{enumerate}

Now the auditor can verify each signature and confirm the validity of $TX^{Cov}$. Note that the auditor has no guarantee that the covenant commitments will be enforced, since they have no proof that the enforcement private keys were deleted.  

\section{Covenant Mechanism Comparison}
\label{sec:covenant-comparison}

Here we will consider three plausible soft-fork upgrade paths for Bitcoin which would enable consensus-enforced covenants. By `consensus-enforced' we mean that one or more consensus rules would need breaking to break the enforcement of the covenant. By definition, consensus rules are those followed by all peers in the network. Each node in the network verifies that transactions and blocks adhere to consensus rules, or else they fall out of consensus with the network. On the other hand, deleted-key covenants are enforced by the deletion of enforcement keys, which cannot be verified without interactive participation during the activation of the covenant. Consensus-enforced covenants remove this requirement altogether, they are \textit{verifiable}.

These three plausible soft-fork upgrades each have a code implementation and are actively being debated among Bitcoin developers and enthusiasts; {\tt ANYPREVOUTANYSCRIPT} ({\tt APOAS}) \cite{BIP118-APO}, {\tt OP\_CHECKTEMPLATEVERIFY} ({\tt CTV}) \cite{BIP119}, and {\tt OP\_INSPECT\_X} ({\tt INSPECT\_X}) \cite{tapscript_elements}. {\tt CTV} and {\tt APOAS} are very similar to each other and to deleted-key covenants in the class of possible covenants they enable. There are minor differences in what subsets of transaction data can be committed to. These are summarised in table \ref{tab:class_comparison}. 

The signature checking operations use modified signature message types when {\tt APOAS} is used. The primary differences are that {\tt A1CP} is set by default (only the current input is included) and the output's hash, index, public key script and the TapScript's leaf hash are not included in the message. The downside of this design is that pinning attacks are ever-present. This is becuase adversaries can freely add data-heavy inputs to the transaction and, in the worst case when RBF is enabled for the transaction, adversaries can propagate their replacement transaction across the peer-to-peer network. It has been suggested to make the {\tt A1CP} functionality optional instead \cite{APOAS-darosior}, which would prevent this pinning attack vector. 

The property that is lacking with {\tt CTV} and {\tt APOAS} is \textit{granularity}. {\tt INSPECT\_X} enables granularity for covenant constraints in that any transaction, input or output field can be constrained. This enables a much broader design space for covenants.  

\begin{table}
    \centering
    \begin{tabular}{|l|l|}
    \hline
         \rowcolor{lightgray} \textbf{Mechanism} & \textbf{Class of possible covenants} \\
         \hline
         Deleted-key & Constraints on choice of signature message type: \\ 
         & {\tt ALL}, {\tt SINGLE}, {\tt NONE}, {\tt ALL||A1CP}, {\tt SINGLE||A1CP}, {\tt NONE||A1CP} \\
         \hline
         {\tt APOAS} & Constraints on choice of modified signature message types: \\
         & {\tt APOAS:ALL}, {\tt APOAS:SINGLE}, {\tt APOAS:NONE} \\
         \hline
         {\tt CTV} & Constraints on default transaction template (initial version), \\
         & subset templates (later versions) \\
        \hline 
         {\tt INSPECT\_X} & Constraint on field X \\
         \hline
    \end{tabular}
    \caption{Comparison of the class of possible covenants for several covenant mechanisms. }
    \label{tab:class_comparison}
\end{table}

Another critical difference between the three alternatives is in which dynamic fee allocation methods they support. In comparison with table \ref{tab:FeeMethods}, {\tt APOAS} covenants can make use of the more efficient and intuitive strategies `Fee input(s)' and `Change output(s)' but with the benefit that they work with dependent chains of transactions. It does not matter if transaction IDs change with intermediate transactions in a dependent chain since references to previous transactions are omitted from signature messages. Recall that the covenant transactions should require custodial signatures, $sig_i^{cust}$, with type {\tt ALL} to prevent pinning and malleability attacks, while the enforcement signatures, $sig_l^{enf}$ can choose any type. Covenants with {\tt INSPECT\_X} can use the same fee allocation strategies as {\tt APOAS} covenants. However, {\tt CTV} covenants commit to all parts of the transaction, and must therefore not modify any intermediate transactions in a dependent chain. The best available approach in this case is to use `CPFP output(s)' or to prepare multiple covenant transactions with a range of fees and to commit to each one in a separate TapScript branch. 

The enforcement subscripts are shown for each mechanism in table \ref{tab:cov_mechanism_scripts}. These mechanisms have different relative performances measured by the size of their scripts and in the verification cost of the script. Both {\tt APOAS} and {\tt CTV} commit to a large subset of transaction data efficiently (via a signature or a hash), encompassing numerous constraints on the transaction. The primary cost of verifying the {\tt APOAS} enforcement subscript is one signature verification. The primary cost of verifying the {\tt CTV} enforcement subscript is producing one transaction message hash. For {\tt INSPECT\_X} to commit to the same subset as either {\tt APOAS} or {\tt CTV}, it would require several constraints to be added in the enforcement subscript and would be a rather large size. This is a downside for having primitive operations instead of procedural-macro operations, and it reveals a trade-off for granularity. However, {\tt INSPECT\_X} constraints do not require signature or hash operations, meaning that the verification cost is significantly lower. 

\begin{table}
    \centering
    \begin{tabular}{|l|l|}
        \hline
         \rowcolor{lightgray} \textbf{Soft-fork name} &\textbf{ Enforcement subscript}  \\
         \hline
         {\tt APOAS} & {\tt <sig> <PK>} {\tt CHECKSIG}  \\
         \hline
         {\tt CTV} & {\tt <tx hash>} {\tt CTV} \\
         \hline
         {\tt INSPECT\_X} & {\tt INSPECT\_X <target\_value> EQUAL\_VERIFY} \\
         \hline
    \end{tabular}
    \caption{For each proposed soft-fork upgrade, the covenant enforcement subscripts, which are part of the lock script, are shown.}
    \label{tab:cov_mechanism_scripts}
\end{table}

Concerns have been raised about recursive covenants making Bitcoin contracts `Turing-complete' \cite{ZmnOnReccursion}. It is broadly accepted that Script programs must terminate, or else there would be high risk of denial-of-service attacks on nodes which run those programs, and ultimately on the entire peer-to-peer network. The development community prefer that Script is not Turing-complete since the halting problem \cite{HaltingProblem} (knowing if a program will terminate or not) is unsolvable for Turing-complete languages. However, with recursive covenants, multi-transaction Bitcoin contracts may become Turing-complete. In our opinion, this is not a problem because the throughput of transactions is limited by available block space and each transaction requires a fee. Thus a denial-of-service attack in this case amounts to paying significant amount in fees to overwhelm the memory pool with transactions. Such an attack is very costly to maintain. An attacker can create an infinite loop (a recursive covenant without an exit-path), but the progress of this infinite loop will be incremental, transaction by transaction. A node which is validating transactions in this infinite loop will not need to perform infinite computation in a finite time period. The node's resource limits will not be surpassed because it will automatically restrict the number of transactions it retains in memory to protect itself. It is not necessary to prove that the infinite chain of transactions will terminate, as is prevented by the halting problem.

\section{Conclusion}
\label{sec:del-key-conclusion}

In this chapter we introduced the deleted-key covenant protocol. A series of security objectives for the protocol were proposed and subsequently analysed to discern the precise circumstances in which they can be accomplished, given a set of predefined assumptions and a threat model outlining the potential capabilities of an attacker.

The study revealed that the resilience of the deleted-key covenant protocol against formidable attacks can be substantially enhanced with proper parameterisation of the multi-signature thresholds within both the covenant enforcement subscript and the custodial subscript. This insight serves as a crucial foundation for designing and implementing secure and efficient applications based on deleted-key covenants.

Additionally, this chapter offered a detailed examination of the critical aspects of engineering applications that rely on deleted-key covenants, including the safe composability of covenant transactions and the maintenance of the described security model. A classification of possible covenant commitments was presented, drawing attention to pertinent safety concerns related to modifying covenant transactions. This analysis was instrumental in our evaluation of dynamic fee allocation methods for deleted-key covenant applications, a summary of which was encapsulated in table \ref{tab:FeeMethods}. The potential of extending the basic proof-of-reserves protocol \cite{BIP127} for deleted-key covenants was explored, although the requisite interactivity with the verifier during the protocol was identified as a significant drawback.

Finally, through a baseline comparison with deleted-key covenants we revealed insights about alternative covenant proposals; comparing the class of possible covenant commitments, differences with dynamic fee allocation methods, and relative size and verification cost of each. 

In line with the objectives outlined in the introductory chapter, this investigation has successfully achieved the goal to `Design and analyze a novel covenant protocol that works with state-of-the-art Bitcoin features', as well as to `Provide a comparative analysis of alternative covenant constructions'. Moving forward, the forthcoming chapter will introduce a custody protocol that leverages the deleted-key covenant protocol, further extending the practical applications of this research. Thus, the exploration and refinement of Bitcoin covenants continues.
\chapter{Ajolote: Custody with Bitcoin Vaults}

\label{ch:vault-custody}

In this chapter we present Ajolote, a custody system that uses Bitcoin vault covenants. We specify the system primarily as a series of message sequence diagrams where the human-interaction is explicitly included. We evaluate the state dynamics of the system and its on-chain privacy properties. We compare the system with industry-standard multi-signature custody systems to explain how vault covenants can improve the state-of-the-art. By following the `ceremony analysis' methodology, we construct a risk model for the system. The specification and analysis include the system setup, operation, and recovery (to handle attack on and failure of system components). Finally, we provide new insights into the ongoing debate about consensus-enforce covenants in Bitcoin. This chapter addresses thesis objective \textbf{O3} (see \ref{sec:thesis-objectives}).

\section{Introduction}

\subsection{Overview}

Ajolote is an autonomous custody system, constructed to provide robust protection for individuals—referred to as the \textit{users}—who hold substantial wealth in bitcoin. It is a theoretical design and no implementation has yet been developed.   The custody system derives its name from the Ajolote, an amphibian renowned for its unparalleled regenerative abilities, including the restoration of its limbs and even its head. This aptly characterizes the inherent resilience of the Ajolote custody system, which preserves its operational integrity even in the face of the failure or compromise of individual signing devices. Like the amphibian, the system is designed to `regenerate', continuing to function effectively amidst adverse conditions and thereby reinforcing its reliability as a bitcoin custody solution. 

A comprehensive custody system facilitates the holding, management, and spending of bitcoin. This system should have rigorously defined procedures for its setup, operation, and recovery in case of a compromise or failure. Ajolote embodies such a system with multiple defensive layers, thereby conferring resilience to both failure and compromise. It builds upon the foundations of standard multi-signature custody systems but distinguishes itself through the innovative application of the novel deleted-key covenant protocol, as detailed in the previous chapter.

\begin{figure}
    \centering
    \includegraphics[width=0.9\textwidth]{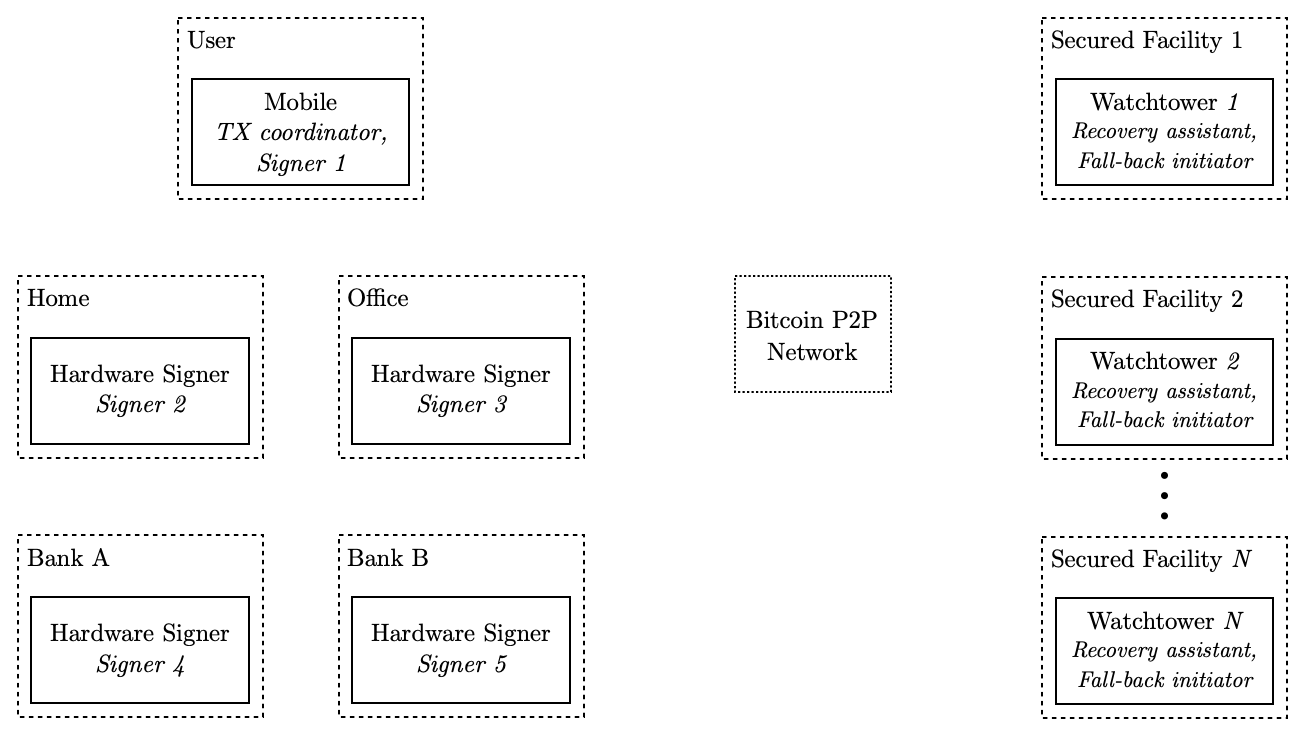}
    \caption{Entity diagram for the Ajolote custody system. Dashed boundaries represent distinct physical security environments. Solid boundaries represent distinct devices. Inside, their type and primary system function is noted (e.g. type: Hardware signer and function: \textit{signer 2}). The `User' environment represents the user's physical location, which often will overlap with their Home or Office environment, and on rare occasions may overlap with the Bank A and Bank B environment. The Bitcoin peer-to-peer network is represented with the dotted box. There are $W$ watchtowers, each in a distinct physical location and physical security environment.}
    \label{fig:entity}
\end{figure}

\begin{figure}
    \centering
    \includegraphics[width=0.8\textwidth]{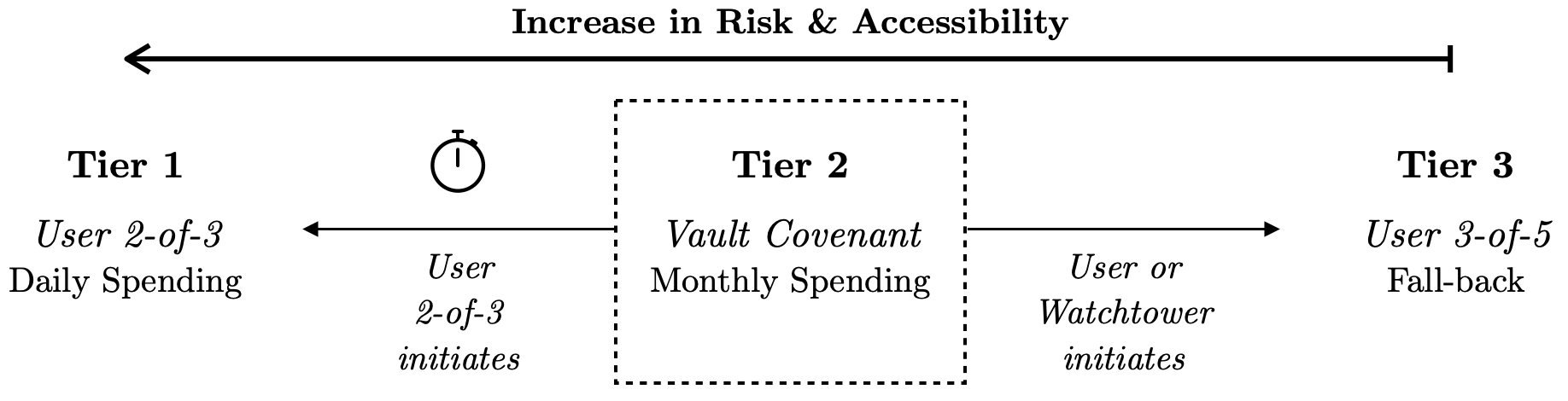}
    \caption{Three distinct risk-accessibility tiers in Ajolote. \textbf{Tier 1}; $2$-of-$3$ multi-signature controlled by the user. \textbf{Tier 2}; Vault covenant which can only pay to Tier 2 or Tier 3. \textbf{Tier 3}; $3$-of-$5$ multi-signature controlled by the user.}
    \label{fig:tiers}
\end{figure}

As operational prerequisites, the user must have access to four separate safe locations and be in possession of a mobile and four distinct hardware signing devices. A simplified entity diagram is shown in figure \ref{fig:entity}. The Ajolote system classifies accessibility into three distinct tiers, as depicted in figure \ref{fig:tiers}.

{}

{}

With each increase in accessibility comes an increase in risk of compromise. Tier 1 is utilised for a minor proportion of the total balance, with the majority of funds sequestered in Tier 2. A streamlined procedure is in place to transition to Tier 3, if necessary. Tier 1 adopts a $2-$of$-3$ signature access policy, while Tier 2 can also be accessed with a $2-$of$-3$ signature, provided a time-lock has expired. Tier 3 employs a $3-$of$-5$ signature access policy and is reserved as a fall-back system for failure and compromise scenarios.

A \textit{watchtower} is an automated server for network monitoring and response. By default, the user engages one or more third-party watchtower providers\footnote{Proficient users with expertise in server administration may opt to deploy their own watchtower(s).} to enforce a withdrawal policy from Tier 2. If that policy is breached, the watchtower or the user can trigger a fallback procedure from Tier 2 to Tier 3, provided they act before the time-lock expires. Accessibility and risk of compromise would be reduced, thus enhancing security. The watchtowers do not possess the ability to access funds directly, they only have the power to trigger the fallback procedure from Tier 2 to Tier 3.

Ajolote obviates the need for seed backups. Instead, the system offers resilience to a loss or theft of up to two signers. Furthermore, Ajolote comprises a set of recovery procedures to invalidate a lost signer and initialise a new one. This allows the user to regain strong safety and security if up to two signers are lost or stolen.

Though Ajolote is designed to function within the current consensus constraints of Bitcoin Core \textit{v24.0}, its efficiency and security could be enhanced if granular, consensus-enforced covenants were introduced to Script via a soft-fork upgrade.

\subsection{Covenant Pattern}
\label{subsec:cov-pattern}

The fall-back procedure is implemented with two types of covenant, which have a distinct transaction pattern and functional behaviour. The first type is a \textit{pay-to covenant}, a constraint on the redeeming transaction to create an output with a pre-defined amount and destination. 

The second type is a \textit{vault covenant}, a constraint on the redeeming transaction to create  an output with two spend paths: one time-locked path, and one immediate path. In Ajolote, the time-locked path can be used to send bitcoin to Tier 1, while the immediate path uses a pay-to covenant with the fall-back address of Tier 3 as its destination. The time-lock grants the user and its watchtowers time to observe the withdrawal attempt and to react by allowing or disallowing it. Watchtowers enforce withdrawal constraints composed of arbitrary rate-limits or permitted transacting hours. The fall-back address has a stricter access policy and is thus used to enhance security. Simultaneously, the fall-back address has a larger key-set and is thus used to enhance safety. 

\subsection{Related Work}

Besides the more general work on custody specifications and risk modelling discussed in chapter \ref{ch:custody}, there are several related works which influenced the design of Ajolote. The idea of a Bitcoin vault covenant was introduced by  M\"{o}ser \textit{et al.} \cite{moeser2016bitcoin}. The focus was on the technical mechanism for how to achieve the vault pattern, but lacked a deeper discussion of a complete custody system design and the benefits it yields. The technical mechanism required powerful new Script commands. 

In other work, McElrath introduced the basic idea of using a specific type of deleted-key covenant to enforce a vault \cite{P2TST}. The sketched mechanism had a weakness in that the key deletion process was a single-point of compromise. The primary benefit here was compatibility with the existing consensus rules in Bitcoin. However, there was no formal description nor analysis of the mechanism. Moreover, a custody system design was not present.

Hommel implemented a prototype of this deleted-key mechanism for a hardware signing device \cite{HommelVaultmbed}. This was a practical study on the data type requirements for message passing between signing devices and transaction coordination applications in the context of deleted-key covenants. Compatibility of hardware signing devices with covenants remains as a barrier for a real-world deployment.

Bishop developed a prototype implementation for constructing trees of pre-signed transactions with support for deleted-key vault covenants and their Script implementation \cite{BishopVaults}. Bishop noted the importance of compartmentalising funds in several separate `vaults', to prevent putting the user's entire balance at risk once the time-lock expires. Indeed, this is a critical insight for designing a complete custody system based on vaults. 

The foundational concepts and insights presented in this chapter draw inspiration from a collaborative effort between myself, McElrath, Bishop, and Hommel. This collaboration led to the development of two papers focusing on Bitcoin covenants \cite{Swambo2020cov} and vault-based custody \cite{Swambo2020vault}. While these papers provided a foundation, the corresponding chapters in this thesis are not merely reproductions. As noted in section \ref{intro:statement-originality}, with the consent of my collaborators, I have substantially revised, corrected, and extended these earlier works. This involved updating to reflect the current state-of-the-art, enhancing methodology, and conducting new analyses and evaluations. This adaptation serves to refine our collective understanding and to provide a more comprehensive and nuanced perspective on the subject at hand. 

As part of a series showcasing Sapio, a development framework for creating multi-transaction Bitcoin contracts, Rubin implemented a vault covenant (where the precise covenant mechanism is abstracted) \cite{Sapio,RubinVaults}. Rubin introduced a variation on the standard vault covenant, a `structured liquidity vault'. In this case, a sequence of transactions is kept wherein each transaction unlocks a portion of the total balance. Imagine a sequence of $n$ transactions and total balance, $B$. Each transaction releases $B/n$ bitcoin to the user and keeps the rest in the vault sequence. Each transaction has a time-lock $T$. This way, `liquidity' is structured to $B/n$ every $T$ blocks. The practical issue here is that the liquidity structure must be pre-specified, and both user's requirements and the price of Bitcoin can vary significantly in short time frames. 

The vault pattern is typically used as a demonstrative example of a covenant mechanism. O'Connor and Piekarska described how to implement one with their {\tt CHECKSIGFROMSTACK} covenant mechanism \cite{Covenants2}. Bartoletti \textit{et al.} used the vault pattern to demonstrate their formal model of Bitcoin contracts. More vault prototypes where implemented by O'Beirne \cite{ObVault} and Poinsot \cite{PoinsotVault} for the covenant mechanisms CTV and APOAS, respectively. Kirstein \textit{et al.} implemented a vault with Ethereum smart contracts and evaluated it with formal specification and verification methods \cite{PhoenixVault}. However, while useful for understanding the mechanisms, these demonstrations do not necessarily translate into effective custody solutions. 

The Revault custody system uses a type of vault, relying on a strict multi-signature policy and pre-signed transactions rather than a covenant mechanism. Unlike the rest, Revault has a comprehensive design specification \cite{practical-revault} with production-grade implementation libraries and a software application \cite{revault-repos}. This system was designed specifically for enterprise multi-party usage. 

An open-source project `Bitcoin Contracting Primitives Working Group' is building a repository of related resources and consolidating research efforts. Vaults is one of the use cases discussed and analysed \cite{contract-wg}. 

\subsection{Contributions}

The focus of this chapter is to explore the question, `How can the introduction of vaults into a custody system enhance its capabilities and improve upon the state-of-the-art?'. For this, a detailed specification of a custody system is required along with a comparison against its equivalent without a vault. Moreover, limitations and risks are created with the introduction of vaults to custody systems and these have yet to be explored in detail. Finally, while vaults are a typical example of the benefits of consensus-enforced covenants in Bitcoin, no study on precisely what benefits are gained has been carried out. This chapter addresses these issues.

\textbf{Introducing Ajolote}: In section \ref{sec:Ajolote-System} we introduce and specify Ajolote, a vault-based custody system, detailing every component and procedure for the setup (\ref{sec:setup-procedures}), operation (\ref{sec:ops-procedures}), and recovery (\ref{sec:rec-procedures}) phases.

\textbf{Evaluating the Design}: In section \ref{sec:Ajolote-Evaluation} we evaluate this design in several ways. We validate the specification of Bitcoin Scripts by using the Miniscript compiler. We describe an operational model to understand the state dynamics and demonstrate important trade-offs. We analyse the privacy properties associated with Ajolote transaction sequences and our use of taproot trees to encode Bitcoin Script. We compare Ajolote with the closest equivalent custody systems which do not use covenants or the vault pattern and offer clarity on the strengths and weaknesses of the vault pattern. 

\textbf{Risk and Threat Modelling}: With section \ref{sec:Ajolote-risk-model} we present a comprehensive risk model that spans the entire life-cycle of the custody system and propose a realistic threat model which, we argue, captures realistic attacker capabilities. 

\textbf{Exploration of Covenant Mechanisms}: Finally, in section \ref{sec:Consensus-covenants-Ajolote}, we explore some expected fundamental differences that derive from the choice of covenant mechanism to determine how consensus-enforced covenants can improve Ajolote and similar systems. Notably, we present a plausibility argument that only granular covenants can fix coin control issues that arise with using the vault pattern.

\subsection{Objectives}
\label{subsec:ax-objectives}

In designing Ajolote we set several objectives. 

\begin{itemize}
    \item[] \textit{Autonomous:} The user is the only entity that can initiate withdrawals and make external payments. No third-party can prevent this.
    \item[] \textit{No single point of failure (SPOF):} Any single hardware device, sensitive datum, or software program may fail or get lost, but the custody system will remain safe and available to the user. 
    \item[] \textit{No single point of compromise (SPOC):} An attacker may compromise any single hardware device, sensitive datum, or software program, but the custody system will remain secure and available to the user. 
    \item[] \textit{Usable:} The procedures to setup and operate the custody system will be sufficiently specified such that users can correctly follow it without creating unexpected safety or security issues. 
    % \item Openness
    \item[] \textit{Bitcoin compatible:} The custody system will be compatible with the consensus rules of Bitcoin.
\end{itemize}

In designing Ajolote we adhered to the following design patterns so as not to create either a SPOF or SPOC.

\begin{itemize}
    \item[] \textit{Modular:} Hardware and software components in the custody system can be replaced or updated, as needed.
    \item[] \textit{Redundant:} Sets of components can perform the same functions so that if one component fails, the custody system remains functional.
    \item[] \textit{Diverse:} Different supply chains, devices, and software are used across components to avoid fragility to malware, software bugs, hardware faults, and so on.
    \item[] \textit{Verifiable:} Failure or compromise of components must be detectable (where possible) so that recovery procedures can be enacted in time.
\end{itemize}

\subsection{Methodology}

Our methodology is underpinned by the following systematic steps:

\textbf{Specification}: We denote the custody system as Ajolote and outline the entire system, including cryptographic key material and their context of use, Scripts and output types,, transaction types and their format, functional elements, and human-device procedures.

\textbf{Analytical Examination}: We employ multiple methods to uncover design flaws and gain insights into custody with Bitcoin vaults, considering state dynamics, privacy properties, Script implementation, and a comparison with equivalent systems with no vault pattern and no covenants.

\textbf{Risk Analysis}: Conducting a thorough risk analysis, we adopt both asset-centric and attacker-centric approaches. That is to say, we list all sensitive and critical assets and the impact from having them leaked or stolen. Then, we describe explicitly the capabilities of an attacker in a threat model. Utilizing \textit{ceremony analysis} (see \ref{subsec:ceremonies}), we enumerate vulnerabilities, attacks, and risks of the Ajolote custody system.

\textbf{Consideration of Alternative Covenants}: Building on prior work, we investigate the potential of different types of consensus-enforced covenants in addressing issues discovered in vault-based Bitcoin custody.

\section{Ajolote System}
\label{sec:Ajolote-System}

In this section, we introduce the inter-dependent Ajolote data structures, which are organized systematically as follows:

\begin{equation*}
    \text{Cryptographic} \text{ keys} \rightarrow \text{High-level} \text{ access} \text{ policies} \rightarrow \text{Transaction} \text{ types} \rightarrow \text{Output} \text{ types} \rightarrow \text{TapTrees}
\end{equation*}

Our system specification needs to address several questions. What cryptographic keys are necessary? What physical and operational boundaries should those keys be subject to? Which keys are used for the access policies of the three risk-accessibility tiers from figure \ref{fig:tiers}? What transaction types can implement these conceptual tiers and the transitions between them? What constraints are encoded by covenant transactions? How should the access policy be implemented in TapScript? What are the functional elements (devices, services, instructions) that manage these data structures? This section provides detailed answers to each of these questions, in preparation for the following sections wherein the procedures are specified. 

\subsection{Bitcoin key boundaries} 

Each private key has well-defined usage boundaries to ensure that single points of failure or compromise are not created. Table \ref{tab:keys} summarises these boundaries. 

The user has 5 master extended private and public key-pairs labelled $(q_n, Q_{n})$ respectively, with $n \in \{1,2,3,4,5\}$. The user generates enforcement key-pairs $(e^v_1, E^v_1)$ and $(e^v_2, E^v_2)$ for each vault $v$. These are not extended because no key derivations are necessary. Each of $N$ watchtowers are labelled by $k$ and have master extended private and public keys $(w_k, W_k)$, where $k \in \{1,2,...,N\}$. 

\begin{table}
    \centering
    \begin{tabular}{|l|l|l|l|l|}
         \hline
         \rowcolor{lightgray} \textbf{Key-pair} & \textbf{Location} & \textbf{Device} & \textbf{Role} \\
         \hline
         $(q_1, Q_1)$ & With User & Mobile & Active \\
         \hline
         $(q_2, Q_2)$ & Home & Hardware signer & Active \\
         \hline
         $(q_3, Q_3)$ & Office & Hardware signer & Active \\
         \hline
         $(q_4, Q_4)$ & Bank A Safe Deposit & Hardware signer & Fall-back \\
         \hline
         $(q_5, Q_5)$ & Bank B Safe Deposit & Hardware signer & Fall-back \\
         \hline
         $(e^v_1, E^v_{1})$ & With User & Mobile & Enforcement \\
         \hline
         $(e^v_2, E^v_{2})$ & Home & Hardware signer & Enforcement \\
         \hline
         $(w_k, W_{k})$ & / & Server & Response Key \\
         \hline
    \end{tabular}
    \caption{The location, storage device, and operational role are shown for each type of key-pair. Non-capital and capital letters represent private and public keys respectively.}
    \label{tab:keys}
\end{table}

Recall from figure \ref{fig:entity} that the user keeps hardware signers in 4 distinct secured locations such as their home, office, bank A and bank B. The user has 4 distinct non-networked hardware signers on which to generate and store master key-pairs (ideally, different manufacturers and distributors). The user keeps another master key-pair on their mobile device which we assume is on their person at all times. Including the mobile device, there should never be more than two signers in the same location at any time. The enforcement private keys will be contained on the user's mobile and their primary hardware signer (in this case, the home signer). The watchtowers should be housed across several data centers or dedicated facilities managed by a service provider. Each watchtower keeps their own key-pair. 

The user has 3 \textit{active} master key-pairs and 2 \textit{fall-back} master key-pairs. Typically the mobile signer will be the \textit{initiator}, and the home signer will be the \textit{finaliser}, as per the roles in the partially-signed bitcoin transaction specification \cite{BIP174}. However, technically any key from the active and fall-back key-sets can fulfill these roles for outputs they control. The \textit{enforcement} private keys are ephemeral, and only used to enforce deleted-key covenants (though the public keys are persistent). The \textit{response} keys are used by watchtowers to initiate the reject procedure. 

\begin{figure}
    \centering
    \includegraphics[width=\textwidth]{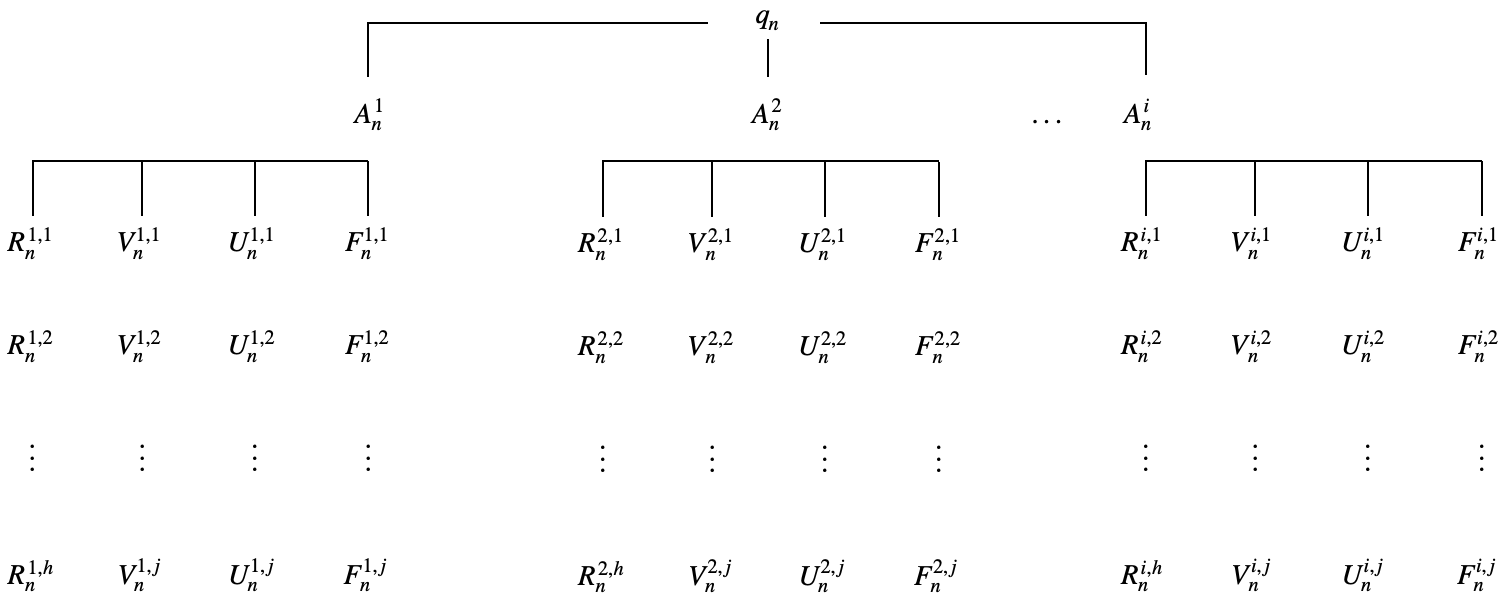}
    \caption{The key-tree for a master key-pair $(q_n, Q_n)$ in Ajolote. The master key-pair is generated and stored on the signer. Accounts, $A^i_n$, are hardened extended public keys to be shared with other signers. The current account is denoted by $i$. The signer may rotate to a new account by deriving a new hardened extended public key, $A^{i+1}_n$. Non-hardened public keys for each Ajolote output type are derived from the $i^{th}$ account and depths $h$ or $j$; $R^{i,h}_n$, $V^{i,j}_n$, $U^{i,j}_n$, $F^{i,j}_n$.}
    \label{fig:key-tree}
\end{figure}

Figure \ref{fig:key-tree} illustrates the key-tree for which the master key-pairs are the root. As per the hierarchical deterministic wallet specification \cite{BIP32}, the master key is comprised of a 32 byte master secret key and a 32 byte chain code. Master public keys are never used explicitly in transaction Scripts. They are confidential, contained to their device for their entire life time. 

To facilitate secure coordination, each signer will store a hardened public key derived from each of the master key-pairs. We call the first hardened extended child public key \textit{account 1}, $A^1_n$. At some point, the signer may need to rotate to a new account. This can be done by generating a new hardened public key as a child of the master. That way, any other entity (e.g. a compromised signer) that knows $A^i_n$ will not automatically know $A^{i'}_n$ where $i\neq i'$. For each signer, an arbitrary number of (non-hardened) public keys can be derived from the account that is in use. Any entity that knows the extended public key for the current account can derive these non-hardened public keys. We label the depth of the derivation with a second superscript $h$ or $j$. At each depth there are keys for different Ajolote output types, which will be described in the following sections; $R^{i,h}_n$, $V^{i,j}_n$, $U^{i,j}_n$, $F^{i,j}_n$. These keys use distinct derivation paths. Notice that $R^{i,h}_n$ can be at a different depth $h$ to the others, which are labelled by $j$. 

{}

\begin{figure}
    \centering
    \includegraphics[width=0.25\textwidth]{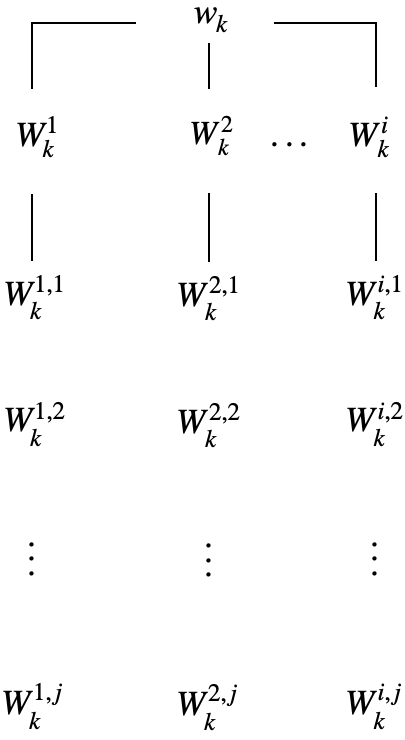}
    \caption{The key-tree for a watchtower's master key-pair $(w_k, W_k)$ in Ajolote. Watchtower accounts, $W^i_k$, are hardened extended public keys to be shared with other signers. A watchtower may rotate to a new account by deriving a new hardened extended public key, $W^{i+1}_k$. Non-hardened public keys are derived from the $i^{th}$ account with depth $j$; $W^{i,j}_k$.}
    \label{fig:WT-key-tree}
\end{figure}

The watchtower key-tree is shown in figure \ref{fig:WT-key-tree}. The watchtower uses the same account rotation mechanism as the user. The watchtowers' accounts are labelled as $W^i_k$. These are hardened extended public keys. From a watchtower's account, their key-chain can be constructed using non-hardened public key derivation. When the user instructs the service provider to re-initialise their watchtowers (during a recovery process), the watchtower will rotate to the next account.

{}

Note that these key-pairs are not backed up in Ajolote as is typically done with single-key custody systems. Backups are not necessary for system-wide resilience. Moreover, backups are cumbersome to use either when performing system-health checks or when in a recovery scenario. The added procedural complexity is not only a usability issue but it increases the attack-surface.

Of course, there are alternatives to the boundaries set here. For example, Ajolote could be deployed with a fiduciary co-signer (controlling an active key), or fiduciary recovery signer (controlling a fall-back key). Or, an active or fall-back key could be stored at the home of a family member or close friend. However, these alternatives would have to account for insider attacks and collusion in their risk model. 

\subsection{High-level Policies}

Let us write general, high-level descriptions of \textit{Miniscript policies} for Ajolote \cite{miniscript}. First we will label and describe some policies and policy fragments. Let the receive policy be 
\begin{equation}
    l_{receive} = thresh(2, pk(R_1^{i,h}), pk(R_2^{i,h}), pk(R_3^{i,h}))
\end{equation}
This represents a $2-$of$-3$ multi-signature policy, with the $3$ public keys derived from $Q_1$, $Q_2$, and $Q_3$, the user's active key-set. 

Similarly, we will use a $2-$of$-5$ multi-signature policy fragment to initiate withdrawals (using the active and fall-back key-set)
\begin{equation}
    l_{wit} = thresh(2, pk(V_1^{i,j}), pk(V_2^{i,j}), pk(V_3^{i,j}), pk(V_4^{i,j}), pk(V_5^{i,j}))
\end{equation} 
We write the covenant enforcement fragment as 
\begin{equation}
    l_{enf} = and(pk(E^{v}_1), pk(E^{v}_2))
\end{equation}
where the vault number, $v$, is an integer. This fragment requires signatures for both enforcement public keys. The custodial subscript for a fall-back output is intended to allow the user or any of their watchtowers to finalise a Fall-back transaction, \TX{fb} (described in section \ref{subsec:tx-design}). It can be satisfied with only $1$ signature. We write this as
\begin{equation}
    l_{pay-to-fb} = thresh(1,pk(U_1^{i,j}),pk(U_2^{i,j}),pk(U_3^{i,j}),pk(U_4^{i,j}),pk(U_5^{i,j}),pk(W^{i,j}_1),pk(W^{i,j}_2),...,pk(W^{i,j}_{N}))
\end{equation}
The spend policy fragment has the same format as the receive policy, but with keys from a different derivation path:
\begin{equation}
    l_{spend} = thresh(2, pk(U_1^{i,j}), pk(U_2^{i,j}), pk(U_3^{i,j}))
\end{equation}
Let us define the user’s fall-back policy as
\begin{equation}
   l_{fb} = thresh(3, pk(F_1^{i,j}), pk(F_2^{i,j}), pk(F_3^{i,j}), pk(F_4^{i,j}), pk(F_5^{i,j})) 
\end{equation}
This represents a $3-$of$-5$ multi-signature policy with all of the user's keys. With these, we can get an intuitive understanding for the meaning of the policies for each output type in Ajolote, as summarised in table \ref{tab:policies}. The length of the delay length is $T$ and it is implemented as a relative time-lock, denoted \textit{older} in policy. 

Finally, we define the test policy as
\begin{equation}
    l_{test} = thresh(5+N, pk(A_1^i), pk(A_2^i), pk(A_3^i), pk(A_4^i), pk(A_5^i), pk(W^i_1), pk(W^i_2),...,pk(W^i_{N}))
\end{equation}

It is used to verify whether or not the signing devices for the custody system are functional. 
\renewcommand{\arraystretch}{1.25}
\begin{table}
    \centering
    \begin{tabular}{|l|l|l|}
    \hline
        \rowcolor{lightgray} \textbf{Output Type} & \textbf{High-level Policy} \\
        \hline
         Receive Output & $l_{receive}$ \\
        \hline
         Vault Output & $and(l_{wit}, l_{enf})$ \\
         \hline
         Unvault Output & $or(and(l_{spend}, older(T)),$\\
         & $and(l_{pay-to-fb}, l_{enf}))$  \\
         \hline
         Fall-back Output & $l_{fb}$  \\
         \hline
         Test Output & $l_{test}$ \\
         \hline 
    \end{tabular}
    \caption{Miniscript policies and policy fragments in the Ajolote custody system.}
    \label{tab:policies}
\end{table}
\renewcommand{\arraystretch}{1}

\subsection{Transaction Design}
\label{subsec:tx-design}

\begin{figure}
    \centering
    \includegraphics[width=0.8\textwidth]{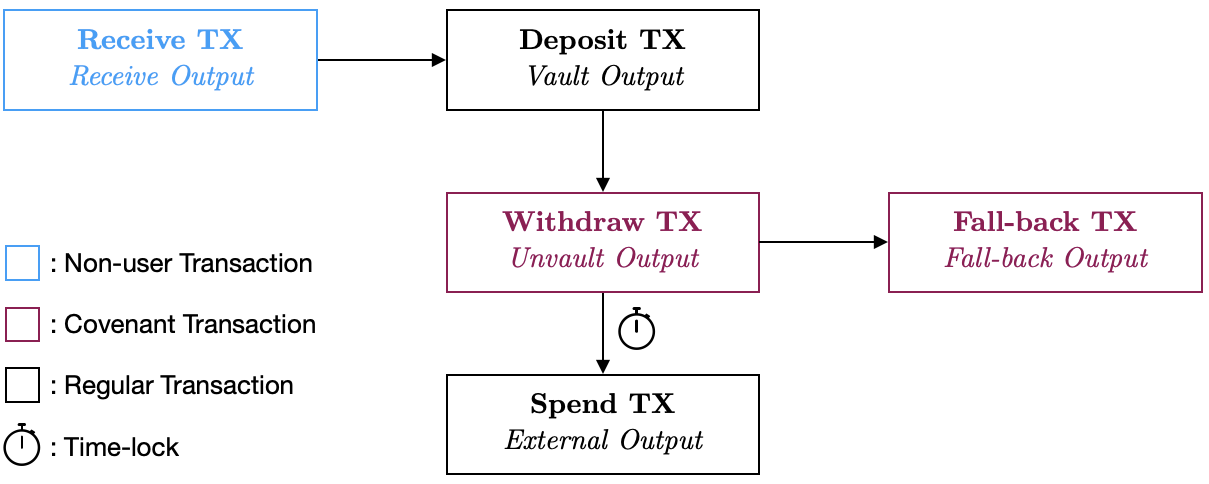}
    \caption{Transaction (TX) diagram of the Ajolote custody system. Covenant transactions are partially pre-signed; their enforcement subscripts are satisfied. The primary output type is written for each named transaction. These outputs are consumed by the proceeding transaction. Change outputs are not shown here for simplicity.}
    \label{fig:AjoloteTxs}
\end{figure}

The transaction diagram for Ajolote is shown in figure \ref{fig:AjoloteTxs}. The receive and vault-deposit are separated into two steps, two transactions, \TX{rec} and \TX{dep}. This is primarily because the vault-deposit procedure requires the user to complete an interactive deleted-key covenant protocol (as detailed in Chapter \ref{ch:bitcoin-covenants}) with their mobile signer, home signer, and watchtowers and bringing an external entity into that protocol creates unnecessary risk. Unfortunately, received funds are not automatically guarded by the vault covenant. Note that \TX{dep} may aggregate several receive outputs in one transaction, creating fewer vault outputs. 

Each vault output is created with a pair of covenant transactions; Withdraw, \TX{wit}, and Fall-back, \TX{fb}. We specify this procedure in section \ref{subsec:vault-dep}. The Spend transaction, \TX{spend}, is unrestricted, except that it is invalid until the time-lock expires ($T$ blocks after \TX{wit} is included in a block). 

The test transaction, \TX{test}, is not part of a sequence (and not shown in the diagram) but it is critical for testing and checking the health of the custody system (detailed in sections \ref{subsec:test-setup} and \ref{subsec:system-hc}). It is formatted for the Bitcoin test network. However, its input is invalid and the transaction should not be broadcast to the network. As we will see, it is sufficient for signers to verify each others' signatures on \TX{test} to check whether or not they have the correct setup data.  

{}

Recall that in table \ref{tab:FeeMethods} in chapter \ref{ch:bitcoin-covenants} we summarised the strategies for allocating fees to deleted-key covenant transactions. For Ajolote, we must choose an appropriate strategy for both \TX{wit} and \TX{fb}. \TX{wit} has a dependent transaction and thus cannot be modified. For now we will select a simple but inefficient strategy and leave optimization for future work. We prepare both \TX{wit} and \TX{fb} with a hard-coded base fee of 40 satoshis per byte. This is very high compared to the the average transaction fee per day for the last year\footnote{According to statistics provided on \href{https://statoshi.info}{statoshi.info}}. A very high fee gives more certainty that the transactions will be processed quickly and reduces reliance on a dynamic fee allocation strategy such as CPFP outputs. Implementing a CPFP output strategy would create procedural burden for the user and possibly create pinning attack vectors by the watchtowers. 

\subsection{TapTrees}
\label{sec:ax-taptrees}

Now, we will describe our TapTree constructions based on the high-level policies. The TapTrees are designed to enhance operational privacy by not requiring public keys, parameters and TapScripts to be revealed unnecessarily. They are equivalent to the high-level policies; the combination of TapScripts in a tree equates to the high-level policy, while the individual TapScripts can be stricter. Formally, the logical disjunction of the policy of all TapScripts in a TapTree is equivalent to the high-level policy.  

For a given output type, the TapTree will be unique from the leafs, through the branches to the root. Why? Because each leaf node is the hash of a unique TapScript which is comprised of newly derived public keys. The uniqueness of TapTrees prevents a network peer from observing simple and definitive clusters of outputs controlled by the user. A deeper analysis of the correlation of outputs is given in section \ref{subsec:tx-seq-corr}. 

Recall that TapTree information is leaked when the output is spent in a transaction, not when the output was created. The information is revealed as part of the witness field of the transaction. It contains signatures, the TapScript used, and the control block that contains branch nodes necessary to prove the inclusion of the TapScript in the TapTree. 

\begin{figure}
    \centering
    \includegraphics[width=0.6\textwidth]{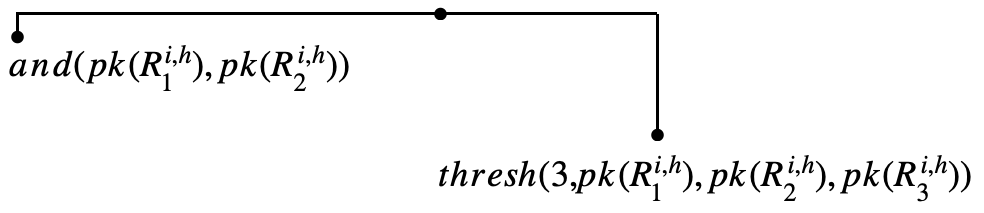}
    \caption{Receive Output TapTree.}
    \label{fig:receive-taptree}
\end{figure}

\textbf{Receive outputs.} In general, receive outputs will be spent using the mobile and home signer. This is the left-most branch of the TapTree depicted in figure \ref{fig:receive-taptree}. Using this branch reveals a 2-of-2 TapScript. Using this branch hides the fact that a 2-of-3 branch was available. In the case that either the mobile or home signer have been lost or compromised, the right-most branch may be used to access coins.

\begin{figure}
    \centering
    \includegraphics[width=0.8\textwidth]{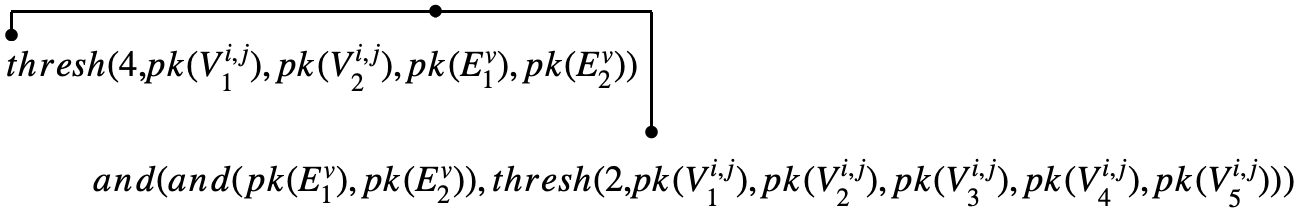}
    \caption{Vault Output TapTree.}
    \label{fig:vault-taptree}
\end{figure}

\textbf{Vault outputs.} Vault outputs will typically be spent using the mobile and home signer. This is the left-most branch of the TapTree in  figure \ref{fig:vault-taptree}. Using this branch reveals a 4-of-4 TapScript. If either the mobile or home signer have been lost or compromised, the right-most branch may be used to withdraw funds from the vault. This branch would reveal a TapScript with 2 required signatures and a 2-of-5 multi-signature. 

\begin{figure}
    \centering
    \includegraphics[width=\textwidth]{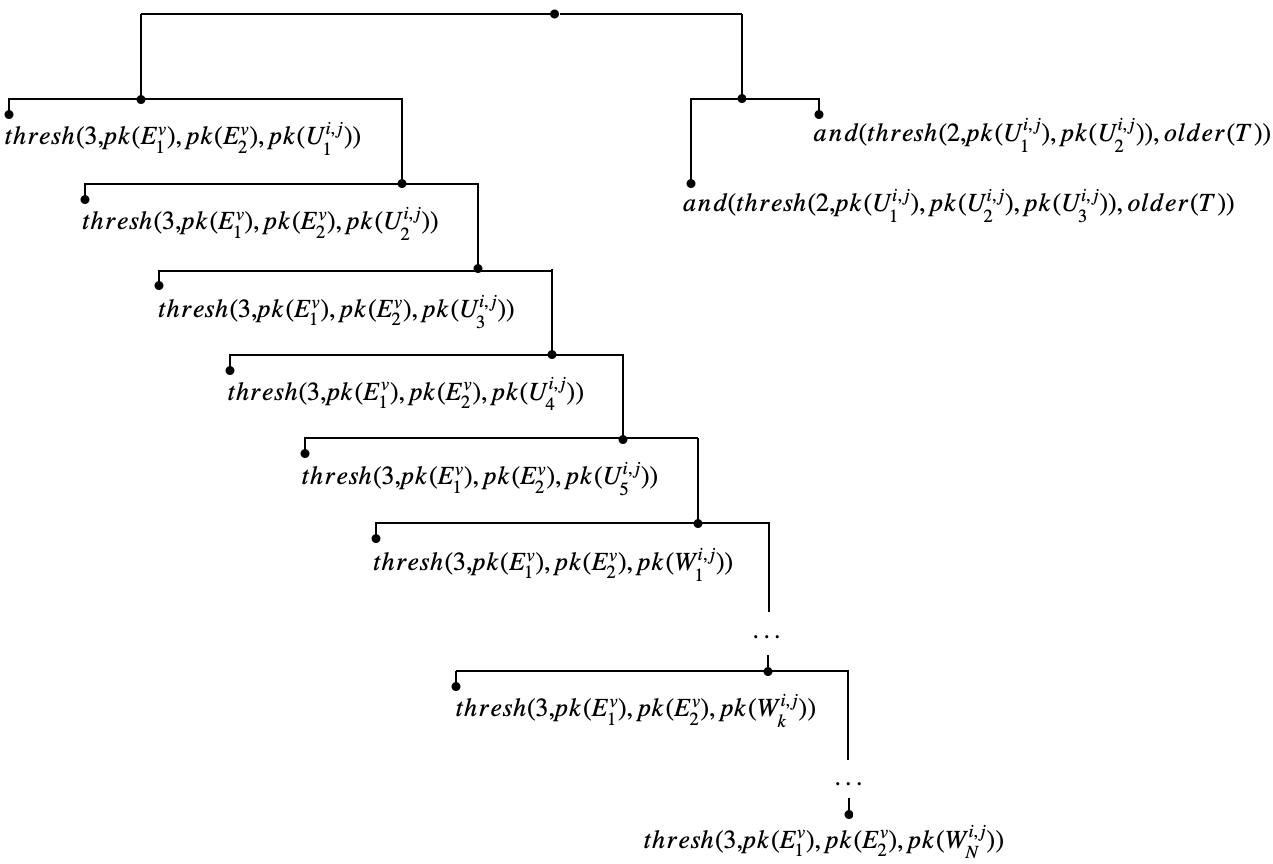}
    \caption{Unvault Output TapTree.}
    \label{fig:unvault-taptree}
\end{figure}

\textbf{Unvault outputs.} This is the most complicated TapTree in Ajolote. It is depicted in figure \ref{fig:unvault-taptree}. For successful withdrawals, the user will typically spend using the 2-of-2 TapScript between the mobile and home signer, after waiting for the time-lock to expire (the right-most branch). This would reveal a 2-of-2 TapScript with a time-lock of length $T$. The user has the option to spend with a 2-of-3 between the mobile, home and office signers. In this case, a 2-of-3 TapScript with a time-lock would be revealed. In both cases, a withdrawal reveals a minimum TapTree depth of 2 (since te control block contains two TapBranch hashes). For reject procedures, a 3-of-3 TapScript will be revealed. A minimum TapTree depth of up to $N+5$ will also be revealed, depending on which watchtower initiates the procedure. The rationale for structuring the tree to have a maximum depth of $N+5$ was to obfuscate how many watchtowers are present to observing network peers. For fall-back procedures, any of the users' signers can be used.  

\begin{figure}
    \centering
    \includegraphics[width=0.65\textwidth]{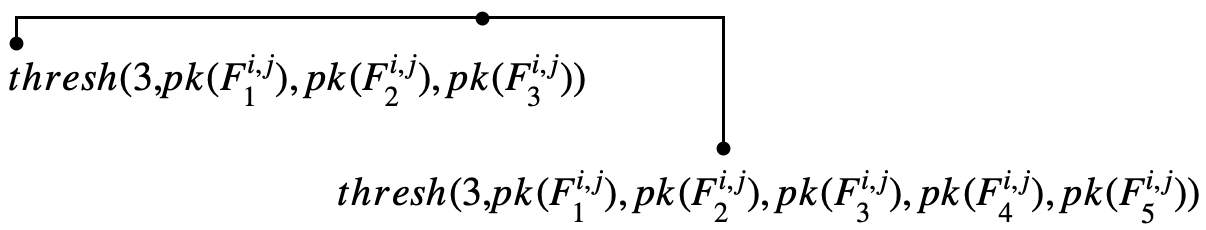}
    \caption{Fall-back Output TapTree.}
    \label{fig:fb-taptree}
\end{figure}

\textbf{Fall-back outputs.} For withdrawals that are rejected, the left-most branch of the TapTree depicted in figure \ref{fig:fb-taptree} will be used. This will reveal a 3-of-3 TapScript. For fall-back scenarios, a 3-of-5 multi-signature will be revealed. 

In table \ref{tab:Ajolote-scripts} we summarise the result of compiling the Miniscript policies (for each TapScript leaf in each TapTree) to Bitcoin Script. Note that the compiler is for version 0 segregated witness output types and therefore supports $\tt OP\_CHECKMULTISIG$ and $\tt OP\_CHECKMULTISIGVERIFY$. As specified in BIP-342 \cite{BIP-Tapscript}, we can replace these disabled instructions with $\tt OP\_CHECKSIGADD$ and $\tt OP\_NUMEQUAL$\footnote{A Script $2$ $P_1$  $P_2$  $P_3$  $3$ $\tt OP\_CHECKMULTISIG$ can be implemented as $P_1$ $\tt OP\_CHECKSIG$ $P_2$ $\tt OP\_CHECKSIGADD$ $P_3$ $\tt OP\_CHECKSIGADD$ $2$ $\tt OP\_NUMEQUAL$ with TapScript.}. We have adjusted the output of the compiler to reflect this. 

\renewcommand{\arraystretch}{1.5}
\begin{table}
    \centering
    \begin{tabular}{|l|p{6.5cm}|>{\raggedright\arraybackslash}p{7cm}|}
        \hline
        \rowcolor{lightgray} \textbf{Output} & \textbf{Miniscript Policy} & \textbf{Bitcoin Script} \\
        \hline
        Receive & $and(pk(R^{i,m}_1),pk(R^{i,m}_2))$ & $R^{i,m}_1$ ${\tt OP\_CHECKSIGVERIFY}$ $R^{i,m}_2$ ${\tt OP\_CHECKSIG} $  \\ 
% Script: 70 WU
% Input: 146.000000 WU
% Total: 216.000000 WU
        \hline
         Receive & $thresh(3,pk(R^{i,h}_1),pk(R^{i,h}_2),pk(R^{i,h}_3))$ & $R^{i,h}_1$ $\tt OP\_CHECKSIGVERIFY$ $R^{i,h}_2$ $\tt OP\_CHECKSIGVERIFY $ $R^{i,h}_3$ $\tt OP\_CHECKSIG$ \\
% Script: 105 WU
% Input: 219.000000 WU
% Total: 324.000000 WU
        \hline
        Vault&$thresh(4,pk(V^{i,j}_1),pk(V^{i,j}_2),$ $pk(E^{v}_1),pk(E^{v}_2))$ & $V^{i,j}_1$ $\tt OP\_CHECKSIGVERIFY$ $V^{i,j}_2$ $\tt OP\_CHECKSIGVERIFY$ $E^{v}_1$ $\tt OP\_CHECKSIGVERIFY$ $E^{v}_2$ $\tt OP\_CHECKSIG$ \\
% Script: 140 WU
% Input: 292.000000 WU
% Total: 432.000000 WU
%         \hline
%         Vault&$and(and(pk(E^{v}_1),pk(E^{v}_2)),thresh(2,$ $pk(V^{i,j}_1),pk(V^{i,j}_2),pk(V^{i,j}_3),$ $pk(V^{i,j}_4),pk(V^{i,j}_5))$ & $E^{v}_1$ $\tt OP\_CHECKSIGVERIFY$ $E^{v}_2$  $\tt OP\_CHECKSIGVERIFY$ $2$  $V^{i,j}_1$
% $V^{i,j}_2$ $V^{i,j}_3$ $V^{i,j}_4$ $V^{i,j}_5$ $5$ $\tt OP\_CHECKMULTISIG$ \\
        \hline
        Vault&$and(and(pk(E^{v}_1),pk(E^{v}_2)),thresh(2,$ $pk(V^{i,j}_1),pk(V^{i,j}_2),pk(V^{i,j}_3),$ $pk(V^{i,j}_4),pk(V^{i,j}_5))$ & $E^{v}_1$ $\tt OP\_CHECKSIGVERIFY$ $E^{v}_2$  $\tt OP\_CHECKSIGVERIFY$  $V^{i,j}_1$ $\tt OP\_CHECKSIG$ $V^{i,j}_2$ $\tt OP\_CHECKSIGADD$ $V^{i,j}_3$ $\tt OP\_CHECKSIGADD$ $V^{i,j}_4$ $\tt OP\_CHECKSIGADD$ $V^{i,j}_5$ $\tt OP\_CHECKSIGADD$ $2$ $\tt OP\_NUMEQUAL$ \\
% Script: 243 WU
% Input: 293.000000 WU
% Total: 536.000000 WU
        \hline
        Fall-back&$thresh(3,pk(F^{i,j}_1),pk(F^{i,j}_2),pk(F^{i,j}_3))$& $F^{i,j}_1$ $\tt OP\_CHECKSIGVERIFY$ $F^{i,j}_2$ $\tt OP\_CHECKSIGVERIFY$ $F^{i,j}_3$ $\tt OP\_CHECKSIG$ \\
% Script: 105 WU
% Input: 219.000000 WU
% Total: 324.000000 WU
        \hline
        Fall-back&$thresh(3,pk(F^{i,j}_1),pk(F^{i,j}_2),$ $pk(F^{i,j}_3),pk(F^{i,j}_4),pk(F^{i,j}_5))$ & $F^{i,j}_1$ $\tt OP\_CHECKSIG$ $F^{i,j}_2$ $\tt OP\_CHECKSIGADD$ $F^{i,j}_3$ $\tt OP\_CHECKSIGADD$ $F^{i,j}_4$ $\tt OP\_CHECKSIGADD$ $F^{i,j}_5$ $\tt OP\_CHECKSIGADD$ $3$ $\tt OP\_NUMEQUAL$ \\
% Script: 173 WU
% Input: 220.000000 WU
% Total: 393.000000 WU
        \hline
        Unvault&$and(thresh(2,pk(U^{i,j}_1),$ $pk(U^{i,j}_2)),older(T))$& $U^{i,j}_1$ $\tt OP\_CHECKSIGVERIFY$ $U^{i,j}_2$ $\tt OP\_CHECKSIGVERIFY$ $T$ 
$\tt OP\_CHECKSEQUENCEVERIFY$ \\
% Script: 74 WU
% Input: 146.000000 WU
% Total: 220.000000 WU
        \hline
        Unvault&$and(thresh(2,pk(U^{i,j}_1),pk(U^{i,j}_2),$ $pk(U^{i,j}_3)),older(T))$& $U^{i,j}_1$ $\tt OP\_CHECKSIG$ $U^{i,j}_2$ $\tt OP\_CHECKSIGADD$ $U^{i,j}_3$ $\tt OP\_CHECKSIGADD$ $2$ $\tt OP\_NUMEQUALVERIFY$ $T$ $\tt OP\_CHECKSEQUENCEVERIFY$ \\
% Script: 109 WU
% Input: 147.000000 WU
% Total: 256.000000 WU
        \hline
        Unvault&$thresh(3,pk(E^v_1),pk(E^v_2),pk(U^{i,j}_n))$& $E^v_1$ $\tt OP\_CHECKSIGVERIFY$ $E^v_2$ $\tt OP\_CHECKSIGVERIFY$ $U^{i,j}_n$ $\tt OP\_CHECKSIG$ \\
% Script: 105 WU
% Input: 219.000000 WU
% Total: 324.000000 WU
        \hline
        Unvault&$thresh(3,pk(E^v_1),pk(E^v_2),pk(W^{i,j}_k))$& $E^v_1$ $\tt OP\_CHECKSIGVERIFY$ $E^v_2$ $\tt OP\_CHECKSIGVERIFY$ $W^{i,j}_k$ $\tt OP\_CHECKSIG$ \\
% Script: 105 WU
% Input: 219.000000 WU
% Total: 324.000000 WU
        \hline
        % & & \\
    \end{tabular}
    \caption{For each Ajolote output type, the Miniscript policy is shown with its compiled Bitcoin Script.}
    \label{tab:Ajolote-scripts}
\end{table}
\renewcommand{\arraystretch}{1}

\subsection{Functional Elements}

\subsubsection{Mobile}

The mobile device acts as a coordinator and signer for the custody system. The user primarily interacts with the system through the user interface on the mobile. The watchtowers communicate with the mobile. The hardware signers communicate with the mobile. However, the mobile is not trusted to behave correctly. It can behave in arbitrarily malicious ways. It can tamper with messages or display false information to the user. To alleviate this, the procedures which comprise the setup, operation and recovery often involve verifying that the mobile data is consistent with the hardware signers and watchtowers. 

The mobile runs a (hypothetical) Ajolote application. It supports everything a generic bitcoin transaction coordinator and signer should support, including; hardware signer support, consensus-library, key-management, signature generation and verification, transaction validation, partially signed bitcoin transactions, hierarchical deterministic derivation support, output script descriptors, address utilities, and Miniscript policies. The application also runs a bitcoin-core full-node but prunes most of the history, retaining only recent blocks. Alternatively, it could connect to a full-node server run by the user. However, for simplicity we consider that the mobile itself is operating a full-node. 

Moreover, the Ajolote application includes an implementation to enact every procedure (as described in the following sections). It has a library of Ajolote data types, message types, transaction and output types, TapTrees, Miniscript policies and scripts. It has an implementation to track the complete state and history of the custody system; both what is on-chain as unspent transaction outputs and off-chain as covenant transactions. It has a secure-delete implementation for strong enforcement of the covenants. It has support for the Noise protocol and the Onion Router (TOR) protocol. Finally, it manages user authentication with a PIN and bio-metric reading. 

\subsubsection{Watchtower}

A Watchtower (WT) is a server program that performs specific monitor and response functions on behalf of the user. A WT can be containerised (e.g. with Docker) for easy deployment and launched with orchestration services on cloud providers (e.g. Docker swarm) to enable replication with geographical redundancy for reliable up-time. Ideally, the WT implementation will be open-source and verifiable by the user and the industry. 

In Ajolote a WT broadly performs the following functions; 
\begin{itemize}
    \item[] \textbf{Bitcoin full-node: } The WT continuously monitors the Bitcoin blockchain. It establishes connections to the peer-to-peer network using TOR to hide its IP address. It validates each transaction and each block it sees in order to remain in consensus and maintain an accurate view of the current state.
    \item[] \textbf{Onion Service: } The WT operates as an anonymous Onion Service for the user, to keep its IP address private and reduce the network-based attack surface. 
    \item[] \textbf{Covenant storage provider: } The WT maintains a register of user \textit{vaults}, and stores partially-signed (with enforcement signatures) transaction pairs $(TX^{wit}, TX^{fb})$ for each vault. 
    \item[] \textbf{Withdrawal constraint enforcer: } Upon detecting a user's $TX^{wit}$, the WT validates whether or not the withdrawal attempt is permitted according to the \textit{withdrawal constraints} (set during watchtower initialisation). If so, the WT does not respond further. If not, the WT signs and broadcasts the associated $TX^{fb}$.
    \item[] \textbf{Ajolote recovery assistant: } Upon request from the user, the WT will send its registry of vaults and partially-signed transaction pairs $(TX^{wit}, TX^{fb})$.
\end{itemize}  

Messages between a WT and the user are transported through encrypted and authenticated channels. The user and WT establish Noise channels during the setup procedure (detailed in section \ref{subsec:wt-init}). They use Noise key-pairs which are distinct from their Bitcoin key-pairs. We denote the user's and the $k^{th}$ WT's noise key-pairs as $(p^{Noise},P^{Noise})$ and $(w^{Noise}_k,W^{Noise}_k)$ respectively. Moreover, the Watchtower and user communicate with specific messages only; those included in the message sequence diagrams for each procedure specified in the following sections. Other message types are considered to be invalid inputs and are ignored. 

\subsubsection{Hardware signers}
\label{subsubsec:HS}

The Ajolote custody system requires hardware signers that support both generic Bitcoin functionality and Ajolote-specific functionality. The hardware signers are offline devices. 

Generic Bitcoin functionality includes; key-management, signature generation and verification, transaction validation, partially signed bitcoin transactions, hierarchical deterministic derivation support, output script descriptors, address utilities, and  Miniscript policies. 

Ajolote-specific functionality includes; secure-delete functionality (for covenant enforcement keys), Ajolote message types (as in the procedure specifications), Ajolote transaction and output types, Ajolote TapTree support, and Ajolote key-tree support. In particular, there will be strict formatting for message, transaction and output types that the device enforces. 

Hardware signers are PIN protected. Furthermore, the hardware signer wipes itself after more than three invalid PIN entries. The hardware signer can connect to a transaction coordinator via a wire or through a camera-QR code channel (where both devices have a screen and a camera). Hardware signers have sufficient storage capacity, and encrypt sensitive data before storing it. 

\subsubsection{User Manual}

A (hypothetical) user manual guides the user in all aspects of the system and ensures they act as needed to maintain the integrity and security of their custody system. The user manual describes how the user should enact each procedure during setup, operation, and recovery. This is an essential functional element without which the user could be tricked to compromise their own system. Moreover, it makes the system accessible to less experienced users, giving them confidence to operate an advanced custody system. Ideally, the manual would be well-designed and user-friendly, in addition to being correct.  

\subsubsection{Computer}

A computer is used during the setup phase in order to double-check that data received from the service provider and imported to the user's other devices have not been tampered with. The computer runs a minimal software tool to fill in a few data types and transfer these to hardware signers. 

\section{Ajolote Setup Procedures}
\label{sec:setup-procedures}

The setup is required to initialise all system components, generate all key material, and test the system. Without a secure setup, no operation can be secure regardless of how well designed it is. The setup has been decomposed into procedures; mobile initialisation, watchtower initialisation, hardware signer initialisation, setup finalisation, and the setup test. These should be completed sequentially. The user must abort if any verification step fails. Devices must abort if any verification step fails or if they receive an unexpected input. 

\subsection{Pre-requisites}

Here is a list of pre-requisites for the setup.
\begin{itemize}
    \item The user has obtained and verified an Ajolote user manual to guide them through the setup.
    \item The user has prepared a private space to enact the setup. 
    \item The user has a computer, mobile, four hardware signers, paper, pen, and the user manual. 
    \item The user has prepared their four hardware signer storage locations: Home, Office, Bank A and Bank B. 
    \item The user's computer and mobile are connected to different networks; Wi-Fi and mobile data.  
    \item The watchtowers are live and their administrating service provider has a list of their hidden-service onion addresses. 
    \item Hardware signers have Ajolote software installed. Due to differences in how hardware signers handle application management, we do not specify this procedure. Users have already set their PIN on the hardware signers and have backed up that PIN. 
    \item The computer has downloaded the Ajolote setup software. 
\end{itemize}

\subsection{Mobile Initialisation}
\label{subsec:mobile-init}

\begin{figure}
    \centering
    \includegraphics[width=\textwidth]{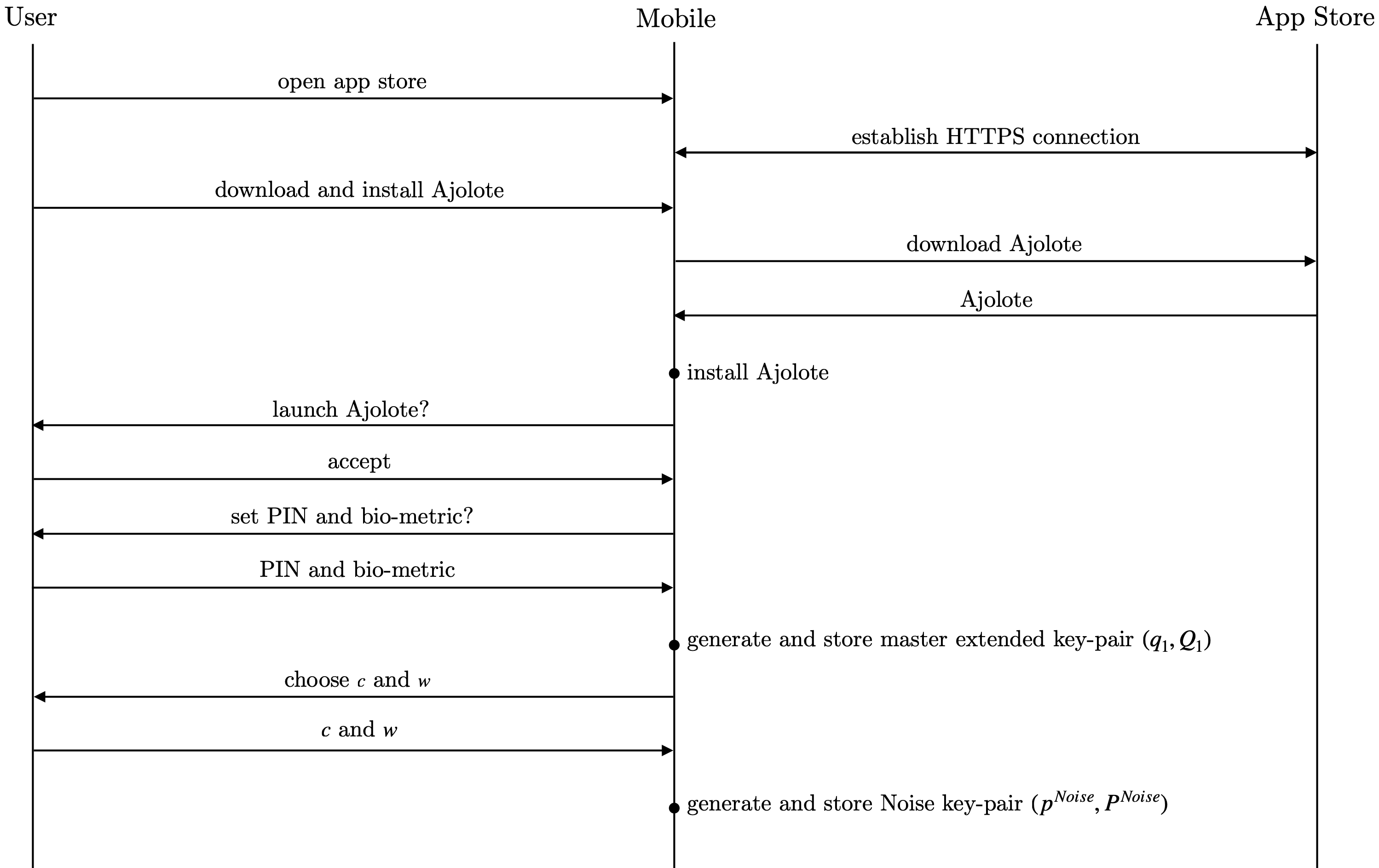}
    \caption{Mobile initialisation. The final mobile state contains: Ajolote software, Bitcoin master extended key-pair, Noise key-pair, configuration parameters $c$, withdrawal constraints $w$, data to verify PIN and bio-metric tokens.}
    \label{fig:mobile-init}
\end{figure}

The mobile initialisation is specified in figure \ref{fig:mobile-init}. Notice that we specify the sub-procedure as a ceremony where the user is included in the protocol specification. The user instructs the mobile to connect with the App store, then to download and install `Ajolote' software from the App store. The mobile and App store communicate using HTTPS. This way, the mobile can authenticate the app store and the communication is encrypted in both directions. The app store serves the requested software and the mobile installs it. Upon launching it, the software prompts the user to set both a PIN and a bio-metric authentication token. The mobile will not store copies of the PIN and bio-metric data, but only sufficient data to verify them (such as their hash). The mobile generates its Bitcoin master extended key-pair and stores it in the mobile's secure element. The software prompts the user to set the configuration parameters $c = \{T, V_{min}, V_{max}, N\}$ and the withdrawal constraints $w$. $T$ is the time-lock length and is constant for all withdrawals. $V_{min}$ and $V_{max}$ are the minimum and maximum amounts of bitcoin the user intends to deposit into each vault. $N$ is the total number of watchtowers the user will hire. Finally, the software generates the Noise key-pair and stores them in the mobile's secure element. 

{}

For the remainder of the setup, the user maintains the same authenticated session on the Ajolote mobile application.

\subsection{Watchtower Initialisation}
\label{subsec:wt-init}

\begin{figure}
    \centering
    \includegraphics[width=\textwidth]{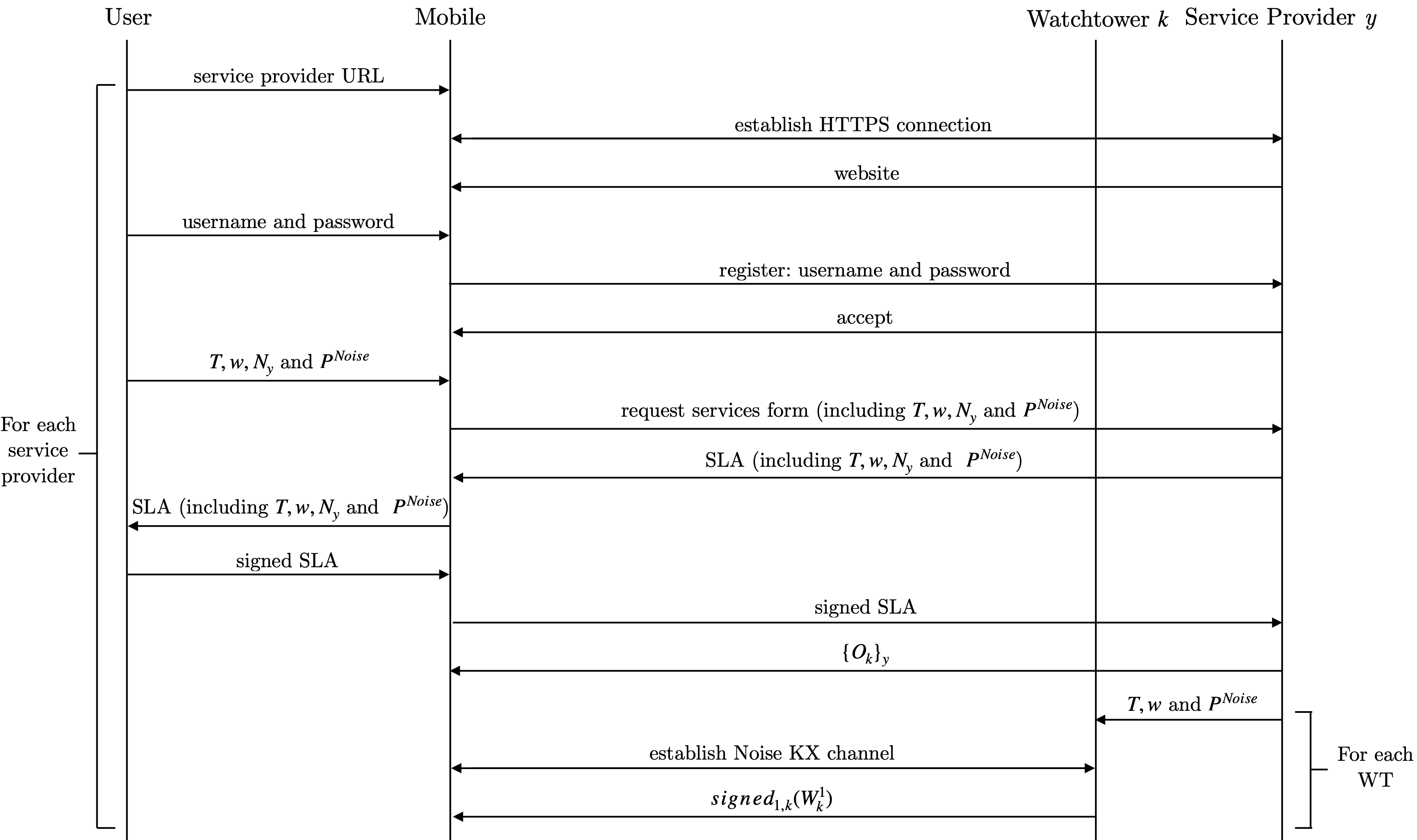}
    \caption{Watchtower (WT) initialisation. The final state of the Ajolote application includes the onion address, Bitcoin public key, and associated service provider for each watchtower; $(O_k, W^1_k, y)$ $ \forall$ $y,k$. The final state of each watchtower includes $(O_k, w_k, W_k, T, w, P^{Noise})$.}
    \label{fig:wt-init}
\end{figure}

In figure \ref{fig:wt-init} we depict the initialisation of the watchtower. The user interacts with the browser application and the Ajolote application, both on their mobile. We use a `Service Provider' abstraction which does not depict the system architecture that enables the service provider to serve their website and control their watchtowers. There are at least two service providers, labelled by $y$. There are $N_y$ watchtowers for each service provider. Each watchtower is labelled by $k$. Note that the total number of watchtowers is $\sum_y N_y = N$. Rather than repeat the same message sequence for each service provider and watchtower, we depict which messages are repeated and show an arbitrary service provider and watchtower. 

The user inputs a URL for the service provider and the mobile and Service provider establish a HTTPS connection. The service provider serves the mobile with their website. The user registers with each service provider by inputting a username and password. These can be used by the service provider to authenticate the user later. The user copies $P^{Noise}$ from Ajolote application and pastes it into their browser application. The user shares relevant setup information ($T, w, N_y, P^{Noise}$) with the service providers. Service providers share a service level agreement (SLA) with the user and the user must evaluate it before signing and making an initial payment. In particular, the user verifies that the setup information is correct. Service providers distribute $(T, w, P^{Noise})$ to each of their watchtowers to store these data. The service provider shares the set of their watchtowers' onion addresses, $\{O_k\}_y$ with the mobile via the browser. The user imports them into the Ajolote application. 

The Ajolote application establishes a Noise KX communication channel with each of the watchtowers \cite{NoiseHS}. `K' means that the user's Noise public key is known by the watchtower. The watchtower can authenticate the user. `X' means the watchtower's Noise public key is transmitted to the user during the handshake. The user's authentication of the watchtower is transitive, relying on secure communication with the service provider. The watchtowers send their first account (a Bitcoin extended public key), $W^{1}_k$, signed with their associated private key, denoted as $signed_{1,k}(W^1_k)$. The mobile can thus verify that the watchtower has control of the public key. The Ajolote application stores $(O_k, W^1_k, y)$ $ \forall$ $y,k$.  

\subsection{Hardware Signer Initialisation}

\begin{figure}
    \centering
    \includegraphics[width=0.8\textwidth]{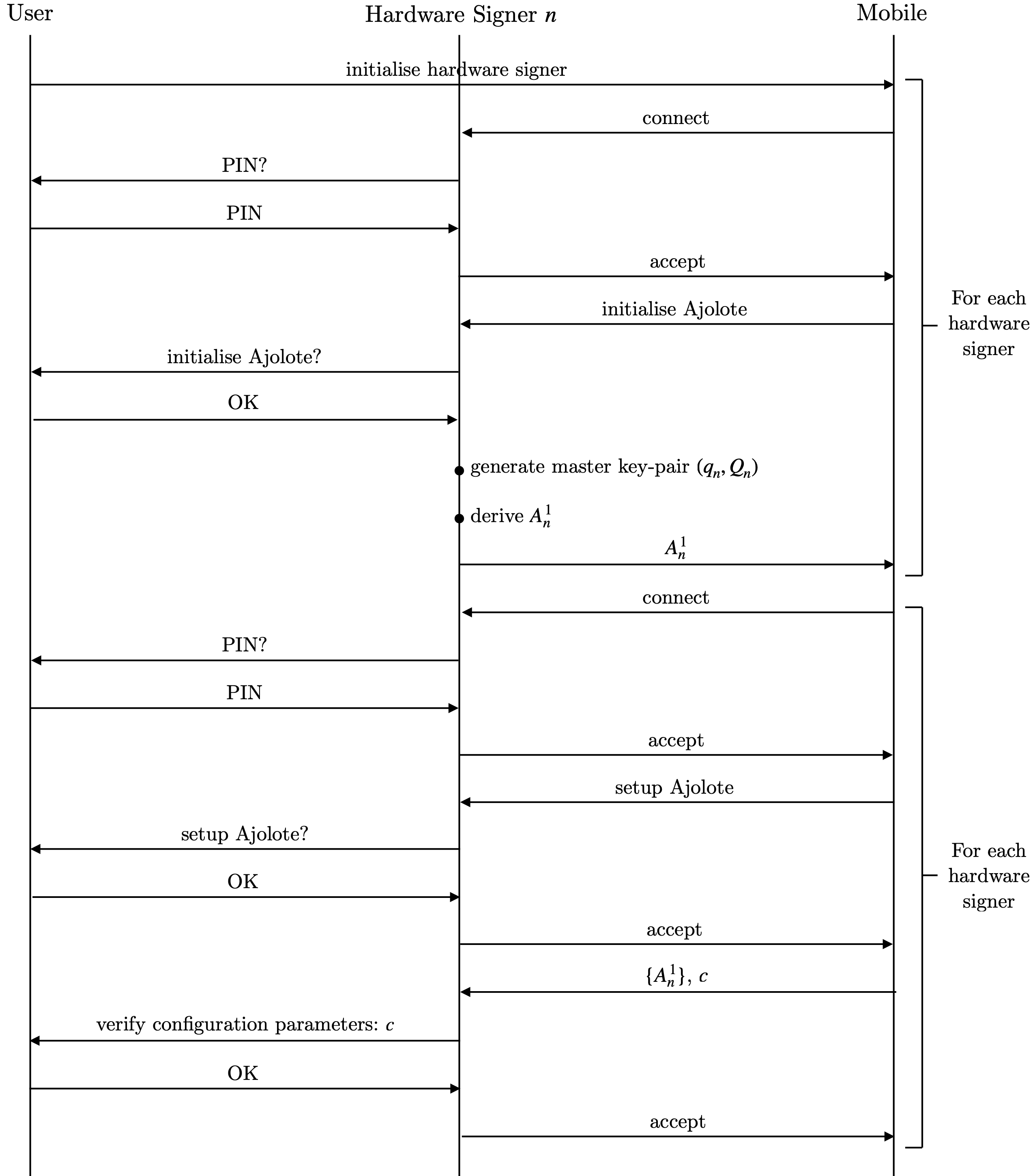}
    \caption{Hardware signer initialisation. The state of the Ajolote application (on mobile) now includes the complete set of first accounts for hardware signers, $\{A^1_n\}$. The state of the hardware signers includes $\{A^1_n\}$ and $c$.}
    \label{fig:hs-init}
\end{figure}

Figure \ref{fig:hs-init} depicts the initialisation procedure for all four hardware signers. It involves initial key generation and sharing relevant public keys across all signers.  Rather than repeat the message sequence for each of them, we show which sequence is repeated for an arbitrary hardware signer labelled as $n \in \{2,3,4,5\}$. Hardware signers may connect to the mobile using a wire or using a camera and QR codes. Messages sent from the hardware signer to the user are displayed on its screen. Messages sent from the user to the hardware signer use input buttons.

The hardware signer requires the user to enter a PIN before allowing the mobile to send requests to it. The first request is to initialise Ajolote and it also requires user confirmation. In response, the hardware signer generates its Bitcoin master extended key-pair $(q_n,Q_n)$ and computes its first account $A^1_{n}$. Then it sends $A^1_{n}$ to the mobile. After enacting this with each signer, the user must again connect the mobile to each hardware signer in turn to import the complete set of first accounts. This set is written as $\{A^1_{n}\}$. For now, this data remains unverified and could be inconsistent across hardware signers. The mobile also sends the configuration parameters $c$ to the hardware signer. The user verifies that what is written on the hardware signer's display matches their decision. 

{}

\subsection{Setup Finalisation}

\begin{figure}
    \centering
    \includegraphics[width=\textwidth]{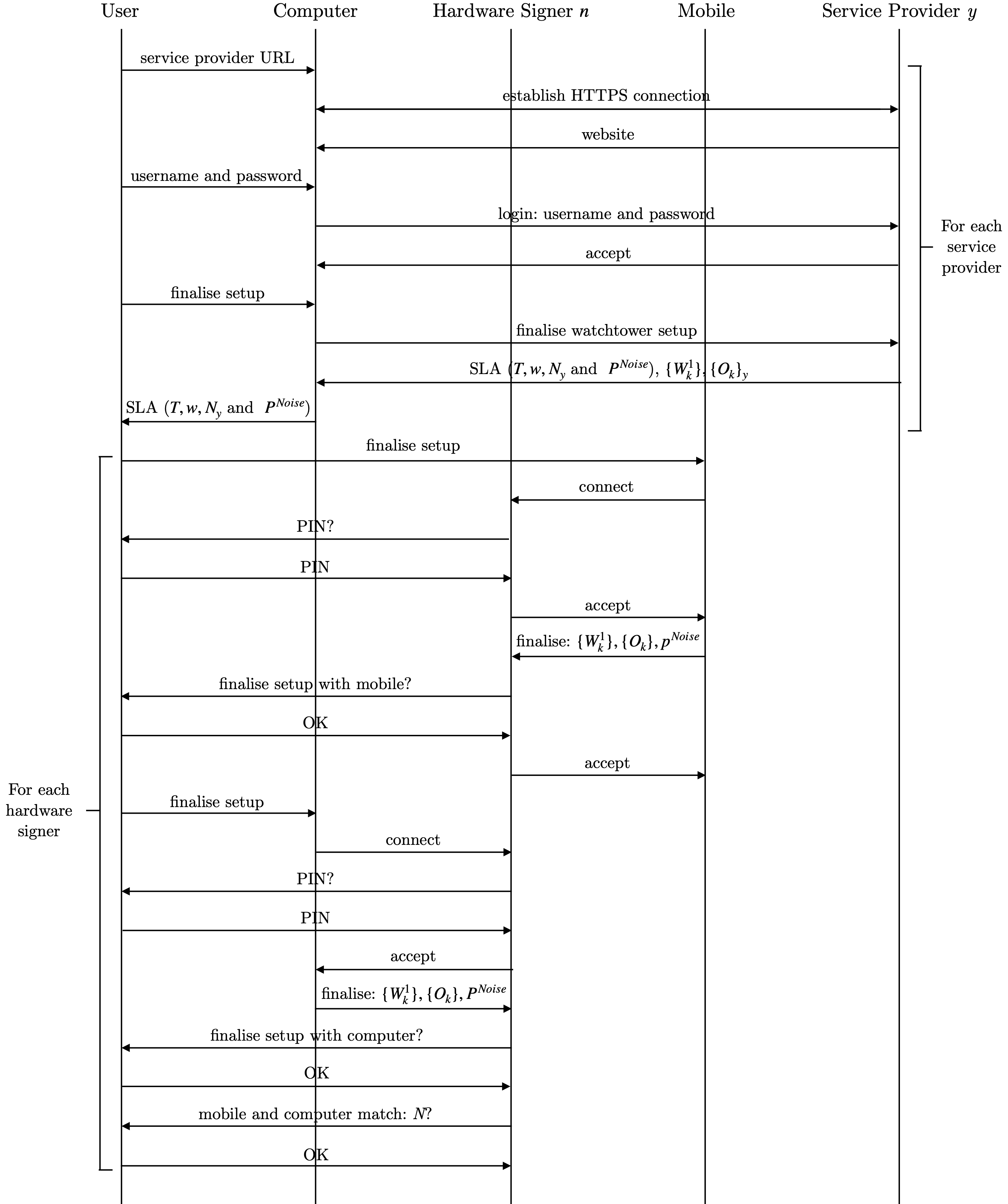}
    \caption{Setup finalisation. The state of the hardware signers now includes $\{W^1_k\}$, $\{O_k\}$, and $p^{Noise}$.}
    \label{fig:setup-fin}
\end{figure}

Figure \ref{fig:setup-fin} depicts how to finalise the setup, which involves transporting watchtower credentials to each hardware signer. To achieve this while preventing a potentially malicious mobile device from tampering with messages, the user communicates with the service provider through a separate channel using their computer. 

First, the user inputs a URL for the service provider and the computer and Service provider establish a HTTPS connection. The service provider serves their website. Then, the user enters their login details to authenticate them-self to the service provider. Once the service provider accepts, the user requests to finalise the setup. The service provider responds with a copy of the SLA including the relevant setup configuration data and watchtower data. The user will verify that the SLA matches the previous one and that $T,w,$ and $N_y$ are correct. This protocol is enacted for each service provider.

Next, the user connects each hardware signer to both the mobile and the computer and retrieves the setup data. From the mobile, it receives the set of watchtower accounts, $\{W^1_k\}$, the set of onion addresses, $\{O_k\}$, and $p^{Noise}$. From the computer, it receives $\{W^1_k\}, \{O_k\}, P^{Noise}$. It verifies that the $p^{Noise}$ it was given by the mobile is the private key corresponding to $P^{Noise}$, as received from the computer. To prevent the mobile or computer from omitting messages to them, the hardware signers display how many of the watchtower public keys and onion addresses were consistent from the mobile and the computer, and prompts the user to verify the total number of watchtowers $N$.

\subsection{Test Setup}
\label{subsec:test-setup}

\begin{figure}
    \centering
    \includegraphics[width=0.8\textwidth]{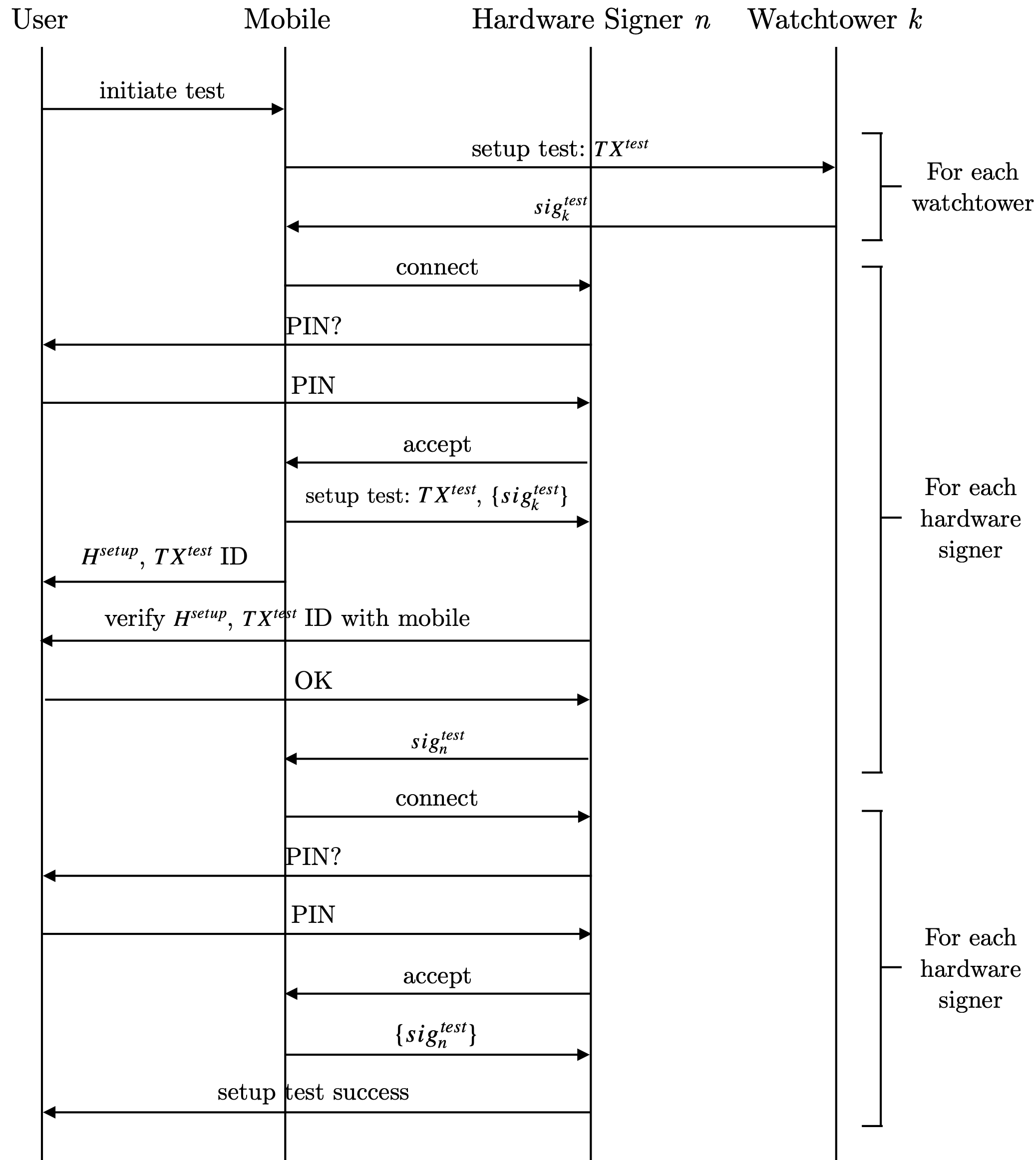}
    \caption{Setup test.}
    \label{fig:setup-test}
\end{figure}

Figure \ref{fig:setup-test} depicts the protocol to test the finalised setup. In this procedure, one goal is to have each signer (mobile, hardware signer, watchtowers) provide a signature for a test transaction \TX{test}, and to have each of the user's devices verify the complete set of signatures. Each of the signers uses their knowledge of $\{W^i_k\}$ and $\{A^i_n\}$ to generate $l_{test}$ independently. Another goal is to have each of the user's devices compute the hash of their setup state,

\begin{equation}
    H^{setup} = hash(\{W^i_k\}||\{A^i_n\}||\{O_k\}||P^{Noise}||c)
\end{equation}
and for the user to verify that these match by comparing what the mobile and hardware signers display. 

The user initiates the test on the mobile. The mobile provides \TX{test} to each watchtower, which generates a signature $sig_k^{test}$ with $w_k$ and sends this to the mobile. The mobile verifies the complete set of watchtowers' signatures $\{sig_k^{test}\}$. The mobile connects to a hardware signer, the user authenticates the connection with their PIN, and the mobile requests a setup test. The mobile provides \TX{test}, and the hardware signer verifies whether the expected $l_{test}$ was used. The mobile provides $\{sig^{test}_k\}$, and the hardware signer verifies each of the signatures. Both the mobile and the hardware signer display $H^{setup}$ and the transaction ID for \TX{test}, prompting the user to verify if they match. If so, the user inputs OK on the hardware signer. The hardware signer generates its own signature $sig^{test}_n$ and sends it to the mobile. This sequence is repeated for each hardware signer. 

Finally, the user reconnects each hardware signer with the mobile (one at a time). The mobile sends the complete set of signatures from the user's signers, $\{sig^{test}_n\}$. The mobile and hardware signer verifies these signatures. If everything is correct, the hardware signers display `setup test success' to the user. 

Note that \TX{test} does not need to be broadcast to allow each of the user's signers to verify that the others have knowledge of their first account, and to prove to others that they have the associated private key. It is sufficient to check that they are consistent among each other. In fact, broadcasting \TX{test} can reveal operational information (such as public keys and the number of watchtowers).

\subsection{Conclusion}

After successfully completing and testing the setup, the user should move hardware signers to their safe and secured locations. Each device will have the state summarised in table \ref{tab:setup-summary}. The computer will not be required again. It should be purged of all information related to the setup. 

\renewcommand{\arraystretch}{1.25}
\begin{table}
    \centering
    \begin{tabular}{|l|l|}
        \hline 
        \rowcolor{lightgray} \textbf{Device} & \textbf{State Contents} \\
         \hline
         Mobile & $\{A^1_n\}$, $\{W^1_k\}$, $\{O_k\}$, $c=(T,V_{min},V_{max})$,$(q_1,Q_1)$, $(p^{Noise},P^{Noise})$, $w$ \\
         \hline
         Home signer & $\{A^1_n\}$, $\{W^1_k\}$, $\{O_k\}$, $c=(T,V_{min},V_{max})$,$(q_2,Q_2)$, $(p^{Noise},P^{Noise})$ \\
         \hline
         Office signer & $\{A^1_n\}$, $\{W^1_k\}$, $\{O_k\}$, $c=(T,V_{min},V_{max})$,$(q_3,Q_3)$, $(p^{Noise},P^{Noise})$ \\
         \hline
         Bank A signer & $\{A^1_n\}$, $\{W^1_k\}$, $\{O_k\}$, $c=(T,V_{min},V_{max})$,$(q_4,Q_4)$, $(p^{Noise},P^{Noise})$ \\
         \hline
         Bank B signer & $\{A^1_n\}$, $\{W^1_k\}$, $\{O_k\}$, $c=(T,V_{min},V_{max})$,$(q_5,Q_5)$, $(p^{Noise},P^{Noise})$ \\
         \hline
         Watchtower $k$ & $w_k, W_k$, $O_k$, $T$, $P^{Noise}$, $w$ \\
         \hline
    \end{tabular}
    \caption{Summary of device state after a successful setup.}
    \label{tab:setup-summary}
\end{table} 
\renewcommand{\arraystretch}{1}

\section{Ajolote Operational Procedures}
\label{sec:ops-procedures}

After a successful setup, the user can begin using the custody system. The following procedures are not necessarily sequential. Though, of course, the user needs to receive bitcoin first. 

In the procedure specifications, when transactions are sent to `Bitcoin Network', assume they have satisfactory witness fields (signatures and Scripts included). Assume also that the user authenticates them-self to the mobile (with their PIN and bio-metric data) before initiating each procedure. 

\subsection{Receive}

In this procedure the user receives bitcoin with a Receive transaction, \TX{rec}. Any entity may authorise a \TX{rec}. A \TX{rec} has no constraints other than creating a \textit{receive output}, i.e. one which is controlled by the user's active key-set, $l_{receive}$. By separating this procedure from the vault-deposit, we ensure that external entities are not participants in the deleted-key covenant protocol. This way trust in counter-parties is minimised. 

A secure procedure for the user to generate an address and share with a counter-party was described by Arapinis \textit{et al.} \cite{FormalHardware}. The basic idea is to verify the address generated on the mobile with one generated on a hardware signer, and to send this to the counter-party through a channel that's not controlled by the mobile. This prevents a compromised mobile device from replacing the address with one controlled by an adversary; an \textit{address generation} attack. 

\begin{figure}
    \centering
    \includegraphics[width=0.8\textwidth]{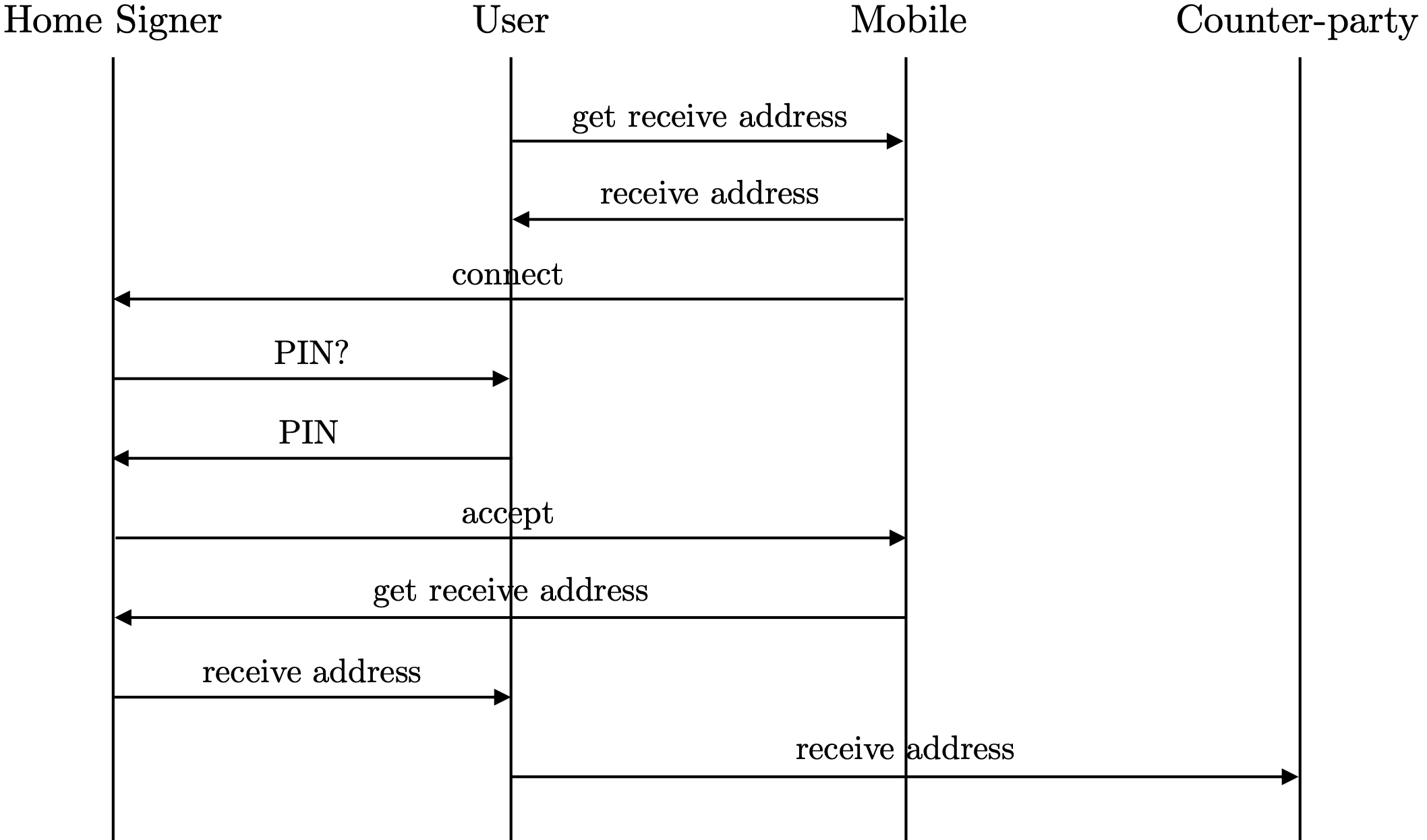}
    \caption{Receive. Critically, the communication between the user and counter-party is not mediated by the mobile.}
    \label{fig:receive}
\end{figure}

\subsection{Vault-deposit}
\label{subsec:vault-dep}

\begin{figure}
    \centering
    \includegraphics[width=\textwidth]{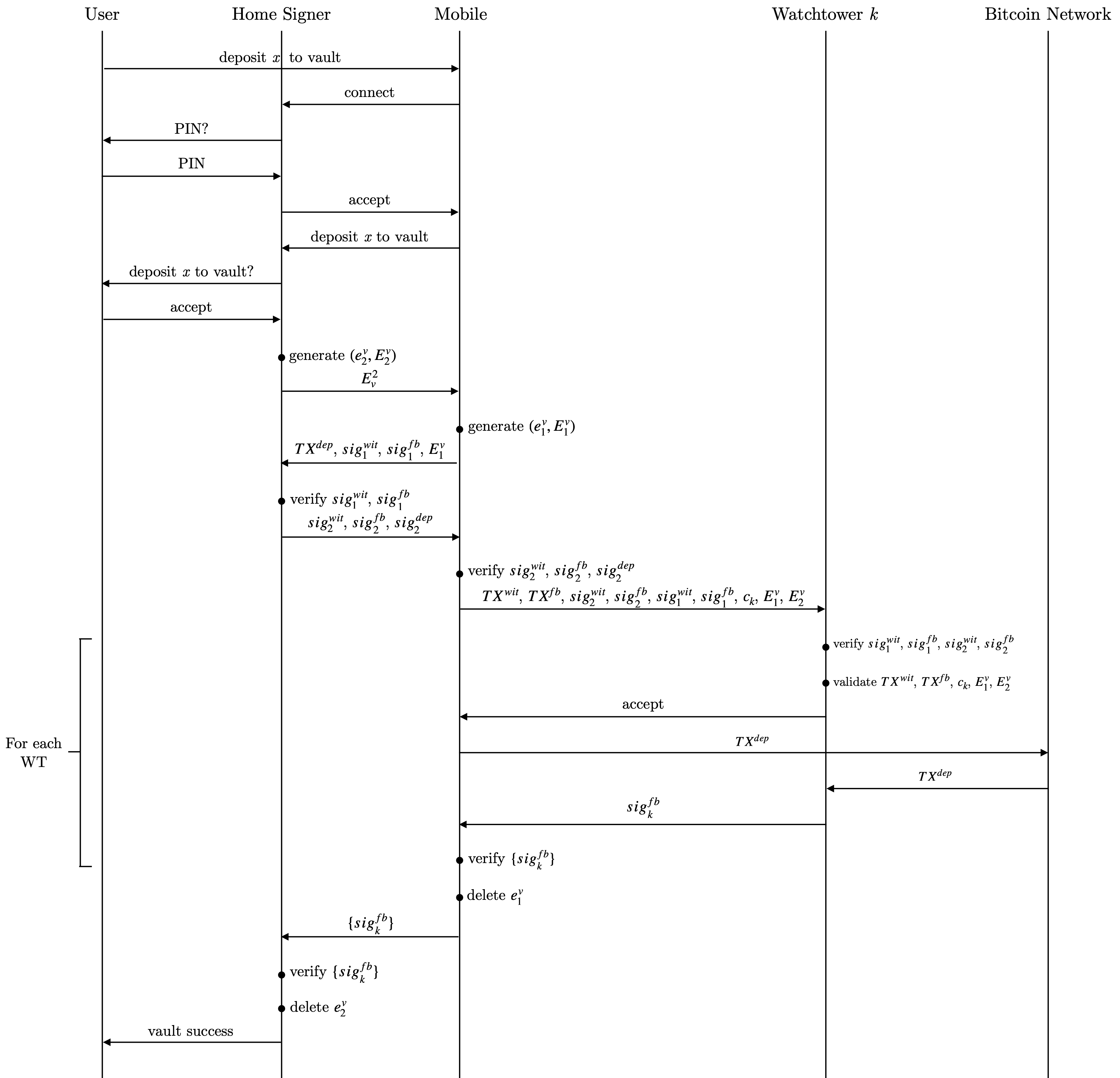}
    \caption{Vault-deposit. Protocol to deposit funds into a deleted-key vault covenant in the Ajolote custody system. The mobile and home signer use ephemeral keys to create enforcement signatures for the withdrawal and fall-back transactions that implement `vault' and `pay-to-fallback' covenants.}
    \label{fig:Ajolote-deposit-vault}
\end{figure}

Conceptually, in this procedure the user adds a layer of security to previously received funds by depositing them into a bitcoin vault. This procedure is where the user constructs the two covenant transactions, \TX{wit} and \TX{fb} (withdrawal and fall-back) and pays to them with the vault-deposit, \TX{dep}. Recall from the previous chapter, section \ref{subsec:protocol-spec}, how a deleted-key covenant should be implemented and the required constraints on transactions. In this case the user self-imposes the covenant. The user enacts each role with their mobile device and home signer; `depositor', `enforcer', and `custodian'. The user also sends the covenant transactions, \TX{wit} and \TX{fb} to their watchtowers. Recall that watchtowers participate in the custodial sub-script for \TX{fb}, so they too enact the `custodian' role. The communication protocol for the procedure is depicted in figure \ref{fig:Ajolote-deposit-vault}. In addition, important actions by each device are shown between messages. 

{}

The user initially selects an amount $x$ to deposit into a vault. This amount can be any number that is the sum of a subset of amounts held in receive outputs. An example user interface could be a slider, with intervals fixed to valid amounts. The user authenticates the connection between the mobile and home signer. The mobile requests to deposit $x$ from the home signer, which relays the request to the user for verification.

The additional data stored by the home signer, mobile, and watchtowers after a successful instance of the vault-deposit protocol is summarised in table \ref{tab:vault-dep-summary}. Each entity has sufficient data to fulfil their duties. The covenant transactions and their enforcement signatures are redundantly stored by the mobile, home signer, and all watchtowers. Signatures with subscript 1 and 2 were generated by the mobile and home signer respectively, using ephemeral private keys $e_1^v$ and $e_2^v$. These keys are subsequently deleted to enforce the covenants. Notice that the home and mobile signer wait for confirmation from the watchtowers before deleting.  

Spending taproot outputs also requires knowledge of the TapScript and the control block which form part of the witness. This is a necessary subset of the TapTree required to spend. The public keys $E^v_1$ and $E^v_2$ are part of the vault and unvault output TapTrees and are stored by the mobile, home signer, and watchtowers. The mobile and home signer can derive the TapTree with these, however the watchtowers do not know the accounts of the user's signers nor of each other watchtower. This is why they are also given $c_k$, the control block needed to authorise \TX{fb} in a reject procedure.

The mobile stores \TX{dep} in its transaction history records. The home signer and mobile store $j$, the derivation index variable which is incremented with each new vault-deposit. This is used to derive new public keys from the current accounts of all signers. 

The home signer only provides signatures for an acceptable \TX{wit} and \TX{fb}. How? By constructing them from its own data; this includes constructing the descriptors, setting the transaction fee, and setting the outputs. The home signer trusts the mobile when setting the input in \TX{wit} (using the vault output information from \TX{dep}). While the home signer can not verify if its \TX{wit} and \TX{fb} will be valid with respect to the state of the blockchain, it can verify if they are valid with respect to the consensus rules and the Ajolote system rules. As well as ensuring appropriate output types and a transaction fee, the home signer can enforce that the vault-deposit amount is within the acceptable range of $(V_{min},V_{max})$. Moreover, the home signer will be a storage device for enforcement signatures. It must verify the ones received from the mobile before accepting and storing them. The home signer must abort and warn the user if the verification fails.  

The mobile will verify all signatures received from the home signer and notify the user if there is ever a failure. Similarly, the watchtowers will verify each signature (received from the mobile) and check the validity of the covenant transactions it receives. The watchtower validates against the withdrawal constraints $w$ and the Bitcoin consensus library. Furthermore, the watchtower must check that $c_k, E_1^v$ and $E_2^v$ are used to derive the Merkle root of the TapTree by matching it against the taproot output key for the unvault output. The mobile waits for acknowledgement from each watchtower before broadcasting \TX{dep} to the Bitcoin network. 

The watchtowers monitor the network for \TX{dep}. Once it has been included in a block, they generate a signature for the fall-back transaction and send this to the mobile, which will forward it to the home signer. This signature is verified by both mobile and home signer, and interpreted as an authenticated message that \TX{dep} has been confirmed, and that the watchtowers have verified and validated the covenant transactions and associated data. If the home signer verifies these signatures, it notifies the user that the vault-deposit was successful. 

\renewcommand{\arraystretch}{1.5}
\begin{table}
    \centering
    \begin{tabular}{|l|l|}
        \hline 
        \rowcolor{lightgray} \textbf{Device} & \textbf{New State Contents} \\
         \hline
         Home signer & \TX{dep}, \TX{wit}, \TX{fb}, $sig^{wit}_1,sig^{wit}_2,sig^{fb}_1,sig^{fb}_2,E^v_2,E^v_1, j$\\
         \hline
         Mobile & \TX{dep}, \TX{wit}, \TX{fb}, $sig^{wit}_1,sig^{wit}_2,sig^{fb}_1,sig^{fb}_2,E^v_2,E^v_1,j$\\
         \hline
         Watchtower $k$ & \TX{dep}, \TX{wit}, \TX{fb}, $sig^{wit}_1,sig^{wit}_2,sig^{fb}_1,sig^{fb}_2, c_k, v$ \\
         \hline
    \end{tabular}
    \caption{New state entries after a successful vault deposit.}
    \label{tab:vault-dep-summary}
\end{table} 
\renewcommand{\arraystretch}{1}

\subsection{Withdrawal}
\label{sec:procedures-withdrawal}

The withdrawal procedure is restricted by the covenant enforcement. In the typical case, where the withdrawal is honest, the user must wait for the time-lock to expire before being able to spend funds. In the event that a compromise is detected or the withdrawal policy is breached, the reject procedure will be initiated before the time-lock expires, resulting in a failed withdrawal attempt. 

\begin{figure}
    \centering
    \includegraphics[width=0.8\textwidth]{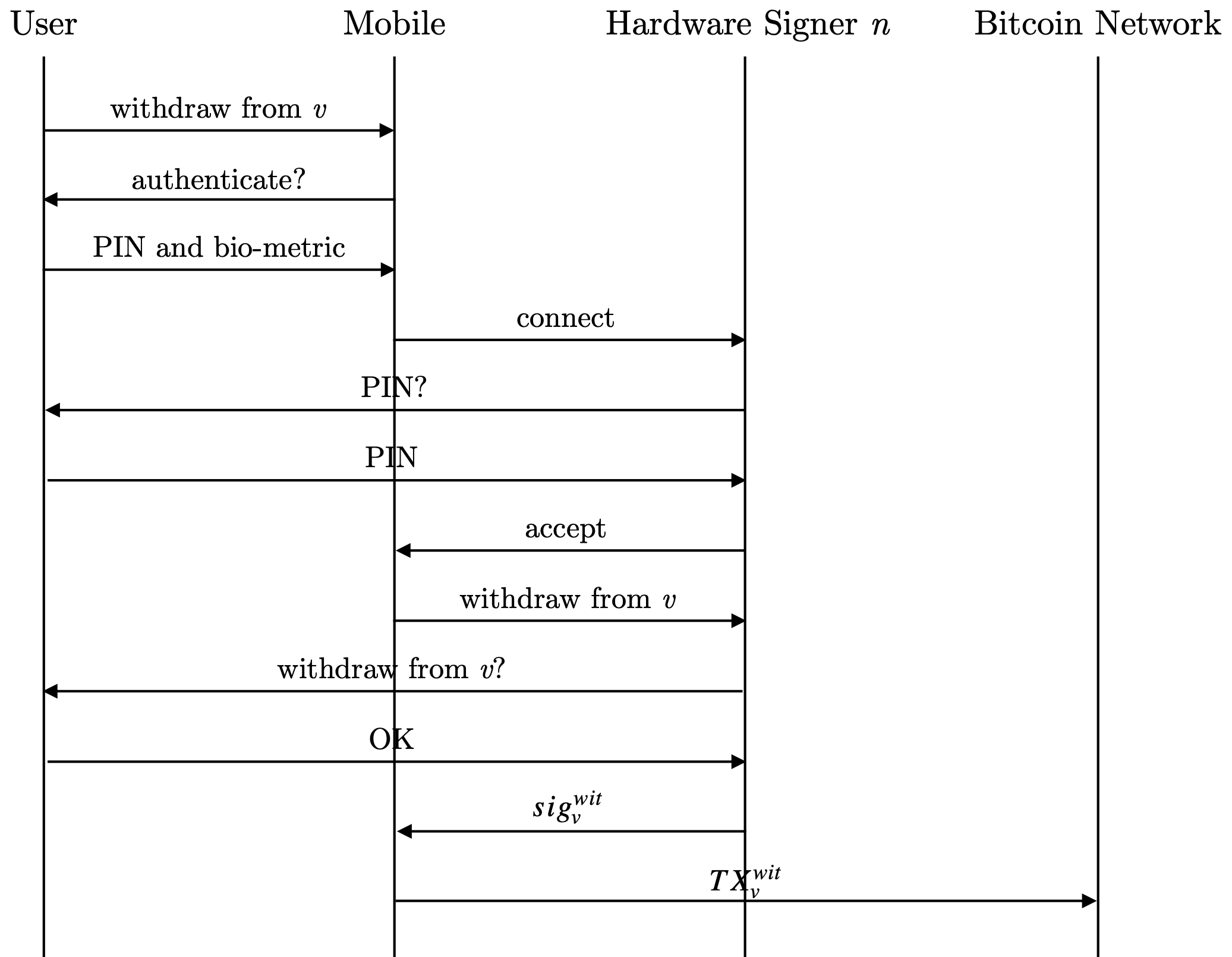}
    \caption{Withdrawal.  }
    \label{fig:withdrawal}
\end{figure}

The user must satisfy $and(l_{wit},l_{enf})$ to finalise \TX{wit}. In addition to the enforcement signatures $sig_{1}^{enf}$ and $sig_{2}^{enf}$ which will have been stored on the mobile device, the user must generate signatures to satisfy the $l_{wit}$ fragment, $sig_{1}^{wit}$ and $sig_{2}^{wit}$. Withdrawals can be initiated with user's mobile and any of the user's hardware signers, as depicted in figure \ref{fig:withdrawal}. Under normal operating scenarios the mobile and home signer will be used. The user selects which vault $v$ to withdraw. The relative time-lock for the spend path begins once the associated withdrawal transaction, $TX_v^{wit}$, is included in a valid block.

{}

\subsection{Spend}

\begin{figure}
    \centering
    \includegraphics[width=\textwidth]{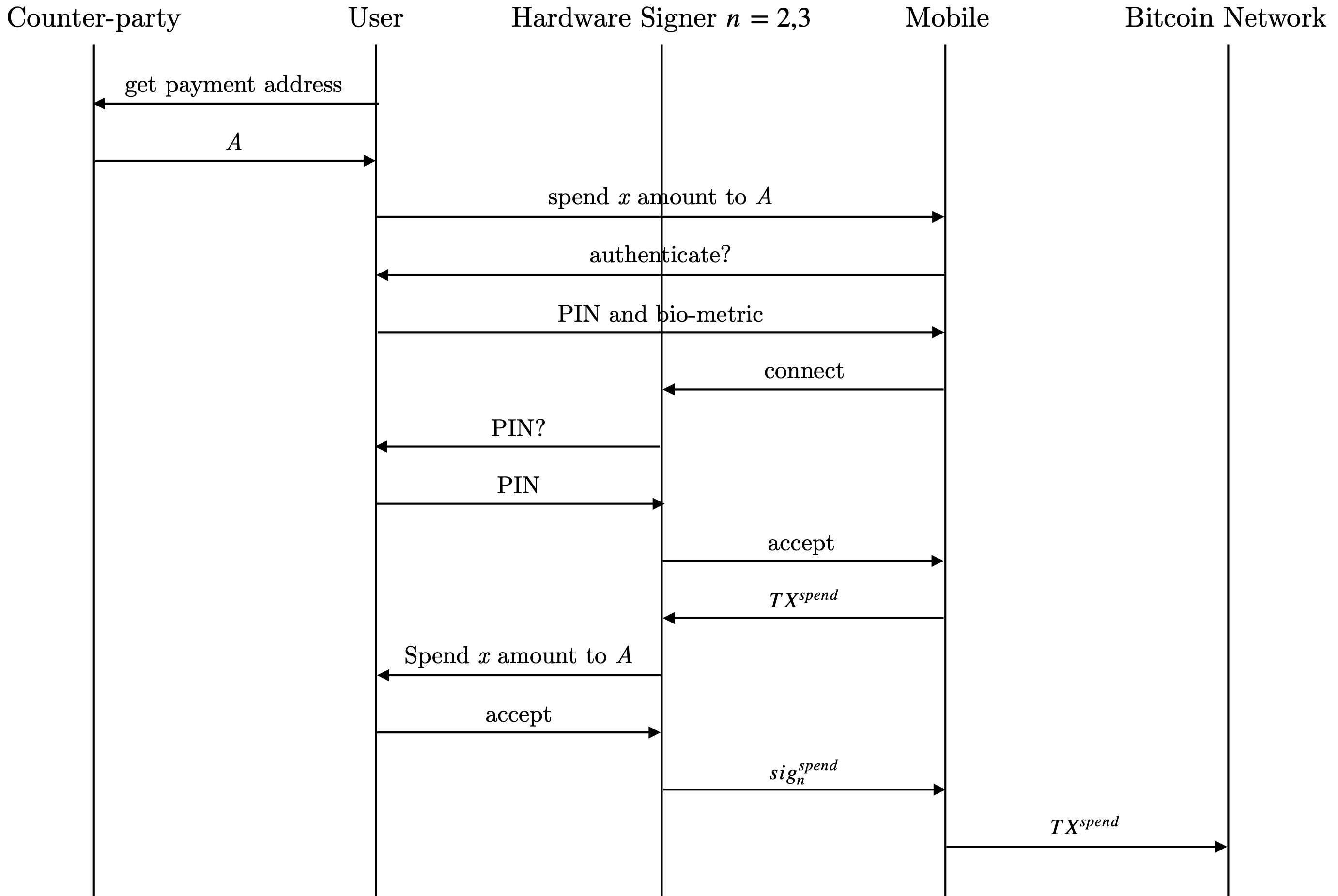}
    \caption{Spend. }
    \label{fig:spend}
\end{figure}

The spend procedure, depicted in figure \ref{fig:spend}, allows the user to make arbitrary payments to external entities with funds that have successfully been withdrawn and any other outputs that are currently locked to $l_{receive}$. The user first acquires a payment address, $A$, from the counter-party they intend to pay. The user initiates the spend by specifying an amount $x$ and $A$ in the Ajolote application on the mobile. The mobile and home (or office) signer are used to authorize the spend transaction, \TX{spend}. The user verifies that the mobile sent the appropriate spend details to the home signer by checking the amount, $x$, and destination address, $A$, on the hardware signer's display. 

{}

\subsection{Reject}

\begin{figure}
    \centering
    \includegraphics[width=0.35\textwidth]{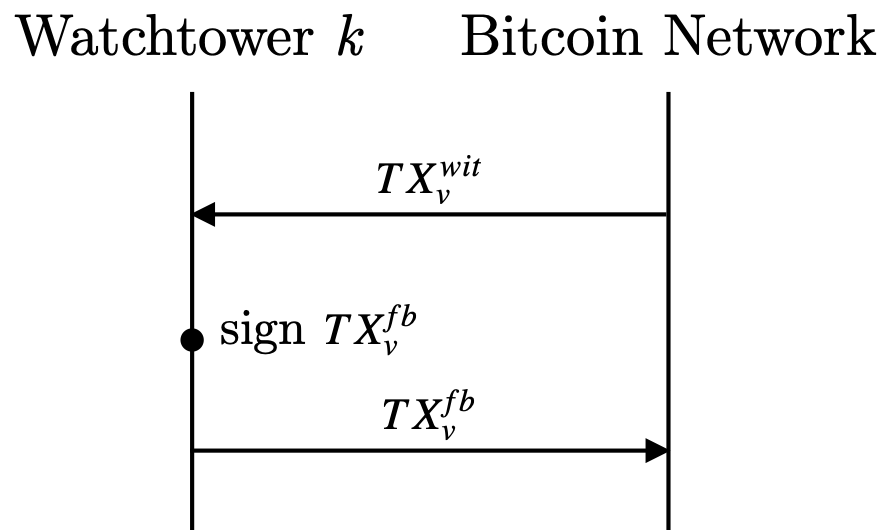}
    \caption{Reject.}
    \label{fig:reject}
\end{figure}

A reject procedure should be initiated automatically by at least one watchtower if the broadcast of a withdrawal transaction conflicts with the pre-determined withdrawal constraints, $w$. This scenario is depicted in figure \ref{fig:reject}. This procedure must occur before the expiration of the relative time-lock.

{}

\subsection{System health check}
\label{subsec:system-hc}

\begin{figure}
    \centering
    \includegraphics[width=\textwidth]{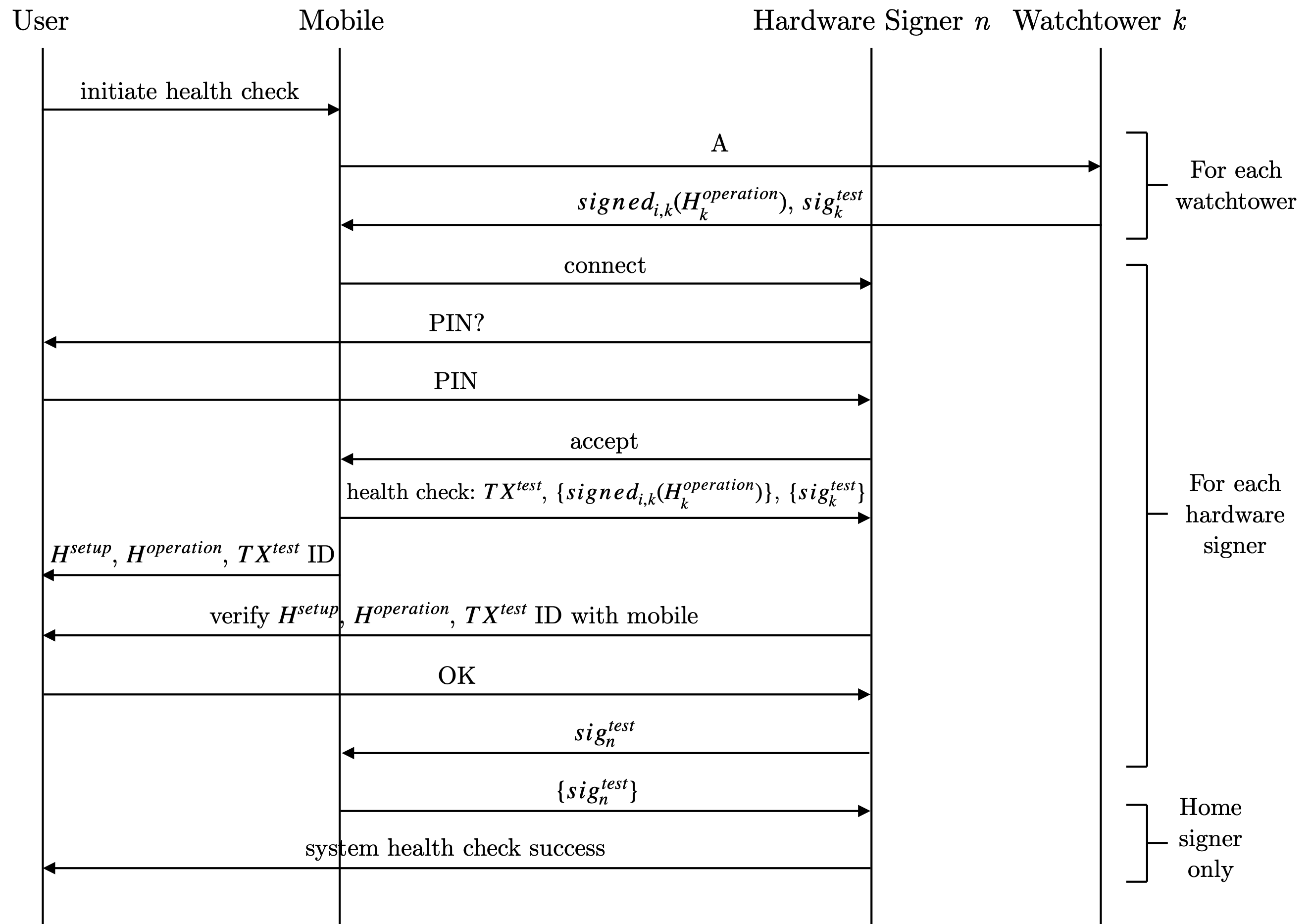}
    \caption{System health check.}
    \label{fig:system-health-check}
\end{figure}

A complete system health check is important to carry out periodically. In figure \ref{fig:system-health-check} we depict the system health check ceremony. This is an extended version of the setup test. It includes checking the consistency of operational state across devices. The public and private Bitcoin keys, covenant transactions and enforcement signatures, public and private Noise keys, and configuration parameters. In this procedure, we define an additional hash value for the user to compare,

\begin{equation}
    H^{operation} = hash(\{TX^{wit}\}||\{TX^{fb}\}||\{sig^{wit}_{enf}\}||\{sig^{fb}_{enf}\}||\{(E_1^v,E_2^v)\})
\end{equation}
which is a hash of the complete set of withdrawal and fall-back transactions, their enforcement signatures and public keys. Together, $H^{setup}$ and $H^{operation}$ allow the user to match the consistency of the majority of data across signers and watchtowers. 

To begin, the user initiates the system health check on the mobile. The mobile sends a request to the watchtower, including the test transaction to sign. The watchtower responds with its signature, and $H^{operation}$ which it has signed with $w^i_k$ so that the hardware signers can authenticate that the message came from the watchtower. 

Then, the user must connect to each hardware signer. To do this, they must travel to each one's safe location. The mobile provides the hardware signer with the test transaction and the data from each watchtower. The user matches the values displayed on the screen of the hardware signer with the mobile. Any discrepancy demonstrates that the device is not consistent with the rest of the system. The hardware signer verifies the signed $H^{operation}$ from each watchtower, and their test transaction signatures. From the hardware signers, only the home signer can compute $H^{operation}$ itself and match against what it received from the watchtowers. The others merely check that all watchtowers are consistent with each other. If successful, the hardware signers shares its test signature with the mobile.  

Finally, the user connects to the home signer, which receives the test signatures from all other signers. The home signer verifies the signatures and notifies the user that the system health check was a success. 

{}

\section{Ajolote Recovery Procedures}
\label{sec:rec-procedures}

Recovery procedures are as critical to the security of continued operations as is the setup. Recovery and setup procedures are similar. However, after a period of operation, the state of the custody system is not empty (as with the initial setup). Instead, there is a history of transactions, funds controlled by the custody system across its three tiers, and the hardware signers are distributed across different locations. 

Whether or not the custody system is under attack or is experiencing component failures, the user should follow the same process (albeit with more urgency in the case of attack). If a device has been lost or stolen, we assume that the data it contained has been compromised. An attacker could use setup information to disrupt the custody system in various ways or simply to break the user's privacy. These are considered in more detail in the risk model \ref{sec:Ajolote-risk-model}.  

\subsection{Decision Tree}

\begin{figure}
    \centering
    \includegraphics[width=0.85\textwidth]{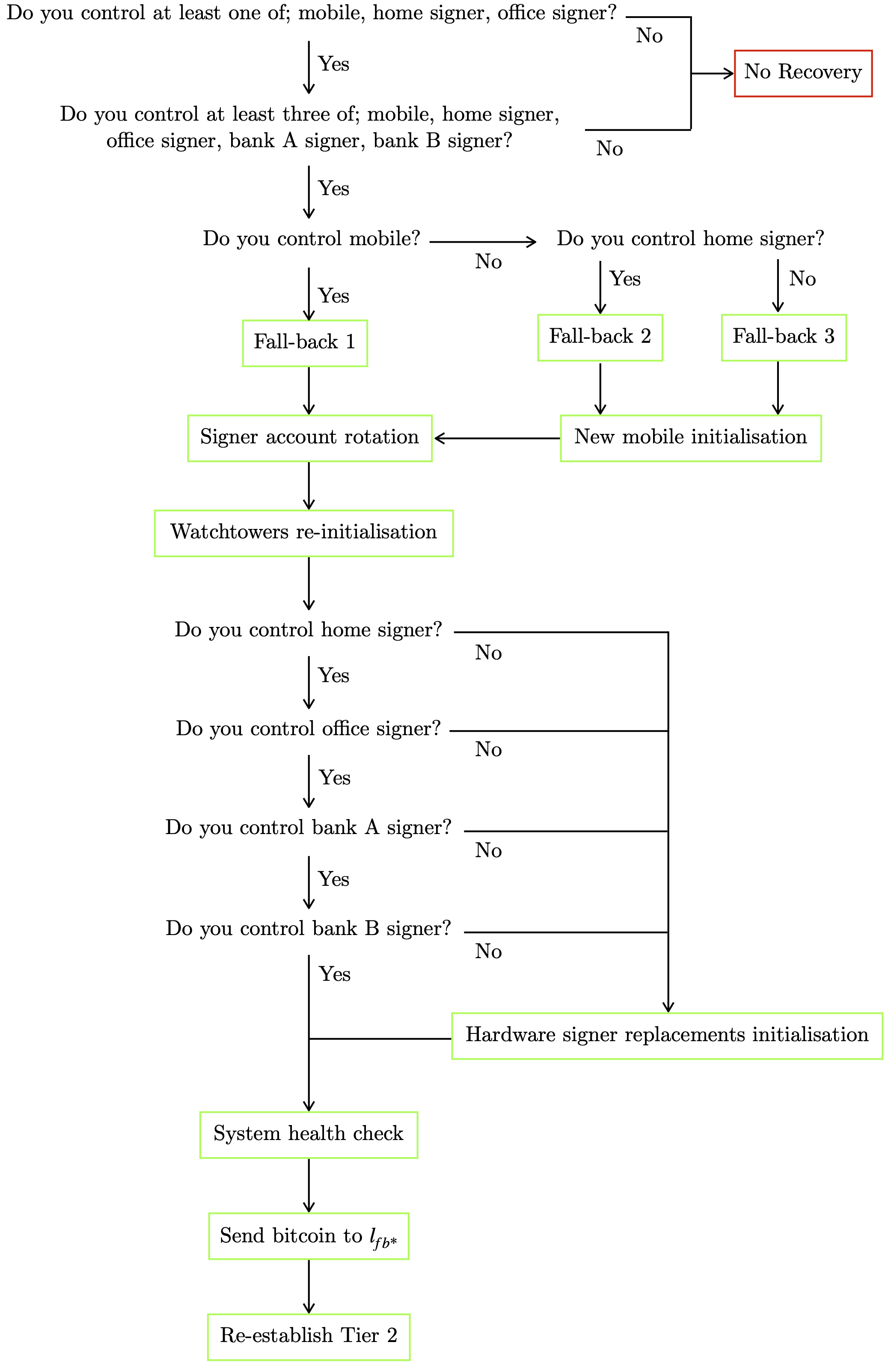}
    \caption{Decision tree for recovery procedures in Ajolote.}
    \label{fig:decision-tree}
\end{figure}

With an advanced custody system, we must have a plan of action to prevent failures and attacks from becoming catastrophic. To simplify the process for a user, they should follow the decision tree shown in figure \ref{fig:decision-tree}. It will be included as part of the user manual. The decision tree guides the user to enact a series of procedures which are specified in the following subsections. 

In addition, the user must exercise some situational awareness and pursue actions that are beyond the scope of these procedures. For example, was their home or office intruded? Has Bank A turned malicious? Has the user received coercive threats? It may be necessary for the user to contact law enforcement. The user should not re-use compromised locations or rely on corrupt counter-parties. 

{}

\subsection{Fall-back}

\begin{figure}
    \centering
    \includegraphics[width=0.8\textwidth]{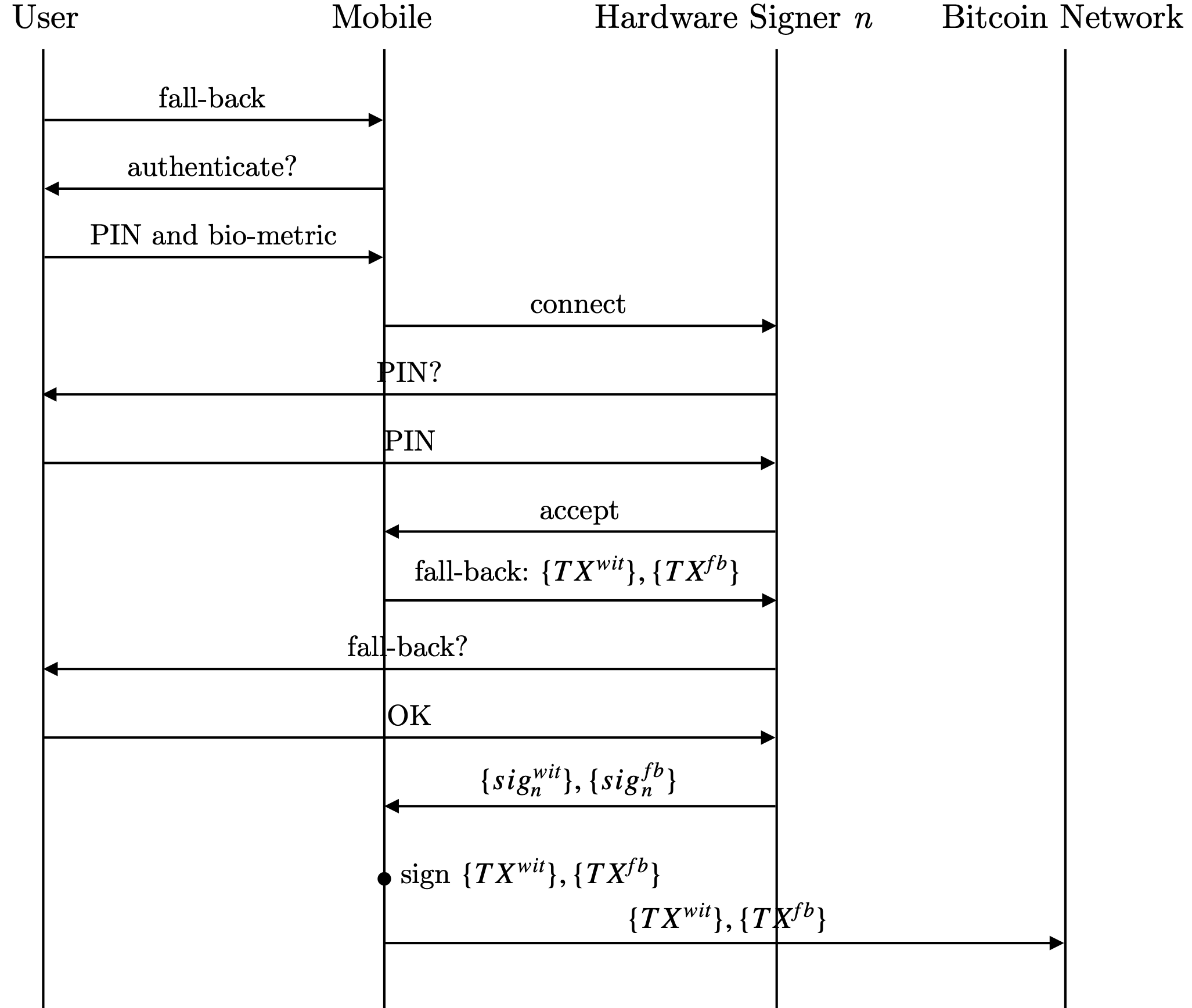}
    \caption{Fall-back using the mobile.}
    \label{fig:fb1}
\end{figure}

\begin{figure}
    \centering
    \includegraphics[width=0.9\textwidth]{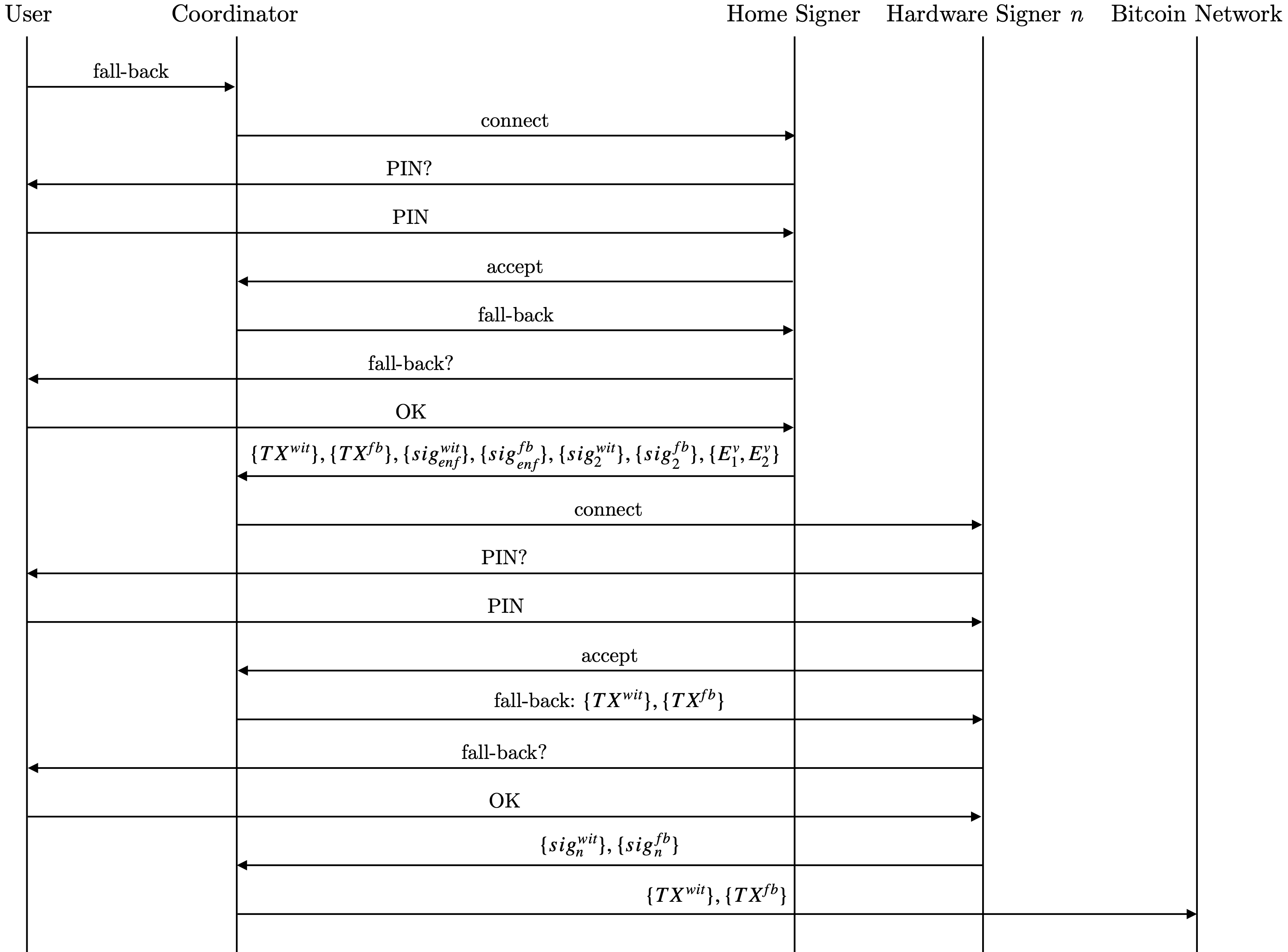}
    \caption{Fall-back using home signer (if mobile and one other hardware signer are not available). The coordinator entity is any mobile or desktop running Ajolote application in emergency mode.}
    \label{fig:fb2}
\end{figure}

\begin{figure}
    \centering
    \includegraphics[width=\textwidth]{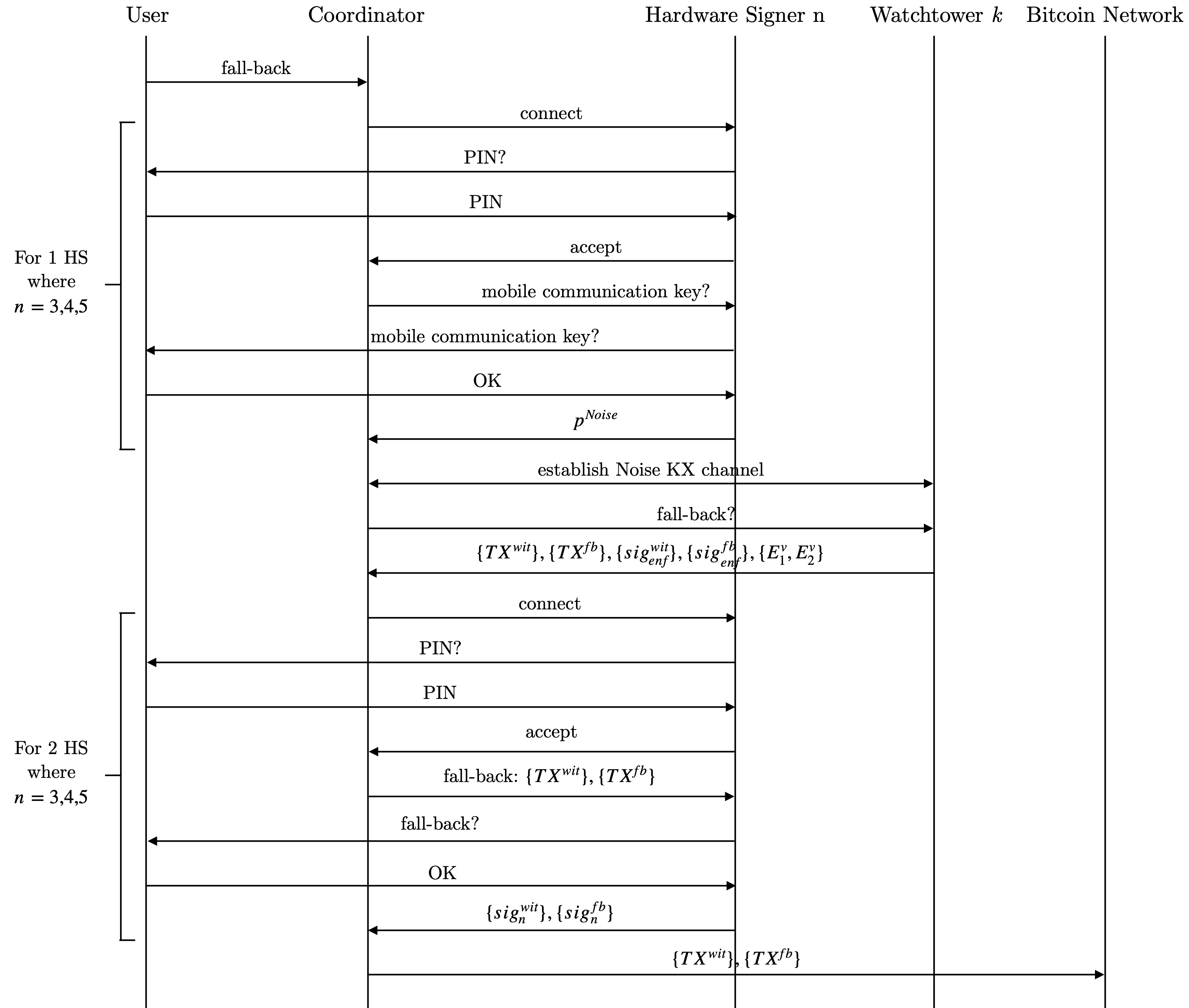}
    \caption{Fall-back without mobile and home signer. The coordinator entity is any mobile or desktop running Ajolote application in emergency mode.}
    \label{fig:fb3}
\end{figure}

The fall-back procedure is a response based on a perceived threat, compromise, or significant failure of the custody system. It is not expected to occur often. It is for an emergency scenario, such as the compromise of one or two of the user's devices. The fall-back procedure depends on which devices have been compromised. There are three variations depicted in figures \ref{fig:fb1}, \ref{fig:fb2}, and \ref{fig:fb3}.

The goal of this procedure is to authorise and broadcast the complete set of withdrawal and fall-back transactions, \{\TX{wit}\} and \{\TX{fb}\}. The private keys controlled by any two of the user's signing devices (including mobile) can be used to authorise \TX{wit}, and any one of their signers can authorise \TX{fb}. 

If one of the hardware signers has been compromised, the procedure can be enacted with the mobile and any remaining signer as in figure \ref{fig:fb1}. 

However, if the mobile has been compromised, then the user must acquire a new device to act as an emergency coordinator. The Ajolote application can be downloaded and launched in `emergency mode'. It requires no additional setup data. It can run on desktop and mobile. The user must access a desktop or mobile (a spare device they own or temporarily using one belonging to a friend). 

In figure \ref{fig:fb2}, the user's mobile has been compromised but they have access to their home signer. The home signer contains the complete set of covenant transactions and enforcement signatures. 

In figure \ref{fig:fb3}, the user's mobile and home signer have been compromised. In this case, the emergency coordinator needs to acquire the user's Noise keys and onion addresses from one of the remaining hardware signers, and request the complete set of covenant transactions and enforcement signatures from one of the watchtowers. 

{}

{}

{}

\subsection{New Mobile Initialisation}

\begin{figure}
    \centering
    \includegraphics[width=0.8\textwidth]{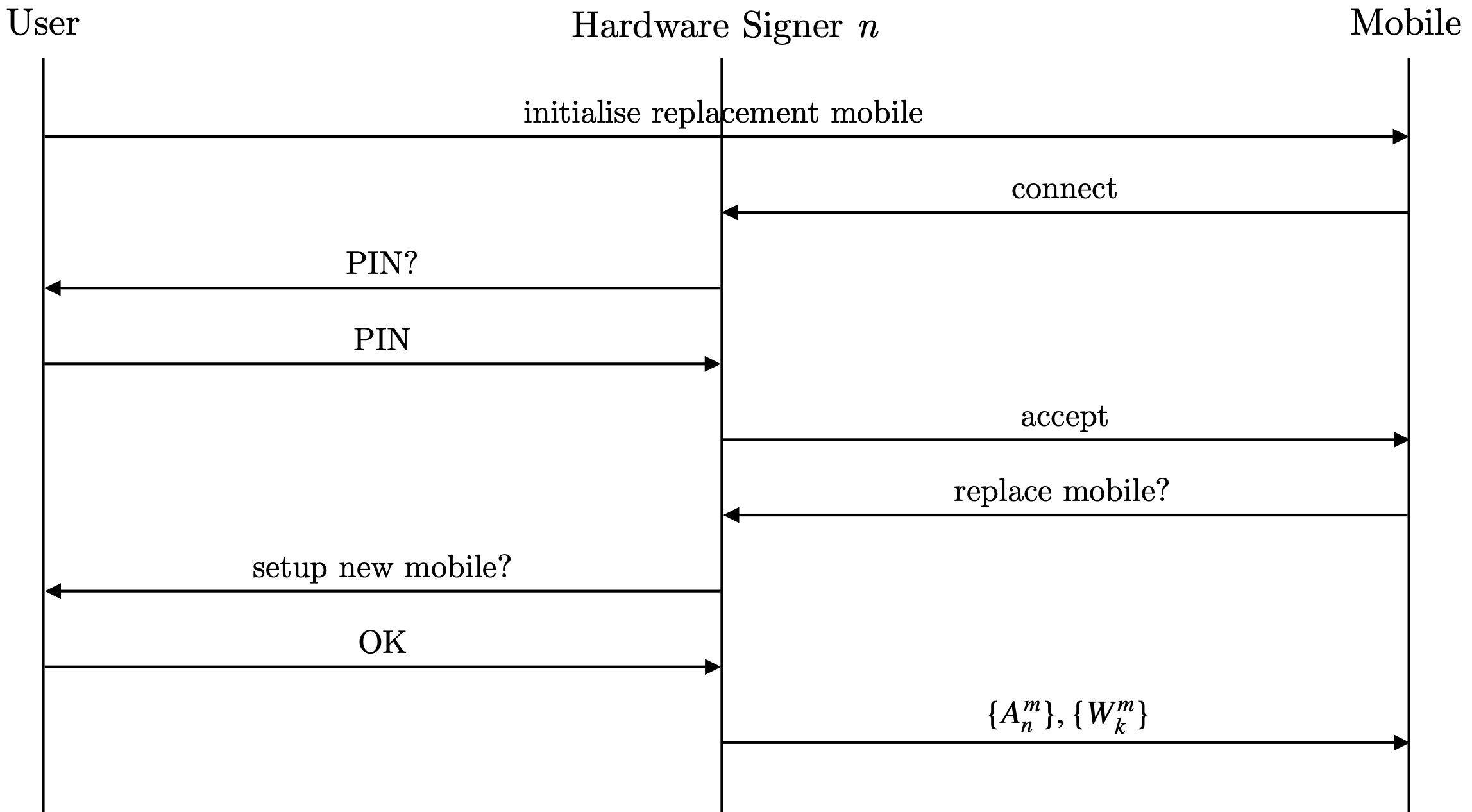}
    \caption{Mobile replacement initialisation.}
    \label{fig:new-mob}
\end{figure}

This procedure describes how to initialise a replacement mobile if the original has been lost, damaged or stolen. The user should begin by enacting the mobile initialisation procedure in section \ref{subsec:mobile-init}. Then, they should enact the steps in figure \ref{fig:new-mob}. Here, the mobile communicates with a hardware signer to retrieve the original setup information. By $\{A^m_n\}$ and $\{W^m_k\}$ we denote the complete set of accounts that have been used previously by signers and watchtowers, respectively. This enables the Ajolote application to deduce the current state and history of the custody system. During blockchain synchronisation, the mobile can derive addresses based on $l_{receive}$ and $l_{fb}$ and by reference to the blockchain can determine all currently available coins. 

{}

\subsection{Signer Account Rotation}

\begin{figure}
    \centering
    \includegraphics[width=0.9\textwidth]{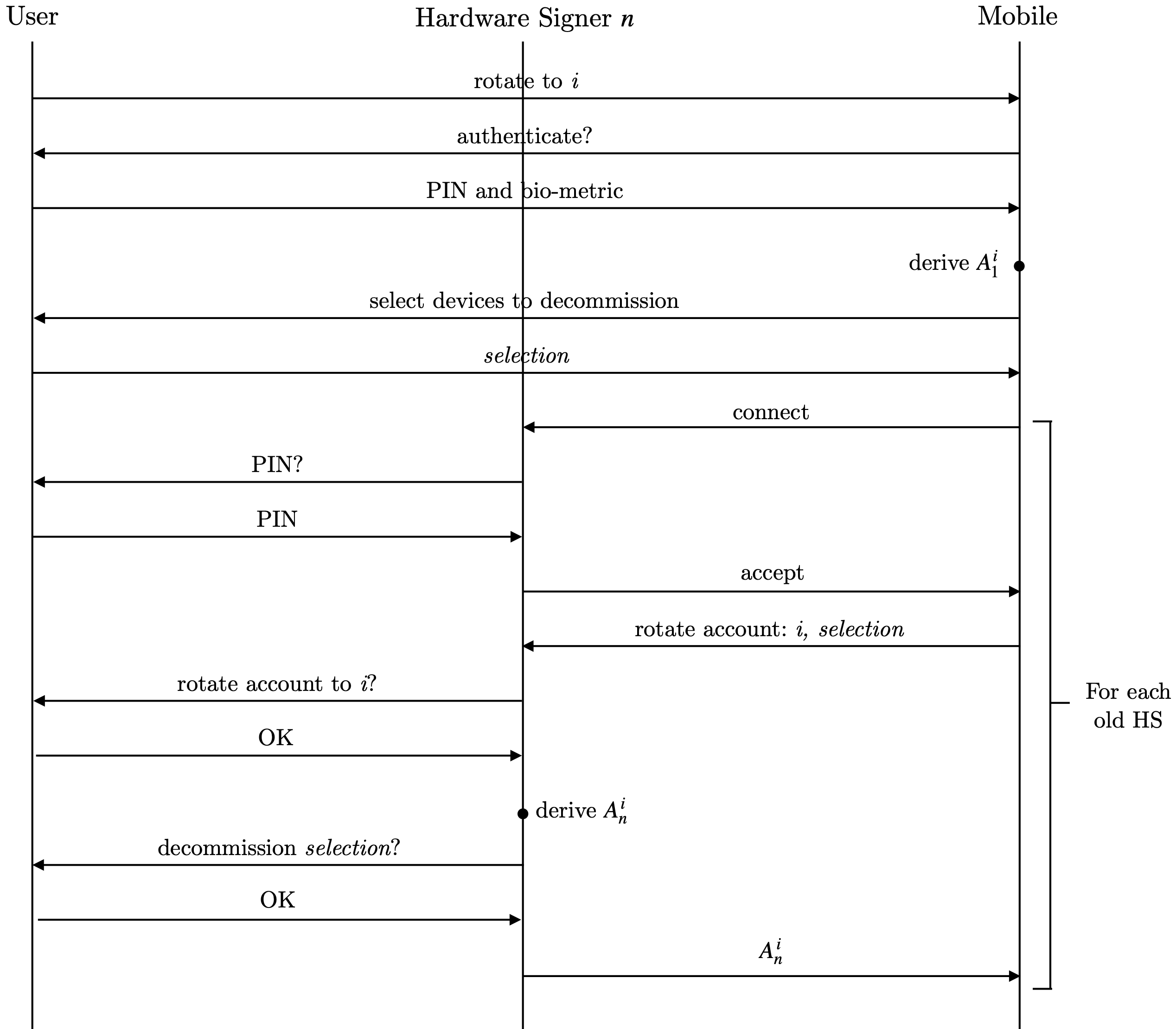}
    \caption{Mobile and Hardware Signer (HS) account rotation.}
    \label{fig:account-rotation}
\end{figure}

In figure \ref{fig:account-rotation} we depict the procedure to rotate the accounts of all `old' (i.e. not compromised or lost) hardware signers. The user instructs the mobile application to initiate the rotation, and inputs \textit{i}, for the $i^{th}$ rotation that they have enacted. The user authenticates them-self to the mobile, which derives a new account $A^i_1$ from their master key-pair. The mobile prompts the user to select which devices to decommission (which have been compromised or lost). Then, for each old hardware signer, the mobile connects and requests to rotate to the $i^{th}$ account. The user verifies both $i$ and the \textit{selection} of devices to decommission on the hardware signers' display. The hardware signers share their new account with the mobile. Note that during this procedure, devices have not tested nor verified the consistency of accounts. This will happen during the system health check. 

{}

\subsection{Watchtower Re-initialisation}

\begin{figure}
    \centering
    \includegraphics[width=\textwidth]{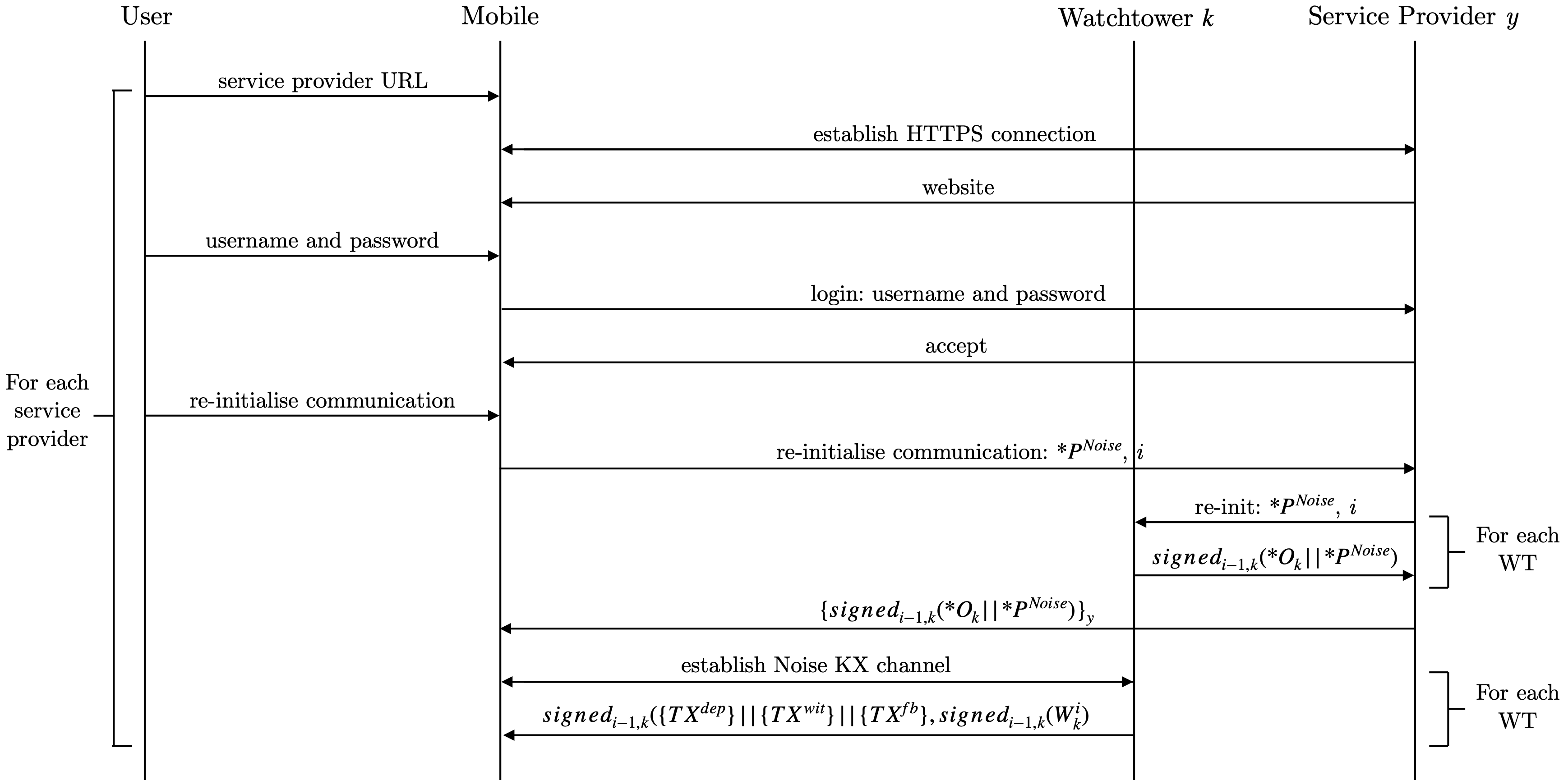}
    \caption{Watchtower (WT) re-initilaisation.}
    \label{fig:wt-reinit}
\end{figure}

If a user's device was lost, damaged or stolen, the communication channels with watchtowers should be re-initialised, as shown in figure \ref{fig:wt-reinit}. The user must have their username and password for this. Service providers re-initialise watchtowers by sending them the user's new Noise public key $\textbf{}^*P^{Noise}$, and the current account index $i$. Watchtowers start new hidden services and generate new onion addresses, $^*O_k$. Watchtowers stop their old hidden services (making old onion addresses unusable). Critically, watchtowers continue to monitor all registered vaults (if any remain). Watchtowers rotate to a new extended Bitcoin public key $W^i_k$.

The watchtower sends a signed message to its managing service provider, containing $^*O_k$ and $\textbf{}^*P^{Noise}$. The watchtowers sign their message with their previous account's private key, now labelled by $i-1$. The service provider sends a batch of these signed messages (one for each of its watchtowers) to the user. 

The mobile establishes a new Noise KX communication channel with each watchtower. The mobile obtains two signed message from the watchtowers. The first contains the complete history of covenant transactions and deposit transactions, The second contains the current Bitcoin account $W^i_k$. Again, these message were signed with the previous account's private key, labelled by $i-1$. This data can be authenticated by old devices. 

{}

\subsection{Hardware Signer Replacements Initialisation}

New hardware signers are initialised with this procedure. The new hardware signers generate their Bitcoin master key-pair $(q_n,Q_n)$ and derive their current account, $A^i_n$. The mobile receives $A^i_n$ from each new hardware signer. Following this, the user must connect with each hardware signer (travelling to their location if necessary). Hardware signers are sent all of the new setup information. Data from the watchtowers are signed. Account information will be verified later, in the system health check procedure. The hardware signers display configuration parameters, $c$ to the user for verification. When connected to the home signer, the mobile also sends the complete history of covenant transactions and deposit transactions, signed by each watchtower with their previous account key. 

\begin{figure}
    \centering
    \includegraphics[width=\textwidth]{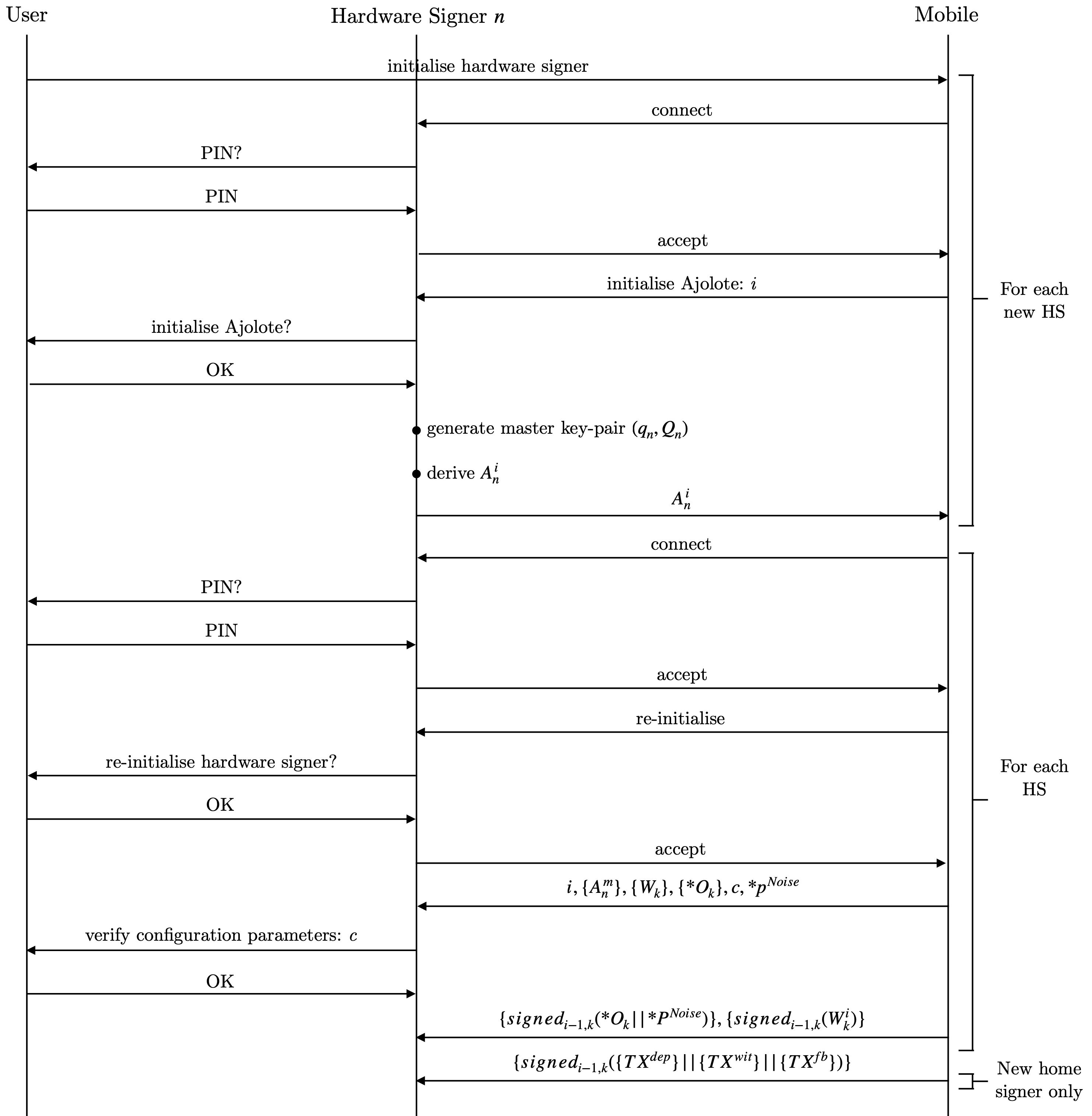}
    \caption{Hardware signer (HS) replacements.}
    \label{fig:new-hs}
\end{figure}{}

\subsection{Send Bitcoin to New Fall-back}

\begin{figure}
    \centering
    \includegraphics[width=\textwidth]{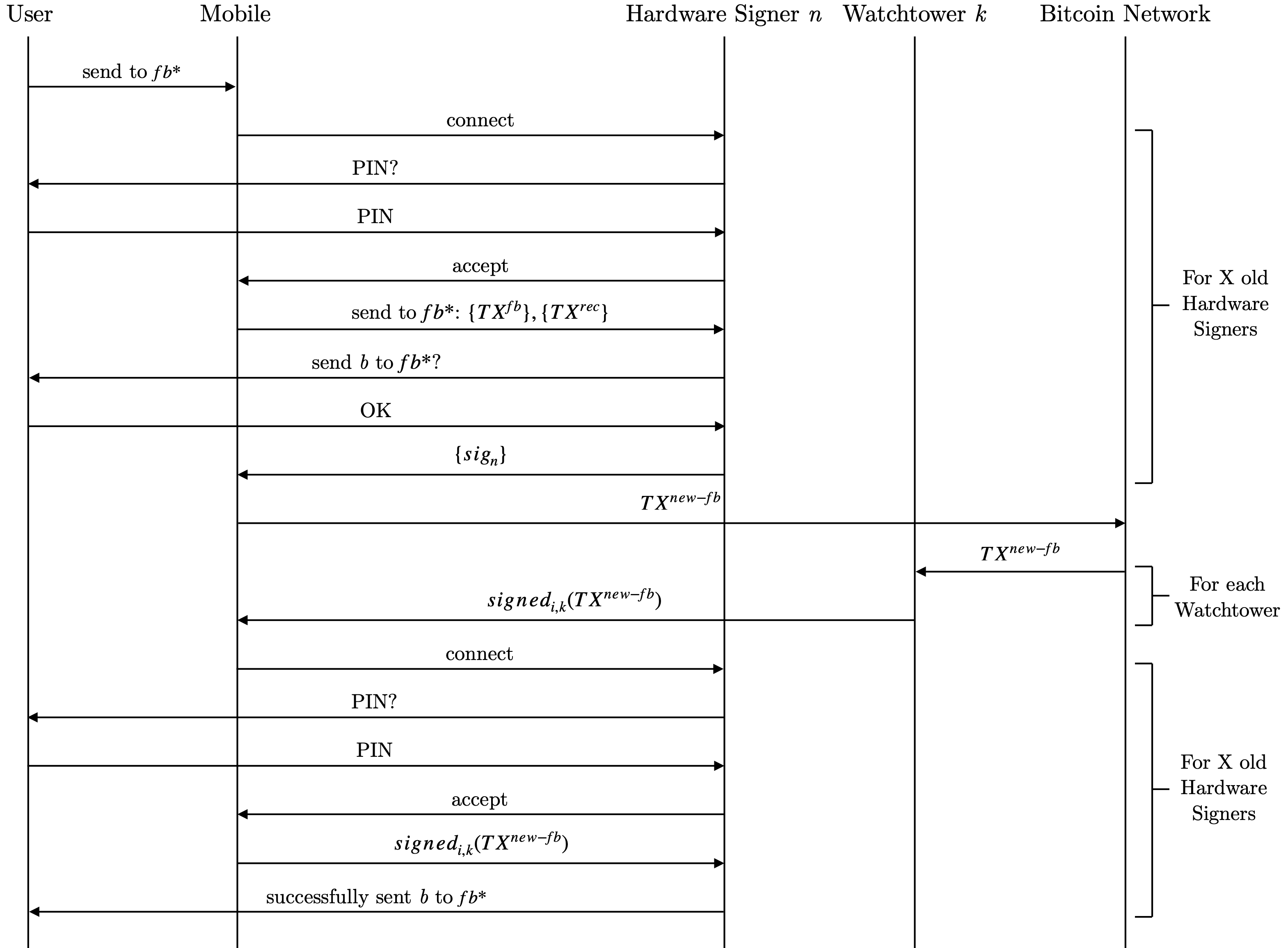}
    \caption{Send bitcoin to new fallback address, $fb^*$.}
    \label{fig:send-to-new-fb}
\end{figure}

In this procedure, depicted in figure \ref{fig:send-to-new-fb}, the user sends all bitcoin that's available in their custody system to the fall-back policy with the new accounts. The user must sign a transaction, \TX{new-fb}, consuming bitcoin from the old fall-back outputs and receive outputs. This single transaction consumes all available outputs, with total balance $b$. 

To satisfy the old $l_{fb}$, the user requires any 3 of the old signers (perhaps including the mobile). The (old or new) mobile will coordinate the transaction. If the mobile is new, then $X=3$ old hardware signers will be used, otherwise $X=2$ old hardware signers will be used. To satisfy the old $l_{receive}$, the user requires any 2 of the old active signers. Signers need not be brought to the same location. The user should travel to each signer to interact with them.

Each participating hardware signer enacts the following, one by one. The mobile sends the complete set of available fall-back and receive transactions, $\{TX^{fb}\}$ and $\{TX^{rec}\}$, to the hardware signer. The hardware signer requests authentication from the user to send $b$ bitcoin to $fb^*$. The user confirms the amount and instructs the hardware signer to proceed. Based on the available outputs, the hardware signer constructs \TX{new-fb} according to a deterministic algorithm. For example, lexicographic ordering of inputs' previous transaction ID and a fixed high fee of 40 satoshis per kilobyte. \TX{new-fb} has one output, which can only be spent by satisfying $l_{fb^*}$, the fall-back policy with the latest accounts. The hardware signer generates signatures to satisfy each input in the transaction and sends these back to the mobile. 

Once signatures are received from each hardware signer, the mobile broadcasts the signed \TX{new-fb} to the Bitcoin Network. Watchtowers observe \TX{new-fb}, wait for its inclusion in a block, sign the transaction message and send this to the mobile. The mobile forwards this signed message to each hardware signer. The hardware signer authenticates the messages from watchtowers and interprets them as a confirmation that it was included in the blockchain. The hardware signer notifies the user that the procedure was successful. The user considers the procedure successful if each signer displays this notification.  

{}

\subsection{Re-establish Tier 2}

\begin{figure}
    \centering
    \includegraphics[width=\textwidth]{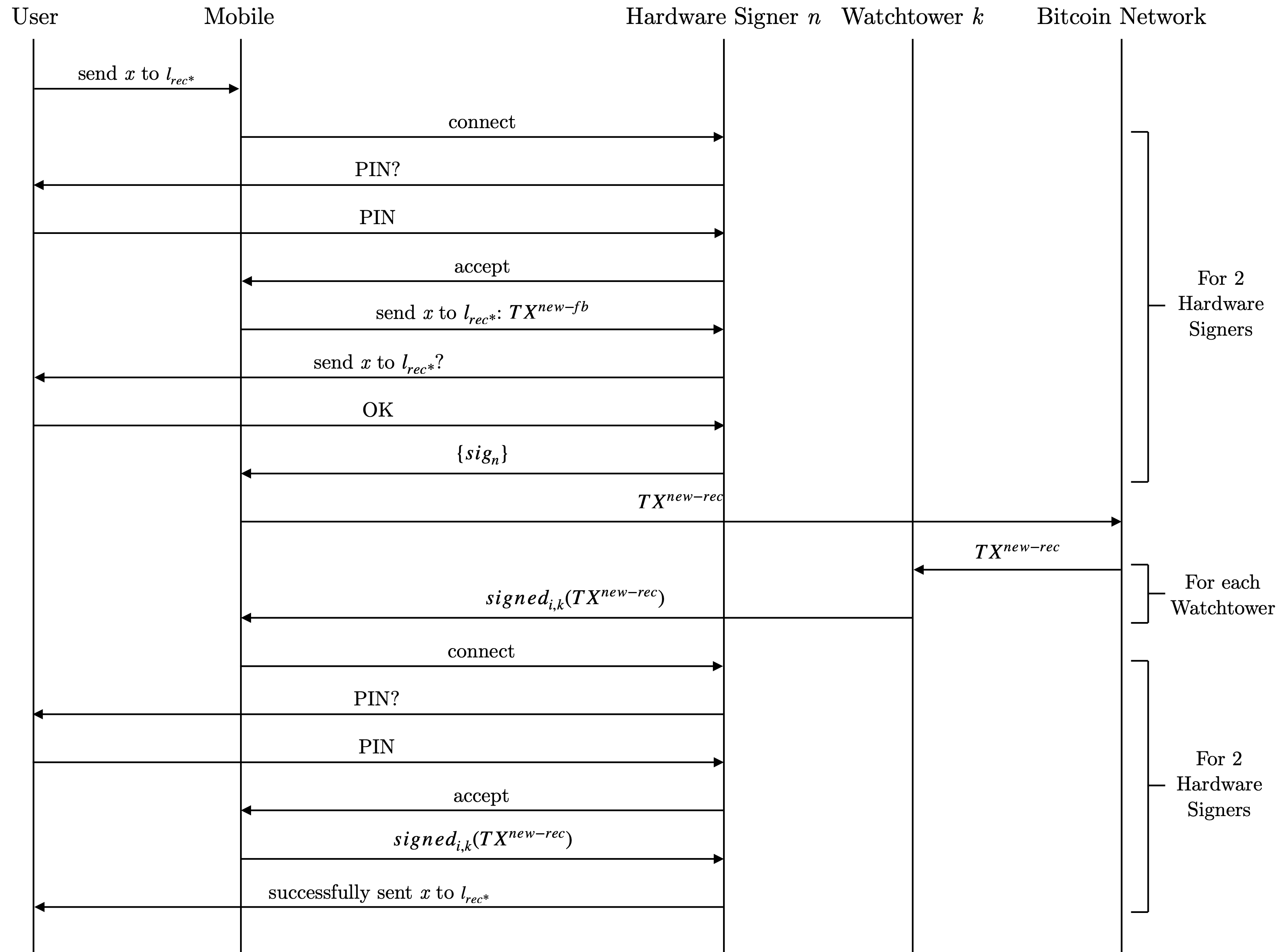}
    \caption{Send bitcoin to new receive outputs, $l_{rec^*}$.}
    \label{fig:send-to-new-rec}
\end{figure}

Once the bitcoin in custody is safely and securely controlled by the fall-back key-set, the user can begin to re-establish the balance of Tier 2. Incrementally, the user should transact from the new fall-back policy (Tier 3) to the new receive policy (Tier 1) and enact vault-deposit procedures. It is critical to do this incrementally so that too much bitcoin is not put at risk at any time. The procedure depicted in figure \ref{fig:send-to-new-rec} specifies the initial step from Tier 3 to Tier 1. The vault-deposit procedure should follow.

The procedure involves a transaction \TX{new-rec}, which consumes an output locked to $l_{fb^*}$  (e.g. the output generated by \TX{new-fb}) and creates two new outputs. The first output has amount $x$ and is locked to $l_{rec^*}$. The second output controls the remaining bitcoin (minus a fee) and is locked to $l_{fb^*}$.  

{}

\section{Ajolote Evaluation}
\label{sec:Ajolote-Evaluation}

\subsection{Operation Dynamics}
\label{subsec:eval-ops-dynamics}

Consider now the low-level state tracking and processing the Ajolote software must perform. At time $t$, there are $N_r$ unspent receive outputs, $N_v$ unspent vault outputs, and $N_u$ unspent unvault outputs. 

Consider first the simple case, where the user has a maximum of 1 unspent vault output at any time. That is, $0 \geq N_v \leq 1$ $\forall$ $t$. At time $t = 0$, $N_r = N_v = N_u = 0$. 

\begin{enumerate}
    \item $t=1$, User receives payment: $N_r = 1$
    \item $t=2$, User enacts vault-deposit: $N_r = 0$, $N_v = 1$
    \item $t=3$, User enacts withdrawal: $N_v = 0$, $N_u = 1$
    \item $t=T$, User enacts spend: $N_u = 0$
\end{enumerate}

This simple example demonstrates a critical issue in the vault pattern. At time $t=T$, the unspent unvault output is susceptible to theft if the attacker has compromised any two of $(q_1,q_2,q_3)$. A cunning attacker would ex-filtrate these private keys without being detected and wait for the time-lock to expire before broadcasting a theft transaction. In order to overcome this critical issue, the user should keep funds in several unspent vault outputs, with a distribution of value that accounts for the threat shown here. The maximum value of an unspent vault output will determine the maximum risk the user must take to access their funds.

For example, the user decides to limit the value of all vault outputs to $5\%$ of their initial total balance. The user only withdraws from one vault output at a time, thus putting at risk a maximum of $5\%$ at any time. This operational limitation significantly mitigates the inherent issue in the vault pattern, at the expense of increasing the users time to access funds. The user must wait (in this case) $T$ blocks for each $5\%$ of value they intend to spend. A custody system using the vault pattern is inherently subject to this \textit{security-accessibility trade-off}. 

Each vault also requires an extra transaction in order to spend it; \TX{wit} and \TX{spend}. Each transaction must pay a fee, and so if the user wants to limit their risk by distributing their balance across, say $100$ vault outputs at $1\%$ each, they must pay more in fees. We find that $N_v$ is proportional to the \textit{cost of operations} and to the \textit{storage space} for the covenant transactions (which is particularly limited on hardware signing devices). Moreover, significant operational complexity is introduced when $V_{max}$ is set. To understand why, we must explore a more detailed example. 

Note that a \TX{dep} may spend several receive outputs at once, and may create several vault outputs.
As detailed in the vault-deposit procedure, each vault output $v$ is controlled by a pair of covenant transactions $(TX_v^{wit}, TX_v^{fb})$. These transactions are partially signed with enforcement private keys $e_1^v, e_2^v$, and those private keys are subsequently deleted. Consequently, no alternative means to spend $v$ is possible. 

Consider the case where at time $t=0$, $N_r = N_u = 0$ and $N_v = 20$. The user has distributed their total balance across $20$ vault outputs each with $5\%$ in them. The user wants to spend $1\%$ of their balance. They select any vault output and enact the withdrawal procedure. At time $t=T$ the user can broadcast a \TX{spend} paying $1\%$ to someone and returning the remaining $4\%$ as change as a new receive output. This process of generating change is unavoidable in real world scenarios where payment values are unpredictable. Now, the user has a choice to either enact a vault-deposit or to keep the new receive output as it is, in a less secure state. Enacting a vault-deposit after each spend is a usability burden. Thus, a  \textit{security-usability} trade-off is present.

To strike a balance for the trade-off, the user can define a minimum value for vault outputs, $V_{min}$. If the set of unspent receive outputs do not have an aggregate value greater than $V_{min}$, then they will not be deposited in a vault. 

The Ajolote software will determine the values of created and selected vault outputs. Together, a \textit{vault creation} algorithm and a \textit{vault selection} algorithm operate to 
\begin{quote}
    1. minimise fee payments,\\
    2. minimise the usability burden,\\
    3. minimise the required storage space, and\\
    4. adhere to the risk control parameterised by $V_{max}$.
\end{quote}
These \textit{coin control} issues are largely a consequence of committing to the complete templates for \TX{wit} and \TX{fb} during the vault-deposit procedure. Were it not for the explicit reference from \TX{fb} to \TX{wit}, then less restrictive signature message types (e.g. {\tt A1CP||SINGLE}) could be used for the enforcement signatures. This would allow for safely modifying \TX{wit} to consume a choice of vault outputs, and to add a change output. Consequently, the required storage space and the fee payments would be less. Further still, any bitcoin that is redirected as change could be a vault output, rather than a receive output. Unfortunately, the usability burden would persist for each vault output, since the vault-deposit procedure requires specifying and verifying specific (though less restricted) transaction templates and generating enforcement signatures.

\subsection{Transaction Sequence Correlation}
\label{subsec:tx-seq-corr}

Operations with Ajolote produce consistent sequences of transactions that are observable on-chain. Let us consider the typical usage pattern. A user's bitcoin goes through output type transitions; 
\begin{equation}
    r_s \rightarrow v_s \rightarrow u_s \rightarrow (Ext, c_s)
\end{equation}
with output types labelled $r$ for receive, $v$ for vault, $u$ for unvault, $Ext$ for external and $c$ for change. We use an integer $s$ to label the sequence. When each output is consumed, the TapScript that was chosen will be revealed. In this typical use scenario, the sequence will appear on-chain as;
\begin{equation}
\begin{split}
    & and(pk(R_1^{i,h}),pk(R_2^{i,h})) \\
    \rightarrow \text{ } & and(and(pk(V_1^{i,j}),pk(V_2^{i,j})),and(pk(E^v_1),pk(E^v_2))) \\
    \rightarrow \text{ } & and(and(pk(U_1^{i,j}),pk(U_2^{i,j})),older(T))
\end{split}
\end{equation}
Public keys are not re-used within this sequence. However, the format of the TapScripts have a consistent pattern. An observer would have high confidence that this sequence is part of an Ajolote custody system. The external output may not be a taproot output. If not, since $c_s$ is a taproot output, it would be obvious which of $Ext$ and $c_s$ is the change output. 

The change output is actually another receive output; $c_s = r_t$. If $r_t$ follows the same sequence, a cross-sequence correlation will be created. An observer would see $c_s = r_t$ in two transaction sequences and would have high confidence that both are part of the same Ajolote custody system. We would typically expect to see change aggregated into a single vault-deposit;
\begin{equation}
    (c_1, c_2, c_3) \rightarrow v_s \rightarrow u_s \rightarrow (Ext, c_s)
\end{equation}
We can observe, however, that this would correlate this sequence with the three sequences that produced $c_1$, $c_2$, $c_3$. Coin-control thus not only has to account for fees, usability, storage, and risk of theft, but also for operational privacy. In this case, if $V_{min}$ is smaller, then less of the available change outputs would need to be grouped together to deposit to a vault. While this creates less cross-sequence correlations, it would cost more in transaction fees and storage. This example demonstrates a \textit{privacy-cost} trade-off.

Let us also consider the sequence involving a rejected withdrawal,
\begin{equation}
    r_s \rightarrow v_s \rightarrow u_s \rightarrow fb_s \rightarrow r_t 
\end{equation}
where $fb$ is a fall-back output type and $t \neq s$ is an integer. The TapScripts revealed with each consumed output are,
\begin{equation}
\begin{split}
    & and(pk(R_1^{i,h}),pk(R_2^{i,h})) \\
    \rightarrow \text{ } & and(and(pk(V_1^{i,j}),pk(V_2^{i,j})),and(pk(E^v_1),pk(E^v_2))) \\
    \rightarrow \text{ } & and(pk(W^{i,j}_k),and(pk(E^v_1),pk(E^v_2))) \\
    \rightarrow \text{ } & and(and(pk(F_1^{i,j}),pk(F_2^{i,j})),pk(F_3^{i,j}))
\end{split}
\end{equation}
Here, the enforcement public keys are re-used within the sequence (for consuming $v_s$ and $u_s$). This creates a strong correlation and a consistent pattern for the reject procedure that can be identified by an observer. The resulting receive output $r_t$ will be consumed in another sequence, either a successful spend or failed withdrawal. In either case, that sequence will be correlated with the reject sequence that preceded it. The next version of Ajolote could instead use distinct enforcement keys for the vault covenant and the pay-to covenant. This would reduce an observer's confidence that this sequence is part of an Ajolote custody system. 
 
Not only is the TapScript revealed when an output is consumed, but so too is the control block. Recall that the control block contains a proof that the chosen TapScript was a leaf of the TapTree that was committed to. If identical branches were used across TapTrees, this would be revealed as one of the branch hashes (TapBranch hash) in the control block. However, in Ajolote, each branch in a TapTree for any given output type ($r$,$v$,$u$,$fb$) contains a unique combination of public keys, resulting in a unique hash. 

The one consistent data point across all TapTrees is the internal public key $P$ that is used. This will be present in the control block of every witness that satisfies an Ajolote output type. This correlation factor can be removed by using different internal public keys for each vault sequence. We believe this is feasible, and briefly sketch a solution but defer a more rigorous study for future work. We must define a function with which all signers can derive a deterministic but pseudo-random number, $r$, from some part of their shared state. This can be used to generate a new internal public key, $P'$, without a known private key simply by computing $P'=P + r.G$. By having it be deterministic and based on existing shared state, we can avoid an interactive protocol among signers to compute $P'$.   

Next, let us consider a fall-back scenario. If the complete set of \TX{wit} and \TX{fb} are broadcast simultaneously then several sequences will appear in the same (or a few successive) blocks. Even if there were no correlations revealed through public key re-use, there will be an observable cross-sequence correlation. In fact, this is a more revealing correlation since an observer could estimate the entire balance that was deposited by the user across vaults, and estimate the preferred range of amounts $(V_{min},V_{max})$. 

Finally, a recovery process includes a fall-back procedure. Therefore, the user's operation after a recovery would be observable on-chain due to the system-wide correlation created with the fall-back. If the user was not being attacked, but merely lost or damaged their device, then they may instead enact a slower fall-back where \TX{wit} and \TX{fb} are incrementally broadcast at random block heights over a period (a week or month). This would largely obfuscate a system-wide correlation. In future work, an integration with privacy services (such as mixers) could be helpful to break the system-wide correlation of outputs after a recovery process.

This analysis demonstrates how using different public keys for each vault greatly reduces correlation among outputs, but also reveals other ways that correlations are introduced. We can see that more work is needed to construct coin control algorithms to adequately account for correlation in vault-based custody systems. Over time, if the set of Ajolote users increases then their operations would obfuscate each other's. More privacy comes from more users. 

\subsection{Comparison with Multi-signature}

Here we compare Ajolote with other multi-signature custody systems. Let us denote a $2-$of$-3$ multi-signature custody system as $C_2^3$, assuming the keys are distributed identically to the Ajolote active key-set. Let us denote a $3-$of$-5$ multi-signature custody system as $C_3^5$, assuming keys are distributed identically to those in the active and fall-back key-set. With Ajolote, a user can utilise the full feature set of $C_2^3$ (Tier 1) and $C_3^5$ (Tier 3), with an additional mid-way feature set with accessibility and security somewhere between the two (Tier 2). Consider how Tier 2 is different. 

Tier 2 is more resilient to loss and to compromise than $C_2^3$. Only one signer can be lost or stolen with $C_2^3$, and if an attacker successfully obtains two signing keys, they can steal all funds. On the other hand, $C_2^3$ is more accessible, not requiring waiting for time-locks to expire and rate-limits (encoded as withdrawal constraints). 

Tier 2 is more accessible than $C_3^5$. Accessing funds in $C_3^5$ requires accessing 3 signers (2 locations and mobile), whereas Tier 2 can be accessed with only 2 signers (1 location and mobile) along with waiting for the expiration of a time-lock. Both Tier 2 and $C_3^5$ remain protected after having 2 signing devices lost or stolen. However, for this to be true with Tier 2, a functional watchtower system needs to be operating in order to reject fraudulent withdrawals or assist with a total fall-back. 

\textit{Can Tier 2 be replicated without the vault pattern? }

One could imagine a custody system that uses a combination of $C_2^3$ and $C_3^5$. The user can store long-term funds in $c_3^5$ and intermittently withdraw (requiring 3 signers) to $C_2^3$ for their more regular payments. They could create partially pre-signed withdrawal transactions (with 1 of the required 3 signatures) from $C_3^5$ to $C_2^3$ and fall-back transactions (with 1 of the required 2 signatures) from $C_2^3$ to $C_3^5$ and keep them on their mobile. This way, they could access funds from $C_3^5$ at any time with 2 other signers (e.g. mobile and home). However, they have no recourse to prevent withdrawals in the event that two signers have been compromised. They can attempt to regain resilience to compromise by quickly broadcasting the fall-back transaction with a high fee, but there's no reliable way to gain priority over theft transactions. The likely outcome would be a bidding war for inclusion in a block by replacing transactions with successively higher fees. Whereas, in Ajolote, the user and watchtowers have consensus-enforced priority until the time-lock expires. This clarifies the unique proposition of a vault-pattern in Bitcoin custody. 

\textit{Can the vault-pattern be replicated without the use of covenants?}

One could imagine a custody system that combines $C_2^3$ and $C_3^5$ with a `vault-deposit' process. As with Ajolote, we will use a \TX{dep} creating a vault output that is consumed with \TX{wit} which encodes the vault pattern (including both the time-locked and immediate access path). However, \TX{wit} is not enforced by a covenant. It is possible to take a path other than those encoded by the vault if you can satisfy the locking conditions for the vault output. For this `Tier 2' to retain a resilience to 2 compromised (stolen) signers, the vault output locking conditions must require at least 3 signatures. Thus, the vault output should be locked with a 3-of-5 multi-signature script. In order to make \TX{wit} accessible (requiring only 2 active signers), the user must pre-sign \TX{wit} with a third signer. They can keep this partially-signed \TX{wit} on their mobile. This way, the vault-pattern has been replicated without the use of covenants. However, notice the critical point that the user has to preemptively access a third signer in order to make \TX{wit} accessible on-demand. To enact the equivalent of the vault-deposit without a covenant, an additional signer must be accessed. This clarifies the unique proposition of a covenant for the vault-pattern; it has improved accessibility. 

To summarise, Tier 2 offers a novel balance for the security-accessibility trade-off present with standard multi-signature custody systems. Tier 2 can not be replicated without a covenant-enforced vault pattern. 

\section{Ajolote Risk Model}
\label{sec:Ajolote-risk-model}

In the following, we consider risk as being comprised of two components; impact and likelihood. A high impact attack is irrelevant if there is no likelihood of it occurring. A low impact attack with high likelihood or a high impact attack with low likelihood should both be considered and mitigated. A high impact attack with high likelihood should not be tolerated and warrants a re-design. Qualitatively, likelihood can be characterised by the difficulty of attack, the potential payoff, and awareness of both. An attacker will not attack if they are not aware that bitcoin is held by the user. However, if the attacker knows how much bitcoin the user is holding and has sufficient expertise to exploit vulnerabilities, then the attack is likely to happen.  

\subsection{Assumptions}

Numerous assumptions underpin our risk model. Each assumption is realistic. However, if they do not hold in a particular context it would invalidate the basis of the following risk model. The first set of assumptions we make are about the user. The user;

\begin{itemize}
    \item understands they must abort setup if there are discrepancies between the software prompts and the user manual since these could cause critical security or safety issues
    \item understands they must abort setup if they fail to verify that hardware signers are consistent with each other
    \item is reliable at matching a few strings of alphanumeric text
    \item understands how to choose an appropriate $T, V_{min}, V_{max}$ and $w$ for their personal context
    \item knows how many watchtowers they will hire from each service provider, $N_y$, as well as the total number of watchtowers, $N$
    \item knows what account number they are using, $i$
    \item backs up their mobile PIN discretely and not digitally
    \item backs up their hardware signer PIN discretely and not digitally
    \item (for each service provider) backs up their username and password discretely and not digitally
\end{itemize}

Next we have assumptions about the context the setup, operation and recovery procedures:

\begin{itemize}
    \item A different vendor and manufacturer were used for each hardware signer.
    \item The user has physical control of their devices (mobile, hardware signers, computer) unless they have been lost or stolen.
    \item The user enacts procedures in private. Device displays are not visible to anyone else. The user is not in the presence of other people. Nobody knows the user is enacting procedures for their custody system. 
    \item Hardware signers have Ajolote compatibility (all the features described in section \ref{subsubsec:HS}).
    \item The user has the correct user manual.
\end{itemize}

We make the following assumptions about the service providers:

\begin{itemize}
    \item There exists more than one watchtower service provider with a track record of good service and a strong reputation.
    \item They have verified their watchtowers' software using standard techniques to verify open-source software. 
    \item They have distinct physical locations for each watchtower server.
    \item They have authenticated and encrypted communication channels with each of their watchtowers.
    \item They follow strict privacy preservation practices with user information. 
\end{itemize}

Finally, we assume that cryptographic mechanisms have the following security properties, unless the device on which they are performed is compromised.

\begin{itemize}
    \item Random number generator: truly random
    \item Hash functions: pre-image resistance, second pre-image resistance, collision resistance
    \item Data encryption: confidentiality
    \item Noise KX channels: encrypted with mutual authentication (if the Noise public key was securely pre-shared with the watchtower)
    \item HTTPS channels: encrypted with authentication of server (if the certificate authority has not been compromised)
\end{itemize}

\subsection{Risk Tiers}

The primary asset of the custody system is \textbf{bitcoin}. Although there is no bitcoin in the system until operations begin, an attacker may discretely intervene during or prior to the setup and wait for the unsuspecting user to deposit bitcoin into the system. The tiers of risk in Ajolote are relevant in all phases; setup, operation, recovery.  

The Ajolote custody system has three tiers of risk for the bitcoin it controls. Tier 1 bitcoin (receive outputs) can be accessed with 2 signers from the active set. Tier 2 bitcoin (vault outputs) can be accessed with 2 signers from the active set after waiting time $T$ and are subject to withdrawal constraints enforced by the watchtowers. Tier 3 bitcoin (fall-back outputs) require 3 signers from the user's active and fall-back set. Access control for the user's bitcoin depends on private master keys $q_n$ for $n \in \{1,2,3,4,5\}$. An attacker can only access $q_n$ if they compromise the signing device. Their attack will not be significantly profitable unless they compromise two or more signing devices. Compromising one signer is an intermediate attack step towards an attacker stealing a user's bitcoin. Typically, the majority of bitcoin will be controlled in Tier 2 and can be pushed to Tier 3 with the reject or fall-back procedures.

These risk tiers are a basic overview and are not comprehensive. They encapsulate the main risk for the primary asset (bitcoin) but do not capture the risk associated with other assets (privacy and operational continuity). Moreover, they do not describe the risk associated with enacting the procedures; vulnerabilities that can be exploited to affect the processes for moving bitcoin up and down through tiers of risk. In the following, we discuss in depth the known risks for procedures in each phase.

\subsection{Impact}
\label{subsec:impact}

In table \ref{tab:stolen-data-impact} we summarise the impact of stolen data. Note that data types are shown individually but, since data will be grouped together on devices, the impact must be considered in aggregates. Refer to table \ref{tab:setup-summary} for data aggregates after the setup and to table \ref{tab:vault-dep-summary} for data aggregates during operation. To this end, we consider several assets and determine the impact of stolen data aggregates towards compromising those assets. 

The major impact of storing system-wide bitcoin public key data and configuration data ($\{A^i_n\},\{W_k\},T, V_{min},V_{max}$) redundantly on all signing devices is that the privacy of the custody system can be broken by compromising a single signing device. Critically, private master keys $q_n$ are compartmentalised across signing devices, making it difficult for an attacker to attempt to steal bitcoin once it is deposited into the system (which happens only after a successful setup). 

The mobile and home signer both store the complete set of covenant transactions and their enforcement signatures. With these and the complete set of \TX{dep}, an attacker could break the user's operational privacy. Worse, \TX{fb} can be authorised with an additional signature from one of their master private keys $q_1$ or $q_2$ which are stored on the same device. So an attacker could disrupt the custody operation by triggering reject procedures as a user attempts withdrawals. This is a temporary denial-of-service attack. 

Data that enables an entity to reach and authenticate themselves to the watchtowers ($p^{noise}, \{O_k\}$) is redundantly stored on all signing devices. Compromising a single signing device would enable an attacker to spoof the user identity in communication with all watchtowers. They could; register fake vaults (by enacting vault-deposit procedure), obtain signatures for test transactions (by enacting setup test or system health check), obtain signed hash of operational data (by enacting system health check), or obtain the complete set of covenant transactions and their enforcement signatures (by enacting fall-back procedure). 

However, reaching a watchtower is only the first step towards compromising it. The watchtowers' IP addresses remain hidden (since they run as hidden services) so no physical location data is revealed. Moreover, watchtowers validate each input they receive and only respond if a message is an Ajolote message type. While there may be an exploit in the implementation of the watchtower, in theory the attack surface is extremely limited. The possibility of malware injection or denial-of-service should be a primary consideration for the implementation. 

Another aggregate of data is stored by each watchtower. If an attacker compromised a watchtower, they would learn private operational behaviour of the custody system such as the length of the time-lock, the withdrawal constraints, the complete set of vault-deposits, and the complete set of covenant transactions (revealing the bitcoin deposited in vaults) and their enforcement signatures. They would have access to the bitcoin key-pair $(w_k,W_k)$ which can be used to initiate reject procedures (once the system is operational) as a temporary denial-of-service attack. They would be able to spoof the identity of the watchtower in communication with the mobile (through controlling the service at $O_k$) and with the hardware signers (through producing signed messages).

In summary, the Ajolote custody system has minor risk associated with `single points of compromise'. The impact of stealing data from the home signer or mobile is; the attacker would have completely broken operational privacy, could launch a temporary denial-of-service attack, or use this as an intermediate attack step towards theft of bitcoin. The impact of stealing data from the office, bank A or bank B signer is; an intermediate attack step towards theft of bitcoin and partially broken operational privacy. The impact of stealing data from a watchtower is; enabling a temporary denial-of-service attack and completely broken operational privacy.

\renewcommand{\arraystretch}{1.25}
\begin{table}
    \centering
    \begin{tabular}{|c|p{10cm}|}
        \hline
        \rowcolor{lightgray} \textbf{Stolen Data} & \textbf{Impact} \\
        \hline
        $\{A^i_n\}$ & On-chain privacy compromised; receive outputs and fall-back outputs observable \\
        \hline
        $\{W^i_k\}$ & On-chain privacy compromised; reject sequences correlated \\
        \hline
        $\{O_k\}$ & Watchtowers become reachable  \\
        \hline
        $w$ & Operational behaviour revealed; withdrawal/reject conditions determined \\
        \hline
        $T$ & Vault behaviour revealed; time-lock length  \\
        \hline
        $(V_{min}, V_{max})$ & Vault behaviour revealed; min and max vault size  \\
        \hline
        $(q_n,Q_n)$ & Reduction in safety (resilience to loss) and security (resilience to theft) and privacy (can derive accounts $A^i_n$). \\
        \hline
        $(p^{Noise},P^{Noise})$ & Can authenticate as user to watchtowers \\
        \hline
        $(w_k, W_k)$ & Can initiate reject procedure (given \TX{fb}) \\
        \hline
        mobile PIN & Can access the Ajolote application (given bio-metric and access to mobile) \\
        \hline
        bio-metric & Can access the Ajolote application (given mobile PIN and access to mobile) \\
        \hline
        user and password for service provider & Can authenticate as user to service provider \\
        \hline
        PIN for hardware signer & Can connect to hardware signer (given physical access to hardware signer) \\
        % \hline
        \hline
        $\{TX^{wit}\}$ & User balance deposited in vaults revealed. Required to initiate withdrawals.  \\
        \hline
        $\{sig^{wit}_{enf}\}$ & Required to initiate withdrawals. \\
        \hline
        $\{TX^{fb}\}$ & User balance deposited in vaults revealed. Required to initiate reject or fall-back.  \\
        \hline
        $\{sig^{fb}_{enf}\}$ & Required to initiate reject or fall-back \\
        \hline
        $\{E_v^1,E_v^2\}$ & Can deduce some information about the TapTrees for vault and unvault output types. Useful for initiating withdrawal, reject or fall-back with $q_n$. \\
        \hline
        $\{c_k\}$ & Useful for initiating reject or fall-back with $w_k$ \\
        \hline
        $v$ & Number of vaults revealed. \\
        \hline 
        $i$ & Number of accounts revealed. \\
        \hline
        $\{TX^{dep}\}$ & Historical and current vault-deposit information revealed. \\
        \hline
        $\{TX^{spend}\}$ & Historical spend information revealed. \\
        \hline
        $\{TX^{rec}\}$ & Historical incoming payments information revealed. \\
        % \hline 
        \hline
        $\{A^m_n\}$ & On-chain privacy compromised for previous custody system setups (prior to recovery procedures). \\
        \hline
    \end{tabular}
    \caption{Summary of impact of stolen data.}
    \label{tab:stolen-data-impact}
\end{table}
\renewcommand{\arraystretch}{1}

\subsection{Threat model}
\label{subsec:ax-threat-model}

Rather than analyse the custody system as set of network security protocols, we consider it more broadly as a set of security ceremonies. To this end, we include the user as a node and messages written and read by the user. We propose a threat model which is more realistic than Dolev-Yao for our analysis. We define the capabilities of an attacker with respect to different types of channels, e.g. $\tt human \xleftrightarrow \tt offline \text{ } device$, $\tt human \xleftrightarrow \tt networked \text{ } device$, $\tt networked \text{ } device \xleftrightarrow \tt networked \text{ } device$, and $\tt networked \text{ } device \xleftrightarrow \tt offline \text{ } device$. In general, an attacker can completely control compromised networked devices, has limited power with compromised offline devices, and has no control over the human. A summary of attacker capabilities per-channel is given in table \ref{tab:attacker-capabilities-summary}. Herein, we expand on our rationale for the threat model.

For our purposes, we further adapt the attacker capabilities as presented by Martina \textit{et al.} \cite{AdaptiveThreatCeremony} to reflect controls built into the Ajolote software. Despite a successful active attack (where the attacker either compromises a device or engages in a man-in-the-middle attack), the messages accepted by honest Ajolote software (mobile application, hardware signer application, or watchtower application) is strictly limited to the Ajolote message types. Furthermore, Ajolote software will expect a particular ordering of messages and halt if this ordering is not respected. To reflect this, we denote the limited capability with a `$^*$'. $I^*$ and $F^*$ mean that an attacker can initiate and fabricate only Ajolote message types, respectively. $B^*$ and $O^*$ mean that an attacker can block and re-order Ajolote message types only, in keeping with the procedure specifications. 

If the mobile/computer has not been compromised, the user's communication with it is not observable to an attacker and cannot be modified in transit. Similarly, if the hardware signer has not been compromised (e.g. with malware or a hardware implant), the user's communication with it is not observable to an attacker and cannot be modified in transit.

The user physically controls the mobile and hardware signers. Since the mobile is a networked device, it can be remotely compromised. Once compromised, it can send messages to the attacker. The hardware signers have no network interfaces. They may only communicate directly with the user (via their display and input buttons) or with the mobile via a wire or a camera-QR code system. The hardware signer cannot send a message to the attacker without the assistance of the mobile signer. In the case where the hardware signer has been compromised but the mobile is honest, the hardware signer cannot get a message to the attacker. However, if both the hardware signer and the mobile have been compromised, the attacker has full read-write control for these devices. The wired and camera-QR code based communication channels between the mobile and hardware signer cannot be remotely `tapped' without compromising the mobile itself. Control of the wire or camera/QR code channels would require the physical presence of an attacker. The same rationale for the mobile to hardware signer channels applies to channels between the computer and hardware signers.

An attacker can remotely compromise a channel between two networked devices and as a result will have full control over the messages between the two devices. 

\begin{table}
    \centering
    \begin{tabular}{|p{5.5cm}|p{5cm}|}
         \hline
         \rowcolor{lightgray} \textbf{Channel} & \textbf{Attacker Capabilities} \\
         \hline
         $\tt User \xleftrightarrow \tt Mobile$,\newline $\tt User \xleftrightarrow \tt Computer$ & \\
         \hline
         $\tt User \xleftrightarrow \tt Hardware\text{ } Signer$ & \\
         \hline
         $\tt User \xleftrightarrow \tt Mobile \text{ } (compromised)$ & $E$, $I^*$, $B^*$, $A$, $C$, $F^*$, $S$, $O^*$  \\
         \hline
         $\tt User \xleftrightarrow \tt Hardware \text{ } Signer \text{ } (compromised)$ & $I^*$, $B^*$, $A$, $C$, $F^*$, $S$, $O^*$ \\
         \hline
         $\tt Mobile \text{ } (compromised) \xleftrightarrow \tt Hardware \text{ } Signer$, $\tt Computer \text{ } (compromised) \xleftrightarrow \tt Hardware \text{ } Signer$ & $E$, $I^*$, $B^*$, $A$, $C$, $F^*$, $S$, $O^*$ \\
         \hline
         $\tt Mobile \xleftrightarrow \tt Hardware \text{ } Signer \text{ }(compromised)$, $\tt Computer \xleftrightarrow \tt Hardware \text{ } Signer \text{ }(compromised)$ & $I^*$, $B^*$, $A$, $C$, $F^*$, $S$, $O^*$  \\
         \hline
         $\tt Mobile \text{ } (compromised) \xleftrightarrow \tt Hardware \text{ } Signer \text{ } (compromised)$, $\tt Computer \text{ } (compromised) \xleftrightarrow \tt Hardware \text{ } Signer \text{ } (compromised)$ & $E$, $I$, $B$, $A$, $C$, $F$, $S$, $O$ \\
         \hline
         $\tt Networked \text{ } Device \xleftrightarrow \tt Networked \text{ } Device$ & $E$, $I$, $B$, $A$, $C$, $F$, $S$, $O$ \\
         \hline
    \end{tabular}
    \caption{A summary of attacker capabilities per channel type in the Ajolote custody system.}
    \label{tab:attacker-capabilities-summary}
\end{table}

\subsection{Authentication}

Entities authenticate each other during the setup procedures using a variety of methods. In some cases, communication is encrypted too. Authentication methods generally rely on either what you know (e.g. password), what you have (e.g. physical device), or who you are (e.g. bio-metric data). All authentication data must be sufficiently random to avoid brute-force attacks. In table \ref{tab:auth-summary} we show which entity authenticates another with which data. We also show what entities the authentication process depends on. If the dependency is compromised, the authentication will be too. Authentication is used to ensure that communication channels are established among the correct entities. An attacker that compromises authentication data (directly or via a dependency) can spoof the identity of the authenticated entity, and engage in man-in-the-middle attacks without being detected.  

\begin{table}
    \centering
    \begin{tabular}{|p{3cm}|p{3cm}|p{3cm}|p{4cm}|}
         \hline
         \rowcolor{lightgray} \textbf{Authenticating Entity} & \textbf{Authenticated Entity} & \textbf{Authentication Information} & \textbf{Dependencies }\\
         \hline
         Mobile & User & PIN and bio-metric & \\
         \hline
         Mobile & App Store & certificate & Certificate Authority \\
         \hline
         Mobile & Service Provider & certificate & Certificate Authority \\
         \hline
         Computer & Service Provider & certificate & Certificate Authority \\
         \hline
         Hardware Signer & User & PIN & \\
         \hline
         Service Provider & User & username and password & \\
         \hline
         Watchtower & Mobile & $P^{Noise}$ & Service Provider \\
         \hline
         Mobile & Watchtower & $O_k$ & Service Provider \\
         \hline
         Hardware Signer & Watchtower & $W_k$ & Service Provider and 1 of (Mobile, Computer) \\
         \hline
         User & Hardware Signer & Physical Display & \\
         \hline 
         User & Mobile & Physical Display & \\ 
         \hline 
    \end{tabular}
    \caption{Table summarising data used for authentication of entities. In some cases the data depends on additional entities being honest.}
    \label{tab:auth-summary}
\end{table}

\subsection{Setup Security}

The goals of the setup are to; 
\begin{itemize}
    \item[a)] prepare each devices' state (as summarised in table \ref{tab:setup-summary}), 
    \item[b)] to verify setup state consistency across devices,
    \item[c)] to verify correctness of data where possible, 
    \item[d)] to demonstrate the ability of each signer to create valid signatures for Ajolote outputs.
\end{itemize}
If an entity (device or user) detects unexpected behaviour from any other entity, the user is instructed to discard and replace the misbehaving component before beginning again. The user can detect unexpected behaviour by following the user manual alongside the software prompts. On one hand, this makes denial-of-service attacks relatively easy and could be frustrating for the user. On the other hand, this acts as a forcing function for correct behaviour and since a denial-of-service attack would be unprofitable (in the setup phase) there is little motivation for an attacker to do this. 

By studying the message sequence diagrams for each setup procedure it should be clear that the final state of each device will contain the correct information if the protocol is enacted by honest participants. Progress through the procedure is forced since blocked messages would be detected (by the user or by honest devices) and the culprit subsequently replaced. Goal \textit{a} is satisfied if the entire process completes without halting.  However, the possibility of compromise requires mechanisms to verify the \textit{consistency} of data across devices and the correctness of data (where possible). 

Unfortunately, it is not possible to determine if any device has been compromised if they enact the protocol as an honest entity would. A consistency verification merely assures the user that the devices are functional. Not every data type can be verified to be correct. This is why it is critical to design the custody system with tolerance to failure and compromise (this is as relevant during setup as it is in operation). Data is only demonstrably correct if it is legible to the user. Other data can be checked by software for correct formatting but it will be indistinguishable whether or not they are controlled by an attacker. This includes Bitcoin key-pairs, onion addresses and Noise key-pairs.

\subsubsection{User's Device Generated State}

During the mobile initialisation the mobile downloads the Ajolote application from the App store. Here we rely on the software verification process by the vendor (e.g. Apple app store or Google Play store) and the Certificate Authority to authenticate the vendor. These vendors have multiple layers of countermeasures to prevent false software from being provided to users \cite{appleAppSecurity,googleAppSecurity}. The risk is that an attacker overcomes these countermeasures and serves malicious Ajolote software to the user's mobile without the user knowing. The compromised mobile would have to appear to adhere to the ceremony specification or cause the setup to be aborted and the device discarded. The result of a mobile running malware would be a custody system that becomes operational with less resilience to failure and attack. Recall the more detailed discussion of impact from section \ref{subsec:impact}. When initialising the mobile, the user directly inputs $c = (T, V_{min}, V_{max})$, and $w$.

During the hardware signer initialisation the user verifies requests from the mobile to `initialise' and `setup' Ajolote. This prevents a compromised mobile from triggering the hardware signers to perform unwanted actions. The user directly verifies the correctness of $c$ on the hardware signers (and so they are consistent with the mobile). Thus, goal \textit{c} is satisfied.

% Consistency of accounts
The design of the setup test enables the user to observe inconsistencies in accounts, $\{A^1_n\}$ and $\{W^1_k\}$, and setup data caused by misbehaving devices. Each honest hardware signer will construct the output in \TX{test} using $l_{test}$ based on its own knowledge of accounts. An honest hardware signer will include its own public key as part of $l_{test}$. The user verifies that each hardware signer and the mobile constructed the same transaction by matching the transaction ID displayed during the test. Moreover, the user verifies that $H^{setup}$ is consistent across the mobile and signing devices. Goal \textit{b} is satisfied for state generated on user's devices as long as one of them is honest. 

The user's signers verify signatures from all watchtowers, demonstrating that the watchtowers have access to the associated private keys. The user's mobile and home signer verify all hardware signers' signatures for \TX{test}, demonstrating that they have access to the associated private keys. Goal \textit{d} is thus satisfied if the user's mobile and home signer remain honest and do not deceive the user.

The setup procedure does not prevent a compromised signer from participating with Bitcoin keys controlled by an attacker. It merely enables honest signers to ensure their inclusion and that they have consistent state among each other. 

\subsubsection{Service Providers' Device Generated State} 

The setup finalisation is used to mitigate risks that emerge during the watchtower initialisation. For the data types generated by the service providers, there is a risk of man-in-the-middle attacks. This type of attack is where the attacker simultaneously spoofs the identity of the user to the service provider, and of the service provider to the user. Recall from table \ref{tab:auth-summary} that user authentication depends on the username and password, while authentication of the service provider depends on the certificate authority. 

The user initially sends data to the service provider via the mobile (during watchtower initialisation). Later, the service provider sends these data back to the user via the computer (during setup finalisation). This increases the difficulty of a successful man-in-the-middle attack because the attacker must spoof the identity of the user to the service provider in two separate sessions on distinct devices (the mobile and the computer) which are operating on distinct networks (e.g. 4g and WiFi). The mobile and hardware signers check the consistency of $\{O_k\}$, $\{W^1_k\}$ and $p^{Noise}$ from both routes. 

The hardware signer displays $N$ to the user. This demonstrates to the user that the hardware signer matched that many $O_k$ and $W^1_k$ values from both the mobile and the computer. It prevents a compromised mobile or computer from omitting $O_k$ or $W^1_k$ from the set without being detected.

A successful attack may compromise both the mobile and computer (or their networks) or compromise each of the service providers. In the later case, the attacker can target service providers' websites and the certificate authority, or can target every watchtower. Given that the service providers are trained security professionals whose business and reputation depend on resisting these types of attack, we expect an attacker to attack the weaker target; the user's mobile and computer. 

If successful, the attacker can eavesdrop or tamper with the setup state values sent to the service provider ($T$, $w$, $N_y$ and $P^{Noise}$) and sent to the user ($\{O_k\}$,$\{W^1_k\}$). Let us consider some of the attacks to build an intuition for how difficult they are and thus for their likelihood. Refer back to section \ref{subsec:impact} for the impact associated with each data type. 

Eavesdropping requires no further action from the attacker.

Tampering $T$, $w$ would go undetected and could result in watchtowers enforcing weak or overly restrictive withdrawal policy, or of believing they have more time to react than they really do. 

Tampering $P^{Noise}$ will require the attacker to operate a false entity mimicking the mobile behaviour. The attacker could replace the value with one for which they have the private key and spoof the mobile identity to the watchtowers. 

Tampering $\{O_k\}$ would require the attacker to also operate hidden services running Ajolote watchtower software (or at least mock implementations). Why? Because the user will try to connect with those services and progress with the setup and they would detect if the watchtowers are not operational. An attack that successfully convinces the user to connect with malicious watchtowers would degrade the reliability of reject and fall-back procedures. 

If these attacks were achieved \textit{for each} service provider, then the user would not have any honest watchtowers and their operation would be vulnerable. Specifically, the reject procedure and fall-back (version 3) would be unreliable. The bitcoin in Tier 2 would only have the security and safety features of Tier 1. 

\subsubsection{Summary}

In summary, the user's device generated state is consistent if at least one of their devices is honest. The user legible data is consistent and correct on each of the user's devices. However, the service provider generated state is less reliable. The user can ensure consistency of the data, but cannot know if the watchtowers are honest. The highest risk (impact and likelihood) is that a sophisticated attacker compromises the user's mobile and computer (or their networks), and completes the setup procedure with malicious watchtowers, remaining undetected. This compromises the additional security of Tier 2 bitcoin compared with Tier 1. All that remains for an attacker is to compromise another of the user's signer. 

Our analysis indicates aspects that can be strengthened with simple countermeasures. First, the user should keep knowledge of their bitcoin holdings confidential. They will be much more secure by not making them-self a target. Second, the user should use a new and dedicated mobile and computer. This will significantly reduce the likelihood of being compromised. Third, their mobile and computer should have industry-standard VPN and anti-virus software. 

\subsection{Operational Security}

The operational security depends on whether or not the setup was enacted correctly. We will assume it was completed but that some signers may have been compromised prior to the system becoming operational. 

Throughout this analysis we will present vulnerabilities, describe the attack, note the impact and suggest mitigating countermeasures. 

\subsubsection{Receive and Spend}

The receive procedure was designed to reveal inconsistency between the mobile and home signer before sharing the receive address with the counter-party who will send bitcoin to the user. A formal treatment of this was given by Arapinis \textit{et al.} \cite{FormalHardware}. However, through ceremony analysis we can see clearly that there are still risks within the procedure. 

There are two vulnerabilities to exploit via two attacks, resulting in the same impact. A general countermeasure can mitigate both attacks. 
\begin{table}[H]
    \centering
    \begin{tabular}{|>{\columncolor{lightgray}\raggedleft\arraybackslash}p{3.5cm}|p{12cm}|}
    \hline
         \textbf{Impact} & Payment is sent to an attacker, not to the user.\\
         \textbf{Countermeasure} & Limit the size of amounts being sent and make incremental payments. That way, no single receive procedure will carry too much risk. \\
    \hline
    \end{tabular}
    \label{risk:receive}
\end{table}

\begin{table}[H]
    \centering
    \begin{tabular}{|>{\columncolor{lightgray}\raggedleft\arraybackslash}p{3.5cm}|p{12cm}|}
    \hline
        \textbf{Vulnerability A}& Mobile is subject to broad remote exploits. Home signer is subject to malware injection and hardware implants (particularly while in the supply-chain). \\
        \textbf{Attack A}& Compromise the mobile and home signer and replace the receive address with one controlled by an attacker.\\
        \textbf{Countermeasure A}& Verify the receive address on another user device (e.g. office signer).  \\
    \hline
    \end{tabular}
\end{table}

\begin{table}[H]
    \centering
    \begin{tabular}{|>{\columncolor{lightgray}\raggedleft\arraybackslash}p{3.5cm}|p{12cm}|}
    \hline
\textbf{Vulnerability B}& Weak mutual authentication method between the user and counter-party. \\
\textbf{Attack B}& Man-in-the-middle attack between the user and counter-party to replace the receive address with one controlled by an attacker.\\
\textbf{Countermeasure B}& Use strong mutual authentication between the user and counter-party. \\
    \hline
    \end{tabular}
\end{table}

The channel between the user and the counter-party is unspecified here. It could, for example, be; face-to-face, via a chat app, via a web-based payment portal, or something else entirely. The point is that if the channel is not \textit{securely} mutually authenticated then an attacker could launch a man-in-the-middle attack between the user and counter-party and trick the counter-party to pay them instead of the user.

The same type of vulnerabilities, attacks and countermeasures are present for the spend procedure, where instead the user pays the counter-party. Address replacement attacks, and worse, transaction replacement attacks are possible if either; the channel between the user and the counter-party is compromised, or both the mobile and home signer are compromised. However, the impact is different. In this case, the attacker can authorise any kind of theft transaction from Tier 1 bitcoin. 

\subsubsection{Vault-deposit}
\label{subsec:ops-risk-vault-deposit}

The vault-deposit procedure is subject to two categories of attack; either prevent the covenant enforcement from happening or break the covenant enforcement after a successful vault-deposit. As we will see, neither is feasible or realistic if the appropriate countermeasures are used. 
\begin{table}[H]
    \centering
    \begin{tabular}{|>{\columncolor{lightgray}\raggedleft\arraybackslash}p{3.5cm}|p{12cm}|}
    \hline
\textbf{Vulnerability C}& Deceive the user to think vault-deposits are successful, but they are not. \\
\textbf{Attack C}& Compromise the hardware signer (either \textit{(i)} when in the supply-chain, or \textit{(ii)} by penetrating the user's physical security system at home without being detected). Compromise the mobile (remotely). Instruct both devices to pretend to enact vault-deposits. Steal their master private key-pairs $(q_n,Q_n)$ for $n=1,2$. \\
\textbf{Impact C}& The longer the deception lasts, the more bitcoin will accumulate in Tier 1. Attacker is able to steal all bitcoin controlled by Tier 1. \\
\textbf{Countermeasures C}& \textit{(i)} Buy direct from producer, device authentication with producer, use tamper-evident packaging, and purchase privately.\\
& \textit{(ii)} Prepare a strong physical security system at home with alarms, cameras, a strong safe, locked doors, and tamper-evident packaging. \\
    \hline
    \end{tabular}
\end{table}

The major difficulty to overcome here is to attack the home signer without being detected, whether it is scenario \textit{(i)} or \textit{(ii)}. Could an attacker reproduce tamper-evident packaging? Could an attacker compromise the hardware producer's packaging facility without being detected? Could the attacker collude with the producer (e.g. a state-level attacker colluding with the producer) without the producer being caught and losing their reputation? Could an attacker breach the user's home, by-pass each lock, alarm, camera, and safe without leaving a trace? Maybe, but these are unlikely attacks. 

If the user is suspicious and has detected some evidence of an attack on the home signer, they should fall-back and investigate the situation more thoroughly.

There is an important take-away point here for integrating a vault pattern into a custody system. \textit{The procedure depends on the user's verification of the state of their custody system.} In this case, the user depends on both their mobile and home signer to notify them whether or not the vault-deposit was a success. These devices must both have a mechanism to verify if it was successful. The mobile has its own connectivity to the Bitcoin network and can monitor the status independently. The home signer depends on signed messages from the watchtowers. If the user had another method to verify their custody system's state, it would be harder-still to deceive them. For example, the user could verify signed messages from the watchtowers using the office signer. However, this negates the usability of Tier 2. The user could use their computer, connect to one or more block explorers, and manually verify that \TX{dep} was confirmed on-chain. However, again, this is a cumbersome process that reduces the usability of Tier 2. Notice that this is independent of the choice of covenant mechanism.

By considering the enforcement conditions for activating a deleted-key covenant (as described in section \ref{subsec:enf-cond}) we describe under what conditions the vault-deposit is successful. Briefly, the vault-deposit procedure depends on a) construction of \TX{wit} and \TX{fb}, b) key-deletion, and c) verifying the inclusion of \TX{dep}. An attacker may deceive the user about any of these conditions in attack \textbf{C}. 

Notice too that condition b) is subject to another vulnerability, as follows.
\begin{table}[H]
    \centering
    \begin{tabular}{|>{\columncolor{lightgray}\raggedleft\arraybackslash}p{3.5cm}|p{12cm}|}
    \hline
\textbf{Vulnerability D}& Key-deletion process leaves residual information on the mobile and home signer. The `deleted' keys can be recovered through advanced physical inspection attacks (as discussed in section \ref{subsec:key-deletion}). \\
\textbf{Attack D}& Breach the physical environment where the mobile and home signer are located and steal the devices. By-pass the access control for both devices (PINs and bio-metric). Extract private key material and state. Determine the deleted-keys for each vault with physical inspection of residual charge with advanced instruments. \\
\textbf{Countermeasure D}& Trigger a fall-back before the attacker succeeds in theft. \\
    \hline
    \end{tabular}
\end{table}

The countermeasure to attack \textbf{D} is highly effective. The attack requires specialist equipment, expertise, and time. Since it would need to be completed without alerting the user (who can trigger a fall-back), it is highly unlikely to be successful. 

Finally, let us consider a denial-of-service attack. 
\begin{table}[H]
    \centering
    \begin{tabular}{|>{\columncolor{lightgray}\raggedleft\arraybackslash}p{3.5cm}|p{12cm}|}
    \hline
\textbf{Vulnerability E}& The vault-deposit procedure will not complete if any one of the; mobile, home signer, or watchtowers does not behave as expected. \\
\textbf{Attack E}& Compromise or shutdown any of the devices. Ensure that it does not respond to the others.\\
\textbf{Impact E}& User can not move bitcoin from Tier 1 to Tier 2. The attack would be detected since the procedure is failing. This is an intermediate attack, increasing the payoff of a Tier 1 theft.\\
    \hline
    \end{tabular}
\end{table}

\subsubsection{Withdrawal}

As with the vault-deposit, the primary vulnerability related to the withdrawal procedure emerges from the user relying on the mobile and home signer for their view on what is their custody system state. The attack \textbf{F} is an extended version of \textbf{C}, depending on deceiving the user, and the countermeasures for \textbf{C} apply here to.
\begin{table}[H]
    \centering
    \begin{tabular}{|>{\columncolor{lightgray}\raggedleft\arraybackslash}p{3.5cm}|p{12cm}|}
    \hline
\textbf{Vulnerability F}& User's devices trigger withdrawals without the user's permission.\\
\textbf{Attack F}& Compromise both the mobile (remotely) and the home signer (physically) without being detected. Deceive the user, with malware on mobile and home signer, about the balance deposited to vaults. Gradually withdraw bitcoin from Tier 2 to Tier 1, adhering to the withdrawal conditions enforced by watchtowers. Use the two compromised signers to authorise a theft transaction.\\
\textbf{Impact F}& Theft of (up to) all of the bitcoin initially controlled by Tier 2 and by Tier 1. \\
    \hline
    \end{tabular}
\end{table}

Adding another signer to $l_{wit}$ is not a viable countermeasure to this problem, since that would negate the usability of Tier 2. The enforcement of withdrawal constraints by the watchtower and the consensus-enforced time-lock are a useful countermeasure in limiting the rate of potential theft. However, theft can still occur. Critically, if the user lacks an accurate view into the ongoing state of their custody system, an attacker can cause significant impact despite strict withdrawal constraints and long time-locks. 

\subsubsection{Reject}

The reject procedure is only reliable before the time-lock expires because \TX{fb} only has priority over \TX{spend} during that period. During the time-lock period, a watchtower has to detect \TX{wit}, detect any violations of withdrawal constraints, sign \TX{fb} and broadcast it to the network. At a minimum, the length of the time-lock should be sufficiently long for that process. However, another aspect to consider is network fee dynamics. The network fee exhibits non-linear changes. It intermittently jumps between values that can be orders of magnitude apart. While \TX{fb} has been prepared with a high fee (40 satoshis per kilobyte), this will not be sufficient in some edge case scenarios. Moreover, if $T$ is short, the network fee dynamics could be gamed by an attacker. 
\begin{table}[H]
    \centering
    \begin{tabular}{|>{\columncolor{lightgray}\raggedleft\arraybackslash}p{3.5cm}|p{12cm}|}
    \hline
\textbf{Vulnerability G}& \TX{fb} is not included in a block before the time-lock in \TX{wit} expires.\\
\textbf{Attack G}& Attacker broadcasts a batch of high fee-rate (satoshis per kilobyte) transactions to drive up the network fee to the extent that \TX{fb} loses priority and is not included in the next block. Attacker maintains the attack until the time-lock expires.\\
\textbf{Impact G}& \TX{fb} loses priority over \TX{spend}. Expensive to enact this attack.\\
\textbf{Countermeasure G}& Increase $T$, thus increasing the expense of the attack. \\
    \hline
    \end{tabular}
\end{table}

Note that this is a partial attack. To profit, the attacker would also need to compromise two of the user's signers to be able to satisfy $l_{wit}$ with a theft transaction. 

Next, let us consider a denial-of-service attack using the reject procedure.
\begin{table}[H]
    \centering
    \begin{tabular}{|>{\columncolor{lightgray}\raggedleft\arraybackslash}p{3.5cm}|p{12cm}|}
    \hline
\textbf{Vulnerability H}& Reject procedure blocks a valid withdrawal attempt from the user.\\
\textbf{Attack H}& Compromise either a watchtower or the user's mobile or home signer. Wait for a user to initiate a withdrawal. Sign and broadcast the associated \TX{fb}.\\
\textbf{Impact H}& The user is temporarily denied access to the amount controlled by \TX{wit}. The bitcoin is pushed to Tier 3, making it less accessible to the user. \\
\textbf{Countermeasure H}& Replace compromised devices with the recovery procedures. Replace service providers that violate their service level agreement. \\
    \hline
    \end{tabular}
\end{table}

\subsubsection{System Health-check}

The system health-check is the same as the test procedure, except for the additional operational data that is verified by the user and the fact that the user must travel to the location of each hardware signer to interact with them. It grants the same consistency guarantees about watchtowers' data and about the user's device data as the setup test. It demonstrates that signers have access to their private keys. 

\subsection{Recovery Security}

\subsubsection{Fall-back}

Critically, each signer must receive an authenticated message from the user before participating in the fall-back procedure. This prevents a malicious coordinator from tricking a signer into participating.

A malicious watchtower can not authorise \TX{wit} and so it has no ability to trigger a fall-back scenario unilaterally. 

As with the vault-deposit and withdrawal procedures, the user may be vulnerable to deception.
\begin{table}[H]
    \centering
    \begin{tabular}{|>{\columncolor{lightgray}\raggedleft\arraybackslash}p{3.5cm}|p{12cm}|}
    \hline
\textbf{Vulnerability I}& User is deceived to think that a fall-back procedure has been completed, but in reality it has not. \\
\textbf{Attack I}& Compromise the coordinator (mobile or emergency coordinator) with malware. Instruct coordinator to pretend to the user that a fall-back has been completed. It will enact the procedure with the hardware signer(s), but will not broadcast \TX{wit} or \TX{fb}. \\
\textbf{Impact I}& User's bitcoin remains in a less secure and less safe state, in Tier 2.\\
\textbf{Countermeasure I}& Send watchtower-signed acknowledgements of the fall-back procedure to the hardware signer(s).\\
    \hline
    \end{tabular}
\end{table}

In this case, the user verification of their custody system state is weaker than with the vault-deposit because they are only using the coordinator. The proposed countermeasure would enhance that verification process, requiring the attacker to compromise an additional signer to prevent the user from noticing the fall-back procedure failed. 

The user would realise during the recovery process, when they attempt to send bitcoin to the new fall-back key-set.

Recall that there are three variations of the procedure which depend on which device(s) have been lost or stolen. If the user does not have access to their mobile and home signer (as in figure \ref{fig:fb3}), they are relying on at least one watchtower being honest to assist in the fall-back procedure. Otherwise, they cannot fall-back. 

As with the reject procedure, the reliability of a fall-back depends on the length of the time-locks used. The longer the time-lock $T$, the less risk there is for \TX{fb} to lose priority to potential theft transactions. During a fall-back procedure, the risk is magnified for several reasons.

First, since all \TX{wit} are broadcast at once, the impact of the time-lock expiring results in the total balance within Tier 2 moving to Tier 1. Second, one or more of the devices may have been compromised and in that case, Tier 1 controls are less secure. Third, the network fee will increase when a batch of high fee-rate transactions are broadcast (the complete set of \TX{wit} and \TX{fb}), perhaps causing a non-linear reaction that leads to a sustained high fee-rate period. This reason is already mitigated by preparing both \TX{wit} and \TX{fb} with the same fee-rate because they will be prioritised equally by (non-colluding) miners. Finally, and most importantly, the user must enact the procedure. It is not automatic. The user will need time to travel to the necessary signers. 
\begin{table}[H]
    \centering
    \begin{tabular}{|>{\columncolor{lightgray}\raggedleft\arraybackslash}p{3.5cm}|p{12cm}|}
    \hline
\textbf{Vulnerability J}& The user is unable to initiate the fall-back procedure. \\
\textbf{Attack J}& Wait for the user to be far away (for example, in a different country or continent). Compromise two of the user's signing devices. Attempt withdrawals and theft transactions. \\
\textbf{Impact J}& The rate of theft must not violate the withdrawal constraints enforced by watchtowers. \\
\textbf{Countermeasure J}& The mobile notifies the user of withdrawal attempts. If the user has not initiated withdrawals, they can manually trigger reject procedures by signing \TX{fb} with the mobile. \\
    \hline
    \end{tabular}
\end{table}

Of course, this countermeasure only works if the user has access to their mobile and it has not been remotely compromised.  

Now, let us consider how an attacker can take advantage of the fall-back procedure to launch a temporary denial-of-service attack.  
\begin{table}[H]
    \centering
    \begin{tabular}{|>{\columncolor{lightgray}\raggedleft\arraybackslash}p{3.5cm}|p{12cm}|}
    \hline
\textbf{Vulnerability K}& Trigger the user to enact a fall-back procedure.\\ 
\textbf{Attack K}& Alert the user to some danger. For example, send a threatening message, steal one of their devices, or attempt to break into their home. Wait for the user to trigger the fall-back procedure.\\
\textbf{Impact K}& The user's total balance that was in Tier 2 is pushed to Tier 3. It is now less accessible for the user. \\
    \hline
    \end{tabular}
\end{table}

This temporary denial-of-service attack demonstrates another trade-off that arises when trying to design a custody system that uses the vault pattern. If it is easy to trigger a fall-back procedure, it will be more secure and safe but it will be more susceptible to denial-of-service attacks. This is a typical \textit{security-accessibility} trade-off.

\subsubsection{Device Replacements}

Once the fall-back is complete, the goals of the recovery are the same as those of the setup, except that the state required on each device may include operational data. The recovery procedures are different. The state of a new device can be `bootstrapped' by an old device, and subsequently verified with the remaining old devices. Consequently, if we assume that data on old devices was correct, then we can use these devices to verify correctness of data for new devices. This assumption is reasonable since those devices have been operating and would have failed otherwise. Following that, we can ensure that all devices (old and new) are consistent with each other. At the end, we rely on the system health check procedure.

The verification of correctness requires authenticating the device originating the data, demonstrating that it has not been tampered with while being transported across one or more channels. Then, we can say that the data was not tampered unless all of the originating entities have been compromised (either the device itself or with man-in-the-middle attacks). Refer to table \ref{tab:auth-summary} to recall how entities authenticate each other. In tables \ref{tab:mobile-replacement-data}, \ref{tab:hs-replacement-data}, \ref{tab:home-replacement-data} we summarise the data acquired via `bootstrapping'. The `confirmation with' column states which entities are relied on for verification. Only one of those entities needs to be honest to detect tampered data.

\begin{table}
    \centering
    \begin{tabular}{|p{2cm}|p{3cm}|p{3cm}|p{3cm}|p{3cm}|}
         \hline
         \rowcolor{lightgray} \textbf{Data} & \textbf{Source} & \textbf{Received in Procedure} & \textbf{Confirmation in Procedure} & \textbf{Confirmation with} \\ %\textbf{Consistency Set} \\
         \hline
         $\{A_n^m\}$ & old hardware signer & mobile replacement initialisation & system health check & old hardware signers  \\ %mobile, all hardware signers \\
         \hline
         $\{W^m_k\}$ & old hardware signer & mobile replacement initialisation & system health check & old hardware signers  \\ %mobile, all hardware signers, watchtower $k$ \\
         \hline
         $\{\text{}^*O_k\}$ & watchtower $k$ & watchtower re-initialisation & system health check & old hardware signers \\%& mobile, all hardware signers, watchtower $k$ \\
         \hline
         $w$ & user & mobile initialisation & mobile initialisation & user \\%mobile, all watchtowers \\
         \hline
         $c$ & user & mobile initialisation & mobile initialisation & user \\
         \hline
         $\{TX^{dep}\}$, $\{TX^{wit}\}$, $\{TX^{fb}\}$ & watchtowers & watchtower re-initialisation & system health check & all watchtowers (authenticated with $\{W^m_k\}$ from old hardware signers) \\
         \hline 
    \end{tabular}
    \caption{Summary of new mobile setup and operational state data received across channels. For each data type we show; the source of data, the procedure wherein the data is received, the procedure wherein the data is confirmed with trusted entities and which entities those are.}
    \label{tab:mobile-replacement-data}
\end{table}

\begin{table}
    \centering
    \begin{tabular}{|p{2cm}|p{3cm}|p{3cm}|p{3cm}|p{3cm}|}
         \hline
         \rowcolor{lightgray} \textbf{Data} & \textbf{Source} & \textbf{Received in Procedure} & \textbf{Confirmation in Procedure} & \textbf{Confirmation with} \\ %\textbf{Consistency Set} \\
         \hline
         $\{A_n^m\}$ & old hardware signer & hardware signer replacement initialisation & system health check & old hardware signers  \\
         \hline
         $\{W_k\}$ & old hardware signer & hardware signer replacement initialisation & system health check & old hardware signers  \\ 
         \hline
         $\{\text{}^*O_k\}$ & watchtower $k$ & hardware signer replacement initialisation & system health check & old hardware signers \\%& mobile, all hardware signers, watchtower $k$ \\
         \hline
         $c$ & user & hardware signer replacement initialisation & hardware signer replacement initialisation & user \\%& mobile, all watchtowers \\
         \hline
    \end{tabular}
    \caption{Summary of new hardware setup state data received across channels. For each data type we show; the source of data, the procedure wherein the data is received, the procedure wherein the data correctness is confirmed with trusted entities and which entities those are.}
    \label{tab:hs-replacement-data}
\end{table}

\begin{table}
    \centering
    \begin{tabular}{|p{2cm}|p{3cm}|p{3cm}|p{3cm}|p{3cm}|}
         \hline
         \rowcolor{lightgray} \textbf{Data} & \textbf{Source} & \textbf{Received in Procedure} & \textbf{Confirmation in Procedure} & \textbf{Confirmation with} \\
         \hline
         $\{TX^{dep}\}$, $\{TX^{wit}\}$, $\{TX^{fb}\}$ & mobile & hardware signer replacement initialisation & hardware signer replacement initialisation & watchtowers \\
         \hline
    \end{tabular}
    \caption{Summary of home signer operational state data received across channels. For each data type we show; the source of data, the procedure wherein the data is received, the procedure wherein the data is confirmed with trusted entities and which entities those are.}
    \label{tab:home-replacement-data}
\end{table}

\subsubsection{User's Device Generated Data}

In the signer account rotation procedure, the user authenticates them-self to the mobile and old hardware signers. The user can then verify that hardware signers are rotating to the correct new account \textit{i} and decommissioning the correct \textit{selection} of devices. 

The user's devices each receive the accounts of each of the others ($\{A^i_n\}$) from the mobile. These are verified in the system health check. 

\subsubsection{Service Providers' Device Generated Data}

Thanks to the `bootstrapping' of new devices from old devices, the recovery process removes the need to rely on a computer as a secondary means to verify data for interacting with the watchtowers. Instead, since the user's devices already contain the previous accounts for each watchtower $\{W^{i-1}_k\}$, these can be used to authenticate messages from the watchtower, including  $\{\text{}^*O_k\}$, $\{W^{i}_k\}$ and $\text{}^*P^{Noise}$. The home signer and mobile also receive $\{TX^{dep}\}$, $\{TX^{wit}\}$, $\{TX^{fb}\}$ signed by the watchtowers. Assuming that the initial setup resulted in correct and consistent data from service providers, and that the `bootstrapping' resulted in correct and consistent data across the user's new devices, then it follows that this aspect of the recovery is secure. Why? Because each of the user's devices can verify the authenticity of messages independently, and can verify the consistency of messages across all watchtowers.  

\subsubsection{Transition to New Setup}

With the `send to new fall-back' procedure, the user's bitcoin is sent to Tier 3. With the `re-establish Tier 2' procedure, bitcoin is sent in increments to Tier 1 (receive outputs) and then to Tier 2 (vault-deposits). This process ensures that the custody system retains strong security and safety for the majority of funds, and does not put the entire balance in the riskier Tier 1 at the same time.    

Although the `send to new fall-back' procedure follows on from a system health check, there is no guarantee that the user's devices remain honest. If the mobile is compromised, it can be used to deceive the user. 
\begin{table}[H]
    \centering
    \begin{tabular}{|>{\columncolor{lightgray}\raggedleft\arraybackslash}p{3.5cm}|p{12cm}|}
    \hline
\noindent \textbf{Vulnerability L}& Deceive the user about their available balance, $b$.\\
\textbf{Attack L}& Compromise the mobile. Omit transactions from the sets $\{TX^{fb}\}$ and $\{TX^{rec}\}$ from the message sent to hardware signers. \\
\textbf{Impact L}& \TX{new-fb} does not send the \textit{entire} balance to $fb^*$. The remaining balance is controlled by the previous set of accounts, which had been partially compromised or had partially failed.\\ 
\textbf{Countermeasure L}& User keeps track of their balance once their custody system is operational, for example, in a personal ledger of records. \\
    \hline
    \end{tabular}
\end{table}

Once again, we find that the user's view of the custody system state is a critical function which if inaccurate, creates vulnerabilities an attacker can exploit. In this case, the countermeasure is manual and simple.

The procedure has been designed so that the user will not be deceived to think that \TX{new-fb} was successful, when it was not. The idea is to use multiple independent channels to notify the user of the successful inclusion of \TX{new-fb} in the blockchain. The first channel is between the mobile and the user. The other channels are between the watchtowers and the user, authenticated by their hardware signers. The mobile, if compromised, can not forge signatures for the watchtowers. A compromised mobile can not tamper with messages signed by the watchtower. The hardware signers can authenticate messages from the watchtower despite the fact that they were forwarded from the mobile. Thus, an attacker would need to compromise the mobile and all watchtowers, or the mobile and all participating hardware signers in order to deceive the user about the success of \TX{new-fb}.  

The signers each independently construct \TX{new-fb} based on their setup data. This enforces \TX{new-fb} to have a correct output; paying the total amount to the new accounts with policy $l_{fb^*}$. The output key in \TX{new-fb} cannot be tampered with if at least one of the participating signers remains honest.  

The `re-establish Tier 2' procedure is almost identical to `send to new fall-back'. To summarise; the user knows if \TX{new-rec} has been included in a block and if the amount $x$ was correct if at least one of the participating signers is honest, and each honest signer enforces the correctness of $l_{rec^*}$. 

\subsection{Summary}

Our analysis demonstrates how the `ceremony analysis'  methodology can help to identify vulnerabilities in custody systems and build a model of risk. Typically, the security of Bitcoin custody systems is characterised by the Script that enforces access-control. However, when each procedure in the setup, operation and recovery are considered step-by-step, including human actions, it becomes clear that more detailed risk models for custody systems must be constructed. As shown here, each procedure is subject to vulnerabilities that are separate from the `Risk Tiers' encapsulated by Script-based access-controls. 

Our analysis is not comprehensive. It deals with only one threat model and one set of assumptions. Attacks are described briefly and could be expanded on in more detail. Countermeasures are described briefly and introducing them into the design creates new risks to add to the model. Our analysis is exploratory; it provided insights into the design process for custody systems and into the vault pattern. Our analysis sets a foundation for future work. 

By systematically constructing a risk model for Ajolote we found weaknesses inherent to the vault pattern as well as design failures that can be addressed in future work. First let us review the former.

The user-initiated fall-back procedure creates temporary denial-of-service attacks that are relatively simple to perform (for example, anonymously threaten the user). This would be frustrating for the user. 

The vault-pattern depends on data-availability (for covenant transactions, enforcement signatures, and other witness data) or else access to bitcoin will be lost. As a consequence, data must be redundantly stored and consistency must be verified across devices. In this, procedural burden is created for the user. Relying on watchtowers for data-availability alleviates part of the procedural burden but creates third-party risk.  
 
Verifying watchtower-generated data is less reliable than user-generated data. However, the user depends on these data for reliable movement of bitcoin form Tier 1 to Tier 2 (vault-deposit) and from Tier 2 to Tier 3 (reject and fall-back). This holds despite the fact that watchtowers have no ability to permanently block access to or steal bitcoin. Though movement from Tier 2 to Tier 3 is streamlined and tolerant to compromise and failure, one can not assume that Tier 2 is as secure as Tier 3.

Now let us review the latter.

The user's live view of the custody system typically depends on only the mobile, or on a combination of the mobile and a hardware signer with one honest watchtower. The user's view of the custody system is critical to ensure that procedures are completed successfully. In particular, deceiving the user can result in severe attacks on the vault-deposit, withdrawal, and fall-back procedures. Strengthening the user's view of the custody system requires creating additional independent communication channels between the user and the Bitcoin network. How to do this without significantly degrading usability remains an open question. An example might be to use the mobile only as the coordinator (not a general signer) and have another hardware signer device that the user carries with them at all times, with their mobile. If setup correctly, this device could verify watchtower messages and communicate directly with the user, with less risk of being compromised remotely than the mobile has. 

\section{Consensus Enforced Covenants}
\label{sec:Consensus-covenants-Ajolote}

It is worth exploring how consensus enforced covenants can solve issues for vault-based custody systems. In this section we will discuss whether or not three separate proposals can fundamentally improve Ajolote by solving some of its issues and better achieving some of its objectives. The proposals we're interested in here are; {\tt OP\_CHECKTEMPLATEVERIFY} (henceforth {\tt CTV}), {\tt ANYPREVOUTANYSCRIPT} (henceforth {\tt APOAS}), and {\tt INSPECT\_X}. These were discussed in the previous chapter in sections \ref{subsec:related_covenants} and \ref{sec:covenant-comparison}. 

\subsection{Two-step Vault-deposit}

Recall that in Ajolote, to avoid an interactive protocol with their recipient, the user receives payments as receive outputs. Then, they enact a vault-deposit, typically using their mobile and home signers. By design, the hardware signer stores Ajolote policies and verified extended public keys so that it can verify if vault-deposits are correct before they are irrevocably committed to. The signer is also used to enforce the covenant (by generating an enforcement signature and deleting the associated private key). The \textit{verification} and \textit{enforcement} functions are separate, and must be distinguished to gain clarity on the benefits of consensus-enforced covenants in vault-based custody systems. 

With {\tt CTV and APOAS}, lock Scripts commit to covenant transaction templates. These templates must be specified in advance, including output Scripts in the covenant transactions. In Ajolote, these Scripts contain public keys controlled by various hardware signers. Rather than rely on the mobile signer alone to generate the template and specify the public keys, the user should verify these public keys and Script policies with a second air-gapped device such as the home signer. Consequently, the \textit{verification} function still requires action on behalf of the user, and the vault-deposit procedure should still be completed in two steps (first receive, then deposit). Address generation attacks prevent a non-interactive, one-step vault-deposit. 

With {\tt INSPECT\_X}, covenant enforcement Scripts commit to specific constraints on fields of the covenant transaction. The constraints are granular, and can, for example, partially constrain an output Script in the covenant transaction. This means that the precise public keys do not need to be specified in advance, while the format of the policy remains fixed. Can we make use of this to safely defer \textit{verification} to the future, and enable a one-step, non-interactive vault-deposit? Unfortunately, no. By not committing to the public keys in the enforcement Script, the constraints are meaningless. Any public keys can be inserted to satisfy the policy. The inserted keys are not physically constrained to user's hardware signers kept in secured locations. If an attacker can compromise the custodial Script in order to authorise the covenant transaction, they must adhere to the enforced policy format but can insert public keys where they control the associated private keys. Thus, verification of the enforcement Script must happen during the vault-deposit and the same two-step process is required even with this granular covenant mechanism. 

\subsection{Enforcement Security}

Enforcement for deleted-key covenants depends on the process of key deletion. Recall attack \textbf{D} from section \ref{subsec:ops-risk-vault-deposit} which describes how the enforcement can be broken by recovering deleted keys. With consensus-enforced covenants, this attack would not be possible. Instead, breaking the covenant would require re-writing the blockchain from the latest block to the block that includes \TX{dep}. Otherwise, the commitment in the output script of \TX{dep} is fixed. Re-writing the blockchain is a very expensive attack and becomes exponentially more expensive with every new block that is produced. Thus, consensus-enforced covenants are arguably much harder to break than deleted-key covenants. 

\subsection{Verifying Covenant Enforcement}

Recall from section \ref{sec:covenant-comparison} in the previous chapter that consensus-enforced covenants are \textit{verifiable} after the covenant has been activated. The verification requires a proof that the public key script commits to covenant constraints. The proof requires the covenant transaction, commitment values (signatures or hash values), the TapScript and control block. To construct the proof, one must show that the taproot output key, which is observable as an unspent-output in the blockchain, can be derived from these data. This can be useful for system health checks, or for auditing purposes. With deleted-key covenants, there is no way to verify that the enforcement private keys were deleted.

\subsection{Covenant Template Management}

The vault pattern depends on data-availability among the user's devices. The data includes covenant transactions and witness data. A safe custody system using the vault pattern needs sufficient redundancy of these data. A secure custody system using the vault pattern needs to restrict access to these data from potential attackers. This is independent of the covenant mechanism that is used. The management of covenant transaction templates and enforcement signatures is an extra burden that is inherent to the custody systems that use the vault pattern. 

\subsection{Coin Control}

As noted in section \ref{subsec:eval-ops-dynamics}, coin control issues arise in Ajolote as a consequence of committing to complete transaction templates for \TX{wit} and \TX{fb}, and because \TX{fb} is dependent on an immutable \TX{wit}. While {\tt CTV} does not offer any new functionality that may solve these issues, we will show that {\tt APOAS} can yield minor improvements to the design and {\tt INSPECT\_X} can plausibly yield major improvements. 

First, recall the basic model, summarised by the sequence of output types created by Ajolote transactions,
\begin{equation}
\begin{split}
      r_s \rightarrow v_s \rightarrow u_s & \rightarrow (Ect, c_s) \\
      & \rightarrow fb_s
\end{split}
\end{equation}
Each \TX{wit} consumes a single $v_s$ and creates an $u_s$. If the user intends to withdraw an amount more than what is contained in $v_s$, they must consume another vault output, $v_t$ and broadcast several \TX{wit}. The user cannot withdraw an amount less than what is contained in $v_s$. In standard custody systems (with no covenants), Bitcoin transactions are constructed at the time they will be used and they may consume or create several outputs at once. This grants the user flexibility. If we utilise the features of {\tt APOAS} and {\tt INSPECT\_X}, we may commit to partial transaction templates and enable \TX{wit} to be mutable without invalidating \TX{fb}. Let us consider how these features can improve coin control. 

Recall from table \ref{tab:class_comparison} that {\tt APOAS} covenants can be enforced with various signature message types that define what transaction fields the signature message includes. We will not consider {\tt APOAS:NONE} because it does not commit to an output and thus cannot enforce constraints on the output. We will not consider {\tt APOAS:ALL} because this commits to all outputs (including their amounts and scripts) and thus would not allow useful modifications to the covenant transactions (such as adding a change output). {\tt APOAS:ALL} enforcement signatures provide essentially the same functionality as deleted-key covenants whose enforcement signature types are {\tt ALL}. 

Let us consider the {\tt APOAS:SINGLE} enforcement signature type. The signature covers the input that is being signed and the output at the same index. It allows new inputs or outputs to be added. The signature will enforce that \TX{wit} contains a $v_s$ input and $u_s$ output at the same index. Multiple $(v_s, u_s)$ pairs can be added to \TX{wit} at the time the user wants to make a withdrawal.

What about the dependent \TX{fb}? The enforcement signature for $u_s$ as an input in \TX{fb} is valid for any previous output that can be satisfied with the same witness. It will enforce that \TX{fb} contains $u_s$ as an input and $fb_s$ as an output. The $u_s$ could be an output in any transaction. This way, modifying \TX{wit} at withdrawal time by aggregating $(v_s, u_s)$ pairs will not invalidate the enforcement signatures in \TX{fb}. \TX{fb} can be modified to include a $(u_s, fb_s)$ pair for each $(v_s, u_s)$ pair in \TX{wit}. 

This would reduce the number of kilobytes used per withdrawal and rejected withdrawal. Operational costs would be less for the user. So, we see that {\tt APOAS} can create this minor improvement. Recall that modifications of covenant transactions can only be authorised by entities that can satisfy the custodial subscript. For \TX{wit}, that means the two of the user's signers can be used to authorise modifications. For \TX{fb}, that means any of the user's signers, or any of the watchtowers can authorise modifications. This minor efficiency improvement creates potential pinning attack vectors from watchtowers, or from an attacker that has compromised one of the user's signers. A more thorough analysis is required to determine vulnerabilities created by this feature. 

Unfortunately, the constraints that can be enforced by {\tt APOAS} signature message types do not enable the flexibility required to significantly improve coin control issues. For that, more granular covenant constraints are necessary. 

With {\tt INSPECT\_X}, we can inspect transaction fields within the execution of a TapScript. We can make a covenant constraint such as ``check that the input amounts sum to $x$ and the output amounts sum to $x-f$" for some pre-defined fee amount $f$. We can make a constraint such as ``check that the output at index $1$ has a public key script equal to $o$". With these in mind, let us consider how to improve on coin control with the vault-pattern.

What we would like is to have a \TX{wit} with inputs $(v_1,v_2,...,v_y)$ and outputs $(u_t,v_t)$, where the user can define the amount of the unvault output at the time of withdrawal, and re-direct excess to a new vault output. Written informally, the following constraints (committed to in the vault output policy) should be enforced;

\begin{itemize}
    \item[a)] $v_1.amount + v_2.amount +...+ v_y.amount = u_t.amount + v_t.amount + y\cdot f$ 
    \item[b)] output at index $0$ has public key script $o_u$ (which encapsulates the unvault output policy)
    \item[c)] output at index $1$ has public key script $o_v$ (which encapsulates the vault output policy) 
    \item[d)] number of outputs $=2$
    \item[e)] all inputs have public key script $o_v$ (which encapsulates the vault output policy)
\end{itemize}

With this, any number of vault outputs, $y$, can be selected at the time of withdrawal. The total amount must be used to create an unvault output at index 0 and a vault output at index 1, and to pay a pre-defined fee $f$ for each of the $y$ inputs. The user must authorise \TX{wit} with two of their signers. The modification of \TX{wit} is protected by its custodial script $l_{wit}$. 

This design is much more practical for coin control because it allows the user to dynamically determine how much they want to withdraw from vault deposits without being restricted to pre-defined amounts. Moreover, excess amounts are deposited directly into a new vault output, rather than a change output. How about \TX{fb}? The following constraints (committed to in the unvault output policy) should be enforced;

\begin{itemize}
    \item[a)] output at index $0$ has public key script $o_{fb}$ (which encapsulates the fall-back policy)
    \item[b)] number of outputs $=1$
    \item[c)] $u_1.amount + u_2.amount + ... + u_z.amount = fb_t.amount + z\cdot f$
    \item[d)] all inputs have the public key script $o_u$ (which encapsulates the unvault output policy)
\end{itemize}

With this, any number of unvault outputs, $z$, can be selected at the time of rejection. The total amount must be used to create a fall-back output at index 0 and to pay a pre-defined fee $f$ for each of the $z$ inputs. The modification of \TX{fb} is protected by its custodial script, $l_{pay-to-fb}$, which any of the user's devices or watchtowers can satisfy.

Consider this as a plausibility argument for how granular covenants can improve custody systems that use the vault pattern. This is not exhaustive. A thorough specification and risk analysis is needed to make a conclusive statement. 

\section{Conclusion}

This chapter introduced a new Bitcoin custody system, Ajolote, based on covenants that enforce the vault pattern. The specification presented herein captures human-device interaction. We demonstrate that this is critically important because often, human-interaction can be a powerful tool and sometimes can be the source of vulnerabilities. We constructed a threat model, a specification of attacker capabilities, that accounts for the differences between humans, networked devices, and offline devices and the ways in which they communicate. We considered the impact associated with each data type in the system being lost or stolen. Together, our understanding of the attacker and the system assets were integrated in a model of risk. This is a novel application of `ceremony analysis' to custody in Bitcoin. 

From our analysis we offer an insight relevant for Bitcoin custody in general. Despite having a distributed key-management system (across hardware signers in different locations), the system may only be as secure as the user's accurate view of their custody system state. If the user solely relies on their transaction coordinator for information about their state, they can be deceived in multiple ways. The user can be deceived about having made a transaction, about having received a transaction, or can be deceived that their payment was sent to their intended counter-party when in reality it was sent to an attacker.   

It is critical to consider the entire life-cycle of keys, from setup, to operation, to rotation to new keys. To this end, we specify and analyse how private keys are generated, how key-trees are used, how public keys are shared, and how keys are rotated. Our analysis also goes beyond a simplistic view of key-management. Custody risk cannot be described purely from a key-management perspective. Resilience to loss and resistance to attack must be considered for all setup state data, all operational data, as well as physical environments that store hardware.   

We have explored in detail the advantages and disadvantages of the vault pattern in custody. On one hand, it enables an access control layer (Tier 2) that is more safe and secure than an equivalent, standard multi-signature access control layer (Tier 1) because it enables a swift movement of funds into a more strict access control layer (Tier 3). This comes at the cost of accessibility. At the same time, it offers more accessibility than the strict access control layer, but with a broader attack surface. The vault pattern offers a new balance on the security-accessibility trade-off. On the other hand, the vault pattern creates operational complexity which can be difficult for the user to understand and cumbersome to use. Also, the vault pattern creates privacy risk based on repeating transaction sequences.  

With Ajolote we have demonstrated a realistic use case for deleted-key covenants. It is interesting to note that the key-deletion process created significantly less risk (according to our model) than the general difficulty for the user to maintain an accurate view of the state of their custody system. If this latter risk can be better addressed, it is plausible that deleted-key covenants will be suitable for realistic custody systems for individual users. This is compatible with the current version of Bitcoin, and does not depend on a soft-fork upgrade to the consensus rules. 
 
Finally, we evaluated how different consensus-enforced covenants could improve vault-based custody. The coin control issues observed in Ajolote cannot be helped with {\tt CTV} or {\tt APOAS}. However, we provided a plausibility argument that granular covenants such as those enabled with {\tt INSPECT\_X} can solve the coin control issues and make vault-based custody much more practical. How? By enabling the user to adjust how much bitcoin to withdraw from a vault and re-direct excess bitcoin into a new vault. This could reduce the system complexity significantly because managing the size of vaults and distributing bitcoin across several vault transactions is no longer necessary.   
   
While this chapter presents a significant move towards a viable vault-based custody system for Bitcoin, there is much work to do to improve on the design and analysis and to create an implementation. Design variations may include an integrated inheritance scheme, a fiduciary co-signer, a fiduciary recovery assistant, self-managed watchtowers, or a fifth hardware signer (so that the mobile acts only as a coordinator). Analysis must be a part of an ongoing cyclic process. Alternative methods can be used to attempt to enumerate risks and design countermeasures to address them. Regarding implementation, building Ajolote will require experts in hardware construction, operating systems, full-stack software engineering, and user experience design.
\chapter{Conclusion}
\label{ch:conclusions}

In this chapter we conclude the thesis by summarising the main contributions, evaluating the achievement of the objectives stated in section \ref{sec:thesis-objectives}, and discussing directions for future work.

\section{Contributions}
\label{sec:conclusion-contributions}
The contributions of this thesis are as follows:

\begin{itemize}
    \item[\textbf{C1}] An overview of the design space for Bitcoin custody and approaches to modelling risk with a review of academic literature, product documentation, and articles from the industry (in chapter \ref{ch:custody}).
    \item[\textbf{C2}] The first detailed study on the deleted-key covenant mechanism and its security model (in chapter \ref{ch:bitcoin-covenants}).
    \item[\textbf{C3}] A comparative analysis of the deleted-key covenant mechansim with three alternative covenant mechanisms proposed in other work (in chapter \ref{ch:bitcoin-covenants}).
    \item[\textbf{C4}] A detailed specification for a custody system called \textit{Ajolote} which uses deleted-key covenants to implement the vault pattern (in chapter \ref{ch:vault-custody}).
    \item[\textbf{C5}] An evaluation of the state dynamics and privacy of Ajolote (in chapter \ref{ch:vault-custody}).
    \item[\textbf{C6}] A realistic threat model for arbitrary custody systems involving humans, offline hardware devices, and networked devices (in section \ref{subsec:ax-threat-model}).
    \item[\textbf{C7}] A risk model for Ajolote using a ceremony analysis methodology and the realistic threat model, covering the setup, operation, and recovery phases (in chapter \ref{ch:vault-custody}).
    \item[\textbf{C8}] Analysis to differentiate how alternative covenant mechanisms support Bitcoin custody using the vault pattern (in chapter \ref{ch:vault-custody}).
    \item[\textbf{C9}] A risk framework for Bitcoin custody operation with the Revault protocol. Validation of the attack-tree methodology and publication of attack-trees (in Appendix \ref{ch:risk-framework-revault}).
\end{itemize}

\section{Achievement of Objectives}

We re-state the objectives from section \ref{sec:thesis-objectives} here.

\begin{itemize}
    \item[O1] Establish categorical definitions within the domain of Bitcoin custody. Evaluate existing technology, documentation, standards and analyses. Write an informed overview to bring coherence to this emerging field of study.
    \item[O2] Design and analyze a novel covenant protocol that works with state of the art Bitcoin features. Provide a comparative analysis of alternative covenant constructions. 
    \item [O3] Design and analyse a novel, self-managed custody system for an individual.
\end{itemize}

Objective \textbf{O1} was broadly achieved  with the literature, documentation, and industry product review in chapter \ref{ch:custody}. Therein, we set the premise for the remainder of the thesis by introducing a selection of categories, definitions, and design principles from the industry and literature about Bitcoin custody. We considered the technology stack, particularly focusing on key-management and privacy. We reviewed the leading open-source custody systems and emerging standards for custody. We summarised methods for constructing and using risk models, and showed how various methods have been used to-date for Bitcoin custody systems and their components. 

In chapter \ref{ch:bitcoin-covenants} we presented novel research on Bitcoin covenants (restrictions on transaction sequences), contributing to \textbf{O2}. Covenants are a technological mechanism often claimed to enable enhanced custody systems, although no detailed study had yet been presented to verify that claim. We presented the first detailed protocol specification for the deleted-key covenant mechanism. We analysed it against a set of security objectives. We described several factors that application designers should consider when using covenants, including; composability, the class of possible covenants, fee allocation strategies, safety concerns, and requirements for a proof-of-reserves protocol. Towards the end of the chapter, in section \ref{sec:covenant-comparison}, we compared deleted-key covenants with three other proposals of covenant mechanisms which require a soft-fork to the consensus rules of Bitcoin. We offer new insights on their advantages and disadvantages, thereby achieving objective \textbf{O2}.

Chapter \ref{ch:vault-custody} addresses questions raised in chapter \ref{ch:bitcoin-covenants} about the value of covenants in Bitcoin custody, contributing further to \textbf{O2}. In particular, the \textit{vault pattern} (see section \ref{subsec:cov-pattern}) enables an access control layer with intermediate accessibility, safety and security; between the standard multi-signature access control layers with equivalent key-sets. It affords a new expression of an inherent security-accessibility trade-off. 

We offered a detailed specification of a novel custody system, Ajolote, in chapter \ref{ch:vault-custody}, contributing to objective \textbf{O3}. The setup, operation, and recovery phases were detailed and the human-interaction was made explicit. We reviewed and applied the ceremony analysis methodology and put forth a realistic threat model suitable for custody systems. Notably, the threat model regards the differences between humans, networked devices, and offline devices, and the ways in which they communicate. Ajolote was a newly conceived case-study and we evaluated its operational features with a simple model of its state dynamics. We constructed a risk model for the system, and revealed several vulnerabilities that may emanate from deception of the user. This class of attack is generally applicable to custody systems, and should be considered by designers when specifying the operational processes.  Altogether, chapter \ref{ch:vault-custody} satisfied objective \textbf{O3}.

In Appendix \ref{ch:risk-framework-revault} we exhibited a risk framework for the Revault custody system as a set of attack-trees. We put forth requirements for our attack model, and concluded that the attack-tree formalism was suitable. These attack-trees included vulnerabilities to exploit for both the technology and operational processes underpinning Revault. By using the model, organisations may quantify risks in their specific context for deploying Revault. Crucially, the model is modular and extensible, and can be adapted as the threat landscape evolves. 

\section{Future work}

In addition to the individual contributions in section \ref{sec:conclusion-contributions}, this thesis can function as an instructive pedagogical resource for custody system designers and analysts. This thesis is the first body of work to synthesise concepts and methods from the widely dispersed academic literature, industry articles, risk-management methodologies, and technological practice into a cohesive whole that illuminates a path forward for research in Bitcoin custody, and which commences down this path. \textit{Why is this path important?} To expand the economic function of Bitcoin as a store of value (and consequently, as a medium of exchange). \textit{What does this path look like?} A continuous effort to build open-source systems and collaborate on the construction of associated risk models. This effort would require a community of designers, engineers, and industry players with incentives aligned around improving the state-of-the-art. Transparency, modularity, extensibility, and a common theoretical basis each align with the philosophy of open security \cite{OpenSecurity} and enable \textit{Evolving Bitcoin Custody}. 

\subsection{Custody Systems}

Bitcoin custody systems require several advancements in user experience to mitigate the risks associated with user deception. This can involve new mental models for users to better understand the state of their system, and new representation formats that are legible to inexperienced users. This can involve well-designed user-manuals and guides for risk-management, such as those provided by Allen and Appelcline \cite{SmartCustodyBook,SmartCustody-multisig}. 

Bitcoin hardware devices require further development and an expansion of functionality in order to assist users in maintaining resilient custody systems. In addition to the points about user experience raised above, this includes better compatibility across devices from different manufacturers to support custody systems with a diversity of hardware devices and to mitigate the risk of compromise in the supply-chain, of the operating system, and of applications. Similarly, client software for mobile and desktop with broad support across hardware device types is critical to enabling more resilient custody systems. 

As noted in section \ref{subsec:custody-access-control}, threshold signature schemes that support distributed key generation and distributed signing have seen significant development in recent years, for both Schnorr signatures and ECDSA signatures. These schemes can enable more privacy for users of custody systems. These schemes can enable key-rotation (sometimes referred to as `key-refreshment') procedures without initiating transactions. Some of these schemes are in use commercially, and are thus being tested in reality. We expect to see more progress and innovation. For example, the types of access control structure that are supported can develop beyond quorum authorisation and eventually include timed-signatures \cite{timelocked-signatures}. Also, these schemes could become non-interactive, and support for them could be added to leading hardware signer devices. 

Methods and tools to enhance the usability and security of the setup phase of custody systems are critical to develop. As we have seen with Ajolote (recall section \ref{sec:setup-procedures}), when several hardware signers are in use, the process to verify the correctness and consistency across custody system components can be arduous. 

\subsection{Risk Models}

Regarding risk models, any contribution towards better understanding of vulnerabilities, attacks, and countermeasures would be valuable. Ideally, the research community will generate a public library of risk models for various custody systems using several methods, as described in section \ref{sec:custody-modelling-risk}. They may integrate feedback from experienced security analysts and a variety of perspectives. This will improve not only the confidence of users in systems, but also the ability for insurance underwriters to assess and to price risk. In turn, this will facilitate the growth of the emerging economy using Bitcoin.

The threat model presented in section \ref{subsec:ax-threat-model} is a promising baseline for risk analysis with Bitcoin custody systems because it captures realistic attacker capabilities. Further investigation of and with the threat model is justified. As exemplified in chapter \ref{ch:vault-custody}, a custody system should be considered in its entirety, from setup, to operation and recovery. In particular, we suggest moving towards constructing a formal specification of the threat model and custody procedures \textit{in each phase}, and attempting to formally verify relevant properties. Formal verification of security ceremonies (which include human interaction) has been investigated in recent work \cite{Formal-ceremonies,Xmen-ceremony}. There is little work in the existing literature which specify and analyse the setup and recovery phases of custody systems. 

\subsection{Vault-Based Custody and Bitcoin Covenants}

Until a consensus-enforced covenant mechanism is enabled for Bitcoin, hardware signing device manufacturers may consider implementing secure deletion functions to support deleted-key covenants. That would contribute towards developing vault-based custody systems like Ajolote as well as other types of application.

There are questions about which covenant-based applications are valuable and worth enabling or improving. There are questions about which covenant mechanisms are best suited for these applications. Based on the contributions of this thesis, we emphasise that the application of vault-based custody systems stand to benefit most from \textit{granular} covenant mechanisms, since these can plausibly improve the coin control issues observed with Ajolote. We expect this to be a fruitful area of research. 

In that direction, researchers must consider systemic constraints such as maintaining consensus and mitigating network-wide denial-of-service attacks. In search of answers, researchers can pursue formal modeling and verification of Script and its modifications to answer definitively if new operations are safe or not, as exemplified in other work \cite{CovenantsUnchained,btc-spendable}. Moreover, developers can test live deployments on other blockchains or side-chains to get practical experience learning from real attacks and user feedback before attempting to change the consensus rules of Bitcoin. This is a complex topic and any upgrade is likely to be stalled until sufficient consensus is realised among Bitcoin developers about a viable path forward. 

\appendix

\chapter{Risk Framework for Bitcoin Custody Operation with Revault Protocol}

\label{ch:risk-framework-revault}

\section{Connection with Previous Chapters}

In chapter \ref{ch:vault-custody}, the risk analysis of Ajolote was particularly focused on the specification of procedures and the system's tolerance to compromise and failure. While several vulnerabilities were enumerated, the attacks that could exploit them were described only briefly. As a complementary methodology, we demonstrate herein how constructing a formal model of attacks can deepen the analysis of a custody system and afford a clearer understanding of the cost, difficulty, and ultimately the likelihood of attack. 

This appendix explores a similar custody system called Revault \cite{practical-revault,revault-pdf}. It has been implemented and tested. The Revault system works in a multi-party setting and still requires managing pre-signed transactions, but it does not use deleted-key covenants. Instead, the transactions can only be re-written with the agreement (or compromise) of all the stakeholders' signing devices. 

Revault uses the vault pattern in two slightly different ways. In the first, funds can be immediately re-directed (`re-vaulted') to the active key-set. In the second, funds can be immediately pushed to an emergency output type. With Revault, the receive policy is an $N-$of$-N$ multi-signature Script among the active key-set and this is what makes the `re-vault' procedure viable. In Ajolote, a `re-vault' mechanism would only be as secure as a $2-$of$-3$ access structure. 

Coin control is still a primary issue, but the ability to recreate pre-signed transactions presents an opportunity to overcome it. Moreover, since the operational burden for the custody system is distributed across users, the per-user burden may be less than with Ajolote. Note that Revault does not (yet) use taproot output types because it was implemented before they were added to Bitcoin.

Note that some of the terminology in this chapter may differ slightly from the rest of the thesis. Herein a `custody system' is called a `custody protocol'. The terms `self-custody' and `third-party custody' are used in place of `autonomous' and `heteronomous', respectively. `Wallet' is used here to describe a user's (local) transaction coordinator and Bitcoin network server, while the `Coordinator' is a server that mediates communication among users. 

\section{Summary}

Our contributions with this chapter are twofold. First, we elucidate the methodological requirements for a risk framework of custodial operations and argue for the value of this type of risk model as complementary with cryptographic and blockchain security models. Second, we present a risk model in the form of a library of attack-trees for Revault -- an open-source custody protocol. The model can be used by organisations as a risk quantification framework for a thorough security analysis in their specific deployment context. Our work exemplifies an approach that can be used independent of which custody protocol is being considered, including complex protocols with multiple stakeholders and active defence infrastructure.

\section{Introduction}

While mainstream acceptance of Bitcoin as an asset appears to be increasing, advanced tools and methods for secure custody of bitcoins are slow to develop. Bitcoin custody encompasses the protection of assets through software, hardware, and operational processes. The foundation of Bitcoin custody is key-management, a well understood topic in the academic literature and in practice. However, Bitcoin custody, in particular multi-stakeholder custody, involves human processes, communication protocols, network monitoring and response systems, software, hardware and physical security environments. Given a secure cryptographic layer, there are still vulnerabilities introduced at the application layer by software developers, at the hardware layer throughout the supply chain, and at the operations layer by users. Without adequate risk management frameworks for custodial operations, Bitcoin users are likely to suffer unexpected losses whether they self-custody funds or employ a third-party custodian.  

Open-source custody protocols are emerging \cite{practical-revault,Glacier,Subzero,Swambo2020vault} and are a critical ecosystem component for improving security standards. If a custody protocol stands to public scrutiny and offers a high-level of security without relying on proprietary processes, users, insurance companies and regulators can have more confidence in it. The emerging custody protocols are trying to reconcile the needs of traditional businesses and banking  with Bitcoin's novel identity-less and irreversible transaction properties. A lack of available and accepted open-source custody protocols means that organisations are heavily relying on third-party custodians, or deploying their own custody protocol.

We propose an attack modelling technique as the basis for a risk framework for Bitcoin custody operations, using the Revault protocol\footnote{Specifically, the version identified as 609b40dda07155abe5cd4a5af77fc2211d11fbc1 which can be found on the open-source repository hosted on Github \cite{practical-revault}.} as a case-study \cite{revault-pdf,practical-revault}. While the process of model construction is intensive, the resultant framework is extensible and modular and some of its components can be re-used with different custody protocols. It is intended to be readily comprehensible, and, given sufficient validation, the framework can be used by any organisation intending to deploy Revault to better understand their risk posture. 

Risk quantification frameworks address several ecosystem problems. Organisations that control bitcoins or other digital assets need accurate models to engage in realistic risk-management.
The complexity of custodial risks leaves insurance companies guessing rather than systematically estimating when pricing their insurance offerings or assessing particular solutions for digital custody. 
Finally, emerging regulatory standards for custody \cite{DACS,CryptoassetCustody} are simple and fail to capture advanced custody architectures or enable context-specific risk analyses that acknowledge the full security environment of a custody operation.    

The remainder of this chapter is structured as follows.
Section \ref{sec:RevaultOverview} summarises the components and processes of the Revault protocol. Section \ref{sec:Methodology} discusses our evaluation criteria for an operational risk framework, and introduces the attack-tree formalism on which our risk model is based. Section \ref{sec:Threat Model} presents our operational risk model for Revault. Section \ref{sec:Conclusion} concludes this chapter. 

\section{Overview of Revault Custody Protocol}
\label{sec:RevaultOverview}

Revault is a multi-party custody protocol that distinguishes between \textit{stakeholders} and fund \textit{managers}.
The primary protection for funds is a high-threshold multi-signature Script controlled by the stakeholders. 
The day-to-day operational overhead of fund management is simplified by enabling portions of funds to be delegated to fund managers.  Stakeholders define spending policies in-line with traditional controls of expenses, and have automated servers to enforce their policies. In addition, a deterrent is withheld by each stakeholder to mitigate incentives to physically threaten the stakeholders. To achieve this, Revault makes use of sets of pre-signed transactions coupled with an active defence mechanism for detecting and responding to attempted theft transactions. In the following, we will outline the components of the Revault architecture, the transaction set,
 the stakeholders' routine signing process and the managers' spend process. Refer to \cite{practical-revault} for the detailed specification of the open-source protocol.

\subsection{Revault Architecture Components}
\label{subsec:Components}

Each stakeholder and manager has a \textit{hardware security module} (HSM) to manage their private keys and generate signatures for transactions. A backup of private keys is stored for each HSM in a separate protected physical environment. 

Each stakeholder and manager uses a \textit{wallet} software to track their co-owned bitcoins, craft transactions, store transaction signatures and communicate with each other through a \textit{coordinator}. The coordinator is a proxy server that simplifies communication for the multi-party wallet. All communication uses Noise KK encrypted and authenticated channels \cite{NoiseProtocolFramework}.

Stakeholders each have one or more \textit{watchtower}, an online server that enforces the stakeholder's spending policy limitations. 
Stakeholders each have an \textit{anti-replay oracle} server.

\subsection{Revault Transaction Set}
\label{sec:TransactionSet}

The use of hierarchical deterministic wallets means that each participant in the Revault protocol has a tree of public and private keys \cite{BIP32}. To discuss ownership of bitcoins, we refer to a generalisation of a locking Script, called a \textit{descriptor}. The wallet will have multiple addresses that correspond to a single abstracted descriptor. Funds are deposited into the multi-party wallet through a Deposit transaction (Tx) output that pays to the deposit descriptor, describing $N-$signatories locking Scripts derived from the stakeholders' (\textit{stk}) extended public keys (\textit{xpub}). In descriptors language formalisation \cite{ScriptDescriptors} it is defined  as: \begin{equation}
    thresh(N,\text{ } stk\_1\_xpub,\text{ } stk\_2\_xpub,\text{ } ...,\text{ } stk\_N\_xpub)
\end{equation}

\begin{figure}[t]
\centering
    \includegraphics[width=0.8\linewidth]{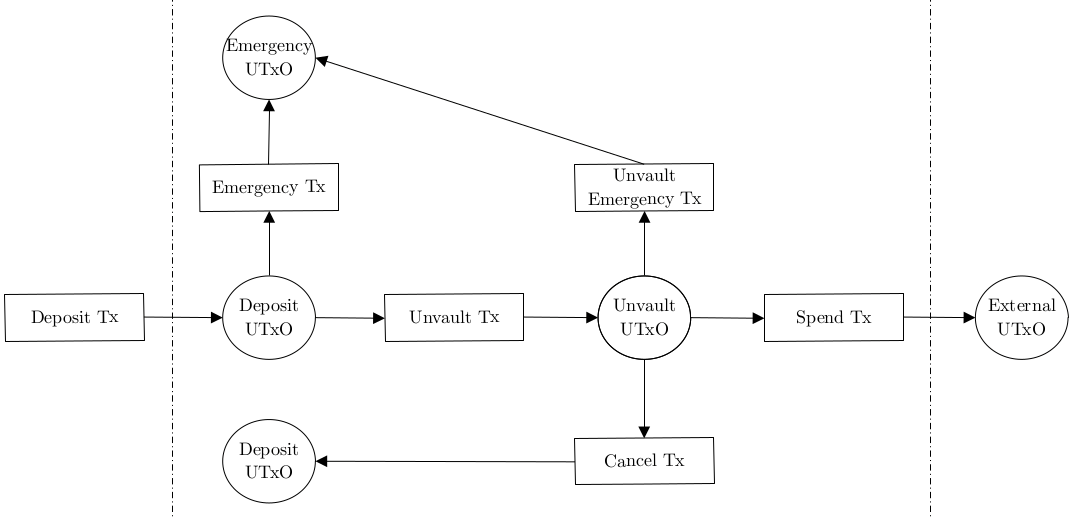}
    \caption{Diagram of the transaction (Tx) set structure in the Revault protocol. An Unspent-transaction-output (UTxO) is created by a preceding Tx and is consumed by an input in a proceeding Tx.}
    \label{fig:tx_diagram}
\end{figure}

The set of transactions prepared with stakeholders' wallets and signed using their hardware security modules (HSM) include the Emergency Tx, Unvault Tx, Unvault-Emergency Tx and Cancel Tx. The managers can only prepare and sign a Spend Tx type. Figure \ref{fig:tx_diagram} depicts these transactions and the essential unspent-transaction-outputs (UTxOs) they create or consume.

An Emergency Tx locks funds to an emergency descriptor which is unspecified by the Revault protocol and is kept private among stakeholders. The descriptor must however be harder to unlock than the deposit descriptor. This is the deterrent for physical threats to the stakeholders. 

An Unvault Tx consumes the deposit UTxO and creates an unvault UTxO locked to the unvault descriptor, 
\begin{equation}
    \begin{split}
    &or(  thresh(N, stk\_1\_xpub, stk\_2\_xpub, ..., stk\_N\_xpub), \\
    &\text{ }\text{ }\text{ }\text{ }and( thresh(K, man\_1\_xpub, ..., man\_M\_xpub), \\
    &\text{ }\text{ }\text{ }\text{ }\text{ }\text{ }\text{ }\text{ }\text{ }\text{ }and(  thresh(N, oracle\_1\_xpub, ..., oracle\_N\_xpub),\\
    &\text{ }\text{ }\text{ }\text{ }\text{ }\text{ }\text{ }\text{ }\text{ }\text{ }\text{ }\text{ }\text{ }\text{ }\text{ }\text{ }older(X) ) ) ),\\
    \end{split}
\end{equation}

that is redeemable by either the $N$ stakeholders \textit{or} the $M$ managers (\textit{man}) along with $N$ automated anti-replay oracles after $X$ blocks.

A Cancel Tx consumes the unvault UTxO and creates a new deposit UTxO. The watchtowers' role is to broadcast the Cancel Tx if a fraudulent spend attempt is detected (either through an unauthorised attempt at broadcasting an Unvault Tx or if a Spend Tx does not abide by the spending policy).  The time-lock gives watchtowers $X$ blocks worth of time to broadcast a Cancel Tx. 
An Unvault-Emergency Tx consumes the unvault UTxO and locks funds to the emergency descriptor. It has the same purpose as the Emergency Tx, only it consumes the unvault UTxO rather than the deposit UTxO.
A Spend Tx is used by managers to pay to external addresses.

\subsection{Stakeholders' Signing Routine}

\begin{figure}[t]
\centering
    \includegraphics[width=0.6\linewidth]{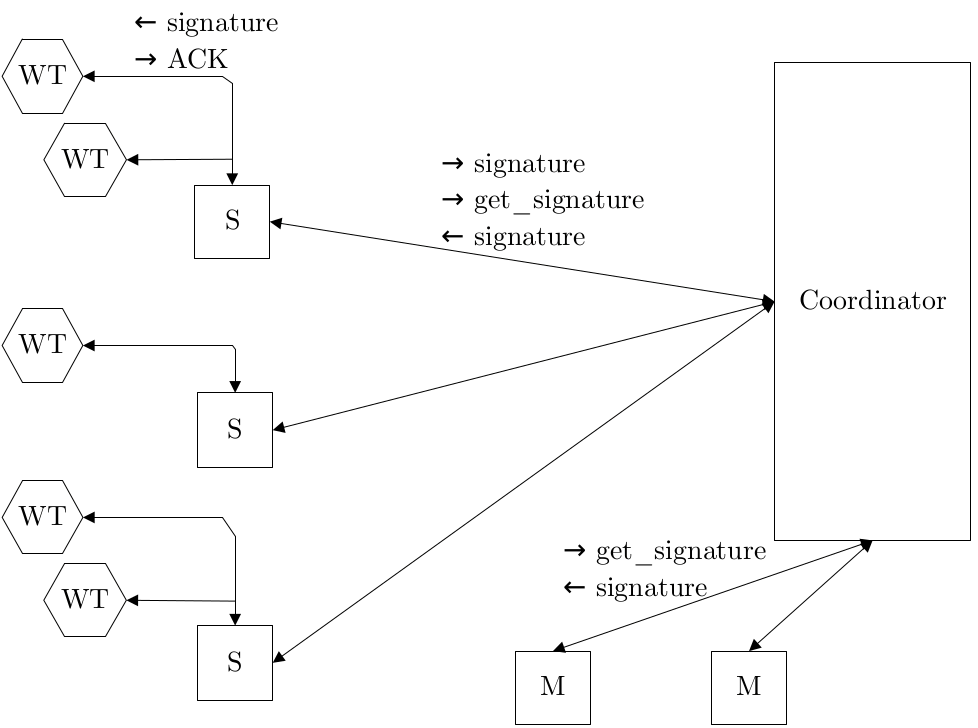}
    \caption{Data flow diagram for the communication of the \textit{stakeholders' signing routine} with an example Revault deployment. There are three stakeholders (S) who each have one or two watchtowers (WT). There are two managers (M) and a coordinator. Signature messages, signature requests and watchtower acknowledgements (ACK) are only shown once per connection type but apply to each connection of that type (e.g. there is \{$\leftarrow$ signature, $\rightarrow$ ACK\} between each WT and S).}
    \label{fig:sig_exchange_diagram}
\end{figure}

Stakeholders' wallets routinely check for new deposits 
%UTxOs locked to their jointly controlled deposit descriptor. 
and each one triggers a signing routine. Figure \ref{fig:sig_exchange_diagram} shows the connections and message types for an example Revault deployment enacting the signing routine. The wallet crafts an Emergency Tx and requests the stakeholder to sign it using their HSM. The stakeholder will verify the emergency descriptor on the HSM before authorising the signature generation\footnote{This feature is not available with current HSMs, but integrating compatibility with descriptors (along with other security features) would improve the human-verification component of HSM security and is being discussed on the \textit{bitcoin-dev} mailing list \cite{AdvancedHMfeatures}.}. 
The wallet then connects to the coordinator to push its signature and will fetch other stakeholders' signatures.

Optionally, stakeholders may also sign the Cancel, Unvault-Emergency, and Unvault Txs to securely delegate funds to the managers. In this case the signing process is the same but is carried out in two steps: first, the signatures for the Cancel and Unvault-Emergency Txs are exchanged with the other stakeholders through the coordinator and then shared with the watchtower(s), and only then are the Unvault Tx signatures shared with managers.

\subsection{Managers' Spending Process}

Most spending policies cannot be inferred from the Unvault Tx alone and so the Spend Tx must be known to the watchtower to validate an unvaulting attempt. In these cases the Spend Tx must be advertised to the watchtowers before unvaulting, otherwise it will be cancelled. The anti-replay oracle is required to avoid the Spend Tx being modified by the managers \textit{after} the unvault time-lock expires and thus by-passing enforcement of the watchtowers' spending policies.

\begin{figure}[t]
\centering
    \includegraphics[width=0.6\linewidth]{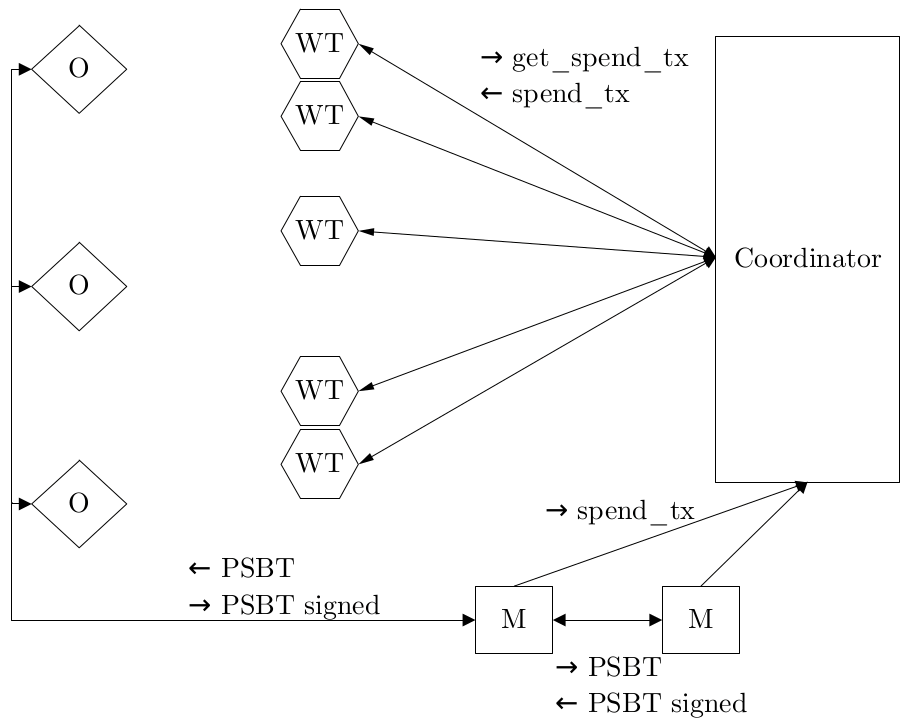}
    \caption{Data flow diagram for the communication of the managers' spend process. In this example there are two managers (M), three anti-replay oracles (O), five watchtowers (WT) and a coordinator. A Partially-signed Bitcoin Tx (PSBT) is exchanged among managers and between a manager and the anti-replay oracles. A fully signed SpendTx is shared with the WTs through the coordinator.}
    \label{fig:spend_tx_exchange}
\end{figure}

Any manager can initiate a spend. Figure \ref{fig:spend_tx_exchange} depicts the spend process. The initiator creates a Spend Tx, verifies and signs it using their HSM and passes it back to the wallet in the partially-signed Bitcoin Tx (PSBT) format \cite{BIP174}. It's  exchanged with a sufficient threshold of the other managers to add their signatures and hand it back to the initiator. The initiator requests a signature from each of the anti-replay oracles and pushes the fully-signed Spend Tx to the coordinator.  
The initiator broadcasts the Unvault Tx, triggering a lookup from the watchtowers to the coordinator for the Spend Tx. If the Spend Tx is valid according to \textit{all} of the watchtowers policies and none of them cancel this unvaulting attempt, the manager waits $X$ blocks and broadcasts the Spend Tx. 

If, during the unvaulting process, there's a significant increase in the fee-level required for a Spend Tx to be mined, a manager needs to bump the fee. Managers use a dedicated single-party fee wallet for this. Similarly, watchtowers use a fee wallet in the case there is high demand for block space to bump the fee for Cancel or emergency Txs. 

\section{Methodology}
\label{sec:Methodology}

To see where this research fits in to the big picture, consider the key life-cycle of a custodial operation. There are three phases; initialization, operation, and termination. Initialization is where wallet and communication keys are generated, where software integrity is verified, hardware security modules are checked, and relevant public information is shared among participants. Operation encompasses the active fund management. Termination is the phase wherein the wallet is de-commissioned and all sensitive information destroyed. Initialization and termination are out of scope for this chapter. Our risk model covers the operations phase. In the following we present our rationale for our chosen attack modeling formalism and explain how this can be used as a risk framework.

\subsection{Operational Security Models}

A framework for high-level risk analyses for the integration of custody into a multi-stakeholder context has not yet been presented. To-date the literature has focused primarily on cryptographic security modeling, dealing with low-level risks associated with cryptographic primitives, key-management protocols, HSMs and single-party wallets. The underlying cryptographic security is fundamental but should be complemented by an operational security model, which is much more likely to be the domain where participants create vulnerabilities for an attacker. Advanced custody protocols that use multi-layer access control with both static and active defences for insider and external attackers demand a whole-system approach to security analysis.

We present now several requirements for our modelling formalism: a) the ability to represent complex processes with numerous components and sequential events; b) supports qualitative risk analysis; c) supports automated quantitative methods for multi-attribute risk analysis; d) readily comprehensible and visual models that are more amenable to open-source intelligence; and e) extensible and modular models to support differential analysis and re-use of modules.  

The two most popular attack modeling techniques in cyber-security literature are attack-trees and attack graphs \cite{AGraphATreeReview}. In short, tools for attack graphs tend to produce graphs that aren't readily comprehensible due to the complexity of real-world attack scenarios \cite{SurveyAttackModeling}. That is, attack graphs don't scale well \cite{AssetCentricAnalysis}. On the other hand, attack-trees seem to meet all of our requirements, at least when considered with the right structure and semantics (as described in section \ref{sec:Attack-treeFormalism}) and thus we construct our risk model using this formalism.

While a statement such as `our custody solution is based on an $m-$of$-n$ security model' can entail a lot for simple multi-signature custody protocols, it doesn't capture the reality nearly as well as our proposed methodology would. It is certainly not sufficient for a more complex custody protocol like Revault. What is the physical environment for those $n$ private keys? Are any of those keys online? Are there key backups and, if so, what protections are in place for these? Too much depends on the broader security environment of a custody protocol for it to be left without scrutiny.   

Application threat modelling has been used to harden the Revault protocol throughout both its theoretical development and implementation. For each application process (spend, routine signing, emergency, revault) a component-by-component and connection-by-connection analysis has been carried out to determine the consequences of outages, data tampering, component corruption, etc., and has resulted in the design specification \cite{practical-revault} and the transaction flow threat model \cite{revault-pdf}. The application threat modeling approach is complementary and has informed us in enumerating the risks presented with the attack-trees. However, in contrast to attack-trees, it lacks a semantic structure which is amenable for automated risk quantification and thus isn't suitable as the basis of a risk framework.

\subsection{Attack-Tree Formalism}
\label{sec:Attack-treeFormalism}

The risk model is presented using the formalism of attack-trees \cite{AttackTrees,ThreatLogicTrees,ThreatTrees}. Attack-trees have an attack at their root, and branches that capture alternate (OR) and complementary (AND) attack pathways comprised of intermediate attack goals as non-leaf nodes and basic attack steps as leaf nodes. As in numerous other works
\cite{ADTool2,DBLP:journals/corr/JhawarKMRT15,AFTrees,10.1093/cybsec/tyaa020,Duqu2,ADTreesSAND}, we extend the basic attack-tree to support  sequential conjunction of branches (SAND) allowing us to model an attack where some sub-tree of an attack pathway has to occur before and in addition to another sub-tree. For brevity we depict our attack-trees as nested lists. The logical gates (OR, AND, SAND) shown with each node apply to the next node at the same depth. This means that at any given depth, a node with a SAND gate occurs \textit{before} other nodes that are shown below it. Some aspects of the system are built to be resilient to attack and failure through redundancy. For example, an attacker needs to compromise all stakeholders' private keys to steal funds locked to the deposit descriptor. To be concise, rather than having several copies of the same sub-tree we write ($X$ times) to note that the sub-tree has to happen $X$ times. During an analysis, these sub-trees should be considered as $X$ separate AND sub-trees, since they are contextually different (corresponding to different participants, remote and physical environments).

We provide a set of attack-trees, capturing prominent risks that have been enumerated primarily by considering tangible and intangible assets. Tangible assets (bitcoins) are distinguished by the access control structures determined by the set of descriptors. We consider operational privacy and business continuity as intangible assets. 

Our work here is focused on security, rather than safety. In principle, the same methodology could be extended to an integrated security-safety model by constructing attack-fault-trees \cite{AFTrees}. Another common extension to the attack-tree formalism is to include countermeasures, producing attack-defence-trees \cite{FoundationsADTrees,ADTreeValue}. The benefit of our modular modelling technique is that it enables future work to integrate these extensions and re-use results from this work. Hence, we prioritise constructing a strong foundation based only on attacks, and aim to incrementally improve on the model presented as new intelligence emerges.

\subsection{On Risk Analysis}

Our purpose in constructing the risk model presented in section \ref{sec:Threat Model} is to provide a framework to support both qualitative and quantitative risk analyses for specific deployment instances of Revault custody. By determining costs, likelihoods, and other attributes for risks associated with custodial processes, an organisation can perform a differential analysis of countermeasures until their risk-tolerance is satisfied. An explicit framework not only helps an organisation deploying Revault with risk-management but could form a standard by which insurance companies and regulators consider specific deployments. As with any model of complex reality, attack-trees are imperfect and cannot capture every possible attack pathway, but the alternative---complete ignorance---is not better. 

To perform a context-specific risk analysis, a set of estimates are made (using in-house empirical data, public research, and expert opinion) for each basic attack step on different attributes such as monetary cost, execution time, or likelihood. With that, a bottom-up procedure (from leaf nodes to the root) is used to compute aggregated attributes. Bayesian methods can be used to update prior estimates with more refined values as new data sources emerge. The process for generating estimates is critical and should be considered with care. In-depth research-based practical guidance on this topic is given by D. W. Hubbard and R. Seiersen in \cite{Hubbard}.
Given specific contextual information, estimations can be improved by further decomposing basic attack steps (e.g. `steal keys backup') into multiple steps (e.g.  `bribe manager to determine backup location' SAND `break into safe'). If a basic attack step has a highly uncertain estimate, then further decomposition into more explicit steps can be beneficial. On the other hand, decomposing into quantities that are more speculative than the first could compound uncertainty rather than reducing it. 

Various methods for analysis can be used to compute aggregated attributes for attack-trees. Kordy \textit{et al.} gave an overview \cite{KordyReview}. Our purpose here is to provide the framework on which to perform analyses rather than to provide a specific analysis. We have not performed a comprehensive evaluation of analysis methods, but offer some suggestions based on a comparison in \cite{KumarThesis}. Two methods that support evaluating the attributes of cost, probability, and time are stochastic-model checking \cite{AFTrees} and game-theoretic analysis \cite{ADTreeValue}. Whichever methods are used must appropriately capture the constraints of our model (including SAND gates) and should be automated to enable rapid attribute-based queries for security metrics such as; the expected attack pay-off for the most likely attack,  or the possible attack pathways given a budget of \$10,000.   

Our approach to constructing the risk model is centered on assets since these are clearly distinguished through Bitcoin descriptors, as continuity of a custodial process, or as operational privacy. However, when performing the risk analysis it can be insightful to consider attacker personas \cite{Shostack}:  a crime syndicate; an opportunistic burglar; a nation state; a business competitor; or even an insider. If the organisation understands any of these personas well (arguably they should especially understand their competitors and employees) they can reduce the uncertainty in their aggregate risk estimates for these scenarios. Attacker-profiles are a useful way to prune attack-trees \cite{KumarThesis}.

\section{Risk Model}
\label{sec:Threat Model}

We have constructed the risk model with several assumptions that limit the scope of the analysis to the operational aspects of custody. Known risks from other protocol and environment dependencies that are discussed in other works should be considered as complementary but are, for the purpose of clarity, assumed to be benign here. First, we assume that the Bitcoin network is functional, realising its live-ness and availability properties \cite{Garay2015,Garay2017,Pass2017,Badertscher2017,Badertscher2018}. We assume that there is significant hash-rate to prevent blockchain reorganisations of a depth higher than the Unvault Tx's relative lock-time. Next, we assume that Revault's Tx model is robust; with scripts that realise the access control structures we expect, without unintended consequences from Tx malleability and network propagation issues as described in \cite{revault-pdf}. We assume the initialization process was secure and safe; private keys and  backups were correctly and confidentially constructed for each participant, software and hardware integrity were verified, relevant public key information for both the wallet and communication was shared among participants leading to a correct configuration for the wallet clients, watchtowers, anti-replay oracles and the coordinator. We assume that Revault's communication security model as described in \cite{practical-revault} is robust. That is, where messages need to be authenticated or confidential, they are. We assume that the software development life-cycle of Revault is secure, such that any deployment is using an implementation that adheres to the protocol specification. Finally, we assume that entities constructing Deposit Txs don't succumb to a man-in-the-middle attack. That is, they lock funds to the deposit descriptor rather than to an attacker's address.

\subsection{Common Attack Sub-Trees}

These attack sub-trees are common to different attacks on Revault, and \textbf{a}, \textbf{b}, \textbf{c}, \textbf{d}, \textbf{e}, \textbf{f} and \textbf{g} are likely to be common to attacks on other custody protocols. 

{\footnotesize
\begin{itemize}[noitemsep,parsep=0pt,partopsep=0pt, leftmargin=0.7cm]
\item[\textbf{a} :] \textbf{Compromise a participant (stakeholder or manager)}
\begin{itemize}[noitemsep,topsep=0pt,parsep=0pt,partopsep=0pt, leftmargin=0.8cm]
\item[1 :] Coerce participant (OR)
\item[2 :] Corrupt participant
\end{itemize}
\end{itemize}
}

\noindent Coercion and insider threats from corrupt participants must be considered. Legal defences for malicious employee behaviour can be effective deterrents here. 

{\footnotesize
\begin{itemize}[noitemsep,parsep=0pt,partopsep=0pt, leftmargin=0.7cm]
\item[\textbf{b} :] \textbf{Compromise a participant's (stakeholder's or manager's) HSM }
\begin{itemize}[noitemsep,topsep=0pt,parsep=0pt,partopsep=0pt, leftmargin=0.8cm]
\item[1 :] Physical attack of HSM (OR)
\begin{itemize}[noitemsep,topsep=0pt,parsep=0pt,partopsep=0pt, leftmargin=0.9cm]
\item[\textit{1.1} :] Determine location of participant’s HSM (SAND)
\item[\textit{1.2} :] Access the physical security environment of the participant’s HSM (SAND)
\item[\textit{1.3} :] Exfiltrate keys (either on premise or after stealing it) (OR)
\item[\textit{1.4} :] By-pass PIN and make the HSM sign a malicious chosen message
\end{itemize}
\item[2 :] Remote attack of HSM (OR)
\begin{itemize}[noitemsep,topsep=0pt,parsep=0pt,partopsep=0pt, leftmargin=0.9cm]
\item[\textit{2.1} :] Compromise a device that is then connected to the HSM (SAND)
\begin{itemize}[noitemsep,topsep=0pt,parsep=0pt,partopsep=0pt, leftmargin=1cm]
\item[\textit{2.1.1} :] (see \textbf{g}) Compromise the participant's wallet software (OR)
\item[\textit{2.1.2} :] Trick participant into connecting their HSM to a compromised device via social engineering
\end{itemize}
\item[\textit{2.2} :] Exploit a firmware vulnerability (OR)
\item[\textit{2.3} :] Trick participant into compromising their own HSM with the user interface of the compromised device
\end{itemize}
\item[3 :](see \textbf{a})  Compromise a participant
\end{itemize}
\end{itemize}
}

{\footnotesize
\begin{itemize}[noitemsep,parsep=0pt,partopsep=0pt, leftmargin=0.7cm]
\item[\textbf{c} :] \textbf{Compromise a participant’s (stakeholder's or manager's) keys backup}
\begin{itemize}[noitemsep,topsep=0pt,parsep=0pt,partopsep=0pt, leftmargin=0.9cm]
\item[1 :] Physical Attack (OR)
\begin{itemize}[noitemsep,topsep=0pt,parsep=0pt,partopsep=0pt, leftmargin=0.9cm]
\item[\textit{1.1} :] Determine location of the keys backup (SAND)
\begin{itemize}[noitemsep,topsep=0pt,parsep=0pt,partopsep=0pt, leftmargin=1cm]
\item[\textit{1.1.1} :] Watch the participant between the custody initialization and the start of operations (OR)
\item[\textit{1.1.2} :] Watch the participant during a backup check (OR)
\end{itemize}
\item[\textit{1.2} :] Access the physical security environment of the keys backup (SAND)
\item[\textit{1.3} :] Depending on backup format, steal or copy it
\end{itemize}
\item[2 :] (see \textbf{a}) Compromise a participant
\end{itemize}
\end{itemize}
}

{\footnotesize
\begin{itemize}[noitemsep,parsep=0pt,partopsep=0pt, leftmargin=0.7cm]
\item[\textbf{d} :] \textbf{Compromise a server (watchtower, anti-replay oracle, or coordinator)}
\begin{itemize}[noitemsep,topsep=0pt,parsep=0pt,partopsep=0pt, leftmargin=0.8cm]
\item[1 :] Remote attack (OR)
\begin{itemize}[noitemsep,topsep=0pt,parsep=0pt,partopsep=0pt, leftmargin=0.9cm]
\item[\textit{1.1} :] Exploit a software vulnerability (OR)
\begin{itemize}[noitemsep,topsep=0pt,parsep=0pt,partopsep=0pt, leftmargin=0.9cm]
\item[\textit{1.1.1} :] Determine the public interfaces of the server (SAND)
\item[\textit{1.1.2} :] Exploit a vulnerability on one of the softwares listening on these interfaces
\end{itemize}
\item[\textit{1.2} :] Exploit a human vulnerability (e.g. trick participant into performing a malicious update)
\end{itemize}
\item[2 :] Physical attack (OR)
\begin{itemize}[noitemsep,topsep=0pt,parsep=0pt,partopsep=0pt, leftmargin=0.9cm]
\item[\textit{2.1} :] Determine server's location (SAND)
\item[\textit{2.2} :] Access the physical security environment of the server (SAND)
\item[\textit{2.3} :] Compromise the server (e.g. inject malware)
\end{itemize}
\item[3 :] (see \textbf{a}) Compromise the participant managing the server
\end{itemize}
\end{itemize}
}

\noindent An attacker who successfully completes \textbf{d} for a watchtower will be able to steal funds from the watchtower's fee wallet and will be able to force an emergency scenario by broadcasting all Emergency and Unvault-Emergency Txs it has stored. They can also prevent broadcast of a Cancel Tx from this watchtower either passively (\textit{ACK} the secure storage of the signature to the stakeholder, but then drop the signature) or actively.

{\footnotesize
\begin{itemize}[noitemsep,parsep=0pt,partopsep=0pt, leftmargin=0.7cm]
\item[\textbf{e} :] \textbf{Shutdown a watchtower}
\begin{itemize}[noitemsep,topsep=0pt,parsep=0pt,partopsep=0pt, leftmargin=0.8cm]
\item[1 :] Physical attack on the watchtower (OR)
\begin{itemize}[noitemsep,topsep=0pt,parsep=0pt,partopsep=0pt, leftmargin=0.9cm]
\item[\textit{1.1} :] Determine watchtower's location (SAND)
\item[\textit{1.2} :] Sever the internet connection to the building in which the watchtower is located (OR)
\item[\textit{1.3} :] Sever the power-line connection to the building in which the watchtower is located (OR)
\item[\textit{1.4} :] Access the physical security of the watchtower and un-plug the machine
\end{itemize}
\item[2 :] Remote attack on the watchtower
\begin{itemize}[noitemsep,topsep=0pt,parsep=0pt,partopsep=0pt, leftmargin=0.9cm]
\item[\textit{2.1} :] Determine public interfaces of watchtower (SAND)
\item[\textit{2.2} :] Denial of Service attack through one of the public interfaces (OR)
\item[\textit{2.3} :] Eclipse attack on the watchtower's Bitcoin node \cite{EclipseAttack} (OR)
\begin{itemize}[noitemsep,topsep=0pt,parsep=0pt,partopsep=0pt, leftmargin=1cm]
\item[\textit{2.3.1} :] Slowly force de-synchronisation of watchtower with the true block height by delaying block propagation \cite{TimeDilationAttack} (OR)
\item[\textit{2.3.2} :] Prevent outgoing propagation of Cancel or Emergency Txs
\end{itemize}
\item[\textit{2.4} :] Denial of Service attack on the fee-bumping UTxOs pool---not enough funds to pay competitive fees
\end{itemize}
\end{itemize}
\end{itemize}
}

{\footnotesize
\begin{itemize}[noitemsep,parsep=0pt,partopsep=0pt, leftmargin=0.7cm]
\item[\textbf{f} :] \textbf{Get signature from participant to unlock UTxO for Theft Tx}
\begin{itemize}[noitemsep,topsep=0pt,parsep=0pt,partopsep=0pt, leftmargin=0.8cm]
\item[1 :] (see \textbf{a}) Compromise a participant (OR)
\item[2 :] (see \textbf{b}) Compromise participant's HSM (OR)
\item[3 :] (see \textbf{c}) Compromise participant's keys backup
\end{itemize}
\end{itemize}
}

{\footnotesize
\begin{itemize}[noitemsep,parsep=0pt,partopsep=0pt, leftmargin=0.7cm]
\item[\textbf{g} :] \textbf{Compromise a participant's wallet}
\begin{itemize}[noitemsep,topsep=0pt,parsep=0pt,partopsep=0pt, leftmargin=0.8cm]
\item[1 :] Physical attack (OR)
\begin{itemize}[noitemsep,topsep=0pt,parsep=0pt,partopsep=0pt, leftmargin=0.9cm]
\item[\textit{1.1} :] Locate participant's device (SAND)
\item[\textit{1.2} :] Access physical security environment of participant's device
\end{itemize}
\item[2 :] Remote attack (OR)
\begin{itemize}[noitemsep,topsep=0pt,parsep=0pt,partopsep=0pt, leftmargin=0.9cm]
\item[\textit{2.1} :] Determine public interfaces of device (SAND)
\item[\textit{2.2} :] Exploit a vulnerability
\end{itemize}
\item[3 :] (see \textbf{a}) Compromise participant
\end{itemize}
\end{itemize}
}

\noindent Participant's wallet devices are expected to be used for day-to-day activities. With many vulnerabilities to exploit, the likelihood of success for \textbf{g} is high.

{\footnotesize
\begin{itemize}[noitemsep,parsep=0pt,partopsep=0pt, leftmargin=0.7cm]
\item[\textbf{h} :] \textbf{Determine the locking Script for a deposit or unvault UTxO (\textit{Witness Script})}
\begin{itemize}[noitemsep,topsep=0pt,parsep=0pt,partopsep=0pt, leftmargin=0.8cm]
\item[1 :] (see \textbf{g}) Compromise any participant's wallet (OR)
\item[2 :] (see \textbf{d}) Compromise a watchtower (OR)
\item[3 :] (see \textbf{d}) Compromise an anti-replay oracle
\end{itemize}
\end{itemize}
}

\noindent Deposit and unvault descriptors are deterministic, but public keys are needed to derive UTxO locking Scripts. These are stored by all wallets, watchtowers and anti-replay oracles.

{\footnotesize
\begin{itemize}[noitemsep,parsep=0pt,partopsep=0pt, leftmargin=0.7cm]
\item[\textbf{i} :] \textbf{Satisfy an input in a Theft Tx that consumes an identified deposit UTxO or unvault UTxO (through $N-$of$-N$)}
\begin{itemize}[noitemsep,topsep=0pt,parsep=0pt,partopsep=0pt, leftmargin=0.8cm]
\item[1 :] (see \textbf{h}) Determine the UTxO locking Script (\textit{Witness Script}) (SAND)
\item[2 :] Prevent the relevant Emergency Tx from being broadcast until the Theft Tx is confirmed (where $A + B = N$) (AND)
\begin{itemize}[noitemsep,topsep=0pt,parsep=0pt,partopsep=0pt, leftmargin=0.9cm]
\item[\textit{2.1} :] (see \textbf{d}) Compromise a watchtower (A times)
\item[\textit{2.2} :] (see \textbf{e}) Shutdown a watchtower (B times)
\item[\textit{2.3} :] (see \textbf{g}) Compromise stakeholder's wallet ($N$ times)
\end{itemize}
\item[3 :] (see \textbf{f}) Get signature from a stakeholder to unlock UTxO for Theft Tx ($N$ times)
\end{itemize}
\end{itemize}
}

{\footnotesize
\begin{itemize}[noitemsep,parsep=0pt,partopsep=0pt, leftmargin=0.7cm]
\item[\textbf{j} :] \textbf{Satisfy an input in a Theft Tx that consumes an identified unvault UTxO (through $K-$of$-M$, anti-replay oracles and time-lock)}
\begin{itemize}[noitemsep,topsep=0pt,parsep=0pt,partopsep=0pt, leftmargin=0.8cm]
\item[1 :] (see \textbf{h}) Determine the UTxO locking Script (\textit{Witness Script}) (SAND)
\item[2 :] Receive signatures for Theft Tx from all $N$ anti-replay oracles (AND)
\begin{itemize}[noitemsep,topsep=0pt,parsep=0pt,partopsep=0pt, leftmargin=0.9cm]
\item[\textit{2.1} :] Compromise a manager’s private communication keys and the set of anti-replay oracles’ public communication keys (OR)
\begin{itemize}[noitemsep,topsep=0pt,parsep=0pt,partopsep=0pt, leftmargin=1cm]
\item[\textit{2.1.1} :] (see \textbf{g}) Compromise a manager’s wallet (OR)
\item[\textit{2.1.2} :] (see \textbf{a}) Compromise a manager
\end{itemize}
\item[\textit{2.2} :] (see \textbf{d}) Compromise the anti-replay oracle
\end{itemize}
\item[3 :] (see \textbf{f}) Get signature from a manager to unlock UTxO for Theft Tx ($K$ times)
\end{itemize}
\end{itemize}
}

{\footnotesize
\begin{itemize}[noitemsep,parsep=0pt,partopsep=0pt, leftmargin=0.7cm]
\item[\textbf{k} :] \textbf{Satisfy an input in a Theft Tx that consumes an identified emergency UTxO}
\begin{itemize}[noitemsep,topsep=0pt,parsep=0pt,partopsep=0pt, leftmargin=0.8cm]
\item[1 :] Determine the emergency descriptor policy (SAND)
\item[2 :] Satisfy the emergency descriptor's locking conditions (may include waiting for time-locks, giving sufficient signatures, giving hash pre-images, \textit{etc}.)
\end{itemize}
\end{itemize}
}

\noindent The details of the emergency descriptor are intentionally not specified with the Revault protocol, except that it is more difficult to access than the deposit descriptor. Stakeholders may compartmentalise and distribute the descriptor information to afford its privacy some resilience to attack.

\subsection{Attack-Trees}
\label{subsec:attck-trees}

The following attack-trees are the foundation for an operational risk framework for Revault.

{\footnotesize
\begin{itemize}[noitemsep,parsep=0pt,partopsep=0pt, leftmargin=0.7cm]
\item[\textbf{A} :] \textbf{Compromise privacy of the custody operation (determine the set of public UTxOs)}
\begin{itemize}[noitemsep,topsep=0pt,parsep=0pt,partopsep=0pt, leftmargin=0.8cm]
\item[1 :] (see \textbf{d}) Compromise any of the servers (OR)
\item[2 :] (see \textbf{a}) Compromise a participant (OR)
\item[3 :] (see \textbf{g}) Compromise a participant's wallet (OR)
\item[4 :] Traffic analysis of connections between servers and/or wallets (OR)
\item[5 :] Blockchain analysis
\end{itemize}
\end{itemize}
}

\noindent Without privacy support for advanced descriptors (such as by using MuSig2 \cite{MuSig2} or MuSig-DN \cite{MuSig-DN} if the proposed Taproot \cite{BIP-Taproot} upgrade is activated by the Bitcoin network) Revault's operational privacy is brittle. 

{\footnotesize
\begin{itemize}[noitemsep,parsep=0pt,partopsep=0pt, leftmargin=0.7cm]
\item[\textbf{B} :] \textbf{Broadcast Theft Tx(s) that consume all deposit UTxOs}
\begin{itemize}[noitemsep,topsep=0pt,parsep=0pt,partopsep=0pt, leftmargin=0.8cm]
\item[1 :] (see \textbf{A}) Determine $\mathcal{D}$, the set of deposit UTxOs (SAND)
\item[2 :] (see \textbf{h}) Determine the locking Script for deposit UTxO ($|\mathcal{D}|$ times)
\item[3 :] (see \textbf{i}) Satisfy an input in a Theft Tx that consumes an identified deposit UTxO ($|\mathcal{D}|$ times) 
\end{itemize}
\end{itemize}
}

\noindent A Theft Tx that consumes all available deposit UTxOs would be catastrophic since this comprises the majority of funds. We recommend a defence wherein each stakeholder is equipped with a panic button that is directly connected to their watchtower or dedicated emergency service. When triggered, all the signed Emergency and Unvault-Emergency Txs are broadcast, negating the pay-off for an attacker and thus acting as a deterrent. 

{\footnotesize
\begin{itemize}[noitemsep,parsep=0pt,partopsep=0pt, leftmargin=0.7cm]
\item[\textbf{C} :] \textbf{Broadcast Theft Tx(s) that consume as many available unvault UTxOs as watchtower spending policies permit}
\begin{itemize}[noitemsep,topsep=0pt,parsep=0pt,partopsep=0pt, leftmargin=0.8cm]
\item[1 :] Determine spending constraints of all watchtowers' policies (SAND)
\begin{itemize}[noitemsep,topsep=0pt,parsep=0pt,partopsep=0pt, leftmargin=0.9cm]
\item[\textit{1.1} :] (see \textbf{a}) Compromise a participant (OR)
\item[\textit{1.2} :] (see \textbf{g}) Compromise a manager's wallet 
\item[\textit{1.3} :] (see \textbf{d}) Compromise a watchtower ($N$ times)
\end{itemize}
\item[2 :] Determine $\mathcal{U}$, the set of available unvault UTxOs (SAND)
\begin{itemize}[noitemsep,topsep=0pt,parsep=0pt,partopsep=0pt, leftmargin=0.9cm]
\item[\textit{2.1} :] (see \textbf{A}) Compromise privacy of the custody operation (determine the set of public UTxOs) (SAND)
\item[\textit{2.2} :] (see \textbf{h}) Determine the locking Script for unvault UTxO ($|\mathcal{U}|$ times)
\end{itemize}
\item[3 :] (see \textbf{i} OR \textbf{j}) Satisfy  an  input  in  a  Theft  Tx  that  consumes  an  identified  unvault UTxO ($|\mathcal{U}|$ times)
\end{itemize}
\end{itemize}
}

\noindent \textbf{C} can be avoided if watchtowers have a white-list of addresses that Spend Txs can pay to.

{\footnotesize
\begin{itemize}[noitemsep,parsep=0pt,partopsep=0pt, leftmargin=0.7cm]
\item[\textbf{D} :] \textbf{Broadcast Theft Tx(s) that consume all available unvault UTxOs, by-passing watchtowers' spending policies}
\begin{itemize}[noitemsep,topsep=0pt,parsep=0pt,partopsep=0pt, leftmargin=0.8cm]
\item[1 :] Prevent watchtower from broadcasting Cancel or Unvault-Emergency Txs before Theft Tx is confirmed ($N$ times SAND)
\begin{itemize}[noitemsep,topsep=0pt,parsep=0pt,partopsep=0pt, leftmargin=0.9cm]
\item[\textit{1.1} :] (see \textbf{d}) Compromise a watchtower (OR)
\item[\textit{1.2} :] (see \textbf{e}) Shutdown a watchtower
\end{itemize}
\item[2 :] Determine $\mathcal{U}$, the set of available unvault UTxOs (SAND)
\begin{itemize}[noitemsep,topsep=0pt,parsep=0pt,partopsep=0pt, leftmargin=0.9cm]
\item[\textit{2.1} :] (see \textbf{A}) Compromise privacy of the custody operation (determine the set of public UTxOs) (SAND)
\item[\textit{2.2} :] (see \textbf{h}) Determine the locking Script for unvault UTxO ($|\mathcal{U}|$ times)
\end{itemize}
\item[3 :] (see \textbf{i} OR \textbf{j}) Satisfy  an  input  in  a  Theft  Tx  that  consumes  an  identified  unvault UTxO ($|\mathcal{U}|$ times)
\end{itemize}
\end{itemize}
}

{\footnotesize
\begin{itemize}[noitemsep,parsep=0pt,partopsep=0pt, leftmargin=0.7cm]
\item[\textbf{E} :] \textbf{Broadcast a Theft Tx that by-passes watchtowers' spending policies}
\begin{itemize}[noitemsep,topsep=0pt,parsep=0pt,partopsep=0pt, leftmargin=0.8cm]
\item[1 :] Determine $\mathcal{U}$, the set of available unvault UTxOs (SAND)
\begin{itemize}[noitemsep,topsep=0pt,parsep=0pt,partopsep=0pt, leftmargin=0.9cm]
\item[\textit{1.1} :] (see \textbf{A}) Compromise privacy of the custody operation (determine the set of public UTxOs) (SAND)
\item[\textit{1.2} :] (see \textbf{h}) Determine the locking Script for unvault UTxO ($|\mathcal{U}|$ times)
\end{itemize}
\item[2 :] (see \textbf{f}) Get signature from a manager to unlock $U \subseteq \mathcal{U}$, a subset of available unvault UTxOs for a valid Spend Tx ($K$ times)
\item[3 :] (see \textbf{i} OR \textbf{j}) Satisfy  an  input  in  a  Theft  Tx  that  consumes  an  identified  unvault UTxO ($|U|$ times)
\item[4 :] (see \textbf{d}) Compromise an anti-replay oracle to get a signature for the valid Spend Tx which consumes $U$, the UTxOs ($N$ times SAND)
\item[5 :] Advertise the valid Spend Tx to the watchtowers through the coordinator (SAND)
\item[6 :] Broadcast all Unvault Txs that the valid Spend Tx depends on and wait for the time-lock to expire
\end{itemize}
\end{itemize}
}

{\footnotesize
\begin{itemize}[noitemsep,parsep=0pt,partopsep=0pt, leftmargin=0.7cm]
\item[\textbf{F} :] \textbf{Force emergency scenario}
\begin{itemize}[noitemsep,topsep=0pt,parsep=0pt,partopsep=0pt, leftmargin=0.8cm]
\item[1 :] Broadcast the full set of signed Emergency and Unvault-emergency transactions 
\begin{itemize}[noitemsep,topsep=0pt,parsep=0pt,partopsep=0pt, leftmargin=0.9cm]
\item[\textit{1.1} :] (see \textbf{d}) Compromise a watchtower (OR)
\item[\textit{1.2} :] (see \textbf{a}) Compromise a stakeholder
\end{itemize}
\end{itemize}
\end{itemize}
}

\noindent The emergency deterrent results in better security from the most egregious physical threats to participants (particularly stakeholders who control the majority of funds) but also in a fragility to the continuity of operations that could be abused by an attacker. Attacks that rely on \textbf{E} may seek a pay-off other than fund theft, such as damaging the reputation of the organisation for having down-time and taking a leveraged bet on the likely market consequences. However, forced down-time attacks through power or internet outages or detainment of personnel are prevalent threats for organisations who aren't deploying Revault. In any case, with this risk model the consequence of not using an emergency deterrent can be considered by performing an analysis with pruned attack-trees.

{\footnotesize
\begin{itemize}[noitemsep,parsep=0pt,partopsep=0pt, leftmargin=0.7cm]
\item[\textbf{G} :] \textbf{Broadcast a Theft Tx which consumes all available UTxOs locked to the emergency descriptor}
\begin{itemize}[noitemsep,topsep=0pt,parsep=0pt,partopsep=0pt, leftmargin=0.8cm]
\item[1 :] (see \textbf{F}) Force an emergency scenario (SAND)
\item[2 :] Determine $\mathcal{E}$, the set of available emergency UTxOs (SAND)
\begin{itemize}[noitemsep,topsep=0pt,parsep=0pt,partopsep=0pt, leftmargin=0.9cm]
\item[\textit{2.1} :] (see \textbf{A}) Compromise privacy of the custody operation (determine the set of public UTxOs)
\end{itemize}
\item[3 :] (see \textbf{k}) Satisfy an input in a Theft Tx that consumes an identified emergency UTxO ($|\mathcal{E}|$ times)
\end{itemize}
\end{itemize}
}

{\footnotesize
\begin{itemize}[noitemsep,parsep=0pt,partopsep=0pt, leftmargin=0.7cm]
\item[\textbf{H} :] \textbf{Broadcast a Theft Tx which spends from a manager’s fee wallet}
\begin{itemize}[noitemsep,topsep=0pt,parsep=0pt,partopsep=0pt, leftmargin=0.8cm]
\item[1 :] (see \textbf{g}) Compromise a manager’s wallet
\end{itemize}
\end{itemize}
}

\noindent While this is a relatively simple attack, the fee wallet will never hold a significant portion of bitcoins and is considered external to the custody protocol.

{\footnotesize
\begin{itemize}[noitemsep,parsep=0pt,partopsep=0pt, leftmargin=0.7cm]
\item[\textbf{I} :] \textbf{Prevent Emergency, Unvault-Emergency, and Cancel Tx valid signature exchange}
\begin{itemize}[noitemsep,topsep=0pt,parsep=0pt,partopsep=0pt, leftmargin=0.8cm]
\item[1 :] $1$ of $N$ stakeholders doesn't sign (OR)
\begin{itemize}[noitemsep,topsep=0pt,parsep=0pt,partopsep=0pt, leftmargin=0.9cm]
\item[\textit{1.1} :] Prevent stakeholder from accessing their HSM (OR)
\item[\textit{1.2} :] Prevent stakeholder from accessing their wallet (OR)
\item[\textit{1.3} :] (see \textbf{a}) Compromise a stakeholder
\end{itemize}
\item[2 :] Shutdown coordinator (OR)
\item[3 :] (see \textbf{e}) Shutdown a watchtower ($N$ times) (OR)
\item[4 :] Blockchain re-organization and Deposit Tx outpoint malleation.
\end{itemize}
\end{itemize}
}

{\footnotesize
\begin{itemize}[noitemsep,parsep=0pt,partopsep=0pt, leftmargin=0.7cm]
\item[\textbf{J} :] \textbf{Prevent Unvault Tx signature exchange}
\begin{itemize}[noitemsep,topsep=0pt,parsep=0pt,partopsep=0pt, leftmargin=0.8cm]
\item[1 :] $1$ of $N$ stakeholders doesn't sign (OR)
\begin{itemize}[noitemsep,topsep=0pt,parsep=0pt,partopsep=0pt, leftmargin=0.9cm]
\item[\textit{1.1} :] Prevent stakeholder from accessing their HSM (OR)
\item[\textit{1.2} :] Prevent stakeholder from accessing their wallet software (OR)
\item[\textit{1.3} :] (see \textbf{a}) Compromise a stakeholder
\end{itemize}
\item[2 :] Shutdown coordinator (OR)
\item[3 :] Prevent all managers from accessing their wallet software
\end{itemize}
\end{itemize}
}

{\footnotesize
\begin{itemize}[noitemsep,parsep=0pt,partopsep=0pt, leftmargin=0.7cm]
\item[\textbf{K} :] \textbf{Prevent managers from broadcasting a Spend Tx}
\begin{itemize}[noitemsep,topsep=0pt,parsep=0pt,partopsep=0pt, leftmargin=0.8cm]
\item[1 :] Prevent managers from signing the Spend transaction (OR)
\begin{itemize}[noitemsep,topsep=0pt,parsep=0pt,partopsep=0pt, leftmargin=0.9cm]
\item[\textit{1.1} :] (see \textbf{d}) Compromise an anti-replay oracle (OR)
\item[\textit{1.2} :] Prevent sufficient threshold of managers from signing the Spend Tx
(where $A + B + C = M-K+1$) (OR) 
\begin{itemize}[noitemsep,topsep=0pt,parsep=0pt,partopsep=0pt, leftmargin=1cm]
\item[\textit{1.2.1} :] (see \textbf{a}) Compromise a manager ($A$ times)
\item[\textit{1.2.2} :] Prevent manager from accessing their HSM ($B$ times)
\item[\textit{1.2.3} :] Prevent manager from accessing their wallet software ($C$ times)
\end{itemize}
\end{itemize}
\item[2 :] Force broadcast of Cancel Tx (OR)
\begin{itemize}[noitemsep,topsep=0pt,parsep=0pt,partopsep=0pt, leftmargin=0.9cm]
\item[\textit{2.1} :] (see \textbf{d}) Compromise a watchtower 
\end{itemize}
\item[3 :] Prevent broadcast of Unvault Tx
\begin{itemize}[noitemsep,topsep=0pt,parsep=0pt,partopsep=0pt, leftmargin=0.8cm]
\item[\textit{3.1} :] High demand for block space making the Unvault Tx not profitable to mine.\footnote{Manager's fee-bumping wallet can not cover this until a network policy such as Package Relay \cite{PackageRelay} is implemented.} % 
\item[\textit{3.2} :] (see \textbf{g}) Compromise manager's wallet ($M$ times)
\end{itemize}
\end{itemize}
\end{itemize}
 }

\section{Conclusion}
\label{sec:Conclusion}

The rise of Bitcoin has led to a new commercial ecosystem, with market exchanges enabling its sale and purchase, companies and financial institutions offering secure custody services, and insurance brokers and underwriters willing to insure individuals, exchanges and custodians against loss or theft of their assets. In this chapter we first posit that a methodology to better understand risks in custodial operations is needed, something complementary to understanding blockchain and cryptographic security. We put forth requirements of the modelling technique and propose attack-trees as a formalism which satisfies those requirements. We exemplify the approach by presenting a library of attack-trees constructed for a multi-party custody protocol called Revault and explain how this framework can be used as a basis for risk-management in custodial operations. The next steps for this work are to: construct a set of defences to the prominent risks and incorporate them into the model; and to determine or build a suitable tool for automating computations for a specific analysis.

\addcontentsline{toc}{chapter}{Bibliography}

\bibliographystyle{abbrv}

\bibliography{master}

\begin{thebibliography}{100}

\bibitem{contract-wg}
Bitcoin contracting primitives working group.
\newblock \url{https://github.com/ariard/bitcoin-contracting-primitives-wg}.

\bibitem{BlockchainCommons}
{Blockchain Commons}.
\newblock \url{https://github.com/BlockchainCommons}.

\bibitem{gordian-seed-tool}
{Blockchain Commons: Gordian Seed Tool}.
\newblock \url{https://github.com/BlockchainCommons/GordianSeedTool-iOS}.

\bibitem{ccss}
{CryptoCurrency Security Standard}.
\newblock \url{https://cryptoconsortium.notion.site/CryptoCurrency-Security-Standard-e372d9cad52f4615aa3ad0c47c24ea21}.

\bibitem{foundation-passport}
{Foundation Devices, Inc.: Foundation Passport}.
\newblock \url{https://foundationdevices.com/passport/}.

\bibitem{googleAppSecurity}
{Google: Use Google Play Protect to help keep your apps safe and your data private}.
\newblock \url{https://support.google.com/googleplay/answer/2812853?hl=en-GB#zippy=\%2Chow-malware-protection-works\%2Chow-google-resets-permissions-for-unused-apps}.

\bibitem{commercial-HWs}
Hardware wallet.
\newblock \url{https://en.bitcoin.it/wiki/Hardware_wallet#Commercial_hardware_wallets_.28ordered_chronologically.29}.

\bibitem{miniscript}
Miniscript.
\newblock \url{https://bitcoin.sipa.be/miniscript/}.

\bibitem{nunchuck-multisig}
{Nunchuck}.
\newblock \url{https://nunchuk.io/about-us/}.

\bibitem{ScriptDescriptors}
{Output Script Descriptors: a language for abstracting out the spending conditions of a Bitcoin transaction output}.
\newblock \url{https://github.com/bitcoin/bitcoin/blob/master/doc/descriptors.md}.

\bibitem{PackageRelay}
{Package Relay design questions for the Bitcoin P2P network}.
\newblock \url{https://github.com/bitcoin/bitcoin/issues/14895}.

\bibitem{practical-revault}
{Practical Revault}.
\newblock \url{https://github.com/revault/practical-revault}.

\bibitem{revault-repos}
{Revault}.
\newblock \url{https://github.com/revault/}.

\bibitem{rusty-guide}
{Rusty's Remarkably Unreliable Guide To Bitcoin Storage: 2018 Edition}.
\newblock \url{https://github.com/rustyrussell/bitcoin-storage-guide}.

\bibitem{ScriptWiki}
{Script}.
\newblock \url{https://en.bitcoin.it/wiki/Script}.

\bibitem{sparrow-wallet}
{Sparrow wallet}.
\newblock \url{https://sparrowwallet.com/}.

\bibitem{wasabi}
{Wasabi}.
\newblock \url{https://docs.wasabiwallet.io/using-wasabi/}.

\bibitem{yeticold}
{YetiCold.com Bitcoin Storage}.
\newblock \url{https://github.com/JWWeatherman/yeticold/blob/master/README.md}.

\bibitem{STRIDE}
{Microsoft Corporation: Security Design by Threat Modeling}, 2005.
\newblock \url{https://docs.microsoft.com/en-us/previous-versions/commerce-server/ee810542(v=cs.20)}.

\bibitem{secp256k1}
{Certicom Research: SEC 2: Recommended Elliptic Curve Domain Parameters}, 2010.
\newblock \url{http://www.secg.org/sec2-v2.pdf}.

\bibitem{MSC}
{Telecommunication Standardization Sector of ITU: Message Sequence Chart (MSC)}.
\newblock 2011.
\newblock \url{https://www.itu.int/rec/dologin_pub.asp?lang=e&id=T-REC-Z.120-201102-I!!PDF-E&type=items}.

\bibitem{ECDSA}
National institute of standards and technology: Digital signature standard (dss).
\newblock {\em {FIPS PUB 186-4 FEDERAL INFORMATION PROCESSING STANDARDS PUBLICATION}}, 2013.
\newblock \url{https://nvlpubs.nist.gov/nistpubs/FIPS/NIST.FIPS.186-4.pdfhttps://nvlpubs.nist.gov/nistpubs/FIPS/NIST.FIPS.186-4.pdf}.

\bibitem{SHA256}
{National Institute of Standards and Technology: FIPS 180-4: Secure Hash Standard (SHS)}, 2015.
\newblock \url{https://csrc.nist.gov/publications/detail/fips/180/4/final}.

\bibitem{BIP39-diceware}
{Bip39-diceware}, 2017.
\newblock \url{https://github.com/taelfrinn/Bip39-diceware}.

\bibitem{cc-framework}
{Common Criteria Development Board: Common Criteria for Information Technology Security Evaluation Part 1}, 2017.
\newblock \url{https://www.commoncriteriaportal.org/files/ccfiles/CCPART1V3.1R5.pdf}.

\bibitem{Glacier}
{Glacier Design Document}, 2017.
\newblock \url{https://glacierprotocol.org/assets/design-doc-v0.9-beta.pdf}.

\bibitem{blockstream-multisig}
{How does Blockstream Green's multisig security work?}, 2017.
\newblock \url{https://help.blockstream.com/hc/en-us/articles/900001391763-How-does-Blockstream-Green-s-multisig-security-work-}.

\bibitem{electrum-multisig}
{Multisig Wallets}, 2017.
\newblock \url{https://electrum.readthedocs.io/en/latest/multisig.html}.

\bibitem{stonewall}
{Introducing Boltzmann and STONEWALL}, 2018.
\newblock \url{https://blog.samouraiwallet.com/post/173544815052/full-bech32-support-introducing-boltzmann-and}.

\bibitem{NIST-CSF}
{National Institute of Standards and Technology: Framework for Improving Critical Infrastructure Cybersecurity}, 2018.
\newblock \url{https://nvlpubs.nist.gov/nistpubs/CSWP/NIST.CSWP.04162018.pdf}.

\bibitem{Subzero}
{Square, Inc.: Subzero: HSM-backed Bitcoin Cold Storage}.
\newblock 2018.
\newblock \url{https://subzero.readthedocs.io/en/master/}.

\bibitem{DACS}
{Capital Markets and Technology Association: Digital Assets Custody Standard}.
\newblock 2020.
\newblock \url{https://www.cmta.ch/content/272/cmta-digital-assets-custody-standard-v1-public-consultation.pdf}.

\bibitem{Casa-WSP}
{Casa: Wealth Security Protocol}.
\newblock 2020.
\newblock \url{https://docs.keys.casa/wealth-security-protocol/}.

\bibitem{coinmonks-dice-seed}
{Generating Device Seeds Using Dice}, 2020.
\newblock \url{https://medium.com/coinmonks/generating-device-seeds-using-dice-894082d43aea}.

\bibitem{Bitbox-threat-model}
{Shift Crypto AG: BitBox02 threat model}, 2020.
\newblock \url{https://shiftcrypto.ch/bitbox02/threat-model/}.

\bibitem{enno-TM}
{Enno Wallet Threat Model for Mobile Apps}, 2021.
\newblock \url{https://github.com/Enno-Wallet-Enno-Cash/security-public/tree/main/threat-model}.

\bibitem{appleAppSecurity}
Apple: App security overview, 2022.
\newblock \url{https://support.apple.com/en-gb/guide/security/sec35dd877d0/1/web/1}.

\bibitem{offline-bip39-dice}
{Bip39 Offline Mnemonic Generator}, 2022.
\newblock \url{https://github.com/veebch/Bip39-Dice}.

\bibitem{CIS-CSC}
{Center for Internet Security: The 18 CIS Critical Security Controls}, 2022.
\newblock \url{https://www.cisecurity.org/controls/cis-controls-list}.

\bibitem{Coldcard-hsm-threat-model}
{Coinkite Inc.: HSM Security Notes}, 2022.
\newblock \url{https://coldcard.com/docs/hsm/security}.

\bibitem{Ledger-threat-model}
{Ledger Donjon: Threat Model}, 2022.
\newblock \url{https://donjon.ledger.com/threat-model/#security-mechanisms}.

\bibitem{Trezor-threat-model}
{SatoshiLabs: Common security threats}, 2022.
\newblock \url{https://trezor.io/learn/a/common-security-threats}.

\bibitem{Donjon-trezor-attack}
K.~Abdellatif, C.~Guillemet, and O.~Hériveaux.
\newblock {Unfixable Seed Extraction on Trezor - A practical and reliable attack}, 2019.
\newblock \url{https://blog.ledger.com/Unfixable-Key-Extraction-Attack-on-Trezor/}.

\bibitem{ANOSS21}
D.~Abram, A.~Nof, C.~Orlandi, P.~Scholl, and O.~Shlomovits.
\newblock {Low-Bandwidth Threshold ECDSA via Pseudorandom Correlation Generators}.
\newblock {\em Cryptology ePrint Archive, Paper 2021/1587}, 2021.
\newblock \url{https://eprint.iacr.org/2021/1587}.

\bibitem{HW-sc-security}
L.~Abrams.
\newblock {Criminals are mailing altered Ledger devices to steal cryptocurrency}, 2021.
\newblock \url{https://www.bleepingcomputer.com/news/cryptocurrency/criminals-are-mailing-altered-ledger-devices-to-steal-cryptocurrency/}.

\bibitem{DPKSSecureAgainstLeakage}
A.~Akavia, S.~Goldwasser, and C.~Hazay.
\newblock {Distributed Public Key Schemes Secure Against Continual Leakage}.
\newblock In {\em Proceedings of the 2012 ACM Symposium on Principles of Distributed Computing}, PODC '12, pages 155--164. ACM, 2012.

\bibitem{BoEcrypto}
R.~Ali, J.~Barrdear, R.~Clews, and J.~Southgate.
\newblock {The economics of digital currencies}.
\newblock 2014.
\newblock \url{https://www.bankofengland.co.uk/-/media/boe/files/quarterly-bulletin/2014/the-economics-of-digital-currencies}.

\bibitem{SmartCustody-cold}
C.~Allen and S.~Appelcline.
\newblock {Cold Storage Self-Custody Scenario}.
\newblock 2019.
\newblock \url{https://github.com/BlockchainCommons/SmartCustodyBook/blob/master/manuscript/02-scenario.md}.

\bibitem{SmartCustodyBook}
C.~Allen and S.~Appelcline.
\newblock {Smart Custody Book}, 2019.
\newblock \url{https://github.com/BlockchainCommons/SmartCustodyBook/blob/master/manuscript/}.

\bibitem{SmartCustody-multisig}
C.~Allen and S.~Appelcline.
\newblock {Multisig Self-Custody Scenario}.
\newblock 2022.
\newblock \url{https://github.com/BlockchainCommons/SmartCustody/blob/master/Docs/Scenario-Multisig.md}.

\bibitem{ThreatTrees}
E.~G. Amoroso.
\newblock {\em {Fundamentals of Computer Security Technology}}.
\newblock Prentice-Hall, Inc., 1994.

\bibitem{Anderson:2008:SEG:1373319}
R.~J. Anderson.
\newblock {\em {Security Engineering: A Guide to Building Dependable Distributed Systems}}.
\newblock Wiley Publishing, 2nd edition, 2008.

\bibitem{BIP13}
G.~Andresen.
\newblock {Address Format for pay-to-script-hash}, 2011.
\newblock \url{https://github.com/bitcoin/bips/blob/master/bip-0013.mediawiki}.

\bibitem{BIP16}
G.~Andresen.
\newblock {Pay to Script Hash}, 2012.
\newblock \url{https://github.com/bitcoin/bips/blob/master/bip-0016.mediawiki}.

\bibitem{Antonopoulos:2014:MBU:2695500}
A.~M. Antonopoulos.
\newblock {\em {Mastering Bitcoin: Unlocking Digital Crypto-Currencies}}.
\newblock O'Reilly Media, Inc., 1st edition, 2014.

\bibitem{FormalHardware}
M.~Arapinis, A.~Gkaniatsou, D.~Karakostas, and A.~Kiayias.
\newblock {A Formal Treatment of Hardware Wallets}.
\newblock {\em Financial Cryptography and Data Security}, pages 426--445, 2019.

\bibitem{btc-spendable}
M.~Arenas, T.~Reisenegger, J.~Reutter, and D.~Vrgoč.
\newblock Is it possible to verify if a transaction is spendable?
\newblock {\em Frontiers in Blockchain}, 4, 2021.

\bibitem{I2P}
F.~Astolfi, J.~Kroese, and J.~van Oorschot.
\newblock {I2P - The Invisible Internet Project}.
\newblock \url{http://mediatechnology.leiden.edu/images/uploads/docs/wt2015_i2p.pdf}.

\bibitem{bip126}
K.~Atlas.
\newblock {Best Practices for Heterogeneous Input Script Transactions}, 2016.
\newblock \url{https://github.com/bitcoin/bips/blob/master/bip-0126.mediawiki}.

\bibitem{CCC}
A.~Au, M.~Hoffman, J.~Mattingly, P.~McAteer, and Y.~Takahashi.
\newblock {Care, Custody, \& Control (CCC): Identification, quantification, and mitigation of cryptocurrency custodial risk.}
\newblock 2020.
\newblock \url{https://www.paulmcateer.com/wp-content/uploads/2021/01/CCC_Pub_Version.pdf}.

\bibitem{attacking-thresh-wal}
J.-P. Aumasson and O.~Shlomovits.
\newblock Attacking threshold wallets.
\newblock {\em Cryptology ePrint Archive, Paper 2020/1052}, 2020.
\newblock \url{https://eprint.iacr.org/2020/1052}.

\bibitem{Back2002}
A.~Back.
\newblock {Hashcash - A Denial of Service Counter-Measure}.
\newblock 2002.
\newblock \url{http://www.hashcash.org/papers/hashcash.pdf}.

\bibitem{Sidechains101}
A.~Back, M.~Corallo, L.~Dashjr, M.~Friedenbach, G.~Maxwell, A.~Miller, A.~Poelstra, J.~Timón, and P.~Wuille.
\newblock {Enabling Blockchain Innovations with Pegged Sidechains}, 2014.
\newblock \url{https://blockstream.com/sidechains.pdf}.

\bibitem{Badertscher2018}
C.~Badertscher, J.~Garay, U.~Maurer, D.~Tschudi, and V.~Zikas.
\newblock {But Why Does It Work? A Rational Protocol Design Treatment of Bitcoin}.
\newblock In J.~B. Nielsen and V.~Rijmen, editors, {\em Advances in Cryptology -- EUROCRYPT 2018}, pages 34--65. Springer International Publishing, 2018.

\bibitem{Badertscher2017}
C.~Badertscher, U.~Maurer, D.~Tschudi, and V.~Zikas.
\newblock {Bitcoin as a transaction ledger: A composable treatment}.
\newblock {\em Lecture Notes in Computer Science (including subseries Lecture Notes in Artificial Intelligence and Lecture Notes in Bioinformatics)}, 10401 LNCS(Crypto 2017):324--356, 2017.

\bibitem{BSKKC20}
Z.~Bao, W.~Shi, S.~Kumari, Z.-y. Kong, and C.-M. Chen.
\newblock Lockmix: a secure and privacy-preserving mix service for bitcoin anonymity.
\newblock {\em International Journal of Information Security}, 19, 2020.

\bibitem{CovenantsUnchained}
M.~Bartoletti, S.~Lande, and R.~Zunino.
\newblock Bitcoin covenants unchained.
\newblock {\em Leveraging Applications of Formal Methods, Verification and Validation: Applications}, 2020.

\bibitem{SoundBitcoinTokens}
M.~Bartoletti, S.~Lande, and R.~Zunino.
\newblock Computationally sound bitcoin tokens.
\newblock {\em 2021 IEEE 34th Computer Security Foundations Symposium (CSF)}, 2021.

\bibitem{BitML}
M.~Bartoletti and R.~Zunino.
\newblock {BitML: A Calculus for Bitcoin Smart Contracts}.
\newblock {\em Cryptology ePrint Archive, Paper 2018/122}, 2018.
\newblock \url{https://eprint.iacr.org/2018/122}.

\bibitem{SurveyFormalBitcoinContracts}
M.~Bartoletti and R.~Zunino.
\newblock Formal models of bitcoin contracts: A survey.
\newblock {\em Frontiers in Blockchain}, 2019.

\bibitem{BLMS20}
M.~Battagliola, R.~Longo, A.~Meneghetti, and M.~Sala.
\newblock {Threshold ECDSA with an Offline Recovery Party}.
\newblock {\em Mediterranean Journal of Mathematics}, 2020.

\bibitem{Bella2007FormalCO}
G.~Bella.
\newblock {\em Formal Correctness of Security Protocols}.
\newblock 2007.

\bibitem{ControlStructure}
A.~Berentsen and F.~Schar.
\newblock {The Case for Central Bank Electronic Money and the Non-case for Central Bank Cryptocurrencies}.
\newblock {\em Federal Reserve Bank of St. Louis Review, Second Quarter 2018}, pages 97--106, 2018.

\bibitem{DeanonymisationP2P}
A.~Biryukov, D.~Khovratovich, and I.~Pustogarov.
\newblock {Deanonymisation of clients in Bitcoin P2P network}.
\newblock {\em Proceedings of the 2014 ACM SIGSAC Conference on Computer and Communications Security}, abs/1405.7418, 2014.
\newblock \url{http://arxiv.org/abs/1405.7418}.

\bibitem{BitcoinTor}
A.~Biryukov and I.~Pustogarov.
\newblock {Bitcoin over Tor isn't a good idea}.
\newblock {\em CoRR}, abs/1410.6079, 2014.
\newblock \url{http://arxiv.org/abs/1410.6079}.

\bibitem{BishopVaults}
B.~Bishop.
\newblock {On-chain vaults prototype}, 2020.
\newblock \url{https://lists.linuxfoundation.org/pipermail/bitcoin-dev/2020-April/017755.html}.

\bibitem{UR-crypto-commons}
{Blockchain Commons}.
\newblock {Uniform Resources (UR): An Introduction}, 2021.
\newblock \url{https://github.com/BlockchainCommons/crypto-commons/blob/master/Docs/ur-1-overview.md}.

\bibitem{BGG19}
D.~Boneh, R.~Gennaro, and S.~Goldfeder.
\newblock {Using Level-1 Homomorphic Encryption to Improve Threshold DSA Signatures for Bitcoin Wallet Security}.
\newblock {\em {Progress in Cryptology -- LATINCRYPT 2017}}, 2019.

\bibitem{BNMCKF14}
J.~Bonneau, A.~Narayanan, A.~Miller, J.~Clark, J.~Kroll, and E.~Felten.
\newblock Mixcoin: Anonymity for bitcoin with accountable mixes.
\newblock {\em Cryptology ePrint Archive, Paper 2014/077}, 2014.
\newblock \url{https://eprint.iacr.org/2014/077}.

\bibitem{BIP112}
BtcDrak, M.~Friedenbach, and E.~Lombrozo.
\newblock {CHECKSEQUENCEVERIFY}, 2015.
\newblock \url{https://github.com/bitcoin/bips/blob/master/bip-0112.mediawiki}.

\bibitem{gazibtd513088}
Y.~Bulut and I.~Sertkaya.
\newblock Security problem definition and security objectives of cryptocurrency wallets in common criteria.
\newblock {\em Bilişim Teknolojileri Dergisi}, 13(2):157 -- 165, 2020.

\bibitem{Buterin-Deterministic-wallets}
V.~Buterin.
\newblock {Deterministic Wallets, Their Advantages and their Understated Flaws}, 2013.
\newblock \url{https://bitcoinmagazine.com/technical/deterministic-wallets-advantages-flaw-1385450276}.

\bibitem{ERFAON}
R.~Canetti, Y.~Dodis, S.~Halevi, E.~Kushilevitz, and A.~Sahai.
\newblock {Exposure-resilient Functions and All-or-nothing Transforms}.
\newblock In {\em Proceedings of the 19th International Conference on Theory and Application of Cryptographic Techniques}, EUROCRYPT'00, pages 453--469. Springer-Verlag, 2000.

\bibitem{CGGMP21}
R.~Canetti, R.~Gennaro, S.~Goldfeder, N.~Makriyannis, and U.~Peled.
\newblock {UC Non-Interactive, Proactive, Threshold ECDSA with Identifiable Aborts}.
\newblock {\em Cryptology ePrint Archive, Paper 2021/060}, 2021.
\newblock \url{https://eprint.iacr.org/2021/060}.

\bibitem{HumanWeakness}
M.~Carlos and G.~Price.
\newblock Understanding the weaknesses of human-protocol interaction.
\newblock In J.~Blyth, S.~Dietrich, and L.~J. Camp, editors, {\em Financial Cryptography and Data Security}, pages 13--26. Springer Berlin Heidelberg, 2012.

\bibitem{CCLST19}
G.~Castagnos, D.~Catalano, F.~Laguillaumie, F.~Savasta, and I.~Tucker.
\newblock {Two-Party ECDSA from Hash Proof Systems and Efficient Instantiations}.
\newblock {\em Cryptology ePrint Archive, Paper 2019/503}, 2019.
\newblock \url{https://eprint.iacr.org/2019/503}.

\bibitem{CCLST21}
G.~Castagnos, D.~Catalano, F.~Laguillaumie, F.~Savasta, and I.~Tucker.
\newblock {Bandwidth-efficient threshold EC-DSA revisited: Online/Offline Extensions, Identifiable Aborts, Proactivity and Adaptive Security}.
\newblock {\em Cryptology ePrint Archive, Paper 2021/291}, 2021.
\newblock \url{https://eprint.iacr.org/2021/291}.

\bibitem{foundation-TM}
L.~Childs and K.~Carpenter.
\newblock {Passport Security Model}, 2021.
\newblock \url{https://github.com/Foundation-Devices/passport-firmware/blob/main/SECURITY/SECURITY.md}.

\bibitem{BIP174}
A.~Chow.
\newblock {Partially Signed Bitcoin Transaction Format}, 2017.
\newblock \url{https://github.com/bitcoin/bips/blob/master/bip-0174.mediawiki}.

\bibitem{BIP370}
A.~Chow.
\newblock {PSBT Version 2}, 2017.
\newblock \url{https://github.com/bitcoin/bips/blob/master/bip-0370.mediawiki}.

\bibitem{FormalWallet}
D.~Coutts and E.~de~Vries.
\newblock {Formal specification for a Cardano wallet}.
\newblock 2018.
\newblock \url{https://iohk.io/en/research/library/papers/formal-specification-for-a-cardano-wallet/}.

\bibitem{CKM21}
E.~Crites, C.~Komlo, and M.~Maller.
\newblock {How to Prove Schnorr Assuming Schnorr: Security of Multi- and Threshold Signatures}.
\newblock {\em Cryptology ePrint Archive, Paper 2021/1375}, 2021.
\newblock \url{https://eprint.iacr.org/2021/1375}.

\bibitem{HW-SPOF}
A.~Dabrowski, K.~Pfeffer, M.~Reichel, A.~Mai, E.~R. Weippl, and M.~Franz.
\newblock Better keep cash in your boots - hardware wallets are the new single point of failure.
\newblock In {\em Proceedings of the 2021 ACM CCS Workshop on Decentralized Finance and Security}, DeFi '21, page 1–8. Association for Computing Machinery, 2021.

\bibitem{DOKSS20}
A.~Dalskov, C.~Orlandi, M.~Keller, K.~Shrishak, and H.~Shulman.
\newblock {\em Securing DNSSEC Keys via Threshold ECDSA from Generic MPC}, pages 654--673.
\newblock 2020.

\bibitem{DJNPO20}
I.~Damgård, T.~P. Jakobsen, J.~B. Nielsen, J.~I. Pagter, and M.~B. Østergård.
\newblock {Fast Threshold ECDSA with Honest Majority}.
\newblock {\em Cryptology ePrint Archive, Paper 2020/501}, 2020.
\newblock \url{https://eprint.iacr.org/2020/501}.

\bibitem{APOAS-darosior}
Darosior.
\newblock {ANYPREVOUT in place of CTV}.
\newblock 2022.
\newblock \url{https://lists.linuxfoundation.org/pipermail/bitcoin-dev/2022-April/020276.html}.

\bibitem{Exact-bip32}
P.~Das, A.~Erwig, S.~Faust, J.~Loss, and S.~Riahi.
\newblock {The Exact Security of BIP32 Wallets}.
\newblock {\em Cryptology ePrint Archive, Paper 2021/1287}, 2021.
\newblock \url{https://eprint.iacr.org/2021/1287}.

\bibitem{Davidson:1996:SIS:548392}
J.~D. Davidson and W.~Rees-Mogg.
\newblock {\em The Sovereign Individual: How to Survive and Thrive During the Collapse of the Welfare State}.
\newblock Simon \& Schuster, Inc., 1996.

\bibitem{decker2018}
C.~Decker, R.~Russell, and O.~Osuntokun.
\newblock {Eltoo: A Simple Layer 2 Protocol for Bitcoin}.
\newblock 2018.
\newblock \url{https://blockstream.com/eltoo.pdf}.

\bibitem{BIP118-APO}
C.~Decker and A.~Towns.
\newblock {SIGHASH\_ANYPREVOUT for Taproot}, 2021.
\newblock \url{https://github.com/bitcoin/bips/blob/master/bip-0118.mediawiki}.

\bibitem{InfoProp}
C.~Decker and R.~Wattenhofer.
\newblock {Information propagation in the Bitcoin network}.
\newblock In {\em IEEE P2P 2013 Proceedings}, pages 1--10, 2013.

\bibitem{Decker2015}
C.~Decker and R.~Wattenhofer.
\newblock {A fast and scalable payment network with bitcoin duplex micropayment channels}.
\newblock {\em Lecture Notes in Computer Science (including subseries Lecture Notes in Artificial Intelligence and Lecture Notes in Bioinformatics)}, 9212:3--18, 2015.

\bibitem{DELGADOSEGURA2020832}
S.~Delgado-Segura, C.~Pérez-Solà, G.~Navarro-Arribas, and J.~Herrera-Joancomartí.
\newblock A fair protocol for data trading based on bitcoin transactions.
\newblock {\em Future Generation Computer Systems}, 107:832--840, 2020.

\bibitem{ResilientCustody}
V.~Di, R.~Longo, F.~Mazzone, and G.~Russo.
\newblock Resilient custody of crypto-assets, and threshold multisignatures.
\newblock {\em Mathematics}, 8:10, 2020.

\bibitem{DS16}
P.~Dikshit and K.~Singh.
\newblock {Weighted threshold ECDSA for securing bitcoin wallet}.
\newblock {\em ACCENTS Transactions on Information Security}, 2:43--51, 2016.

\bibitem{StrongFederations}
J.~Dilley, A.~Poelstra, J.~Wilkins, M.~Piekarska, B.~Gorlick, and M.~Friedenbach.
\newblock {Strong Federations: An Interoperable Blockchain Solution to Centralized Third Party Risks}.
\newblock {\em CoRR}, abs/1612.05491, 2016.

\bibitem{TOR}
R.~Dingledine, N.~Mathewson, and P.~Syverson.
\newblock {Tor: The Second-Generation Onion Router}.
\newblock {\em 13th USENIX Security Symposium (USENIX Security 04)}, 13, 2004.

\bibitem{RIPEMD-160}
H.~Dobbertin, A.~Bosselaers, and B.~Preneel.
\newblock Ripemd-160: A strengthened version of ripemd.
\newblock In D.~Gollmann, editor, {\em Fast Software Encryption}, pages 71--82. Springer Berlin Heidelberg, 1996.

\bibitem{DKLS18}
J.~Doerner, Y.~Kondi, E.~Lee, and abhi shelat.
\newblock {Secure Two-party Threshold ECDSA from ECDSA Assumptions}.
\newblock {\em Cryptology ePrint Archive, Paper 2018/499}, 2018.
\newblock \url{https://eprint.iacr.org/2018/499}.

\bibitem{DKLS19}
J.~Doerner, Y.~Kondi, E.~Lee, and abhi shelat.
\newblock {Threshold ECDSA from ECDSA Assumptions: The Multiparty Case}.
\newblock {\em Cryptology ePrint Archive, Paper 2019/523}, 2019.
\newblock \url{https://eprint.iacr.org/2019/523}.

\bibitem{DolevYao}
D.~Dolev and A.~Yao.
\newblock On the security of public key protocols.
\newblock {\em IEEE Transactions on Information Theory}, 29(2):198--208, 1983.

\bibitem{DLC}
T.~Dryja.
\newblock {Discreet Log Contracts}.
\newblock \url{https://adiabat.github.io/dlc.pdf}.

\bibitem{Ellison2007CeremonyDA}
C.~M. Ellison.
\newblock Ceremony design and analysis.
\newblock {\em IACR Cryptology ePrint Archive}, 2007:399, 2007.

\bibitem{CoinSelection}
M.~Erhardt.
\newblock {An Evaluation of Coin Selection Strategies}, 2016.
\newblock \url{http://murch.one/wp-content/uploads/2016/11/erhardt2016coinselection.pdf}.

\bibitem{eskandari2015usability}
S.~Eskandari, D.~Barrera, E.~Stobert, and J.~Clark.
\newblock {A first look at the usability of bitcoin key management}.
\newblock In {\em Workshop on Usable Security (USEC)}, 2015.

\bibitem{Eyal2021OnCW}
I.~Eyal.
\newblock On cryptocurrency wallet design.
\newblock {\em IACR Cryptology ePrint Archive}, 2021.

\bibitem{hd-privilege}
C.-I. Fan, Y.-F. Tseng, H.-P. Su, R.-H. Hsu, and H.~Kikuchi.
\newblock Secure hierarchical bitcoin wallet scheme against privilege escalation attacks.
\newblock {\em International Journal of Information Security}, 19, 2020.

\bibitem{Dandelion++}
G.~C. Fanti, S.~B. Venkatakrishnan, S.~Bakshi, B.~Denby, S.~Bhargava, A.~Miller, and P.~Viswanath.
\newblock {Dandelion++: Lightweight Cryptocurrency Networking with Formal Anonymity Guarantees}.
\newblock {\em CoRR}, abs/1805.11060, 2018.

\bibitem{mtGox}
A.~Feder, N.~Gandal, J.~Hamrick, and T.~Moore.
\newblock {The impact of DDoS and other security shocks on Bitcoin currency exchanges: Evidence from Mt. Gox}.
\newblock {\em Journal of Cybersecurity}, 3:137--144, 2017.

\bibitem{RFC2616}
R.~Fielding, J.~Gettys, J.~Mogul, H.~Frystyk, L.~Masinter, P.~Leach, and T.~Berners-Lee.
\newblock {Hypertext Transfer Protocol -- HTTP/1.1}.
\newblock {\em IETF Request for Comments, RFC 2616}, 1999.
\newblock \url{https://www.rfc-editor.org/info/rfc2616}.

\bibitem{DLC-covs}
L.~Fournier.
\newblock {CTV dramatically improves DLCs}.
\newblock 2022.
\newblock \url{https://lists.linuxfoundation.org/pipermail/bitcoin-dev/2022-January/019808.html}.

\bibitem{tordl}
Fresheneesz.
\newblock {Tordl Wallet Protocols}.
\newblock \url{https://github.com/fresheneesz/TordlWalletProtocols}.

\bibitem{HCI-threat-model}
M.~Froehlich, P.~Hulm, and F.~Alt.
\newblock {Under Pressure. A User-Centered Threat Model for Cryptocurrency Owners}.
\newblock {\em Proceedings of the 2021 4th International Conference on Blockchain Technology and Applications}, 2021.

\bibitem{SoK-HCI-crypto}
M.~Froehlich, F.~Waltenberger, L.~Trotter, F.~Alt, and A.~Schmidt.
\newblock {Blockchain and Cryptocurrency in Human Computer Interaction: A Systematic Literature Review and Research Agenda}.
\newblock {\em Proceedings of the 2022 ACM Designing Interactive Systems Conference}, 2022.

\bibitem{ADTool2}
O.~Gadyatskaya, R.~Jhawar, P.~Kordy, K.~Lounis, S.~Mauw, and R.~Trujillo-Rasua.
\newblock {Attack Trees for Practical Security Assessment: Ranking of Attack Scenarios with ADTool 2.0}.
\newblock {\em International Conference on Quantitative Evaluation of Systems}, 2016.

\bibitem{GJDM20}
A.~Gagol, J.~Kula, D.~Straszak, and M.~Swietek.
\newblock {Threshold ECDSA for Decentralized Asset Custody}.
\newblock {\em Cryptology ePrint Archive, Paper 2020/498}, 2020.
\newblock \url{https://eprint.iacr.org/2020/498}.

\bibitem{Garay2015}
J.~Garay, A.~Kiayias, and N.~Leonardos.
\newblock {The Bitcoin backbone protocol: Analysis and applications}.
\newblock {\em Lecture Notes in Computer Science (including subseries Lecture Notes in Artificial Intelligence and Lecture Notes in Bioinformatics)}, 9057:281--310, 2015.

\bibitem{Garay2017}
J.~Garay, A.~Kiayias, and N.~Leonardos.
\newblock {The Bitcoin Backbone Protocol with Chains of Variable Difficulty}.
\newblock In J.~Katz and H.~Shacham, editors, {\em Advances in Cryptology -- CRYPTO 2017}, pages 291--323. Springer International Publishing, 2017.

\bibitem{GKMN21}
F.~Garillot, Y.~Kondi, P.~Mohassel, and V.~Nikolaenko.
\newblock Threshold schnorr with stateless deterministic signing from standard assumptions.
\newblock {\em Cryptology ePrint Archive, Paper 2021/1055}, pages 127--156, 2021.
\newblock \url{https://eprint.iacr.org/2021/1055}.

\bibitem{GG18}
R.~Gennaro and S.~Goldfeder.
\newblock {Fast Multiparty Threshold ECDSA with Fast Trustless Setup}.
\newblock In {\em Proceedings of the 2018 ACM SIGSAC Conference on Computer and Communications Security}, CCS '18, pages 1179--1194. ACM, 2018.

\bibitem{GGN16}
R.~Gennaro, S.~Goldfeder, and A.~Narayanan.
\newblock {Threshold-Optimal DSA/ECDSA Signatures and an Application to Bitcoin Wallet Security}.
\newblock {\em Cryptology ePrint Archive, Paper 2016/013}, 2016.
\newblock \url{https://eprint.iacr.org/2016/013}.

\bibitem{GJKR05}
R.~Gennaro, S.~Jarecki, H.~Krawczyk, and T.~Rabin.
\newblock {Secure Distributed Key Generation for Discrete-Log Based Cryptosystems}.
\newblock {\em Proceedings of the 17th International Conference on Theory and Application of Cryptographic Techniques}, 2005.

\bibitem{Gervais:2014:PPB:2664243.2664267}
A.~Gervais, S.~Capkun, G.~O. Karame, and D.~Gruber.
\newblock {On the Privacy Provisions of Bloom Filters in Lightweight Bitcoin Clients}.
\newblock In {\em Proceedings of the 30th Annual Computer Security Applications Conference}, ACSAC '14, pages 326--335. ACM, 2014.

\bibitem{SoK-privacy-21}
S.~Ghesmati, W.~Fdhila, and E.~Weippl.
\newblock {SoK: How private is Bitcoin? Classification and Evaluation of Bitcoin Mixing Techniques}.
\newblock {\em Cryptology ePrint Archive, Paper 2021/629}, 2021.
\newblock \url{https://eprint.iacr.org/2021/629}.

\bibitem{coinjoin-usability}
S.~Ghesmati, W.~Fdhila, and E.~Weippl.
\newblock {Usability of Cryptocurrency Wallets Providing CoinJoin Transactions}.
\newblock {\em Cryptology ePrint Archive, Paper 2022/285}, 2022.
\newblock \url{https://eprint.iacr.org/2022/285}.

\bibitem{low-level-hw-attacks}
A.~Gkaniatsou, M.~Arapinis, and A.~Kiayias.
\newblock Low-level attacks in bitcoin wallets.
\newblock {\em Information Security}, pages 233--253, 2017.

\bibitem{CookieBlockchain}
S.~Goldfeder, H.~Kalodner, D.~Reisman, and A.~Narayanan.
\newblock {When the cookie meets the blockchain: Privacy risks of web payments via cryptocurrencies}.
\newblock {\em Proceedings on Privacy Enhancing Technologies}, 2018, 2017.

\bibitem{CoinJoin-Maxwell}
{Gregory Maxwell}.
\newblock {Coinjoin: Bitcoin privacy for the real world}, 2013.
\newblock \url{https://bitcointalk.org/index.php?topic=279249.0}.

\bibitem{GS22}
J.~Groth and V.~Shoup.
\newblock {Design and analysis of a distributed ECDSA signing service}.
\newblock {\em Cryptology ePrint Archive, Paper 2022/506}, 2022.
\newblock \url{https://eprint.iacr.org/2022/506}.

\bibitem{Groth2022}
J.~Groth and V.~Shoup.
\newblock {On the Security of ECDSA with Additive Key Derivation and Presignatures}.
\newblock {\em Cryptology ePrint Archive, Paper 2021/1330}, 2022.
\newblock \url{https://eprint.iacr.org/2021/1330}.

\bibitem{Donjon-extracting-seeds}
C.~Guillemet.
\newblock {Extracting seeds from Wallets}, 2019.
\newblock \url{https://blog.ledger.com/Extracting-Seeds/}.

\bibitem{Donjon-software}
C.~Guillemet and J.-B. Bédrune.
\newblock {On the security model of software wallets}, 2021.
\newblock \url{https://blog.ledger.com/software-wallets/}.

\bibitem{Donjon-ellipal-attack}
C.~Guillemet and O.~Hériveaux.
\newblock {Extracting seed from Ellipal wallet}, 2019.
\newblock \url{https://blog.ledger.com/Ellipal-Security/}.

\bibitem{BeatCoin}
M.~Guri.
\newblock {BeatCoin: Leaking Private Keys from Air-Gapped Cryptocurrency Wallets}.
\newblock {\em CoRR}, abs/1804.08714, 2018.
\newblock \url{http://arxiv.org/abs/1804.08714}.

\bibitem{gutoski2015hierarchical}
G.~Gutoski and D.~Stebila.
\newblock {Hierarchical deterministic Bitcoin wallets that tolerate key leakage (short paper)}.
\newblock In {\em Proceedings of the 19th International Conference on Financial Cryptography and Data Security (FC 2015)}. Springer, 2015.

\bibitem{DeletingSecretData}
F.~{Hao}, D.~{Clarke}, and A.~F. {Zorzo}.
\newblock {Deleting Secret Data with Public Verifiability}.
\newblock {\em IEEE Transactions on Dependable and Secure Computing}, 13(6):617--629, 2016.

\bibitem{SurveyAttackModeling}
M.~S. Haque.
\newblock {An Evolutionary Approach of Attack Graphs and Attack Trees: A Survey of Attack Modeling}.
\newblock 2017.
\newblock \url{http://dcsl.cs.ua.edu/papers/SAM9712.pdf}.

\bibitem{BIP125}
D.~A. Harding and P.~Todd.
\newblock {Opt-in Full Replace-by-Fee Signaling}, 2015.
\newblock \url{https://github.com/bitcoin/bips/blob/master/bip-0125.mediawiki}.

\bibitem{HF16}
M.~Harrigan and C.~Fretter.
\newblock {The Unreasonable Effectiveness of Address Clustering}.
\newblock {\em 2016 Intl IEEE Conferences on Ubiquitous Intelligence \& Computing, Advanced and Trusted Computing, Scalable Computing and Communications, Cloud and Big Data Computing, Internet of People, and Smart World Congress}, 2016.

\bibitem{Hayek.1978}
F.~A. Hayek.
\newblock {\em {Denationalisation of Money - The Argument Refined: Second (Extended) Edition}}.
\newblock The Institute of Economic Affairs, 1978.

\bibitem{CovInvestopedia}
A.~Hayes.
\newblock {Covenant}, 2020.
\newblock \url{https://www.investopedia.com/terms/c/covenant.asp}.

\bibitem{Android-wallet-attacks}
D.~He, S.~Li, C.~Li, S.~Zhu, S.~Chan, W.~Min, and N.~Guizani.
\newblock Security analysis of cryptocurrency wallets in android-based applications.
\newblock {\em IEEE Network}, PP:1--6, 2020.

\bibitem{BIP37}
M.~Hearn and M.~Corallo.
\newblock {Connection Bloom filtering}, 2012.
\newblock \url{https://github.com/bitcoin/bips/blob/master/bip-0037.mediawiki}.

\bibitem{EclipseAttack}
E.~Heilman, A.~Kendler, A.~Zohar, and S.~Goldberg.
\newblock {Eclipse Attacks on Bitcoin Peer-to-Peer Network}.
\newblock In {\em 24th {USENIX} Security Symposium ({USENIX} Security 15)}, pages 129--144. {USENIX} Association, 2015.

\bibitem{Delphi}
O.~Helmer.
\newblock An experimental application of the delphi method to the use of experts.
\newblock {\em Management Science}, 9:458--467, 1963.

\bibitem{ADTreeValue}
H.~Hermanns, J.~Krämer, J.~Krcál, and M.~Stoelinga.
\newblock {The Value of Attack-Defence Diagrams}.
\newblock {\em Principles of Security and Trust}, pages 163--185, 2016.

\bibitem{HommelVaultmbed}
S.~Hommel.
\newblock {Vault-mbed}, 2020.
\newblock \url{https://github.com/fmr-llc/Vault-mbed}.

\bibitem{SecurityRefArchitecture}
I.~Homoliak, S.~Venugopalan, Q.~Hum, and P.~Szalachowski.
\newblock A security reference architecture for blockchains.
\newblock In {\em 2019 IEEE International Conference on Blockchain (Blockchain)}, pages 390--397, 2019.

\bibitem{SurveySC-21}
B.~Hu, Z.~Zhang, J.~Liu, Y.~Liu, J.~Yin, R.~Lu, and X.~Lin.
\newblock A comprehensive survey on smart contract construction and execution: paradigms, tools, and systems.
\newblock {\em Patterns}, 2:100179, 2021.

\bibitem{SmartphoneWalletThreats}
Y.~Hu, S.~Wang, G.-H. Tu, L.~Xiao, T.~Xie, X.~Lei, and C.~Li.
\newblock {Security Threats from Bitcoin Wallet Smartphone Applications: Vulnerabilities, Attacks, and Countermeasures}.
\newblock {\em Proceedings of the Eleventh ACM Conference on Data and Application Security and Privacy}, page 89–100, 2021.

\bibitem{Hubbard}
D.~W. Hubbard and R.~Seiersen.
\newblock {\em {How to Measure Anything in Cybersecurity Risk}}.
\newblock 2016.

\bibitem{ISO-IEC}
{International Standards Organization}.
\newblock {ISO/IEC 27001 and related standards}, 2022.
\newblock \url{https://www.iso.org/isoiec-27001-information-security.html}.

\bibitem{ExchangeReservesRisk}
S.~Jain, E.~W. Felten, and S.~Goldfeder.
\newblock Determining an optimal threshold on the online reserves of a bitcoin exchange.
\newblock {\em Journal of Cybersecurity}, 2018.

\bibitem{DBLP:journals/corr/JhawarKMRT15}
R.~Jhawar, B.~Kordy, S.~Mauw, S.~Radomirovic, and R.~Trujillo{-}Rasua.
\newblock {Attack Trees with Sequential Conjunction}.
\newblock {\em CoRR}, abs/1503.02261, 2015.

\bibitem{Crypto-vs-Fiat-Wallet}
S.~Jokić, A.~S. Cvetković, S.~Adamović, N.~Ristić, and P.~Spalević.
\newblock Comparative analysis of cryptocurrency wallets vs traditional wallets.
\newblock {\em Ekonomika}, 2019.

\bibitem{JPS22}
S.~Joshi, D.~Pandey, and K.~Srinathan.
\newblock {\em ATSSIA: Asynchronous Truly-Threshold Schnorr Signing for Inconsistent Availability}, pages 71--91.
\newblock 2022.

\bibitem{JBWD18}
M.~Jourdan, S.~Blandin, L.~Wynter, and P.~Deshpande.
\newblock {Characterizing Entities in the Bitcoin Blockchain}.
\newblock {\em 2018 IEEE International Conference on Data Mining Workshops (ICDMW)}, pages 55--62, 2018.

\bibitem{tapscript_elements}
S.~Kanjalkar.
\newblock {TapScript OP\_CODES documentation for elements}.
\newblock 2021.
\newblock \url{https://github.com/ElementsProject/elements/blob/master/doc/tapscript_opcodes.md#new-opcodes-for-additional-functionality}.

\bibitem{EPoSE}
N.~P. Karvelas and A.~Kiayias.
\newblock {Efficient Proofs of Secure Erasure}.
\newblock In {\em Security and Cryptography for Networks}, pages 520--537. Springer International Publishing, 2014.

\bibitem{Bitmatrix}
B.~Keceli.
\newblock {Bitmatrix: A Constant Product Market Maker Based on Recursive Covenants}.
\newblock 2021.
\newblock \url{https://docs.bitmatrix.app/v1/11_21_21/Bitmatrix_Paper_Early_Preview.pdf}.

\bibitem{SchnorrSecurity}
E.~Kiltz, D.~Masny, and J.~Pan.
\newblock Optimal security proofs for signatures from identification schemes.
\newblock {\em Cryptology ePrint Archive, Paper 2016/191}, 2016.
\newblock \url{https://eprint.iacr.org/2016/191}.

\bibitem{glacier-dice}
P.~Kim.
\newblock {The Glacier Protocol and Using Dice To Generate Keys}, 2021.
\newblock \url{https://blog.keyst.one/the-glacier-protocol-and-using-dice-to-generate-keys-6677550c2b86}.

\bibitem{PhoenixVault}
U.~Kirstein, S.~Grossman, M.~Mirkin, J.~Wilcox, I.~Eyal, and M.~Sagiv.
\newblock {Phoenix: A Formally Verified Regenerating Vault}.
\newblock 2021.
\newblock \url{https://arxiv.org/pdf/2106.01240.pdf}.

\bibitem{NIST}
R.~Kissel, M.~A. Scholl, S.~Skolochenko, and X.~Li.
\newblock {SP 800-88 Rev. 1. Guidelines for Media Sanitization}.
\newblock {\em National Institute of Standards \& Technology}, 2006.

\bibitem{KG21}
C.~Komlo and I.~Goldberg.
\newblock {FROST: Flexible Round-Optimized Schnorr Threshold Signatures}.
\newblock {\em Cryptology ePrint Archive, Paper 2020/852}, 2020.
\newblock \url{https://eprint.iacr.org/2020/852}.

\bibitem{FoundationsADTrees}
B.~Kordy, S.~Mauw, S.~Radomirovic, and P.~Schweitzer.
\newblock {Foundations of Attack–Defense Trees}.
\newblock {\em Lecture Notes in Computer Science}, 2010.

\bibitem{KordyReview}
B.~Kordy, L.~Piètre-Cambacédès, and P.~Schweitzer.
\newblock {DAG-Based Attack and Defense Modeling: Don't Miss the Forest for the Attack Trees}.
\newblock {\em Computer Science Review}, 13, 2013.

\bibitem{AnonymitybitcoinP2P}
P.~Koshy, D.~Koshy, and P.~McDaniel.
\newblock {An Analysis of Anonymity in Bitcoin Using P2P Network Traffic}.
\newblock {\em Financial Cryptography and Data Security}, pages 469--485, 2014.

\bibitem{timelocked-signatures}
S.~A. KrishnanThyagarajan, A.~Bhat, G.~Malavolta, N.~Döttling, A.~Kate, and D.~Schröder.
\newblock Verifiable timed signatures made practical.
\newblock {\em Cryptology ePrint Archive, Paper 2020/1563}, 2020.
\newblock \url{https://eprint.iacr.org/2020/1563}.

\bibitem{KumarThesis}
R.~Kumar.
\newblock {\em {Truth or Dare: Quantitative security risk analysis using attack trees}}.
\newblock PhD thesis, 2018.

\bibitem{AFTrees}
R.~Kumar and M.~Stoelinga.
\newblock {Quantitative Security and Safety Analysis with Attack-Fault Trees}.
\newblock {\em 2017 IEEE 18th International Symposium on High Assurance Systems Engineering (HASE)}, 2017.

\bibitem{kurose2013computer}
J.~Kurose and K.~Ross.
\newblock {\em Computer Networking: A Top-down Approach}.
\newblock Always learning. Pearson, 2013.

\bibitem{PrivacySurvey2018}
M.~C. {Kus Khalilov} and A.~{Levi}.
\newblock {A Survey on Anonymity and Privacy in Bitcoin-Like Digital Cash Systems}.
\newblock {\em IEEE Communications Surveys Tutorials}, 2018.

\bibitem{AGraphATreeReview}
H.~Lallie, K.~Debattista, and J.~Bal.
\newblock {A review of attack graph and attack tree visual syntax in cyber security}.
\newblock {\em Computer Science Review}, 35:100219, 2020.

\bibitem{BIP-PUSHTXDATA}
J.~Lau.
\newblock {OP\_PUSHTXDATA}, 2017.
\newblock \url{https://github.com/jl2012/bips/blob/vault/bip-0ZZZ.mediawiki}.

\bibitem{L21}
Y.~Lindell.
\newblock {Fast Secure Two-Party ECDSA Signing}.
\newblock {\em Advances in Cryptology -- CRYPTO 2017}, pages 613--644, 2017.

\bibitem{L22}
Y.~Lindell.
\newblock {Simple Three-Round Multiparty Schnorr Signing with Full Simulatability}.
\newblock {\em Cryptology ePrint Archive, Paper 2022/374}, 2022.
\newblock \url{https://eprint.iacr.org/2022/374}.

\bibitem{LN18}
Y.~Lindell and A.~Nof.
\newblock {Fast Secure Multiparty ECDSA with Practical Distributed Key Generation and Applications to Cryptocurrency Custody}.
\newblock In {\em Proceedings of the 2018 ACM SIGSAC Conference on Computer and Communications Security}, {CCS '18}, pages 1837--1854. ACM, 2018.

\bibitem{AdvancedHMfeatures}
K.~Loaec.
\newblock {Hardware wallets and ``advanced" Bitcoin features}, 2021.
\newblock \url{https://lists.linuxfoundation.org/pipermail/bitcoin-dev/2021-January/018352.html}.

\bibitem{revault-pdf}
K.~Loaec and A.~Poinsot.
\newblock {Revault: a multi-party Bitcoin vault architecture}, 2020.
\newblock \url{https://revault.dev/assets/revault\_documentation.pdf}.

\bibitem{BIP141}
E.~Lombrozo, J.~Lau, and P.~Wuille.
\newblock {Segregated Witness (Consensus layer)}, 2015.
\newblock \url{https://github.com/bitcoin/bips/blob/master/bip-0141.mediawiki}.

\bibitem{LoppCoinSelection}
J.~Lopp.
\newblock {The Challenges of Optimizing Unspent Output Selection}, 2015.
\newblock \url{https://blog.lopp.net/the-challenges-of-optimizing-unspent-output-selection/}.

\bibitem{lopp-seed}
J.~Lopp.
\newblock {A Treatise on Bitcoin Seed Backup Device Design}.
\newblock 2022.
\newblock \url{https://blog.lopp.net/a-treatise-on-bitcoin-seed-backup-device-design/}.

\bibitem{lopp-metal}
J.~Lopp.
\newblock {Metal Bitcoin Seed Storage Reviews}.
\newblock 2022.
\newblock \url{https://jlopp.github.io/metal-bitcoin-storage-reviews/}.

\bibitem{HaltingProblem}
S.~Lucas.
\newblock The origins of the halting problem.
\newblock {\em Journal of Logical and Algebraic Methods in Programming}, 121:100687, 2021.

\bibitem{thesis-HW-security}
{Lukáš Kozák}.
\newblock {Security Analysis of Hardware Crypto Wallets}, 2020.
\newblock \url{https://dspace.cvut.cz/bitstream/handle/10467/88181/F8-BP-2020-Kozak-Lukas-thesis.pdf}.

\bibitem{arcula}
A.~D. Luzio, D.~Francati, and G.~Ateniese.
\newblock Arcula: A secure hierarchical deterministic wallet for multi-asset blockchains.
\newblock {\em Cryptology ePrint Archive, Paper 2019/704}, 2019.
\newblock \url{https://eprint.iacr.org/2019/704}.

\bibitem{Donjon-laser}
{M. Mouchous}.
\newblock {Mounting a low-cost laser bench}, 2022.
\newblock \url{https://blog.ledger.com/laser-bench-low-price/}.

\bibitem{CryptoassetCustody}
H.~N. M.~Sato, M.~Shimaoka.
\newblock {General Security Considerations for Cryptoassets Custodians}.
\newblock 2019.
\newblock \url{https://tools.ietf.org/html/draft-vcgtf-crypto-assets-security-considerations-05}.

\bibitem{usability-multi-device}
E.~V. Mangipudi, U.~Desai, M.~Minaei, M.~Mondal, and A.~Kate.
\newblock Uncovering impact of mental models towards adoption of multi-device crypto-wallets.
\newblock {\em Cryptology ePrint Archive, Paper 2022/075}, 2022.
\newblock \url{https://eprint.iacr.org/2022/075}.

\bibitem{AdaptiveThreatCeremony}
J.~Martina, E.~Dos~Santos, M.~Carlos, G.~Price, and R.~Custódio.
\newblock An adaptive threat model for security ceremonies.
\newblock {\em International Journal of Information Security}, 14, 2014.

\bibitem{Matetic2018BITEBL}
S.~Matetic, K.~W{\"u}st, M.~Schneider, K.~Kostiainen, G.~O. Karame, and S.~Capkun.
\newblock {BITE: Bitcoin Lightweight Client Privacy using Trusted Execution}.
\newblock In {\em IACR Cryptology ePrint Archive}, 2018.

\bibitem{Duqu2}
P.~Maynard, K.~Mclaughlin, and S.~Sezer.
\newblock {Modelling Duqu 2.0 Malware using Attack Trees with Sequential Conjunction}.
\newblock {\em International Conference on Information Systems Security and Privacy}, 2016.

\bibitem{10.1093/cybsec/tyaa020}
P.~Maynard, K.~McLaughlin, and S.~Sezer.
\newblock {Decomposition and sequential-AND analysis of known cyber-attacks on critical infrastructure control systems}.
\newblock {\em Journal of Cybersecurity}, 6(1), 2020.

\bibitem{PreventingExchangeHeist}
P.~McCorry, M.~Möser, and S.~Ali.
\newblock Why preventing a cryptocurrency exchange heist isn’t good enough.
\newblock {\em Security Protocols XXVI}, pages 225--233, 2018.

\bibitem{P2TST}
B.~McElrath.
\newblock {Re-Imagining Cold Storage with Timelocks}.
\newblock 2016.
\newblock \url{https://medium.com/@BobMcElrath/re-imagining-cold-storage-with-timelocks-1f293bfe421f}.

\bibitem{MPJLMVS16}
S.~Meiklejohn, M.~Pomarole, G.~Jordan, K.~Levchenko, D.~Mccoy, G.~Voelker, and S.~Savage.
\newblock A fistful of bitcoins: Characterizing payments among men with no names.
\newblock {\em Communications of the ACM}, 59, 2016.

\bibitem{Menezes:1996:HAC:548089}
A.~J. Menezes, S.~A. Vanstone, and P.~C.~V. Oorschot.
\newblock {\em {Handbook of Applied Cryptography}}.
\newblock CRC Press, Inc., 1st edition, 1996.

\bibitem{Merkle1987ADS}
R.~C. Merkle.
\newblock A digital signature based on a conventional encryption function.
\newblock In {\em Annual International Cryptology Conference}, 1987.

\bibitem{TrackerBlockerSurvey}
G.~{Merzdovnik}, M.~{Huber}, D.~{Buhov}, N.~{Nikiforakis}, S.~{Neuner}, M.~{Schmiedecker}, and E.~{Weippl}.
\newblock {Block Me If You Can: A Large-Scale Study of Tracker-Blocking Tools}.
\newblock In {\em 2017 IEEE European Symposium on Security and Privacy (EuroS P)}, pages 319--333, 2017.

\bibitem{Exchange-risk-empirical}
T.~Moore and N.~Christin.
\newblock {Beware the Middleman: Empirical Analysis of Bitcoin-Exchange Risk}.
\newblock {\em Financial Cryptography and Data Security}, pages 25--33, 2013.

\bibitem{ExchangeClosure}
T.~Moore, N.~Christin, and J.~Szurdi.
\newblock Revisiting the risks of bitcoin currency exchange closure.
\newblock {\em ACM Transactions on Internet Technology}, 18:1--18, 2018.

\bibitem{moeser2016bitcoin}
M.~M{\"o}ser, I.~Eyal, and E.~G. Sirer.
\newblock {Bitcoin Covenants}.
\newblock In {\em FC '16: Proceedings of the the 20th International Conference on Financial Cryptography}, 2016.

\bibitem{Nakamoto2008}
S.~Nakamoto.
\newblock {Bitcoin: A Peer-to-Peer Electronic Cash System}.
\newblock 2008.
\newblock \url{https://bitcoin.org/bitcoin.pdf}.

\bibitem{CoinPool}
G.~Naumenko and A.~Riard.
\newblock {CoinPool: efficient off-chain payment pools for Bitcoin}.
\newblock 2022.
\newblock \url{https://coinpool.dev/}.

\bibitem{NeedhamSchroeder}
R.~M. Needham and M.~D. Schroeder.
\newblock Using encryption for authentication in large networks of computers.
\newblock {\em Communications ACM}, 21(12):993–999, 1978.

\bibitem{Pinning}
J.~Newbery.
\newblock {What is meant by transaction `pinning'?}, 2018.
\newblock \url{https://bitcoin.stackexchange.com/questions/80803/what-is-meant-by-transaction-pinning/80804#80804}.

\bibitem{BIP129}
H.~Nguyen, P.~Gray, M.~Bencun, A.~Chen, and R.~Novak.
\newblock {Bitcoin Secure Multisig Setup (BSMS)}, 2020.
\newblock \url{https://github.com/bitcoin/bips/blob/master/bip-0129.mediawiki#Security}.

\bibitem{ADTreesSAND}
H.~N. Nguyen, J.~Bryans, and S.~Shaikh.
\newblock {Attack Defense Trees with Sequential Conjunction}.
\newblock {\em 2019 IEEE 19th International Symposium on High Assurance Systems Engineering (HASE)}, 2019.

\bibitem{Liquid}
J.~Nick, A.~Poelstra, and G.~Sanders.
\newblock {Liquid: A Bitcoin Sidechain}, 2020.
\newblock \url{https://blockstream.com/assets/downloads/pdf/liquid-whitepaper.pdf}.

\bibitem{MuSig2}
J.~Nick, T.~Ruffing, and Y.~Seurin.
\newblock {MuSig2: Simple Two-Round Schnorr Multi-Signatures}.
\newblock {\em Cryptology ePrint Archive, Report 2020/1261}, 2020.
\newblock \url{https://eprint.iacr.org/2020/1261}.

\bibitem{MuSig-DN}
J.~Nick, T.~Ruffing, Y.~Seurin, and P.~Wuille.
\newblock {MuSig-DN: Schnorr Multi-Signatures with Verifiably Deterministic Nonces}.
\newblock {\em Cryptology ePrint Archive, Report 2020/1057}, 2020.
\newblock \url{https://eprint.iacr.org/2020/1057}.

\bibitem{ObVault}
J.~O'Beirne.
\newblock {A simple vault structure using OP\_CTV}.
\newblock \url{https://github.com/jamesob/simple-ctv-vault}.

\bibitem{AnonymityTxGraph}
M.~Ober, S.~Katzenbeisser, and K.~Hamacher.
\newblock {Structure and Anonymity of the Bitcoin Transaction Graph}.
\newblock {\em Future Internet}, 5:237--250, 2013.

\bibitem{TXHASH}
R.~O'Connor.
\newblock {TXHASH + CHECKSIGFROMSTACKVERIFY in lieu of CTV and ANYPREVOUT}, 2022.
\newblock \url{https://lists.linuxfoundation.org/pipermail/bitcoin-dev/2022-January/019813.html}.

\bibitem{Covenants2}
R.~O'Connor and M.~Piekarska.
\newblock {Enhancing Bitcoin Transactions with Covenants}.
\newblock In {\em Financial Cryptography and Data Security}, pages 191--198. Springer International Publishing, 2017.

\bibitem{Trezor-glitch-hack}
C.~O'Flynn.
\newblock {Glitching Trezor using EMFI Through The Enclosure}, 2019.
\newblock \url{https://colinoflynn.com/2019/03/glitching-trezor-using-emfi-through-the-enclosure/}.

\bibitem{CompactFilter}
O.~Osuntokun and A.~Akselrod.
\newblock {Compact Client Side Filtering for Light Clients}, 2017.
\newblock \url{https://github.com/Roasbeef/bips/blob/master/gcs_light_client.mediawiki}.

\bibitem{Simplicity}
R.~O’Connor.
\newblock {Simplicity}, 2017.
\newblock \url{https://blockstream.com/simplicity.pdf}.

\bibitem{Understandingcryptography}
C.~Paar and J.~Pelzl.
\newblock {\em Understanding Cryptography: A Textbook for Students and Practitioners}.
\newblock Springer Publishing Company, Incorporated, 1st edition, 2009.

\bibitem{BIP39}
M.~Palatinus, P.~Rusnak, A.~Voisine, and S.~Bowe.
\newblock {Mnemonic code for generating deterministic keys}, 2013.
\newblock \url{https://github.com/bitcoin/bips/blob/master/bip-0039.mediawiki}.

\bibitem{PCCY22}
S.~Pan, K.~Y. Chan, H.~Cui, and T.~H. Yuen.
\newblock {Multi-Signatures for ECDSA and Its Applications in Blockchain}.
\newblock {\em Information Security and Privacy}, page 265–285, 2022.

\bibitem{Pass2017}
R.~Pass, L.~Seeman, and A.~Shelat.
\newblock {Analysis of the Blockchain Protocol in Asynchronous Networks}.
\newblock In J.-S. Coron and J.~B. Nielsen, editors, {\em Advances in Cryptology -- EUROCRYPT 2017}, pages 643--673. Springer International Publishing, 2017.

\bibitem{Donjon-trezor-sc-attack}
M.~S. Pedro, C.~Guillemet, and V.~Servant.
\newblock {Breaking Trezor One with Side Channel Attacks}, 2019.
\newblock \url{https://blog.ledger.com/Breaking-Trezor-One-with-SCA/}.

\bibitem{ProofsOfSecureErasure}
D.~Perito and G.~Tsudik.
\newblock {Secure Code Update for Embedded Devices via Proofs of Secure Erasure}.
\newblock In D.~Gritzalis, B.~Preneel, and M.~Theoharidou, editors, {\em Computer Security -- ESORICS 2010}, pages 643--662. Springer Berlin Heidelberg, 2010.

\bibitem{NoiseProtocolFramework}
T.~Perrin.
\newblock {The Noise Protocol Framework}.
\newblock 2018.
\newblock \url{https://noiseprotocol.org/noise.pdf}.

\bibitem{P21}
M.~Pettit.
\newblock {Efficient Threshold-Optimal ECDSA}.
\newblock {\em Cryptology ePrint Archive, Paper 2021/1386}, 2021.
\newblock \url{https://eprint.iacr.org/2021/1386}.

\bibitem{tapscript_elements_blog}
A.~Poelstra.
\newblock {Tapscript: New Opcodes, Reduced Limits and Covenants}.
\newblock 2022.
\newblock \url{https://blog.blockstream.com/tapscript-new-opcodes-reduced-limits-and-covenants/}.

\bibitem{PoinsotVault}
A.~Poinsot.
\newblock {James O'Beirne CTV vault using ANYPREVOUT in place of CTV}.
\newblock \url{https://github.com/darosior/simple-anyprevout-vault}.

\bibitem{BitcoinMetadata}
L.~Pompianu, M.~Bartoletti, and B.~Bellomy.
\newblock A journey into bitcoin metadata.
\newblock {\em Journal of Grid Computing}, 2019.

\bibitem{Poon2016}
J.~Poon and T.~Dryja.
\newblock {The Bitcoin Lightning Network: Scalable Off-Chain Instant Payments}.
\newblock 2016.
\newblock \url{https://lightning.network/lightning-network-paper.pdf}.

\bibitem{RFC791}
J.~Postel.
\newblock Internet protocol (ip).
\newblock {\em IETF Request for Comments, RFC 791}, 1981.
\newblock \url{https://www.rfc-editor.org/info/rfc791}.

\bibitem{RFC793}
J.~Postel.
\newblock Transmission control protocol (tcp).
\newblock {\em IETF Request for Comments, RFC 793}, 1981.
\newblock \url{https://www.rfc-editor.org/info/rfc793}.

\bibitem{RFC959}
J.~Postel and J.~Reynolds.
\newblock {\em File Transfer Protocol (FTP)}.
\newblock 1985.
\newblock \url{https://www.rfc-editor.org/info/rfc959}.

\bibitem{Trezor-hack}
S.~Rashid.
\newblock {Extracting TREZOR Secrets from SRAM}, 2017.
\newblock \url{https://saleemrashid.com/2017/08/17/extracting-trezor-secrets-sram/}.

\bibitem{SOKSecureDeletion}
J.~{Reardon}, D.~{Basin}, and S.~{Capkun}.
\newblock {SoK: Secure Data Deletion}.
\newblock In {\em 2013 IEEE Symposium on Security and Privacy}, pages 301--315, 2013.

\bibitem{Reid2013}
F.~Reid and M.~Harrigan.
\newblock {An Analysis of Anonymity in the Bitcoin System}.
\newblock {\em Security and Privacy in Social Networks}, pages 197--223, 2013.

\bibitem{rhodes2013information}
M.~Rhodes-Ousley.
\newblock {\em Information Security The Complete Reference, Second Edition}.
\newblock The Complete Reference. Mcgraw-hill, 2013.

\bibitem{Pinning-Riard}
A.~Riard.
\newblock {Pinning : The Good, The Bad, The Ugly}, 2020.
\newblock \url{https://lists.linuxfoundation.org/pipermail/lightning-dev/2020-June/002758.html}.

\bibitem{TimeDilationAttack}
A.~Riard and G.~Naumenko.
\newblock Time-{Dilation} {Attacks} on the {Lightning} {Network}.
\newblock {\em Cryptoeconomic Systems}, 2021.
\newblock \url{https://cryptoeconomicsystems.pubpub.org/pub/riard-lightning-dilation}.

\bibitem{QuantitativeAnalysisTxGraph}
D.~Ron and A.~Shamir.
\newblock {Quantitative Analysis of the Full Bitcoin Transaction Graph}.
\newblock {\em Cryptology ePrint Archive, Paper 2012/584}, 2012.
\newblock \url{https://eprint.iacr.org/2012/584}.

\bibitem{Sapio}
J.~Rubin.
\newblock {Designing Bitcoin Contracts with Sapio}.
\newblock \url{https://learn.sapio-lang.org/}.

\bibitem{BIP119}
J.~Rubin.
\newblock {CHECKTEMPLATEVERIFY}, 2020.
\newblock \url{https://github.com/bitcoin/bips/blob/master/bip-0119.mediawiki}.

\bibitem{RubinVaults}
J.~Rubin.
\newblock {Building Vaults on Bitcoin}, 2021.
\newblock \url{https://rubin.io/bitcoin/2021/12/07/advent-10/}.

\bibitem{RRJSS22}
T.~Ruffing, V.~Ronge, E.~Jin, J.~Schneider-Bensch, and D.~Schröder.
\newblock {ROAST: Robust Asynchronous Schnorr Threshold Signatures}.
\newblock {\em Cryptology ePrint Archive, Paper 2022/550}, 2022.
\newblock \url{https://eprint.iacr.org/2022/550}.

\bibitem{OPTX}
R.~Russell.
\newblock {[PROPOSAL] OP\_TX: generalized covenants reduced to OP\_CHECKTEMPLATEVERIFY}, 2022.
\newblock \url{https://lists.linuxfoundation.org/pipermail/bitcoin-dev/2022-May/020450.html}.

\bibitem{Volokitin-HW-attacks}
{S. Volokitin}.
\newblock {Software Attacks on Hardware Wallets - Black Hat}, 2018.
\newblock \url{https://i.blackhat.com/us-18/Wed-August-8/us-18-Volokitin-Software-Attacks-On-Hardware-Wallets.pdf}.

\bibitem{TransformingWeakestLink}
A.~Sasse, S.~Brostoff, and D.~Weirich.
\newblock Transforming the ‘weakest link’ — a human/computer interaction approach to usable and effective security.
\newblock {\em BT Technology Journal}, 19, 2001.

\bibitem{AssetCentricAnalysis}
C.~Schmitz, A.~Sekulla, and S.~Pape.
\newblock Asset-centric analysis and visualisation of attack trees.
\newblock In {\em Graphical Models for Security: 7th International Workshop, GraMSec 2020, Boston, MA, USA, June 22, 2020, Revised Selected Papers}, page 45–64. Springer-Verlag, 2020.

\bibitem{AttackTrees}
B.~Schneier.
\newblock {Attack Trees}.
\newblock 1999.
\newblock \url{https://www.schneier.com/academic/archives/1999/12/attack_trees.html}.

\bibitem{Schnorr1991}
C.~P. Schnorr.
\newblock {Efficient signature generation by smart cards}.
\newblock {\em Journal of Cryptology}, 4(3):161--174, 1991.

\bibitem{Xmen-ceremony}
D.~Sempreboni and L.~Vigan{\`o}.
\newblock A mutation-based approach for the formal and automated analysis of security ceremonies.
\newblock {\em Journal of Computer Security}, 2022.

\bibitem{Shostack}
A.~Shostack.
\newblock {\em {Threat Modeling: Designing for Security}}.
\newblock 2014.

\bibitem{vault12-dice-seed}
M.~Skibinsky and A.~Krotou.
\newblock {Generate a Seed Phrase using Dice.}, 2022.
\newblock \url{https://vault12.com/securemycrypto/cryptocurrency-security-how-to/dice-crypto-recovery-seed/1-gather-your-pencil-and-paper-your-dice-a-bip39-word-list-and-let-s-get-ready-to-roll}.

\bibitem{SS01}
D.~Stinson and R.~Strobl.
\newblock {Provably Secure Distributed Schnorr Signatures and a (t, n) Threshold Scheme for Implicit Certificates}.
\newblock {\em Information Security and Privacy}, pages 417--434, 2001.

\bibitem{Swambo2020cov}
J.~Swambo, S.~Hommel, B.~McElrath, and B.~Bishop.
\newblock {Bitcoin Covenants: Three Ways to Control the Future}.
\newblock {\em CoRR}, abs/2006.16714, 2020.
\newblock \url{https://arxiv.org/abs/2006.16714}.

\bibitem{Swambo2020vault}
J.~Swambo, S.~Hommel, B.~McElrath, and B.~Bishop.
\newblock {Custody Protocols Using Bitcoin Vaults}.
\newblock {\em CoRR}, abs/2005.11776, 2020.
\newblock \url{https://arxiv.org/abs/2005.11776}.

\bibitem{swanson-dice}
W.~Swanson.
\newblock {Creating Bitcoin Private Keys with Dice}, 2014.
\newblock \url{https://www.swansontec.com/bitcoin-dice.html}.

\bibitem{NoiseHS}
{T. Perrin}.
\newblock {The Noise Protocol Framework: Interactive handshake patterns (fundamental)}, 2018.
\newblock \url{https://noiseprotocol.org/noise.html#handshake-patterns/}.

\bibitem{BIP1}
A.~Taaki.
\newblock {BIP Purpose and Guidelines}, 2011.
\newblock \url{https://github.com/bitcoin/bips/blob/master/bip-0001.mediawiki}.

\bibitem{Pinning-Bastien}
B.~Teinturier.
\newblock {RBF Pinning with Counterparties and Competing Interest}, 2020.
\newblock \url{https://lists.linuxfoundation.org/pipermail/lightning-dev/2020-June/002739.html}.

\bibitem{BIP127}
J.~Timón.
\newblock {Simple Proof-of-Reserves Transactions}, 2015.
\newblock \url{https://github.com/bitcoin/bips/blob/master/bip-0099.mediawiki}.

\bibitem{SurveySC}
P.~Tolmach, Y.~Li, S.-W. Lin, Y.~Liu, and Z.~Li.
\newblock A survey of smart contract formal specification and verification.
\newblock {\em ACM Computing Surveys}, 54:1--38, 2021.

\bibitem{BIP-anyprevout}
A.~Towns.
\newblock {SIGHASH\_ANYPREVOUT for Taproot Scripts}, 2019.
\newblock \url{https://github.com/ajtowns/bips/blob/bip-anyprevout/bip-anyprevout.mediawiki}.

\bibitem{TLKBS18}
M.~Tran, L.~Luu, M.~S. Kang, I.~Bentov, and P.~Saxena.
\newblock {Obscuro: A Bitcoin Mixer using Trusted Execution Environments}.
\newblock {\em Cryptology ePrint Archive, Paper 2017/974}, 2017.
\newblock \url{https://eprint.iacr.org/2017/974}.

\bibitem{CommitmentSchemes}
L.~Trevisan.
\newblock {CS276: Cryptography: Notes for Lecture 27}, 2009.
\newblock \url{http://theory.stanford.edu/~trevisan/cs276/lecture27.pdf}.

\bibitem{InspectionResistantMemory}
J.~Valamehr, M.~Chase, S.~Kamara, A.~Putnam, D.~Shumow, V.~Vaikuntanathan, and T.~Sherwood.
\newblock {Inspection Resistant Memory: Architectural Support for Security from Physical Examination}.
\newblock {\em SIGARCH Comput. Archit. News}, 40(3):130--141, 2012.

\bibitem{VR15}
L.~Valenta and B.~Rowan.
\newblock Blindcoin: Blinded, accountable mixes for bitcoin.
\newblock {\em Financial Cryptography and Data Security}, pages 112--126, 2015.

\bibitem{Forensic-Electrum-Core}
L.~Van Der~Horst, K.-K.~R. Choo, and N.-A. Le-Khac.
\newblock Process memory investigation of the bitcoin clients electrum and bitcoin core.
\newblock {\em IEEE Access}, PP:1--1, 2017.

\bibitem{Dandelion}
S.~B. Venkatakrishnan, G.~C. Fanti, and P.~Viswanath.
\newblock {Dandelion: Redesigning the Bitcoin Network for Anonymity}.
\newblock {\em CoRR}, abs/1701.04439, 2017.

\bibitem{Formal-ceremonies}
L.~Vigan{\`o}.
\newblock Formal methods for socio-technical security: (formal and automated analysis of security ceremonies).
\newblock {\em Coordination Models and Languages (COORDINATION 2022)}, 2022.

\bibitem{bitcoinAnarchist}
Z.~Voell.
\newblock {Sorry, Bitcoin is still Anarchist}.
\newblock 2018.
\newblock \url{https://medium.com/@zackvoell/sorry-bitcoin-is-still-anarchist-3e995d2fbbf1}.

\bibitem{Volokitin}
S.~Volokitin.
\newblock {Glitch in the Matrix: Exploiting Bitcoin Hardware Wallets}, 2019.
\newblock \url{https://www.offensivecon.org/speakers/2019/sergei-volokitin.html}.

\bibitem{MOBT}
H.~Wang, X.~Li, J.~Gao, and W.~Li.
\newblock Mobt: A kleptographically-secure hierarchical-deterministic wallet for multiple offline bitcoin transactions.
\newblock {\em Future Generation Computer Systems}, 101, 2019.

\bibitem{WMDZW21}
H.~Wang, W.~Ma, F.~Deng, H.~Zheng, and Q.~Wu.
\newblock {Dynamic threshold ECDSA signature and application to asset custody in blockchain}.
\newblock {\em Journal of Information Security and Applications}, 61:102805, 2021.

\bibitem{ThreatLogicTrees}
J.~D. Weiss.
\newblock {A system security engineering process}.
\newblock In {\em Proceedings of the 14th National Computer Security Conf.}, 1991.

\bibitem{OpenSecurity}
D.~A. Wheeler.
\newblock {What is open security?}, 2013.
\newblock \url{https://dwheeler.com/essays/open-security-definition.html}.

\bibitem{BIP-Schnorr}
P.~Wuille, , J.~Nick, and T.~Ruffing.
\newblock {Schnorr Signatures for secp256k1}, 2020.
\newblock \url{https://github.com/bitcoin/bips/blob/master/bip-0340.mediawiki}.

\bibitem{BIP-Tapscript}
P.~Wuille, , J.~Nick, and A.~Towns.
\newblock {Validation of Taproot Scripts}, 2020.
\newblock \url{https://github.com/bitcoin/bips/blob/master/bip-0342.mediawiki}.

\bibitem{BIP32}
P.~Wuille.
\newblock {Hierarchical Deterministic Wallets}, 2012.
\newblock \url{https://github.com/bitcoin/bips/blob/master/bip-0032.mediawiki}.

\bibitem{BIP-Taproot}
P.~Wuille, J.~Nick, and A.~Towns.
\newblock {Taproot: SegWit version 1 output spending rules}, 2019.
\newblock \url{https://github.com/bitcoin/bips/blob/master/bip-0341.mediawiki}.

\bibitem{wallets-TM}
{Y. E. Bulut and \.{I}. Sertkaya}.
\newblock {Security Problem Definition and Security Objectives of Cryptocurrency Wallets in Common Criteria}, 2020.
\newblock \url{https://dergipark.org.tr/tr/download/article-file/1081388}.

\bibitem{YCX21}
T.~H. Yuen, H.~Cui, and X.~Xie.
\newblock {Compact Zero-Knowledge Proofs for Threshold ECDSA with Trustless Setup}.
\newblock {\em Cryptology ePrint Archive, Paper 2021/205}, 2021.
\newblock \url{https://eprint.iacr.org/2021/205}.

\bibitem{ParalysisProofs}
F.~Zhang, P.~Daian, I.~Bentov, I.~Miers, and A.~Juels.
\newblock Paralysis proofs: Secure dynamic access structures for cryptocurrency custody and more.
\newblock In {\em Proceedings of the 1st ACM Conference on Advances in Financial Technologies}, AFT '19, page 1–15. Association for Computing Machinery, 2019.

\bibitem{ZmnOnReccursion}
ZmnSCPxj.
\newblock {Speedy covenants (OP\_CAT2)}, 2022.
\newblock \url{https://lists.linuxfoundation.org/pipermail/bitcoin-dev/2022-May/{020434}.html}.

\end{thebibliography}

\end{document}